\documentclass[titlesmallcaps, copyrightpage]{uqthesis}

\usepackage{color}
\usepackage[sans]{dsfont}

\newcommand {\msf}[1] {\mathsf{#1}}

\newcommand {\Ref}[1] {Reference~\cite{#1}}

\newcommand {\etal} {\emph{~et~al.}}

\newcommand {\bul} {$\bullet$ }   

\newcommand {\half} {\frac{1}{2}}

\renewcommand{\H}{\mathcal{H}}
\def\A{\mathcal{A}}
\def\B{\mathcal{B}}
\def\AB{\A\B}
\def\C{\mathcal{C}}
\def\N{\mathcal{N}}
\def\s{\mathcal{S}^+}
\newcommand{\NV}[1]{\mathcal{N}\mathcal{V}^{(#1)}}
\newcommand{\NVg}{\mathcal{N}\mathcal{V}}

\newcommand{\HA}{\H_{\A}}
\newcommand{\HB}{\H_{\B}}
\newcommand{\GHZ}{\ket{\Psi_{\mbox{\tiny GHZ}}}}
\newcommand{\dA}{d_\mathcal{A}}
\newcommand{\dB}{d_\mathcal{B}}

\newcommand{\rw}{\rho_{\mbox{\tiny W$_d$}}}

\newcommand{\rwd}[1]{\rho_{\mbox{\tiny W$_{#1}$}}}
\newcommand{\rI}{\rho_{\text{I}_d}}
\newcommand{\rd}[1]{\rho_{#1}}
\newcommand{\rId}[1]{\rho_{\text{I}_#1}}
\newcommand{\rTA}{\rho_{\mbox{\tiny TA}}}
\newcommand{\rCH}{\rho_{\mbox{\tiny CH}}}
\newcommand{\rCG}{\rho_{\mbox{\tiny CG}}}
\newcommand{\rH}{\rho_{\mbox{\tiny H}}}
\newcommand{\rG}{\rho_{\mbox{\tiny G}}}

\newcommand{\CSLOCC}[1]{\C_{\mbox{\tiny SLOCC}}^{\mbox{\tiny(#1)}}}
\newcommand{\ket}[1]{|#1\rangle}
\newcommand{\bra}[1]{\langle#1|}

\newcommand{\ketbra}[1]{\ket{#1}\!\bra{#1}}
\newcommand{\ME}[1]{\ket{\Phi^+_{#1}}}
\newcommand{\MEd}{\ME{d}}
\newcommand{\AProj}[1]{\ket{#1}_{\A\A}\bra{#1}}
\newcommand{\BProj}[1]{\ket{#1}_{\B\B}\bra{#1}}
\newcommand{\ProjME}[1]{\ketbra{\Phi^+_#1}}
\newcommand{\ProjMEd}{\ProjME{d}}

\newcommand{\en}[1]{{(#1)}}

\newcommand{\tr}{\text{tr}}
\newcommand{\trA}{\text{tr$_\A$}}
\renewcommand{\t}{^{\mbox{\tiny T}}}

\newcommand{\sgn}{{\rm sgn}}
\newcommand{\idx}[1]{#1^\text{th}}
\newcommand{\ii}{{\rm i}}
\newcommand{\dd}{{\rm d}}

\newcommand{\quar}{\frac{1}{4}}

\newcommand{\sqhalf}{\frac{1}{\sqrt{2}}}
\newcommand{\unit}{\mathds{1}}
\newcommand{\zero}{\mathbf{0}}

\newcommand{\spup}{\uparrow}
\newcommand{\spdn}{\downarrow}
\newcommand{\bfr}{\mathbf{r}}
\newcommand{\bfa}{\mathbf{a}}
\newcommand{\bfb}{\mathbf{b}}
\newcommand{\bfc}{\mathbf{c}}
\newcommand{\bfl}{\mathbf{l}}
\newcommand{\bfm}{\mathbf{m}}
\newcommand{\bfn}{\mathbf{n}}
\newcommand{\bft}{\mathbf{t}}

\newcommand{\bfw}{\mathbf{w}}
\newcommand{\bfx}{\mathbf{x}}
\newcommand{\bfy}{\mathbf{y}}
\newcommand{\bfB}{\mathbf{B}}
\newcommand{\bfF}{\mathbf{F}}
\newcommand{\bfp}{\mathbf{p}}
\newcommand{\pLHV}{\bfp_{\mbox{\tiny LHV}}}
\newcommand{\pQM}{\bfp_{\mbox{\tiny QM}}}
\newcommand{\ver}[1]{\text{vert}\left(#1\right)}
\newcommand{\co}[1]{\text{conv}\left(#1\right)}

\newcommand{\mA}{m_{\mbox{\tiny A}}}
\newcommand{\nA}{n_{\mbox{\tiny A}}}
\newcommand{\mB}{m_{\mbox{\tiny B}}}
\newcommand{\nB}{n_{\mbox{\tiny B}}}

\newcommand{\ten}{\otimes}
\renewcommand{\P}{\mathcal{P}}

\newcommand{\Cor}[4]{\mathcal{C}_{#1;#3}^{#2;#4}}
\newcommand{\Corc}[4]{{}^c\mathcal{C}_{#1;#3}^{#2;#4}}
\newcommand{\Cors}[4]{{}^s\mathcal{C}_{#1;#3}^{#2;#4}}

\newcommand{\PLHV}[4]{\mathcal{P}_{#1;#3}^{#2;#4}}
\newcommand{\PLHVc}[4]{{}^c\mathcal{P}_{#1;#3}^{#2;#4}}
\newcommand{\PLHVs}[4]{{}^s\mathcal{P}_{#1;#3}^{#2;#4}}
\newcommand{\PLHVsM}[2]{{}^s\mathcal{P}_{#1}^{#2}}

\newcommand{\Qp}{\mathcal{Q}_{\mA;\mB}^{\nA;\nB}}
\newcommand{\Qs}[2]{\mathcal{Q}_{#1;#2}^{2;2}}

\newcommand{\Cct}{\Corc{\mA}{2}{\mB}{2}}
\newcommand{\Ccs}{\Cors{\mA}{2}{\mB}{2}}
\newcommand{\Cp}{\Cor{\mA}{\nA}{\mB}{\nB}}
\newcommand{\CcM}{{}^s\mathcal{C}_{2;2;\cdots;2}^{2;2;\cdots;2}}

\newcommand{\polycl}{\PLHV{\mA}{\nA}{\mB}{\nB}}
\newcommand{\polyclc}{\PLHVc{\mA}{\nA}{\mB}{\nB}}
\newcommand{\polyclt}{\PLHVc{\mA}{2}{\mB}{2}}
\newcommand{\polycls}{\PLHVs{\mA}{2}{\mB}{2}}
\newcommand{\polycM}{{}^s\mathcal{P}_{2;2;\cdots;2}^{2;2;\cdots;2}}

\newcommand{\oA}{o_a}
\newcommand{\oB}{o_b}
\newcommand{\oAB}{o_ao_b}
\newcommand{\oN}[2]{o_{#2}^{[#1]}}
\newcommand{\sA}{s_a}
\newcommand{\sB}{s_b}

\newcommand{\Coeff}[4]{b_{#1#3}^{#2#4}}

\newcommand{\BBlk}[2]{\bfb_{#1,#2}}
\newcommand{\CoeffG}{b_{\sA\sB}^{\oA\oB}}
\newcommand{\CoeffCnG}{b_{\sA\sB}}
\newcommand{\ProbG}{p_{\sA\sB}^{\oA\oB}}

\newcommand{\ProbTwGJ}{p^{\oA\oB}_{\A\B}(\sA,\sB)}
\newcommand{\ProbTwGMA}{p^{\oA}_{\A}(\sA)}
\newcommand{\ProbTwGMB}{p^{\oB}_{\B}(\sB)}

\newcommand{\ProbTwJ}[4]{p^{#2#4}_{\A\B}(#1,#3)}
\newcommand{\ProbTwMA}[2]{p^{#2}_{\A}(#1)}
\newcommand{\ProbTwMB}[2]{p^{#2}_{\B}(#1)}

\newcommand{\ProbNOAB}[1]{p_{\AB}(#1)}
\newcommand{\ProbNOA}[1]{p_{\A}(#1)}
\newcommand{\ProbNOB}[1]{p_{\B}(#1)}

\newcommand{\bLHV}[1]{\beta_{\mbox{\tiny LHV}}^{(#1)}}

\newcommand{\SLHV}{\mathcal{S}_{\mbox{\tiny LHV}}}
\newcommand{\SLHVBI}[1]{\mathcal{S}_{\mbox{\tiny LHV}}^{\mbox{\tiny (#1)}}}
\newcommand{\Sqm}{\mathcal{S}_{\mbox{\tiny QM}}}
\newcommand{\SqmBI}[1]{\mathcal{S}_{\mbox{\tiny QM}}^{\mbox{\tiny (#1)}}}
\newcommand{\SqmCH}{\mathcal{S}_{\mbox{\tiny QM}}^{\mbox{\tiny (CH)}}}
\newcommand{\SqmCHSH}{\mathcal{S}_{\mbox{\tiny QM}}^{\mbox{\tiny (CHSH)}}}

\newcommand{\Eqm}{E_{\mbox{\tiny QM}}}
\newcommand{\betaLHV}{\beta_{\mbox{\tiny LHV}}}

\newcommand{\Bell}{\mathcal{B}}

\newcommand{\eBell}[2]{\langle\Bell_{#1}\rangle_{#2}}
\newcommand{\eBellCH}{\langle\Bell_{\rm CH}\rangle}

\newcommand{\vvec}[1]{\text{vec}(#1)}
\newcommand{\PiS}{\Pi_+}
\newcommand{\PiA}{\Pi_-}
\newcommand{\Cdg}{\mathbb{C}^d}
\newcommand{\Cd}[1]{\mathbb{C}^{#1}}

\newcommand{\Zp}{\mathbb{Z}^+}

\newcommand{\pLPOVM}[1]{p_{\mbox{\tiny L,#1}}^{\mbox{\tiny POVM}}}
\newcommand{\pLPi}[1]{p_{\mbox{\tiny L,#1}}^\Pi}

\newcommand{\pcLPi}[1]{p_{\mbox{\tiny L,#1}}^{\Pi,c}}
\newcommand{\pS}[1]{p_{\mbox{\tiny S,#1}}}
\newcommand{\Sym}{\msf{Sym}}

\newcommand{\POVMA}[2]{A_{#1}^{#2}}
\newcommand{\POVMB}[2]{B_{#1}^{#2}}
\newcommand{\POVMAg}{\POVMA{\sA}{\oA}}
\newcommand{\POVMBg}{\POVMB{\sB}{\oB}}

\def\Ai{\A_{\rm in}}
\def\Ao{\A_{\rm out}}
\def\Bi{\B_{\rm in}}
\def\Bo{\B_{\rm out}}
\def\E{\mathcal{E}}

\def\Z{\mathcal{Z}}

\newcommand{\prd}{Physical Review D~}

\newcommand{\rmp}{Review of Modern Physics~}

\newenvironment{centre}{\begin{center}}{\end{center}}

\usepackage{play}
\usepackage[grey,times]{quotchap}
\usepackage{makeidx}

\usepackage[square,comma,numbers,sort&compress]{natbib}

\usepackage[nottoc]{tocbibind}

\usepackage{amsmath,amsfonts,amssymb,amsthm} 

\usepackage{graphicx} 

\usepackage{epsfig} 

\usepackage{ifpdf}
\ifpdf
    \DeclareGraphicsExtensions{.pdf}  
\else
    \DeclareGraphicsExtensions{.ps}
\fi

\newtheorem{theorem}{Theorem}
\newtheorem{lemma}[theorem]{Lemma}
\newtheorem{dfn}[theorem]{Definition}

\renewcommand{\exp}[1]{\text{e}^{#1}}

\usepackage[hypertex]{hyperref}

\makeatletter
    \def\adots{\mathinner{\mkern2mu\raise\p@\hbox{.}
    \mkern2mu\raise4\p@\hbox{.}\mkern1mu
    \raise7\p@\vbox{\kern7\p@\hbox{.}}\mkern1mu}}
\makeatother

\begin{document}

\frontmatter

\title{Correlations, Bell Inequality Violation \& Quantum Entanglement}
\author{Yeong-Cherng Liang}
\department{Department of Physics}  
\principaladvisor{Dr. Andrew C.~Doherty}

\titlepage

\chapter{Acknowledgements}

This thesis is a consequence of direct and indirect
contributions from various people, without whom I would not
have gone this far. It is almost inevitable that I would miss
some names in the following enumeration. For that matter, I
would now declare that if you are in doubt, then yes, you must
have been one of the contributors and I thank you for your help
in one way or another. This is, of course, not an excuse for me
to not express my appreciation explicitly and I shall attempt
to do that in what follows.

The successful completion of my PhD candidature as well as this
thesis would have been impossible, if not highly improbable
without the help and guidance from my principal supervisor, Dr.
Andrew C. Doherty. Andrew is always full of ideas and this
means a lot to a junior researcher like me who sometimes lacks
insight into the key issue of a problem. In particular, his
physical intuition has enabled me to see the {\em forest},
instead of {\em trees} at various occasions. His comments on
language usage have always been of great help too. On the other
hand, I must also thank Andrew for his generous support in
regard of me traveling overseas to attend conferences --- these
opportunities have, no doubt, greatly expanded my horizons.
Finally, I owe Andrew a big thank you for going the extra miles
to read through the earlier drafts of this thesis and giving me
his valuable comments.

Many thanks to Guifr\'e Vidal who has effectively acted as my
associate supervisor, giving me his timely advice both in and
out of Physics. This proved to be of upmost importance
especially towards the end of my PhD candidature. Thanks also
to Michael Nielsen, who has offered some critical comments on
my research and who has always tried to made the quantum
information science initiative in the University of Queensland
a wonderful learning environment. His insistence on our active
participation in seminars has undeniably done me a great favor
in these years.

I am also thankful to my other present collaborators ---
Llu\'{\i}s Masanes, Ben Toner, Stephanie Wehner, Valerio
Scarani --- as well as past collaborators --- Dagomir
Kaszlikowski, Leong Chuan Kwek, Berthold-Georg Englert, Ajay
Gopinathan and Choo Hiap Oh --- for giving me the opportunity
to learn from them. Among which, I am above all grateful to
Dagomir Kaszlikowski and Leong Chuan Kwek for bringing me into
this exciting field of quantum information science. Thanks also
to Jing-Ling Chen, Shiang Yong Looi and Meng Khoon Tey for
their helpful discussions.

The fellow PhD students sharing the same office as me are
certainly not to be forgotten. I thank Eric Cavalcanti for
enriching my philosophical understanding of nature, making this
doctorate of philosophy a well-justified one. Of course, his
never ending list of puzzles and paradoxes has also refreshed
my ordinary research life from time to time. Chris Foster is
definitely the most reliable mathematician, technician, and
entertainer in Room 302. His trademark of trying to make sloppy
calculations in Physics rigorous is both enlightening and
inspiring. His help over these years, which often results in
delaying his own research progress, is greatly appreciated.
Next, I would like to thank Paulo Mendon\c ca for always
willing to listen to my complaints and sharing with me the
unusual experience that he has been through. Thanks also to
Paulo for proofreading an earlier draft of this thesis.
Certainly, the random humor from Andy Ferris is much
appreciated too.

I am also grateful to all of my other friends, especially those
coming from the UQ Badminton Club.  You guys have complemented
my academic life in just the right way. Finally, I would also
like to acknowledge the generous financial support from the
International Postgraduate Research Scholarships (IPRS) and the
University of Queensland Graduate School Scholarships (UQGSS).

\chapter{List of Publications}

{\bf Publications by the Candidate Relevant to the Thesis}

\begin{itemize}

\item[\bul] Yeong-Cherng Liang and Andrew C. Doherty, \emph{Better
    Bell-inequality violation by collective measurements}. Physical Review
    A \textbf{73}, 052116 (2006)

\item[\bul] Yeong-Cherng Liang and Andrew C. Doherty,
    \emph{Bounds on quantum correlations in Bell-inequality
    experiments}. Physical Review A \textbf{75}, 042103
    (2007)

\item[\bul] Llu\'{\i}s Masanes, Yeong-Cherng Liang and
    Andrew C. Doherty, \emph{All bipartite entangled states
    display some hidden nonlocality}. Physical Review
    Letters \textbf{100}, 090403 (2008)

\item[\bul] Yeong-Cherng Liang, Llu\'{\i}s Masanes and
    Andrew C. Doherty, \emph{Convertibility between
    two-qubit states using stochastic local quantum
    operations assisted by classical communication}.
    Physical Review A \textbf{77}, 012332 (2008)

\end{itemize}

\noindent {\bf Additional Publications by the Candidate
Relevant to the Thesis but not Forming Part of it}

\begin{itemize}

\item[\bul] Andrew C. Doherty, Yeong-Cherng Liang,
    Stephanie Wehner, and Ben Toner, \emph{The quantum
    moment problem and bounds on entangled multi-prover
    games}, Proceedings of the 23rd IEEE Conference on
    Computational Complexity, pp.~199--210 (eprint
    arXiv:0803.4373)

\end{itemize}

\chapter{Abstract}

It is one of the most remarkable features of quantum physics
that measurements on spatially separated systems cannot always
be described by a {\em locally causal theory}. In such a
theory, the outcomes of local measurements are determined in
advance solely by some unknown (or hidden) variables and the
choice of local measurements. Correlations that are allowed
within the framework of a locally causal theory are termed {\em
classical}. Typically, the fact that quantum mechanics does not
always result in classical correlations is revealed by the
violation of {\em Bell inequalities}, which are constraints
that have to be satisfied by any classical correlations. It has
been known for a long time that {\em entanglement} is necessary
to demonstrate nonclassical correlations, and hence a Bell
inequality violation. However, since some entangled quantum
states are known to admit explicit locally causal models, the
exact role of entanglement in Bell inequality violation has
remained obscure. This thesis provides both a comprehensive
review on these issues as well as a report on new discoveries
made to clarify the relationship between entanglement and Bell
inequality violation. In particular, within the framework of a
standard Bell experiment, i.e., a Bell inequality test that is
directly performed on a single copy of a quantum state $\rho$,
we have derived two algorithms to determine, respectively, a
lower bound and an upper bound on the strength of correlations
that $\rho$ can offer for any given Bell inequality. Both of
these algorithms make use convex optimization techniques in the
form of a {\em semidefinite program}. By examples, we show that
these algorithms can often be used in tandem, in conjunction
with {\em convexity} arguments, to determine if a quantum state
can offer nonclassical correlations and hence violates a given
Bell inequality. On the other hand, since a standard Bell
experiment typically involves measurements over many copies of
the quantum systems, we have also investigated the possibility
of enhancing the strength of nonclassical correlation by,
instead, performing collective  measurements on multiple copies
of the quantum systems. Our findings show that even without
postselection, such joint measurements may also lead to
stronger nonclassical correlations, and hence a better Bell
inequality violation. Meanwhile, previous studies have
indicated that entangled state admitting locally causal models
may still lead to observable nonclassical correlations if,
prior to a standard Bell experiment, the state is subjected to
some appropriate local preprocessing. This phenomenon of {\em
hidden nonlocality} was discovered more than a decade ago, but
to date, it is still not known if all entangled states can
demonstrate nonclassical correlations through these more
sophisticated Bell experiments. A key result in this thesis
then consists of showing that for all bipartite entangled
states, observable nonclassical correlations, in the form of a
Bell-CHSH inequality violation, can indeed be derived if we
allow both local preprocessing and the usage of shared
ancillary state which by itself does not violate the Bell-CHSH
inequality. This establishes a kind of equivalence between
bipartite entanglement and states that cannot be simulated by
classical correlations. In summary, for a standard Bell
experiment where no local preprocessing on a quantum state
$\rho$ is allowed, we have provided two algorithms that can be
used in tandem to determine if $\rho$ can be simulated by a
locally causal theory, whereas in the scenario where local
preprocessing is allowed, we have demonstrated that bipartite
entangled states are precisely those which cannot always be
simulated classically.

\tableofcontents

\listoffigures
\listoftables

\chapter{List of Abbreviations}

\begin{list}{}{%
\setlength{\labelwidth}{24mm} \setlength{\leftmargin}{35mm}}

\item[{\em aka}] also known as
\item[{\em lhs}] left-hand-side
\item[{\em rhs}] right-hand-side
\item[BIV] Bell-inequality-violating
\item[CH] Clauser-Horne
\item[CHSH] Clauser-Horne-Shimony-Holt
\item[CGLMP] Collins-Gisin-Linden-Massar-Popescu
\item[CPM] completely positive map
\item[EPR] Einstein-Podolsky-Rosen
\item[GHZ] Greenberger-Horne-Zeilinger
\item[LB] lower bound
\item[LHV] local hidden variable
\item[LHVM] local hidden-variable model
\item[LHVT] local hidden-variable theory
\item[LMI] linear matrix inequality
\item[LOCC] local quantum operations assisted by classical
    communication
\item[MEMS] maximally entangled mixed states
\item[NBIV] non-Bell-inequality-violating
\item[NSD] negative semidefinite
\item[POVM] positive-operator-valued measure
\item[PPT] positive-partial-transposed
\item[PSD] positive semidefinite
\item[QCQP] quadratically-constrained quadratic program
\item[SLO] stochastic local quantum operations without communication
\item[SLOCC] stochastic local quantum operations assisted by
    classical communication
\item[SDP] semidefinite program
\item[SOS] sum of squares
\item[UB] upper bound
\end{list}

\mainmatter

\chapter{Introduction}

The advent of {\em Quantum Mechanics} is undeniably an
important milestone in our attempt to understand Nature. On the
one hand, quantum mechanics is well-known for giving very
accurate predictions for microscopic phenomena, whereas on the
other, it has also given some counter-intuitive predictions
which seem nonsensical from a classical view point. Among the
many intriguing features of quantum mechanics is {\em
entanglement}~\cite{E.Schroedinger:1935a,E.Schroedinger:1935b}
which, loosely speaking, refers to the situation whereby two or
more spatially separated physical systems are so strongly
correlated that it may become impossible to independently
describe the {\em physical state} of the individual systems.
The significance of entanglement can be seen, for example, in
the following quotation by
Schr\"odinger~\cite{E.Schroedinger:1935a},
\begin{quote}
``\ldots I would not call that one but rather the
characteristic trait of quantum mechanics, the one that
enforces its entire departure from classical lines of thought.
By the interaction the two representatives have become
entangled. \ldots"
\end{quote}

The astonishing features of entanglement were first brought to
our attention in 1935 via the influential work by Einstein,
Podolsky and Rosen (henceforth abbreviated as
EPR)~\cite{EPR:1935}, and subsequently popularized by
Schr\"odinger's thought experiment on an innocent
cat~\cite{E.Schroedinger:1935b}.\footnote{See also the English
translation by Trimmer~\cite{J.D.Trimmer:PAPS:1980}.}
Specifically, in Ref.~\cite{EPR:1935}, EPR considered a pair of
physical systems that are so strongly correlated that it
becomes possible to predict, {\em with certainty}, some
properties of the distant physical system by simply performing
measurements on the local one. Exploiting such bizarre
correlations offered by entanglement, EPR eventually came to
the conclusion that the quantum mechanical predictions of {\em
physical reality} is incomplete~\cite{EPR:1935}, just as
statistical mechanics is incomplete within the framework of
classical
mechanics~\cite{A.Einstein:1949,J.S.Bell:PSFPHEP:1976}.

For a long time after that, discussions arising out of EPR's
paper remained largely a philosophical debate. However, as
Bell~\cite{J.S.Bell:1964,J.S.Bell:RMP:1966} showed in the
1960s, the possibility of completing the quantum mechanical
predictions in the way that EPR sought does lead to
experimentally falsifiable consequences. In particular, by
considering a variant of EPR's argument due to
Bohm~\cite{D.Bohm:Book:1951}, Bell~\cite{J.S.Bell:1964} showed
that quantum mechanical predictions on spatially separated
systems cannot always be described by a {\em locally causal
theory}. In such a theory, the outcomes of measurements are
determined {\em in advance} merely by the choice of local
measurements and some {\em local hidden variable} --- which can
be seen as information exchanged between the subsystems during
their {\em common past}. Bell has thus ruled out the
possibility of providing a locally causal description for all
quantum phenomena --- a brutal fact of life that is now
succinctly called {\em Bell's theorem}.

Typically, the incompatibility between a locally causal
description and the quantum mechanical prediction for a quantum
state due to some choice of observables is revealed by the
violation of {\em Bell inequalities}, which are statistical
constraints that have to be satisfied by all locally causal
theories. Since the early 1980s, there have been numerous
experiments reporting Bell inequality violation in various
physical systems (see, for example,
Refs.~\cite{A.Aspect:Nature:1999,
M.A.Rowe:Nature:2001,P.G.Kwiat:Nature:2001}). While it is clear
that entanglement is necessary to demonstrate a Bell inequality
violation, by generalizing the notion of entanglement to mixed
states, Werner~\cite{R.F.Werner:PRA:1989} has found that not
all entangled states can violate a Bell inequality (see also
Refs.~\cite{RPM.Horodecki:PLA:1995,J.Barrett:PRA:2002,
G.Toth:PRA:2006, M.L.Almeida:PRL:2007}). In fact, it is not
even known if all multipartite pure entangled states are
Bell-inequality-violating~\cite{V.Capasso:IJTP:1973,
N.Gisin:PLA:1991, N.Gisin:PLA:1992, J.L.Chen:PRL:2004}. This
state of affairs has inspired some to consider more
general, nonstandard Bell experiments to reveal the bizarre
correlations hidden in quantum states. In this regard, it was
later shown by Popescu~\cite{S.Popescu:PRL:1995} and
others~\cite{N.Gisin:PLA:1996,A.Peres:PRA:1996} that if a Bell
experiment is preceded with appropriate {\em local
preprocessing}, then a non-Bell-inequality-violating quantum
state may become Bell-inequality-violating --- a phenomenon
that is now known as {\em hidden nonlocality}.

In recent years, the rising field of quantum information
processing has also brought a resurgent interest in the study
of Bell inequality violation. The pioneering work in this
regard is due to Ekert~\cite{A.K.Ekert:PRL:1991}, who
showed that Bell inequality violation can be used to guarantee
the security of a class of quantum key distribution protocols.
Since then, a great deal of work has been carried out in this
regard (see, for example, Refs.~\cite{V.Scarani:PRL:2001,
A.Acin:IJQI:2004,J.Barrett:PRL:2005,A.Acin:PRL:2006,
Ll.Masanes:0606049} and references therein). In fact, recently,
it has even been argued in Refs.~\cite{A.Acin:PRL:2006,
Ll.Masanes:0606049} that Bell-inequality violation is necessary
to guarantee the security of some entanglement-based quantum
key distribution protocols. On the other hand, Bell inequality
violation was also found to be relevant in other quantum
information processing tasks, such as reduction of
communication complexity~\cite{H.Buhrman:SJC:2000,
C.Brukner:PRL:2002, C.Brukner:PRL:2004}. In the context of
quantum teleportation~\cite{C.H.Bennett:PRL:1993},
Horodecki\etal~\cite{RMP.Horodecki:PLA:1996} have shown that
all two-qubit states violating a Bell inequality are useful for
teleportation; Popescu, however, has shown that some two-qubit
states not violating the same Bell inequality are also useful
for teleportation~\cite{S.Popescu:PRL:1994}. Of course, given
that quantum entanglement is an essential ingredient in many
quantum information processing
protocols~\cite{A.Acin:IJQI:2004}, it is by no means accidental
that a verification of entanglement through Bell inequality
violation is carried out daily in many laboratories in the
world.

Given the importance of Bell inequality violation, both from a
foundational point of view and its relevance in quantum
information processing, it is perhaps surprising that there are
still many open problems related to the study of Bell
inequality violation~\cite{N.Gisin:0702021}. In particular,
little is known as to which quantum states can violate a Bell
inequality, both in a standard scenario and in a nonstandard
scenario which also involves local preprocessing. Even when a
quantum state is known to violate a Bell inequality, the extent
of violation is in most cases not well-quantified. On a related
note, the maximal violation that quantum mechanics allows for a
given Bell inequality is also not well-studied beyond some
simple cases~\cite{S.Filipp:PRL:2004, H.Buhrman:PRA:2005,
S.Wehner:PRA:2006, B.F.Toner:quant-ph:0601172, D.Avis:JPA:2006,
M.Navascues:PRL:2007}.

The main goal of this thesis to clarify the relationship
between Bell inequality violation and quantum entanglement by
determining the set of quantum states that can give rise to
nonclassical behavior. The structure of this thesis is as
follows. From Chapter~\ref{Chap:BellThm} --
Chapter~\ref{Chap:Q.Cn:Classical}, we will provide a
comprehensive review of the theoretical background of the
thesis. Specifically, Chapter~\ref{Chap:BellThm} deals with
some of the important concepts relevant to local causality and
the key historical developments leading to Bell's theorem. Then
in Chapter~\ref{Chap:ClassicalCn}, we will give a more
technical introduction to the set of classical
correlations,\footnote{This is the set of correlations allowed
by a locally causal theory.} which includes a formal
introduction to the idea of a tight~\cite{Ll.Masanes:QIC:2003,
D.Collins:JPA:2004}, or facet-inducing Bell
inequality~\cite{D.Avis:JPA:2005}. Some of the well-known tight
Bell inequalities will also be reviewed. After that, we will
proceed to the quantum regime in
Chapter~\ref{Chap:Q.Cn:Classical} and introduce the notion of
quantum correlation following Ref.~\cite{I.Pitowsky:Book:1989}.
Some well-known examples of entangled quantum states admitting
a locally causal description will then be reviewed.

Most of our new research findings can be found in the second
part of the thesis, from Chapter~\ref{Chap:QuantumBounds} --
Chapter~\ref{Chap:Hidden.Nonlocality}, while the rest are left
in the appendices. In Chapter~\ref{Chap:QuantumBounds}, we will
present new findings in relation to the problem of determining
if a given quantum state can violate some fixed but arbitrary
Bell inequality via a standard Bell experiment. In particular,
using convex optimization techniques~\cite{S.Boyd:Book:2004} in
the form of a semidefinite
program~\cite{L.Vandenberghe:SR:1996}, we have derived two
algorithms to determine, respectively, a lower bound and an
upper bound on the strength of correlation that a quantum state
$\rho$ can display in some given Bell experiments. These tools
are also applied in Chapter~\ref{Chap:BellViolation} where we
will look at some of the best known Bell inequality violations
displayed by entangled states. Given that in practice, a Bell
experiment involves measurements on many copies of the same
quantum systems, we also investigated the possibility of
getting a better Bell inequality violation by using collective
measurements without postselection; this is the other subject
of discussion in Chapter~\ref{Chap:BellViolation}. Next, in
Chapter~\ref{Chap:Hidden.Nonlocality}, we will look into the
possibility of deriving nonclassical correlations from all
entangled quantum states. In particular, with the aid of an
ancilla state which does not violate the Bell-CHSH inequality,
we will provide a protocol to demonstrate a Bell-CHSH
inequality violation coming from all bipartite entangled
states. This provides a positive answer to the long-standing
question of whether all bipartite entangled states can lead to
some kind of observable nonclassical correlations. Finally, we
will conclude with a summary of key results and some
possibilities for future research in
Chapter~\ref{Chap:Conclusion}.

\chapter{Bell's Theorem and Tests of Local Causality}
\label{Chap:BellThm}

In this chapter, we will give a brief historical review of the
study of local causality in quantum mechanics. We will begin
with the {\em incompleteness} arguments presented by Einstein,
Podolsky and Rosen~\cite{EPR:1935}, and see how that had led to
the celebrated discovery by
Bell~\cite{J.S.Bell:1964,J.S.Bell:RMP:1966}. After that, some
of the key developments towards an experimental test of local
causality will also be reviewed.

\section{Bell's Theorem}\label{Sec:BellThm}

\subsection{The Einstein-Podolsky-Rosen Incompleteness
Arguments}\label{Sec:EPR}

{\em Quantum mechanics}, as is well-known, only gives
predictions, via the {\em wavefunction} or {\em state vector},
on the probabilities of obtaining a certain outcome in an
experiment (see, for example,
Ref.~\cite{A.Peres:Book:1995,C.J.Isham:Book:1995}). Moreover,
according to {\em Bohr's
complementarity}~\cite{N.Bohr:Naturwissenschaften:1928,N.Bohr:Nature:1928,
M.O.Scully:Nature:1991}, physical quantities described by two
non-commuting observables in the theory are incompatible in
that a complete knowledge of one precludes any knowledge of the
other. This scenario is clearly in discord with the classical
intuition that objective properties of physical systems exist
independent of measurements.

Among those who were unsatisfied with Bohr's complementarity
were Einstein, Podolsky and Rosen (EPR) who together put
forward, in their 1935 paper~\cite{EPR:1935}, the argument that
any {\em complete} physical theory must be such that
\begin{quote}
    ``{\em every element of the physical reality must have a
    counterpart in the physical theory.}"
\end{quote}
A sufficient condition for the {\em reality} of a physical
quantity that they have provided is as follows~\cite{EPR:1935}:
\begin{quote}
    ``{\em if without in any way disturbing a system, we can
    predict with certainty (i.e., with probability equal to unity)
    the value of a physical quantity, then there exists an element
    of physical reality corresponding to this physical quantity.}"
\end{quote}

According to these criteria, Bohr's complementarity implies at
least one of the followings, namely, (1) quantum mechanics is
not a complete theory, or (2) the two physical quantities
corresponding to non-commuting observables cannot have
simultaneous reality. Moreover, by considering local
measurements on two physical systems that have interacted in
the past but are separated at the time of measurements, EPR
came to the conclusion that if (1) is false, so is (2).

As an example, EPR considered a two-particle system described
by the wavefunction
\begin{equation}\label{Eq:Wavefunction:EPR}
    \ket{\Psi(x_1,x_2)}=\int_{-\infty}^{\infty}\dd{p}~\exp{({\ii}p/\hbar)(x_1-x_2+x_0)},
\end{equation}
where $x_1$ and $x_2$ are, respectively, the coordinates
attached to the two particles, $x_0$ is some arbitrary constant
and $p$ is the eigenvalue of the momentum operator for the
first particle. It is not difficult to see that for both
position and momentum measurements on the two particles, the
outcomes derived are always perfectly correlated. In
particular, if Alice and Bob are, respectively, at the
receiving ends of the two particles, their measurement outcomes
on these particles will read:\\
\begin{table}[h!]
    \begin{centre}
    \begin{tabular}{c|cc}
    Measurement & Alice & Bob\\\hline
    Momentum $P$  & $p$ & $-p$\\
    Position $Q$ & $x$ & $x+x_0$
    \end{tabular}
    \end{centre}
\end{table}

Therefore, according to the criterion set up by EPR, should $P$
be measured on the first particle, the momentum of the second
particle is an element of physical reality; whereas if $Q$ is
measured on the first particle, the position of the second
particle is an element of physical reality.
Moreover~\cite{EPR:1935},
\begin{quote}
    {\em  since at the time of measurement the particles no longer
    interact, no real change can take place in the second system in
    consequence of anything that may be done to the first system.}
\end{quote}
Hence, by EPR's criterion of {\em reality}, both $P$ and $Q$ of
the second particle, though corresponding to noncommuting
observables in the theory, can have simultaneous reality,
corresponding to the negation of (2). Since negation of (1)
also led to the negation of (2), while at least one of (1) and
(2) has to be true, EPR concluded that the quantum mechanical
description of physical reality given by wavefunction is
incomplete. Furthermore, at the very end of the
paper~\cite{EPR:1935}, EPR optimistically expressed their
belief that a theory that provides a {\em complete} description
of physical reality is possible.

\subsection{Completeness and Hidden-Variable Theory}
\label{Sec:LHVM}

Although no explicit proposal was given  by EPR, it was
commonly inferred from their arguments and the success of
statistical mechanics that a complete description of physical
reality can be attained if unknown ({\em hidden}) variables are
supplemented to the wavefunction description of physical
reality (see, for example Ref.~\cite{J.S.Bell:PSFPHEP:1976} and
references therein). Indeed, not known to EPR and many other
founding fathers of quantum mechanics, towards the end of
1920s, de Broglie constructed a hidden-variable theory that is
capable of explaining the quantum interference phenomena while
retaining the corpuscular feature of individual
particles~\cite{deBroglie:1927,deBroglie:1928}.

Despite that, the idea of completing the description given by
quantum mechanics with additional variables has received much
criticism over the years (see for example
Ref.~\cite{J.S.Bell:RMP:1966} and references therein). Among
them, von Neumann's proof (pp.~305,
Ref.~\cite{vonNeumann:1932}) of the impossibility of ({\em
noncontextual}) hidden variables probably provided peace in
mind to most of those who were against the proposal. The proof
given by von Neumann in Ref.~\cite{vonNeumann:1932} has,
nevertheless, imposed unnecessary restrictions on the unknown
variables~\cite{J.S.Bell:RMP:1966}. In fact, this was made
blatant after Bohm rediscovered the hidden-variable
theory~\cite{D.Bohm:PR:1952a,D.Bohm:PR:1952b} first formulated
by de Broglie~\cite{deBroglie:1927,deBroglie:1928}.

Nonetheless, {\em Bohmian mechanics} or the {\em pilot-wave
model}, as the de~Broglie-Bohm hidden-variable theory is
currently known, was dismissed by many physicists because of
the explicit ``{\em nonlocal}" flavor in the theory.
Ironically, it was precisely the discovery of this
controversial theory that led
Bell~\cite{J.S.Bell:1987,J.S.Bell:RMP:1966} to consider,
instead, the possibility of a {\em local hidden-variable
theory} and hence his important discovery in
1964~\cite{J.S.Bell:1964}.

For Einstein, he was firmly convinced that (pp~672,
\cite{A.Einstein:1949})
\begin{quote}
    ``{\em\ldots within the framework of future physics, quantum
    theory takes an analogous position as statistical mechanics
    takes within the framework of classical mechanics.}''
\end{quote}
Adhering to the same philosophy, Bell's consideration of a
hidden-variable theory~\cite{J.S.Bell:1964} is such that an
average over some unknown ensemble labeled by the {\em
hidden-variable} gives rise to the statistical behavior of
quantum mechanical prediction.  As Bell emphasized, the
variables are {\em hidden} because they are not known to exist;
they are not even accessible in principle, otherwise ``quantum
mechanics would be observably
inadequate"~\cite{J.S.Bell:RMP:1966,J.S.Bell:1971}.

Clearly, not all hidden-variable theories are welcome in the
physics community~\cite{J.S.Bell:RMP:1966}. For instance, in
the hidden-variable theory formulated by de Broglie and
Bohm~\cite{D.Bohm:PR:1952a,D.Bohm:PR:1952b}, the trajectory of
one particle may depend explicitly on the trajectory as well as
the wavefunction of other particles that it has interacted with
in the past, regardless of their spatial separation. This
``{\em nonlocal}" feature of the theory is in apparent
contradiction with the well-established intuition of causality
that we have learned from special theory of relativity.
Therefore, following EPR's flavor, Bell considered
hidden-variable theories that are {\em local} such that, in
Bell's words~\cite{J.S.Bell:1964}:
\begin{quote}
    ``\ldots{\em  the result of measurement on one system be
    unaffected by operations on a distant system with which it has
    interacted in the past} \ldots"
\end{quote}
In later years, a theory that satisfies Bell's notion of
locality, or more specifically
\begin{quote}
    ``{\em The direct causes (and effects) of events are near by,
    and even the indirect causes (and effects) are no further away
    than permitted by the velocity of light.}"
\end{quote}
is said to be {\em locally causal}~\cite{Bell1990}. Hereafter,
we will use the term local hidden-variable theory (henceforth
abbreviated as LHVT) and the term locally causal theory
interchangeably.\footnote{The other terminology that is also
commonly found in the literature is {\em local realistic
theory}; this is however not as universally accepted, see e.g.
Ref.~\cite{T.Norsen:FP:2007}.} As we shall see below, Bell's
greatest contribution came in by showing that {\em quantum
mechanics is not a locally causal
theory}~\cite{J.S.Bell:1964,J.S.Bell:RMP:1966}.

\subsection{Quantum Mechanics is not a Locally Causal Theory}
\label{Sec:LHVM-QM}

To illustrate this remarkable fact of life,
Bell~\cite{J.S.Bell:1964} has chosen to work within the
framework first presented by Bohm (Sec 15 -- 19, Chap 22,
Ref.~\cite{D.Bohm:Book:1951}) concerning the spin degrees of
freedom of two spin-$\half$ particles, which is the analog of
EPR's scenario for {\em discrete variable quantum
systems}.\footnote{Incidentally, the experimental situation
described in the original EPR paper~\cite{EPR:1935} can indeed
be explained within the framework of a locally casual
theory~\cite{J.S.Bell:ANYAS:1986}.} In this version of EPR's
argument, pairs of spin-$\half$ particles are prepared in the
spin singlet state
\begin{equation}\label{Eq:Dfn:Singlet}
    \ket{\Psi^-}=\sqhalf\left(\ket{\spup}_\A\ket{\spdn}_\B
    -\ket{\spdn}_\A\ket{\spup}_\B\right),
\end{equation}
where $\ket{\spup}_\A$ and $\ket{\spdn}_\A$ are correspondingly
the {\em spin up} and {\em spin down} state of one of the
particles with respect to some spatial
direction\footnote{\label{fn:singlet}Since the spin singlet
state is isotropic, the actual space quantization axis is
immaterial.} (likewise for $\ket{\spup}_\B$ and
$\ket{\spdn}_\B$). After that, particles in each pair are
separated and sent to two experimenters (hereafter always
denoted by Alice and Bob), who can subsequently perform spin
measurements along some (arbitrary) direction $\hat{\alpha}$
and $\hat{\beta}$, respectively, on these particles
(c.f.~Figure~\ref{Fig:TwoPartyBellExperiment}). Now, recall
from quantum mechanics that the expectation value of such
measurements reads
\begin{equation}\label{Eq:Dfn:CorrelationFn:Singlet}
    \Eqm(\hat{\alpha},\hat{\beta})\equiv
    \bra{\Psi^-}\sigma_{\hat{\alpha}}\ten\sigma_{\hat{\beta}}
    \ket{\Psi^-}=-\hat{\alpha}\cdot\hat{\beta},
\end{equation}
where
\begin{gather}
    \sigma_{\hat{\alpha}}\equiv\hat{\alpha}\cdot\vec{\sigma},\quad
    \sigma_{\hat{\beta}}\equiv\hat{\beta}\cdot\vec{\sigma},\\
    \vec{\sigma}\equiv\sum_{l=x,y,z}\sigma_l\hat{e}_l,
    \label{Eq:Dfn:vecSigma}
\end{gather}
$\hat{e}_x$ is the unit vector pointing in the positive $x$
direction (likewise for $\hat{e}_y$ and $\hat{e}_z$) and
\begin{equation}\label{Eq:Dfn:PauliMatrices}
    \sigma_x\equiv\left(
    \begin{array}{cc}
    0 & 1 \\
    1 & 0
    \end{array}\right),\quad
    \sigma_y\equiv\left(
    \begin{array}{cr}
    0 & -\ii\\
    \ii & 0
    \end{array}\right),\quad
    \sigma_z\equiv\left(
    \begin{array}{cr}
    1 & 0 \\
    0 & -1
    \end{array}\right)
\end{equation}
are the Pauli matrices (here, we adopt the convention that
$\sigma_z\ket{\spup}=\ket{\spup}$,
$\sigma_z\ket{\spdn}=-\ket{\spdn}$).  Thus, if
$\hat{\alpha}=\hat{\beta}$, the measurement outcomes on both
sides must be perfectly (anti-) correlated, i.e., if Alice's
measurement outcome reads ``$\spup$", Bob's measurement outcome
must read ``$\spdn$". Since this is true for other pair of
$\hat{\alpha}'$ and $\hat{\beta}'$ such that
$\hat{\alpha}'=\hat{\beta}'$, hence, by virtue of EPR's
original argument, one can conclude that the ``spin" along any
direction for both of these particles must be ``element of
physical reality".

Now, let us follow Ref.~\cite{J.S.Bell:1964} and denote by
$\lambda$ any additional parameters carried by the particles
that could provide a complete specification for these physical
realities. Physically, we can think of $\lambda$ as information
that is exchanged between the particles during the preparation
procedure but which is not completely encoded in the state
vector $\ket{\Psi^-}$. As remarked in
Ref.~\cite{J.S.Bell:1964}, the exact nature of $\lambda$ is
irrelevant, it could refer to a single or a set of random
variables, or even a set of functions and it could take on
continuous as well as discrete values. If we denote by $\oA$
and $\oB$, respectively, the measurement outcome observed at
Alice's and Bob's side, then by Bell's requirement of {\em
locality}, we must have $\oA$ as a function of $\lambda$ and
$\hat{\alpha}$ but not $\hat{\beta}$; likewise for $\oB$.
Furthermore, the measurement outcome at each side is completely
determined by these parameters such that~\cite{J.S.Bell:1964}
\begin{equation}\label{Eq:LHV:ResponseFunction}
    \oA(\hat{\alpha},\lambda)=\pm1,\quad \oB(\hat{\beta}, \lambda)=\pm1;
\end{equation}
here, we adopt the convention that measurement outcomes
``$\spup$" and ``$\spdn$" are assigned the value ``$+1$" and
``$-1$" respectively. Let us now define the {\em correlation
function} as
\begin{equation}\label{Eq:Dfn:CorrelationFn:LHV}
    E(\hat{\alpha},\hat{\beta})\equiv\int_\Lambda\dd{\lambda}~\rho_\lambda~
    \oA(\hat{\alpha},\lambda)~\oB(\hat{\beta},\lambda),
\end{equation}
where $\Lambda$ is the space of hidden-variable and
$\rho_\lambda$ is some normalized probability density such that
\begin{equation}\label{Eq:ProbabilityDensity:Normalization}
    \int_\Lambda\dd\lambda~\rho_\lambda=1.
\end{equation}
Physically, the correlation function,
Eq.~\eqref{Eq:Dfn:CorrelationFn:LHV}, is just the average of
the product of local measurement outcomes over an ensemble of
physical systems characterized by some distribution of
hidden-variable, $\rho_\lambda$. It then follows that a
necessary condition for getting a {\em complete description} of
the above-mentioned physical realities using local
hidden-variable is that for all $\hat{\alpha}$ and
$\hat{\beta}$
\begin{equation}\label{Eq:CorrelationFn:LHVQM}
    E(\hat{\alpha},\hat{\beta})=\Eqm(\hat{\alpha},\hat{\beta})
\end{equation}
for some choice of $\oA(\hat{\alpha},\lambda)$,
$\oB(\hat{\beta},\lambda)$ and some choice of $\rho_\lambda$
that is independent of $\hat{\alpha}$ and $\hat{\beta}$. As we
shall see below, Eq.~\eqref{Eq:CorrelationFn:LHVQM} cannot be
made true in general. Nonetheless, it is interesting to note
that Bell has constructed a specific {\em local hidden-variable
model}\,\footnote{Throughout this thesis, we will use the term
{\em local hidden-variable model} to refer to, say, a set of
rules, that can be used to reproduce some set of experimental
statistics; it is less general than a LHVT, which is supposed
to be able to reproduce all experimental statistics generated
by quantum mechanics.} (henceforth abbreviated as LHVM) that
makes it true for the case when
$\hat{\alpha}\cdot\hat{\beta}=+1,0,-1$~\cite{J.S.Bell:1964}.

To show that Eq.~\eqref{Eq:CorrelationFn:LHVQM} cannot be made
true for all possible choices of measurement parameters, Bell
introduced another unit vector $\hat{\beta}'$ and considered
the following combination of correlation functions:
\begin{align*}
    E(\hat{\alpha},\hat{\beta})-E(\hat{\alpha},\hat{\beta}')
    &=\int_\Lambda\dd{\lambda}~ \rho_\lambda
    \left[\oA(\hat{\alpha},\lambda)~\oB(\hat{\beta},\lambda)-
    \oA(\hat{\alpha},\lambda)~\oB(\hat{\beta}',\lambda)\right].
\end{align*}
From triangle inequality, Eq.~\eqref{Eq:LHV:ResponseFunction}
and Eq.~\eqref{Eq:ProbabilityDensity:Normalization}, it follows
that
\begin{align}
    \left|E(\hat{\alpha},\hat{\beta})-E(\hat{\alpha},\hat{\beta}')\right|
    &\le\int_\Lambda\dd{\lambda}~   \rho_\lambda\left|\oA(\hat{\alpha},\lambda)~\oB(\hat{\beta},\lambda)\right|
    \left[1-\oB(\hat{\beta},\lambda)~\oB(\hat{\beta}',\lambda)\right],\nonumber\\
    &=\int_\Lambda\dd{\lambda}~ \rho_\lambda\left[1-
    \oB(\hat{\beta},\lambda)~\oB(\hat{\beta}',\lambda)\right],\nonumber\\
    &=1-\int_\Lambda\dd{\lambda}~ \rho_\lambda~
    \oB(\hat{\beta},\lambda)~\oB(\hat{\beta}',\lambda).\label{Ineq:Diff:Correlation}
\end{align}
When $\hat{\alpha}=\hat{\beta}$, it follows from
Eq.~\eqref{Eq:Dfn:CorrelationFn:Singlet} that
$\Eqm(\hat{\alpha},\hat{\beta})=-1$. Therefore, Bell further
assumed in Ref.~\cite{J.S.Bell:1964} that if the measurement
parameters chosen by both observers coincide, the outcomes of
measurement, as determined by the hidden variables are also
{\em perfectly correlated}:
\begin{equation}\label{Eq:PerfectCorrelation}
    \oA(\hat{\alpha},\lambda)=-\oB(\hat{\alpha},\lambda).
\end{equation}
With this assumption, the above inequality becomes
\begin{equation}\label{Ineq:Bell:1964}
    \left|E(\hat{\alpha},\hat{\beta})-E(\hat{\alpha},\hat{\beta}')\right|
       -E(\hat{\beta},\hat{\beta}')-1\le0,
\end{equation}
which gives the very first inequality that has to be {\em
satisfied by any LHVT} in the literature~\cite{J.S.Bell:1964}.
In the spirit of Bell's original work, let us introduce the
following definition for a Bell inequality.\footnote{It is
worth noting that among the physics community, the term Bell
inequality, or Bell-type inequality has sometimes been used to
refer to inequality that arises out of an entanglement witness.
To appreciate the distinction between these two kinds of
inequalities, see, for example,
Refs.~\cite{B.M.Terhal:PLA:2000,P.Hyllus:PRA:2005}.}

\begin{dfn}\label{Dfn:BellInequality}
    A Bell inequality is an inequality derived from the assumptions
    of a general local hidden-variable theory.
\end{dfn}

In Ref.~\cite{J.S.Bell:1964}, Bell subsequently gave a formal
proof, based on Eq.~\eqref{Ineq:Bell:1964}, that
$\Eqm(\hat{\alpha},\hat{\beta})$ cannot equal or even be
approximated arbitrarily closely by
$E(\hat{\alpha},\hat{\beta})$. However, to illustrate the point
that quantum mechanics also gives rise to predictions not
allowed by any LHVT, it suffices to show that for some choice
of measurement parameters, the quantum mechanical version of
Eq.~\eqref{Ineq:Bell:1964}, namely,
\begin{equation}\label{Ineq:QM:Bell:1964}
    \left|\Eqm(\hat{\alpha},\hat{\beta})-\Eqm(\hat{\alpha},\hat{\beta}')\right|
    -\Eqm(\hat{\beta},\hat{\beta}')-1\le0,
\end{equation}
is violated. To this end, let us assume that all the spin
measurements are performed on the $x-z$ plane and that
$\hat{\alpha}$ points along the direction of the positive
$z$-axis, i.e., $\hat{\alpha}=\hat{e}_z$. Then, for the choice
of
\begin{equation}
    \hat{\beta}=\frac{\sqrt{3}}{2}\hat{e}_x+\half \hat{e}_z,\qquad
    \hat{\beta}'=\frac{\sqrt{3}}{2}\hat{e}_x-\half \hat{e}_z,
\end{equation}
it can be easily verified using
Eq.~\eqref{Eq:Dfn:CorrelationFn:Singlet} that quantum mechanics
predicts 1/2 for the {\em lhs} of inequality
\eqref{Ineq:QM:Bell:1964}, thereby demonstrating that quantum
mechanical prediction is, in general, incompatible with that
given by any LHVT, c.f. Eq.~\eqref{Ineq:Bell:1964}.

The above finding gives rise to the following important theorem
first derived by Bell~\cite{J.S.Bell:1964}:

\begin{theorem}\label{Thm:Bell}
    No local hidden-variable theory can reproduce all quantum
    mechanical predictions. Equivalently, quantum mechanics is not
    a locally causal theory.
\end{theorem}

\section{Towards an Experimental Test of Local Causality}

\subsection{Bell-Clauser-Horne-Shimony-Holt Inequality}\label{Sec:Bell-CHSH}

The inequality \eqref{Ineq:Bell:1964} derived by
Bell~\cite{J.S.Bell:1964} has clearly demonstrated that {\em
some} quantum mechanical predictions, in the ideal scenario,
cannot be reproduced by any LHVT. However, the assumption of
{\em perfect correlation}, c.f.
Eq.~\eqref{Eq:PerfectCorrelation}, or equivalently,
\begin{equation}\label{Eq:PerfectCorrelation2}
    E(\hat{\alpha}',\hat{\beta})=-1,
\end{equation}
for $\hat{\alpha}'=\hat{\beta}$ is too strong to be justified
in any realistic experimental scenario. The Bell inequality
\eqref{Ineq:Bell:1964} was therefore not readily subjected to
any experimental test. A few years later, in 1969, a resolution
was provided by Clauser, Horne, Shimony and Holt (henceforth
abbreviated as CHSH) who, instead of
Eq.~\eqref{Eq:PerfectCorrelation2}, assumed that for some
$\hat{\alpha}'$~\cite{CHSH:PRL:1969}
\begin{equation}\label{Eq:NonPerfectCorrelation}
    E(\hat{\alpha}',\hat{\beta})=-1+\delta,
\end{equation}
where $0\le\delta\le1$. To conform with the prediction given by
quantum mechanics, one expects that for spin measurement on the
singlet state and when $\hat{\alpha}'$ is (approximately)
aligned with $\hat{\beta}$, $\delta$ is close to  but not
exactly equal to zero.

Now, let's take this {\em imperfect correlation} into account
by dividing the space of hidden-variable $\Lambda$ into
$\Lambda_\pm$ such that
\begin{equation}\label{Eq:HVSpacePM}
    \Lambda_\pm=\{\lambda|\oA(\hat{\alpha}',\lambda)=\pm
    \oB(\hat{\beta},\lambda)\}.
\end{equation}
Then, it follows from Eq.~\eqref{Eq:Dfn:CorrelationFn:LHV},
Eq.~\eqref{Eq:ProbabilityDensity:Normalization},
Eq.~\eqref{Eq:NonPerfectCorrelation} and
Eq.~\eqref{Eq:HVSpacePM} that
\begin{equation}
    2\int_{\Lambda^-}\dd\lambda~\rho_\lambda=2-\delta.
\end{equation}
Instead of inequality \eqref{Ineq:Bell:1964}, inequality
\eqref{Ineq:Diff:Correlation} now leads to
\begin{align*}
    \left|E(\hat{\alpha},\hat{\beta})-E(\hat{\alpha},\hat{\beta}')\right|
    &\le 1-\int_{\Lambda^+}\dd{\lambda}~ \rho_\lambda~
    \oB(\hat{\beta},\lambda)\,\oB(\hat{\beta}',\lambda)-\int_{\Lambda^-}\dd{\lambda}~ \rho_\lambda~
    \oB(\hat{\beta},\lambda)\,\oB(\hat{\beta}',\lambda),\nonumber\\
    &= 1-\int_{\Lambda^+}\dd{\lambda}~ \rho_\lambda~
    \oA(\hat{\alpha}',\lambda)\,\oB(\hat{\beta}',\lambda)+\int_{\Lambda^-}\dd{\lambda}~ \rho_\lambda~
    \oA(\hat{\alpha}',\lambda)\,\oB(\hat{\beta}',\lambda),\nonumber\\
    &= 1-E(\hat{\alpha}',\hat{\beta}')+2\int_{\Lambda^-}\dd{\lambda}~ \rho_\lambda~
    \oA(\hat{\alpha}',\lambda)\,\oB(\hat{\beta}',\lambda),\nonumber\\
    &\le 1-E(\hat{\alpha}',\hat{\beta}')+2\int_{\Lambda^-}\dd{\lambda}~ \rho_\lambda
    \left|\oA(\hat{\alpha}',\lambda)\,\oB(\hat{\beta}',\lambda)\right|,\nonumber\\
    &= 3-E(\hat{\alpha}',\hat{\beta}')-\delta,\nonumber
\end{align*}
which, together with Eq.~\eqref{Eq:NonPerfectCorrelation}, becomes
\begin{equation}\label{Ineq:CHSH:Original}
    \left|E(\hat{\alpha},\hat{\beta})
    -E(\hat{\alpha},\hat{\beta}')\right|+
     E(\hat{\alpha}',\hat{\beta})
    +E(\hat{\alpha}',\hat{\beta}')
    \le 2.
\end{equation}
This is the famous Bell-CHSH inequality that was first derived
in Ref.~\cite{CHSH:PRL:1969}. It is interesting to note that a
few years later~\cite{J.S.Bell:1971}, Bell gave an alternative
derivation\footnote{Strictly, the inequality that was later
derived by Bell reads:
\begin{equation}\label{Ineq:CHSH:Bell}
    \left|E(\hat{\alpha},\hat{\beta})-E(\hat{\alpha},\hat{\beta}')\right|+
    \left|E(\hat{\alpha}',\hat{\beta})+E(\hat{\alpha}',\hat{\beta}')\right|
    \le 2,
\end{equation}
but as we shall see below, we can essentially treat it as the
same inequality as that given by
Eq.~\eqref{Ineq:CHSH:Original}.} of inequality
\eqref{Ineq:CHSH:Original} by respectively replacing
Eq.~\eqref{Eq:LHV:ResponseFunction} and
Eq.~\eqref{Eq:Dfn:CorrelationFn:LHV} with
\begin{equation}
    |\bar{o}_a(\hat{\alpha},\lambda)|\le 1,\quad
    |\bar{o}_b(\hat{\beta},\lambda)|\le 1,
\end{equation}
and
\begin{equation}\label{Eq:Dfn:CorrelationFn:LHV2}
    E(\hat{\alpha},\hat{\beta})\equiv\int_\Lambda\dd{\lambda}~\rho_\lambda~
    \bar{o}_a(\hat{\alpha},\lambda)\,\bar{o}_b(\hat{\beta},\lambda).
\end{equation}
Here, Bell tried to be more general (as compared with his
approach in Ref.~\cite{J.S.Bell:1964}) by assuming that the
measurement apparatuses could also contain hidden-variable that
could influence the experimental results. In the above
expressions, $\bar{o}_a(\hat{\alpha},\lambda)$ is thus used to
denote an average over the hidden-variable associated with
Alice's apparatus when it is set to perform measurements
parameterized by $\hat{\alpha}$; similarly for
$\bar{o}_b(\hat{\beta},\lambda)$.

At this stage, it is worth making a few other remarks. Firstly,
in contrast with Bell's first inequality,
Eq.~\eqref{Ineq:Bell:1964}, that was developed for spin
measurements on the singlet state, the Bell-CHSH inequality is
also relevant to other physical states as well as other
physical systems. In fact, it is applicable, as a constraint
imposed by LHVTs, to {\em any} experimental statistics
involving {\em two} spatially separated subsystems and where
{\em two} {\em dichotomic}\footnote{A dichotomic measurement is
one that yields one out of two possible outcomes.} measurements
--- each giving outcomes labeled by $\pm1$ --- can be performed
on each of the subsystems. Essentially, this means that in the
more general experimental framework, the parameters
$\hat{\alpha}$ etc. are merely labels to distinguish the
different measurements that Alice and Bob may perform on the
subsystem in their possession.

As a result, and for the convenience of subsequent discussion,
let us introduce the following notation for the correlation
function associated with Alice measuring the observable
$A_{\sA}$ and Bob measuring the observable $B_{\sB}$, i.e.,
\begin{equation}\label{Eq:Dfn:CorrelationFn:LHV3}
    E(A_{\sA},B_{\sB})\equiv \int_\Lambda\dd{\lambda}~\rho_\lambda~
    \oA(A_{\sA},\lambda)\,\oB(B_{\sB},\lambda),
\end{equation}
where the outcomes of local measurements $\oA$ and $\oB$ are
now functions of the hidden variable $\lambda$ and,
respectively, the local observables $A_{\sA}$ and $B_{\sB}$. In
particular, if we now make the following associations between
the measurement parameters $\{\alpha,\alpha',\beta,\beta'\}$
and the local observables $\{A_{\sA},B_{\sB}\}_{\sA,\sB=1}^2$:
\begin{equation}
    \hat{\alpha}\to A_2,\quad\hat{\alpha}'\to A_1,\quad \hat{\beta}\to
    B_1,\quad \hat{\beta}'\to B_2,
\end{equation}
it is clear that both inequality \eqref{Ineq:CHSH:Original} and
inequality \eqref{Ineq:CHSH:Bell} imply the following inequality:
\begin{equation}\label{Ineq:CHSH:Convention}
    E(A_1,B_1)+E(A_1,B_2)+E(A_2,B_1)-E(A_2,B_2)\le 2.
\end{equation}
Evidently, if this is a valid constraint that has to be
satisfied by any LHVT, so is any other obtained by relabeling
the local observers (``Alice"~$\leftrightarrow$~``Bob"), local
measurement settings ($A_1\leftrightarrow A_2$,
$B_1\leftrightarrow B_2$) and/or outcomes
($+1\leftrightarrow-1$). For example, if we instead make the
associations $ \hat{\alpha}\to A_1$, $\hat{\alpha}'\to A_2$ and
relabel all the $+1$ outcomes at Alice's site by $-1$ and {\em
vice versa}, then we will arrive at
\begin{equation}\label{Ineq:CHSH:Convention2}
    -2\le E(A_1,B_1)-E(A_1,B_2)+E(A_2,B_1)+E(A_2,B_2),
\end{equation}
which is clearly different from inequality
\eqref{Ineq:CHSH:Convention}. Nonetheless, the difference
between these inequalities, which is due to a different choice
of labels, is physically irrelevant. After all, when testing a
set of experimental data against a Bell inequality, the choice
of these labels is completely arbitrary. As such, let us define
the equivalence class of Bell inequalities as
follows~\cite{Ll.Masanes:QIC:2003,D.Collins:JPA:2004}.
\begin{dfn}\label{Dfn:BI:Equivalence}
    A Bell inequality is equivalent to another if and only if one
    can be obtained from the other by relabeling the local
    observers, local measurement settings and/or measurement
    outcomes.
\end{dfn}
Under this definition, it is straightforward to see that apart
from inequality \eqref{Ineq:CHSH:Convention2}, inequality
\eqref{Ineq:CHSH:Convention} is also equivalent to 6 other
inequalities.  Hereafter, when there is no risk of confusion,
we will refer to inequality \eqref{Ineq:CHSH:Convention} as the
Bell-CHSH inequality and to the entire class of 8 inequalities
that are equivalent to inequality \eqref{Ineq:CHSH:Convention}
as the Bell-CHSH inequalities. In relation to inequality
\eqref{Ineq:CHSH:Original}, it is also not difficult to see
that this inequality is violated if and only if (at least) one
of the Bell-CHSH inequalities is violated; likewise for
inequality \eqref{Ineq:CHSH:Bell}.

As a last remark, we note that the Bell-CHSH inequality is an
example of what is now called a {\em (Bell) correlation
inequality} --- a Bell inequality that only involves linear
combination of {\em correlation functions}. Clearly, a
correlation function, which can be determined experimentally by
averaging over the product of the outcome of local observables,
is not the only quantity that is derivable from a given set of
experimental data; the relative frequency of experimental
outcomes, in the limit of large sample size, gives a good
approximation to the probability of obtaining that particular
outcome. In the next section, we will look at an example of the
other prototype of (linear) Bell inequalities, namely, one that
involves a linear combination of joint and marginal
probabilities of experimental outcomes.

\subsection{Bell-Clauser-Horne Inequality}\label{Sec:Bell-CH}

The earlier work by CHSH is no doubt a big step towards an
experimental test for the feasibility of locally casual
theories. However, due to imperfect detection and other
realistic experimental concerns, the Bell-CHSH inequality
\eqref{Ineq:CHSH:Convention} can only be put into a real
experimental test when supplemented with an auxiliary
assumption on the ensemble of detected
particles~\cite{CHSH:PRL:1969,J.F.Clauser:RPP:1978}.
Specifically, in the context of polarization measurement on
photons, the original assumption made by CHSH is that if a pair
of photons emerges from the respective polarizers located at
Alice's and Bob's side, the probability of their joint
detection is independent of the orientation of the polarizers.

A few years later, work by Clauser and Horne (hereafter
abbreviated as CH) demonstrated that without an auxiliary
assumption, neither the experiment carried out by Freedman and
Clauser~\cite{Freedman:PRL:1972} nor any similar ones with
improved detector efficiency can give a definitive test of
locally causal theories~\cite{CH:PRD:1974}. To remedy the
problem, CH derived, in the same paper~\cite{CH:PRD:1974},
another Bell inequality and showed that when supplemented with
a considerably weaker {\em no enhancement} assumption, the
results obtained by Freedman and Clauser are indeed
incompatible with LHVTs~\cite{CH:PRD:1974}.

\begin{figure}[h!]
    \centering\rule{0pt}{4pt}\par
    \scalebox{1}{\includegraphics{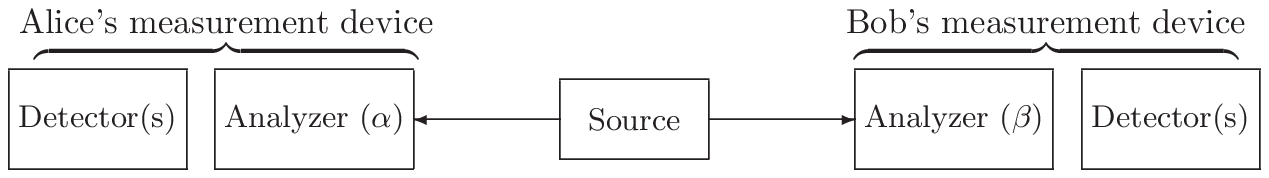}}
    \caption{\label{Fig:TwoPartyBellExperiment}
    Schematic diagram of the experimental setup involved in a
    standard two-party Bell experiment. The source produces pairs
    of physical systems that are subsequently distributed,
    respectively, to Alice and Bob. They then subject the physical
    system that they receive to an analyzer which has an adjustable
    parameter (denoted by $\alpha$ and $\beta$ correspondingly).
    For example, in the case of polarization measurement on
    photons, an analyzer is simply a combination of waveplates and
    a polarizer. The final stage of the measurement process
    consists of detecting the subsystems that pass through each
    analyzer with one or more detectors. In the scenario considered
    by Bell~\cite{J.S.Bell:1964} and
    Clauser\etal~\cite{CHSH:PRL:1969}, there are two detectors at
    each site, whereas in the original experimental scenario
    considered by CH~\cite{CH:PRD:1974}, there is only one detector
    after each analyzer.
    }
\end{figure}

The scenario that CH considered is a familiar one, namely, one
that involves ensembles of two particles being sent to Alice
and Bob respectively. Under the control of each experimenter is
an analyzer with an adjustable parameter (denoted by $\alpha$
and $\beta$ respectively) and a detector. At each run of the
experiment, let us denote by $\lambda$ the state of the
two-particle system and $\ProbNOAB{\alpha,\beta, \lambda}$ the
probability that for this two-particle state, a count is
triggered at both detectors conditioned on Alice setting her
analyzer to $\alpha$ and Bob setting his to $\beta$; the
marginal probabilities of detecting a particle
$\ProbNOA{\alpha,\lambda}$ and $\ProbNOB{\beta,\lambda}$ are
similarly defined. In these terminologies, the no enhancement
assumption states that for a given state $\lambda$, the
probability of detecting a particle with the analyzer removed
is greater than or equal to the probability of detecting a
particle when the analyzer is in place.

Now, note that for a given (normalized) probability density
$\rho_\lambda$ characterizing the ensemble of states emitted,
the observed relative frequencies should correspond to
\begin{gather}
    \ProbNOA{\alpha}=\int_\Lambda\dd{\lambda}~\rho_\lambda~
    \ProbNOA{\alpha,\lambda},\quad
    \ProbNOB{\beta}=\int_\Lambda\dd{\lambda}~\rho_\lambda~
    \ProbNOB{\beta,\lambda},\nonumber\\
    \ProbNOAB{\alpha,\beta}=\int_\Lambda\dd{\lambda}~\rho_\lambda~
    \ProbNOAB{\alpha,\beta,\lambda}.\label{Eq:Prob:AverageOverLambda}
\end{gather}
It is worth noting that as it is, the above formulation could
very well be applied to quantum mechanical prediction, with the
wavefunction $\ket{\psi}$ playing the role of $\lambda$. As
with the correlation function,
Eq.~\eqref{Eq:Dfn:CorrelationFn:LHV}, the condition of local
causality comes in by demanding that the probability of joint
detection factorizes~\cite{CH:PRD:1974}, i.e.,
\begin{equation}\label{Eq:Probability:Factorizes}
    \ProbNOAB{\alpha,\beta,\lambda}=\ProbNOA{\alpha,\lambda}\,
    \ProbNOB{\beta,\lambda}.
\end{equation}

From the definition of probabilities, it follows that
\begin{gather}
    0\le \ProbNOA{\alpha,\lambda}\le1,\quad 0\le
    \ProbNOA{\alpha',\lambda}\le1,\nonumber\\
    0\le \ProbNOB{\beta,\lambda}\le1,\quad 0\le
    \ProbNOB{\beta',\lambda}\le1,\label{Eq:Probabilities}
\end{gather}
where $\alpha'$ and $\beta'$ are some other choice of
parameters for the analyzers. Together,
Eq.~\eqref{Eq:Probability:Factorizes} and
Eq.~\eqref{Eq:Probabilities} imply that~\cite{CH:PRD:1974}
\begin{align*}
    -1\le &\ProbNOA{\alpha,\lambda}\,\ProbNOB{\beta,\lambda}
    + \ProbNOA{\alpha,\lambda}\,\ProbNOB{\beta',\lambda}
    +\ProbNOA{\alpha',\lambda}\,\ProbNOB{\beta,\lambda}\\
    - &\ProbNOA{\alpha',\lambda}\,\ProbNOB{\beta',\lambda}- \ProbNOA{\alpha,\lambda}
    -\ProbNOB{\beta,\lambda} \le 0
\end{align*}
for each given $\lambda$. After averaging over the ensemble
space $\Lambda$, one arrives at
\begin{align}\label{Ineq:CH:Original}
    -1\le \ProbNOAB{\alpha,\beta}+ \ProbNOAB{\alpha,\beta'}
    +\ProbNOAB{\alpha',\beta}- \ProbNOAB{\alpha',\beta'}- \ProbNOA{\alpha}
    -\ProbNOB{\beta} \le 0,
\end{align}
which is the Bell-CH inequality --- the very first Bell
inequality for {\em probabilities} derived in the literature.
Notice that to arrive at the lower limit of inequality
\eqref{Ineq:CH:Original}, we also have to assume that the
probability density $\rho_\lambda$ is normalized,
Eq.~\eqref{Eq:ProbabilityDensity:Normalization}.

Let us now make a few other remarks concerning
inequality~\eqref{Ineq:CH:Original}. To begin with, we note
that although the inequality was derived by considering a
one-output-channel analyzer that is followed by a single
detector, it could very well be applied to measurement devices
equipped with two (or more) detectors, thereby giving rise to
two (or more) possible outcomes.\footnote{Strictly, there are
three possible outcomes when there are two detectors, with the
other possible outcome corresponding to no detection.} In
particular, for the specific case of two possible outcomes,
which we will label as ``$\pm$", the same analysis allows us to
arrive at the inequality~\cite{CH:PRD:1974}
\begin{subequations}\label{Ineq:CH:MoreOutcomes}
\begin{gather}\label{Ineq:Prob:CH}
    \ProbTwJ{1}{\oA}{1}{\oB}+\ProbTwJ{1}{\oA}{2}{\oB} +
    \ProbTwJ{2}{\oA}{1}{\oB}-\ProbTwJ{2}{\oA}{2}{\oB}-\ProbTwMA{1}{\oA}
    -\ProbTwMB{1}{\oB}\le 0,
\end{gather}
and
\begin{gather}\label{Ineq:Prob:CH2}
    -\left[\ProbTwJ{1}{\oA}{1}{\oB}+\ProbTwJ{1}{\oA}{2}{\oB} +
    \ProbTwJ{2}{\oA}{1}{\oB}-\ProbTwJ{2}{\oA}{2}{\oB}-\ProbTwMA{1}{\oA}
    -\ProbTwMB{1}{\oB}\right]\le 1,
\end{gather}
\end{subequations}
where each measurement outcome ${\oA}$ and ${\oB}$ can be
``$\pm$" and $\ProbTwGJ$ is now the probability of Alice
observing outcome ${\oA}$ and Bob observing outcome ${\oB}$
conditioned on her performing the $\idx{\sA}$ measurement and
him performing the $\idx{\sB}$ measurement; the marginal
probabilities $\ProbTwGMA$ and $\ProbTwGMB$ are analogously
defined. Notice that the four inequalities
\eqref{Ineq:Prob:CH2} are actually equivalent to inequalities
\eqref{Ineq:Prob:CH} and can be obtained from the latter, for
example, via the identity $\ProbTwJ{\sA}{+}{\sB}{+}
+\ProbTwJ{\sA}{+}{\sB}{-} =\ProbTwMA{\sA}{+}$.

Let us also remark that the set of 8 inequalities given in
Eq.~\eqref{Ineq:CH:MoreOutcomes} are symmetrical with respect
to swapping $\A$ \& $\B$ and have taken into account all
possible ways of labeling of the outcomes. Nevertheless,
additional equivalent inequalities, such as
\begin{align}\label{Ineq:CH:MoreOutcomes:Convention2}
    -1\le \ProbTwJ{1}{\oA}{2}{\oB}+\ProbTwJ{1}{\oA}{1}{\oB}
    + \ProbTwJ{2}{\oA}{2}{\oB}-\ProbTwJ{2}{\oA}{1}{\oB}-\ProbTwMA{1}{\oA} -\ProbTwMB{2}{\oB}\le 0
\end{align}
can still be obtained by relabeling the local measurement
settings. Hereafter, unless stated otherwise, the term Bell-CH
inequality would refer to Eq.~\eqref{Ineq:Prob:CH} with only
two possible outcomes.

In relation to the Bell-CHSH inequality, we recall that the
correlation function defined in
Eq.~\eqref{Eq:Dfn:CorrelationFn:LHV3} can actually be rewritten
as\footnote{To this end, we are identifying the $\idx{\sA}$
measurement at Alice's site as a measurement of $A_{\sA}$ while
the $\idx{\sB}$ measurement at Bob's site as a measurement of
$B_{\sB}$.}
\begin{equation}\label{Eq:pLHV->ELHV}
    E(A_{\sA},B_{\sB})=
    \ProbTwJ{\sA}{+}{\sB}{+} +\ProbTwJ{\sA}{-}{\sB}{-}
    -\ProbTwJ{\sA}{+}{\sB}{-} -\ProbTwJ{\sA}{-}{\sB}{+},
\end{equation}
i.e., the average value of the product of observables or
\begin{equation}\label{Eq:pLHV->ELHV2}
    E(A_{\sA},B_{\sB})=
    \ProbTwJ{\sA}{\oA}{\sB}{=\oB} -\ProbTwJ{\sA}{\oA}{\sB}{\neq\oB},
\end{equation}
which is the difference between the probability of observing
the same outcomes at the two sides and the probability of
observing different outcomes at the two sides. Thus, by adding
the two inequalities in Eq.~\eqref{Ineq:Prob:CH} with $\oA\neq
\oB$ and subtracting them from the two inequalities with
$\oA=\oB$, one arrives at the Bell-CHSH inequality in the form
of Eq.~\eqref{Ineq:CHSH:Convention}. Conversely, if there are
only two possible outcomes such that
\begin{gather}\label{Eq:Probabilities:TwoOutcome:Normalization}
    \ProbTwMA{\sA}{+}+ \ProbTwMA{\sA}{-}=1~\,\,\forall~\sA,\quad
    \ProbTwMB{\sB}{+}+ \ProbTwMB{\sB}{-}=1~\,\,\forall~\sB,
\end{gather}
then all the four Bell-CH inequalities given in
Eq.~\eqref{Ineq:Prob:CH} can also be obtained from the
Bell-CHSH inequalities via Eq.~\eqref{Eq:pLHV->ELHV} or
Eq.~\eqref{Eq:pLHV->ELHV2}. Hence, when seen as a set of
constraints imposed by LHVTs on two particles, where each of
them is subjected to two alternative dichotomic measurements,
the Bell-CH inequalities are entirely equivalent to the
Bell-CHSH inequalities~\cite{CH:PRD:1974}.

\subsection{Experimental Progress}

Since the late 1960s, many experiments have been carried out,
via the Bell-CH and Bell-CHSH inequalities, to probe the
adequacy of locally causal theories. An account of the early
attempts prior to the 1980s can be found in the excellent
review by Clauser and Shimony~\cite{J.F.Clauser:RPP:1978}.
These early results, however, were not compelling enough to
close the debate due to the various possible loopholes in
experiments~\cite{ShimonySummer2005}.

Among which, the {\em communication loophole} survived happily
till the influential experiment performed by Aspect and
coworkers in 1982 using time-varying
analyzers~\cite{Aspect:PRL:1982b}. Since then, many have
considered the impossibility of a LHVT verified, even though
some still think otherwise (see for
example~\cite{T.W.Marshall:PLA:1983,E.Santos:PRL:1991,
E.Santos:PRA:1992,E.Santos:FP:2004,E.Santos:SHPMP:2005} and
references therein). As of now, the experiment that most
convincingly evades the communication loophole was carried out
by Weihs and collaborators in 1998~\cite{Weihs:PRL:1998}. The
equally notorious {\em detection loophole} has also been closed
quite recently by Rowe and
coworkers~\cite{M.A.Rowe:Nature:2001}. A single experiment that
closes both these loopholes at once is, nevertheless, still
being sought~\cite{A.Aspect:Nature:1999,E.S.Fry:2002}. In this
regard, it is worth noting that some other loopholes such as
those considered in
Refs.~\cite{J.Barrett:PRA:2002b,A.Kent:PRA:2005} exist, but
they are generally considered less compelling. For further
information on recent Bell experiments, see the review by
Genovese~\cite{M.Genovese:PhysicsReports:2005}.

\chapter{Classical Correlations and Bell Inequalities}
\label{Chap:ClassicalCn}

In the last chapter, we have seen two important examples of
Bell inequalities that were developed in the hope of realizing
a convincing test of local causality. Bell inequalities,
nevertheless, can also be understood from a completely
different perspective. Specifically, in this chapter, we will
see that in the space of probability vectors, which we will
call the {\em space of correlations}, the tight Bell
inequalities correspond to hyperplanes that together form the
boundaries of the {\em convex set} of classical
correlations.\footnote{Although our treatment focuses (almost)
exclusively on probability vectors, it should be clear that one
can just as well consider a space of correlations that is
defined in terms of various correlation functions, as in
Eq.~\eqref{Eq:Dfn:CorrelationFn:LHV3}. In that case, a (tight)
Bell correlation inequality similarly defines a closed
halfspace where the convex set of classical correlations
resides.} Froissart is apparently the pioneer of such a
geometrical approach to Bell
inequalities~\cite{M.Froissart:NCB:1981}. Not too long after
that, this approach was discovered independently by Garg and
Mermin~\cite{A.Garg:FP:1984}. A few years later, a general
study along the same lines was also carried out by
Pitowsky~\cite{I.Pitowsky:Book:1989,I.Pitowsky:MP:1991}. A
great advantage of this geometrical approach is that it can be
easily generalized to more complicated experimental scenarios
and hence, allows more complicated Bell inequalities to be
derived in a systematic manner.

\section{Classical Correlations and Probabilities}
\label{Sec:LHVM:Example}

Before we move on to the more general scenario, let us first go
through the following example of a hypothetical Bell experiment
to gain some intuition. In particular, let us consider an
experimental scenario where the Bell-CHSH inequality, or
equivalently the two-outcome Bell-CH inequality, is applicable
(Figure~\ref{Fig:TwoPartyBellExperiment}). Now, let us imagine
that the experimental data collected (Table~\ref{tbl:LHVM}) ---
including those not explicitly shown in the table --- satisfy
the following joint probabilities
\begin{subequations}\label{Eq:JointProbabilities}
\begin{align}
    &\ProbTwJ{1}{+}{1}{+}=1,\,\,\quad \ProbTwJ{1}{+}{1}{-}=0, \quad
    \ProbTwJ{1}{-}{1}{+}=0, \quad \ProbTwJ{1}{-}{1}{-}=0,\label{Eq:Prob:A1B1}\\
    &\ProbTwJ{1}{+}{2}{+}=\frac{1}{2},\quad \ProbTwJ{1}{+}{2}{-}=\frac{1}{2}, \quad
    \ProbTwJ{1}{-}{2}{+}=0, \quad \ProbTwJ{1}{-}{2}{-}=0,\label{Eq:Prob:A1B2}\\
    &\ProbTwJ{2}{+}{1}{+}=\frac{1}{2},\quad \ProbTwJ{2}{+}{1}{-}=0, \quad
    \ProbTwJ{2}{-}{1}{+}=\frac{1}{2}, \quad \ProbTwJ{2}{-}{1}{-}=0,\label{Eq:Prob:A2B1}\\
    &\ProbTwJ{2}{+}{2}{+}=\quar,\quad \ProbTwJ{2}{+}{2}{-}=\quar, \quad
    \ProbTwJ{2}{-}{2}{+}=\quar, \quad \ProbTwJ{2}{-}{2}{-}=\quar,\label{Eq:Prob:A2B2}
\end{align}
and marginal probabilities
\begin{gather}
    \ProbTwMA{1}{+}=1,\quad \ProbTwMA{1}{-}=0,\quad
    \ProbTwMA{2}{+}=\half,\quad \ProbTwMA{2}{-}=\half,\label{Eq:Prob:A1A2}\\
    \ProbTwMB{1}{+}=1,\quad \ProbTwMB{1}{-}=0,\quad
    \ProbTwMB{2}{+}=\half,\quad \ProbTwMB{2}{-}=\half.\label{Eq:Prob:B1B2}
\end{gather}
\end{subequations}
Evidently, we can collect all the 16 joint probabilities
together and think of them as the components of a probability
vector $\bfp$ living in a 16-dimensional space. Let us now make
the following definitions in relation to such a probability
vector.
\begin{dfn}\label{Dfn:ClassicalCn}
    A probability vector is said to be classical if it can be
    generated from some local hidden-variable model.
\end{dfn}

Hereafter, we will also loosely refer to a probability vector
as a correlation. This can be justified by noting that from the
components of a probability vector, we can learn the extent to
which measurement outcomes between subsystems $\A$ and $\B$ are
correlated. For example, if $\A$ and $\B$ involved in the
experiment are totally {\em uncorrelated}, we will expect that
all the joint probabilities factorize and equal to the product
of the corresponding marginal probabilities, i.e.,
\begin{equation}\label{Eq:Probability:Uncorrelated}
    \ProbTwJ{\sA}{\oA}{\sB}{\oB} = \ProbTwMA{\sA}{\oA}\,\ProbTwMB{\sB}{\oB}.
\end{equation}
Moreover, all experimental statistics that could be of
interest, such as the correlation function for given
measurement settings, as well as other higher order moments can
be computed according to the standard procedures.

\begin{table}[!h]
\centering\rule{0pt}{4pt}\par \caption{\label{tbl:LHVM} A
hypothetical set of experimental data gathered in an experiment
to test the Bell-CHSH inequality or the Bell-CH inequality.
Here, $n$ is an index to label each run of the experiment and
$N$ is some very large number such that the data set is
statistically significant. The local measurements that may be
performed by Alice are labeled by $A_1$ and $A_2$ whereas that
for Bob are labeled by $B_1$ and $B_2$. Outcomes of the
experiments are labeled by $\pm1$ and are tabulated under the
respective local measurements that are carried out in each run
of the experiment.}
\begin{tabular}{|c||c|c||c|c|}
    \hline$n$ & $A_1$ & $A_2$ & $B_1$ & $B_2$
    \\ \hline\hline
      1 &  1 &  & 1 & \\ \hline
      2 &  1 &  &  & 1\\\hline
      3 &  & 1 & 1 & \\\hline
      4 & 1 &  & 1 & \\\hline
      5 &  & -1&  & 1\\\hline
      \vdots & \vdots & \vdots & \vdots & \vdots \\ \hline
      1000 &  1 &  &  &-1\\\hline
    \end{tabular}\hspace{1cm}
    \begin{tabular}{|c||c|c||c|c|}
    \hline$n$ & $A_1$ & $A_2$ & $B_1$ & $B_2$
    \\ \hline\hline
      1001 &   & -1 &  & -1\\\hline
      1002 & 1 & & 1 & \\\hline
      1003 &  & 1 &  & 1\\\hline
      1004 & 1 & & 1 &\\\hline
      1005 &  & -1& 1 & \\ \hline
      \vdots & \vdots & \vdots & \vdots & \vdots \\ \hline
      $N$ &  & 1 &  &-1\\ \hline
    \end{tabular}
\end{table}

Now, let us again look at the set of experimental data
presented in Table~\ref{tbl:LHVM}. If there exists a LHVM that
can reproduce this set of data, we will be able to fill in the
blanks corresponding to unperformed measurement results such
that all the joint and marginal probabilities are preserved.
Therefore, if the unfilled entries in the table can be filled
up in such a way that respects all the probabilities listed in
Eq.~\eqref{Eq:JointProbabilities}, we will have got a LHVM that
reproduces all the experimental statistics, and hence
correlations derivable from Table~\ref{tbl:LHVM}. An example of
how this can be done is shown in Table~\ref{tbl:LHVM:filled}.
In this case, we can see $n$ as an index for the local
hidden-variable $\lambda$ that is associated with each run of
the experiment. Then, in each run $n$, once the choice of local
measurement is decided, the outcome of the measurement can be
read off directly from the table (regardless of the other
entries listed in the same row of the table).

\begin{table}[!h]
\centering\rule{0pt}{4pt}\par \caption{\label{tbl:LHVM:filled}
The same set of experimental data as in Table~\ref{tbl:LHVM}
but with the {\em unperformed} measurement results (enclosed
within round brackets) filled in according to some hypothetical
LHVM. In particular, the LHVM works in such a way that the
newly filled entries in the table give rise to the same joint
and marginal probabilities as the original entries listed in
Table.~\ref{tbl:LHVM}, c.f. Eq.~\eqref{Eq:JointProbabilities}.}
\begin{tabular}{|c||c|c||c|c|}
\hline$n$ & $A_1$ & $A_2$ & $B_1$ & $B_2$
\\ \hline\hline
  1 &  1 & {\en 1} & 1 & \en{1}\\ \hline
  2 &  1 & \en{-1} & {\en 1} & 1\\\hline
  3 & {\en 1} & 1 & 1 & \en{1}\\\hline
  4 & 1 & \en{-1} & 1 & \en{-1}\\\hline
  5 & {\en 1} & -1& {\en 1} & 1\\\hline
  \vdots & \vdots & \vdots & \vdots & \vdots \\ \hline
  1000 &  1 & \en{-1} & {\en 1} &-1\\\hline
\end{tabular}\hspace{1cm}
\begin{tabular}{|c||c|c||c|c|}
\hline$n$ & $A_1$ & $A_2$ & $B_1$ & $B_2$
\\ \hline\hline
  1001 & \en{1} & -1 & {\en 1} & -1\\\hline
  1002 & 1 & {\en 1}& 1 & \en{-1}\\\hline
  1003 & {\en 1} & 1 & {\en 1} & 1\\\hline
  1004 & 1 & {\en 1}& 1 & \en{-1}\\\hline
  1005 & {\en 1} & -1& 1 & \en{1}\\ \hline
  \vdots & \vdots & \vdots & \vdots & \vdots \\ \hline
  $N$ & {\en 1} & 1 & {\en 1} &-1\\ \hline
\end{tabular}
\end{table}

That a LHVM can be constructed for the data presented in
Table.~\ref{tbl:LHVM} is not incidental. Simple calculations
using Eq.~\eqref{Ineq:CH:MoreOutcomes} and
Eq.~\eqref{Eq:JointProbabilities} show that none of the Bell-CH
inequalities is violated by the experimental data presented in
Table.~\ref{tbl:LHVM}. Evidently, no-violation of the Bell-CH
inequality is a necessary condition for the existence of a LHVM
for the given experimental data. Nevertheless, as was first
shown by Fine in 1982~\cite{A.Fine:PRL:1982}, fulfillment of
all the Bell-CH inequalities is also sufficient to guarantee
the existence of a LHVM, provided that the experimental data
only involves two dichotomic measurements performed by two
observers~\cite{A.Garg:PRLc:1982,A.Fine:PRLr:1982,A.Garg:PRD:1983}.
Hence, in an experimental scenario involving only two observers
and two dichotomic measurements per site, a complete
characterization of classical correlations can be obtained {\em
solely} using the Bell-CH inequalities. What about experiments
involving more observers, more local measurements per site, or
more outcomes per measurement? These are the questions that we
will discuss in the following sections.

\section{Geometrical Structure of the Set of Classical Correlations}

\subsection{The Spaces of Correlations}
\label{Sec:CorrelationSpace}

For the subsequent discussion, let us consider a more general
scenario whereby a source --- characterized by some physical
state $\rho$ --- distributes pairs of physical systems to Alice
and Bob, and where each of them can perform (on the subsystems
that they receive), respectively, $\mA$ and $\mB$ alternative
measurements  that would each generates $\nA$ and $\nB$
distinct outcomes.\footnote{Of course, one can be more general
than this and allows each measurement to have different number
of possible outcomes. Nevertheless, for brevity, we shall be
contented with a discussion on the case where all local
measurements performed by Alice yield the same number of
possible outcomes (likewise for Bob).} For now, we will
restrict our attention to this bipartite scenario, but most of
the following arguments can be modified easily to cater for the
multipartite scenario. In view of the forthcoming discussion,
let us also introduce the vectors
\begin{equation}
    \bfm\equiv(\mA,\mB),\quad\bfn\equiv(\nA,\nB)
\end{equation}
for, respectively, a compact description of the number of local
measurement settings and the number of possible outcomes for
each local measurement. As with the previous section, the
experimental statistics in such a scenario can be summarized as
a probability vector $\bfp\in\mathbb{R}^{d_p}$ where $d_p=\mA
\mB\nA\nB$ (if one prefers to work in the space of correlations
that is defined only in terms of {\em full} correlation
functions $E(A_{\sA},B_{\sB})$, then we will be working in a
space of dimension $\mA\mB$ ---
Sec.~\ref{Sec:BI:Correlations}). The components of the
probability vectors are the joint probabilities
$\ProbTwGJ$.\footnote{One can, instead, work in a space of
probabilities with dimension $d>d_p$ such that each probability
vector $\bfp$ also has the marginal probabilities $\ProbTwGMA$
and $\ProbTwGMB$ as components. However, this is not necessary,
as the marginal probabilities are not independent from the
joint probabilities.} We will refer to this real vector space
as the {\em space of correlations}, denoted by $\Cp$. Clearly,
for our purpose, not all of the $d_p$ coordinates in $\Cp$ are
independent. For instance, given a particular choice of Alice's
and Bob's measurement, there must be an outcome at Alice's as
well as Bob's site.\footnote{For the purpose of present
discussion, one could treat the possibility of no-detection as
one of the possible outcomes.} Normalization of probability
therefore requires:
\begin{equation}\label{Eq:Probabilities:General:Normalization}
    \sum_{\oA=1}^{\nA}\sum_{\oB=1}^{\nB}
    \ProbTwGJ=1\quad\forall~\sA, \sB,
\end{equation}
where we have labeled the outcomes registered at Alice's site
as $\oA=1,2,\ldots,\nA$ (likewise $\oB=1,2,\ldots,\nB$ at Bob's
site). Moreover, adhering to the principles of relativity, we
shall be contented with correlations that do not allow
faster-than-light signaling. These are correlations that
respect the following equalities~\cite{S.Popescu:FP:1994}:
\begin{equation}\label{Eq:Dfn:No-signaling}
    \sum_{\oA=1}^{\nA} \ProbTwGJ=
    \ProbTwGMB\quad\text{and}\quad
    \sum_{\oB=1}^{\nB} \ProbTwGJ=
   \ProbTwGMA
    \quad\forall~\sA, \sB.
\end{equation}
In words, this means that the marginal probability of Alice
observing local measurement outcome $\oA$, conditioned on her
measuring $\sA$, i.e., $\ProbTwGMA$ is independent of the
choice of measurement $\sB$ made by the spatially separated
observer Bob; likewise for $\ProbTwGMB$. This is now commonly
known as the {\em no-signaling condition} (see, for example,
Refs.~\cite{R.F.Werner:QIC:2001,V.Scarani:PLA:2002}), which was
originally termed the {\em relativistic causality condition} in
Ref.~\cite{S.Popescu:FP:1994}.

By a simple counting argument, one can show that after taking into
account all of these constraints, there are effectively
only~\cite{D.Collins:JPA:2004}
\begin{equation}\label{Eq:Dfn:EffectiveDimension}
    d_p'=\mA \mB (\nA-1)(\nB-1) + \mA (\nA-1)+\mB(\nB-1)
\end{equation}
independent entries in the probability vector, which can be
taken to be all but one of the marginal probabilities
$\ProbTwGMA$ for each $\sA$, likewise for $\ProbTwGMB$, plus
$(\nA-1)(\nB-1)$ of the joint probabilities $\ProbTwGJ$ for
each combination of $\sA$ and $\sB$. Hence, we are essentially
only interested in a subspace of the set of probability vectors
that is of dimension $d_p'$.

\subsection{The Convex Set of Classical Correlations}
\label{Sec:Convexity:ClassicalCn}

Now, let us take a closer look at the set of classical
correlations associated with the experimental scenario
described above. Hereafter, we will denote this set by
$\polycl$ (analogously, we will denote the set of classical
correlations defined in the space of correlation functions as
$\polyclc$). From Eq.~\eqref{Eq:Prob:AverageOverLambda} and
Eq.~\eqref{Eq:Probability:Factorizes}, it follows that a
classical probability vector is one whose entries satisfy
\begin{equation}\label{Eq:Dfn:ClassicalProbVec}
    \ProbTwGJ =
    \int_\Lambda\dd\lambda~\rho_\lambda~
    \ProbTwMA{\sA,\lambda}{\oA}\,\ProbTwMB{\sB,\lambda}{\oB}
\end{equation}
for some choice of $\ProbTwMA{\sA,\lambda}{\oA}$ and
$\ProbTwMB{\sB,\lambda}{\oB}$, and some probability density
$\rho_\lambda$. For any two classical probability vectors
$\pLHV$ and $\pLHV'$, any convex combination of them gives rise
to a probability vector
\begin{equation}
    \bfp''\equiv q\,\pLHV + (1-q)\,\pLHV',
\end{equation}
that is also classical. This is because the resulting
probability vector $\bfp''$ can be realized via a LHVM which
consists of implementing the LHVM associated with $\pLHV$ and
$\pLHV'$ stochastically. Specifically, by tossing a biased coin
with probability $q$ of getting heads and probability $1-q$ of
getting tails, the probability vector $\bfp''$ can be realized
by implementing the LHVM associated with $\pLHV$ whenever the
outcome of the toss is heads, and the LHVM associated with
$\pLHV'$ whenever the outcome of the toss is tails. Therefore,
the set of classical probability vectors $\polycl$ is {\em
convex}.

A natural question that follows is: what are the extreme
points\footnote{An extreme point of a convex set is a point in
the set which cannot be expressed as a nontrivial convex
combination of two or more different points in the
set~\cite{B.Grunbaum:Book:2003,G.M.Ziegler:Book:1995}.} of this
set? With some thought, it is not difficult to see that
probability vectors such that the joint probability factorizes,
i.e.,
\begin{subequations}\label{Eq:Dfn:ExtremePoints:ClassicalCorrelations}
\begin{equation}
    \ProbTwJ{\sA}{\oA}{\sB}{\oB} = \ProbTwMA{\sA}{\oA}\,\ProbTwMB{\sB}{\oB},
\end{equation}
and for which the marginal probabilities are either 0 or 1,
i.e.,
\begin{equation}
    \ProbTwMA{\sA}{\oA}=0,1,\qquad \ProbTwMB{\sB}{\oB}=0,1,
\end{equation}
\end{subequations}
are extreme points of $\polycl$~\cite{D.Collins:JPA:2004}.
These are probability vectors corresponding to deterministic
LHVMs. Physically, each of these probability vectors
corresponds to a scenario where the experimental outcomes for
given local measurement settings are deterministic; once the
local measurement setting is chosen, one and only one of the
local detectors will ever click.\footnote{In the context of
$\polyclc$ and where measurement outcomes are bounded between
$1$ and $-1$ the extreme points correspond to those whereby
$E(A_{\sA},B_{\sB})=\oA(A_{\sA})~\oB(B_{\sB})=\pm1$.}
Conversely, it is also not difficult to see from
Eq.~\eqref{Eq:Dfn:ClassicalProbVec} and
Eq.~\eqref{Eq:Dfn:ExtremePoints:ClassicalCorrelations} that
{\em any} other classical probability vectors can be written as
a nontrivial convex combination of these extremal probability
vectors. In other words, a probability vector is an extreme
point of $\polycl$ if and only if it satisfies
Eq.~\eqref{Eq:Dfn:ExtremePoints:ClassicalCorrelations}.

\begin{figure}[h!]
\centering\rule{0pt}{4pt}\par
\scalebox{1}{\includegraphics{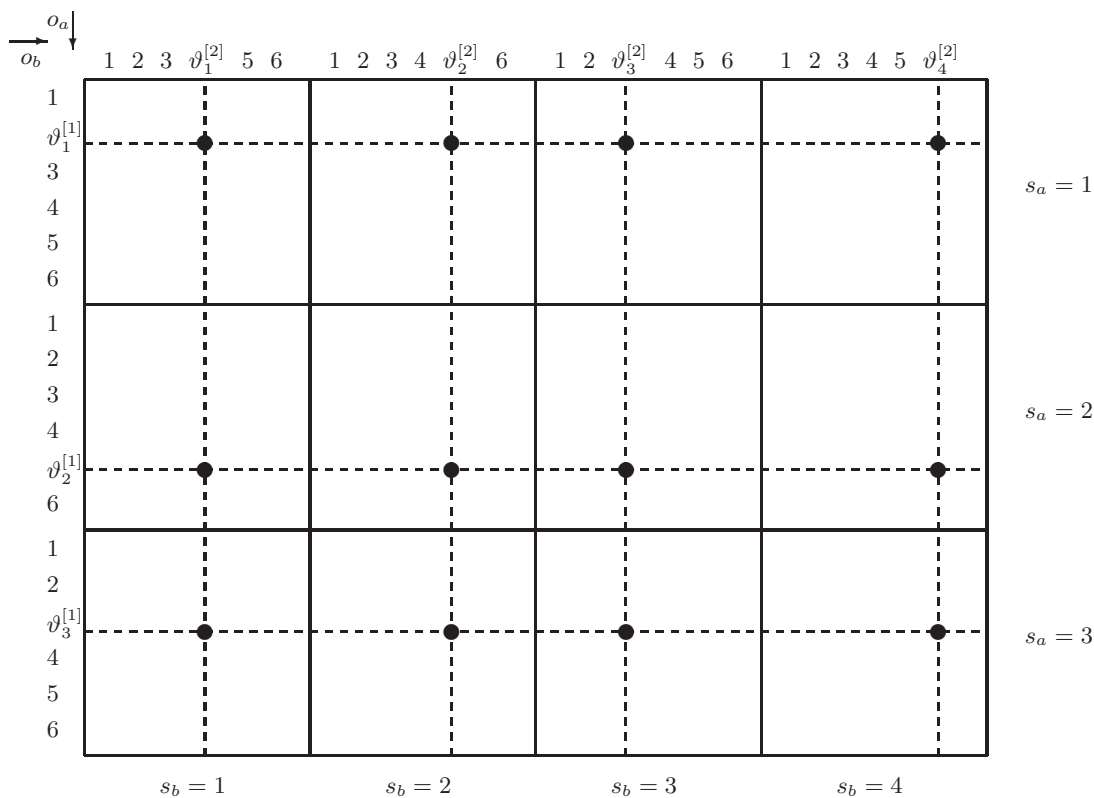}}
\caption{\label{Fig:ExtremalClassicalStrategy}
Schematic representation of the LHVM corresponding to a
particular extreme point of $\PLHV{3}{6}{4}{6}$, denoted by
${}^{\bfa,\bfb}\mathbf{B}_{\A\B}$, where
$\bfa\equiv\left(\vartheta^{[1]}_1,\vartheta^{[1]}_2,\vartheta^{[1]}_3\right)=(2,5,3)$
and $\bfb\equiv\left(\vartheta^{[2]}_1,\vartheta^{[2]}_2,
\vartheta^{[2]}_3,\vartheta^{[2]}_4\right)=(4,5,3,6)$ (adapted
from Figure 1 of Ref.~\cite{A.Peres:FP:1999}). Each row (column),
separated from each other by solid horizontal (vertical)
lines, corresponds to a choice of measurement $s_a$ ($s_b$) for
Alice (Bob). The intersection of a row and a column gives rise
to a sector, which corresponds to particular choice of Alice's
and Bob's measurement. For each extremal LHVM, the outcome of
measurements solely depends on the choice of local measurement.
Hence, once a row (column) is chosen, the measurement outcome
is also determined, and is indicated by a dashed horizontal
(vertical) line. For example, Alice will always observe the
second outcome ($o_a=2$) whenever she chooses to perform the
first measurement ($s_a=1$), regardless of Bob's choice of
measurement.
}
\end{figure}

Given that the physical scenario corresponding to an extreme
point of $\polycl$ is such that the local measurement settings
determine the local measurement outcome with certainty, we
might as well label each of these extreme points by two sets of
indices $\bfa$ and $\bfb$ that are, respectively, associated
with the measurement outcomes observed by Alice and
Bob~\cite{B.M.Terhal:PRL:2003}
(Figure~\ref{Fig:ExtremalClassicalStrategy}). Specifically, let
us denote by ${}^{\bfa,\bfb}\mathbf{B}_{\A\B}$ an extreme point
of $\polycl$, $\vartheta^{[1]}_{\sA}$ Alice's measurement
outcome conditioned on her measuring $A_{\sA}$ and
$\vartheta^{[2]}_{\sB}$ Bob's measurement outcome conditioned
on him measuring $B_{\sB}$. Then the two sets of indices
$\mathbf{a}=(\vartheta^{[1]}_1,\vartheta^{[1]}_2,\ldots,\vartheta^{[1]}_{\mA})$
where $\vartheta^{[1]}_{\sA}=1,2,\ldots,\nA$ and
$\mathbf{b}=(\vartheta^{[2]}_1,\vartheta^{[2]}_2,\ldots,\vartheta^{[2]}_{\mB})$
where $\vartheta^{[2]}_{\sB}=1,2,\ldots,\nB$ will completely
characterize ${}^{\bfa,\bfb}\mathbf{B}_{\A\B}$ in the sense
that its component reads~\cite{B.M.Terhal:PRL:2003}
\begin{equation}\label{Eq:VertPLHV:Boolean}
    {}^{\bfa,\bfb}\mathbf{B}^{\oAB}_{\A\B}
    (\sA,\sB)=\delta_{\oA \vartheta^{[1]}_{\sA}}\delta_{\oB \vartheta^{[2]}_{\sB}}.
\end{equation}

For finite number of local measurement settings and measurement
outcomes, i.e.,
\begin{equation*}
    \mA, \mB, \nA, \nB <\infty,
\end{equation*}
it is possible to enumerate all of these extreme points by
going through all legitimate boolean values of the local
probabilities. In total, there are thus
\begin{equation}
    n_v=\nA^{\mA}\nB^{\mB}
\end{equation}
extremal classical probability vectors, corresponding to $n_v$
extremal deterministic LHVMs. Hereafter, we will also refer to
the extreme points of a convex polytope $\P$ as its {\em
vertices}, denoted as $\ver{\P}$. The fact that there are only
a finite number of extreme points in the (convex) set of
classical correlations immediately implies that $\polycl$ is a
{\em convex polytope}~\cite{B.Grunbaum:Book:2003,
G.M.Ziegler:Book:1995}, which was first called the {\em
correlation polytope} by Pitowsky~\cite{I.Pitowsky:MP:1991}.
Notice that the dimension of the correlation polytope, i.e.,
the dimension of its affine hull\footnote{An affine combination
of a set of points $\mathcal{X}=\{x_1, x_2,\ldots,x_n\}$ is a
linear combination of $x_k$, i.e., $\sum_k q_k\,x_k$ such that
$q_k=1$. The affine hull of $\mathcal{X}$ is the union of all
affine combinations of $\mathcal{X}$.} is $d_p'$.

\subsection{Correlation Polytope and Bell Inequalities}
\label{Sec:Polytope&BI}

A well-established fact about a convex polytope is that it can
equivalently be represented by the intersection of a finite
family of closed
halfspaces~\cite{B.Grunbaum:Book:2003,G.M.Ziegler:Book:1995}.
As is well-known, a closed halfspace in $\mathbb{R}^{d_p}$ can
be represented by an inequality that is {\em linear} in the
$d_p$ coordinates. Let us denote by $I^{(k)}_{\bfm;\bfn}$ the
inequality that is associated with the ``$k$"-th halfspace
``bounding" the polytope $\polycl$, i.e.,
\begin{equation}\label{Ineq:Halfspace}
    I^{(k)}_{\bfm;\bfn}:~ \mathbf{F}^{(k)}\cdot\bfp\le \bLHV{k},
\end{equation}
then the {\em boundary} associated with this halfspace is the
hyperplane
\begin{equation}\label{Eq:Hyperplane}
    \SLHV^{(k)}(\bfm;\bfn;\bfp)\equiv\mathbf{F}^{(k)}\cdot\bfp=\bLHV{k},
\end{equation}
where $\bfp\in\mathbb{R}^{d_p}$ is an arbitrary vector in the space
of correlations, $\bfF^{(k)}$ is a vector defining the ``direction"
of the hyperplane involved, $\bfF^{(k)}\cdot\bfp$ represents the
Euclidean inner product between the two vectors, and $\bLHV{k}$ is
some constant related to the offset of the hyperplane from the
origin.

By definition, a classical probability vector $\pLHV$ is a
member of $\polycl$ and hence must satisfy
inequality~\eqref{Ineq:Halfspace}, i.e.,
\begin{equation}\label{Ineq:Halfspace:LHV}
    \mathbf{F}^{(k)}\cdot\pLHV\le \bLHV{k}.
\end{equation}
The inequality \eqref{Ineq:Halfspace} is therefore a valid
constraint that has to be satisfied by all classical
probability vectors. In other words, it is a {\em Bell
inequality}. It is straightforward to see that any conic
combination\footnote{A conic combination of $n$ points is a
non-negative linear combination of the $n$ points.} of such
inequalities will also give rise to another inequality that has
to be satisfied by all $\pLHV$. There is thus no unique family
of inequalities defining a given correlation polytope
$\polycl$. In principle, one can even write down an infinite
family of Bell inequalities, each associated with a closed
halfspace (and hence hyperplane), which has to be satisfied by
all $\pLHV$. In this regard, it is worth noting that the
smallest family of such closed halfspaces consists of those
whose {\em boundaries} are the affine hull of the
facets\footnote{The intersection of a polytope with a
supporting hyperplane gives rise to a face of the polytope. If
the dimension of a polytope is $d$, then a face of dimension
$d-1$ is known as a facet of the polytope.} of $\polycl$ (pp
31, \cite{B.Grunbaum:Book:2003}). In other words, Bell
inequalities that are associated with this smallest family of
halfspaces are characterized by $\mathbf{F}^{k}$ and $\bLHV{k}$
such that the solution set
$\left\{\bfp^{(k)}\right\}\subset\polycl$ to each of the
corresponding equalities
\begin{equation}
    \mathbf{F}^{(k)}\cdot\pLHV^{(k)}=\bLHV{k},
\end{equation}
is nonempty and whose affine dimension equals $d_p'-1$. For
definiteness, we will refer to them as {\em tight} Bell
inequalities~\cite{Ll.Masanes:QIC:2003,D.Collins:JPA:2004}, or
equivalently facet-inducing Bell
inequalities~\cite{D.Avis:JPA:2005}. It is worth noting that
the coefficients associated with these tight Bell inequalities,
i.e., $\bfF^{(k)}$, when properly normalized, also define a
convex polytope that is {\em dual} to the correlation polytope.
Moreover, a probability vector $\bfp$ is classical if and only
if it satisfies this minimal set of Bell inequalities defining
$\polycl$.\footnote{See, for example Definition 2.10 and
Theorem 2.15 of Ref.~\cite{G.M.Ziegler:Book:1995}.}

For the convenience of subsequent discussion, let us note that
the {\em linearity} of inequality \eqref{Ineq:Halfspace} also
allows us to write the functional form of a generic Bell
inequality, c.f. Eq.~\eqref{Eq:Hyperplane}, in the
following tensorial form
\begin{equation}\label{Eq:SLHV:Generic}
    \SLHV(\bfm; \bfn; \bfp)=\sum_{\sA=0}^{\mA}\sum_{\sB=0}^{\mB}
    {\sum_{\oA=1}^{\nA}}'{\sum_{\oB=1}^{\nB}}' \CoeffG\ProbG +b_{0,0},
\end{equation}
where
\begin{equation}\label{Eq:Prob:Convention}
    \ProbG\equiv\left\{ \begin{array}{c@{\quad:\quad}l}
        \ProbTwGJ  & \sA>0, \sB>0,\\
        \ProbTwGMA & \sA>0, \sB=0,\\
        \ProbTwGMB & \sA=0, \sB>0,
        \end{array} \right.
\end{equation}
is a component of the probability vector $\bfp$ and $\CoeffG$
is the corresponding component of the vector of coefficients
$\bfF$. Notice that the sums over outcomes are restricted in
that when $\sA=0$, there is no sum over $\oA$ and when $\sB=0$,
there is no sum over $\oB$; in these special cases, we shall
write
\begin{equation}
    \CoeffG\equiv\left\{ \begin{array}{c@{\quad:\quad}l}
        b_{\sA0}^{\oA} & \sA>0, \sB=0,\\
        b_{0\sB}^{\oB} & \sA=0, \sB>0.\\
        \end{array} \right.
\end{equation}
We can then write these coefficients in a compact manner via
the following matrix
\begin{equation}\label{Eq:Dfn:b}
    b:\sim\left(
    \begin{array}{c||c|c|c|c}
    b_{0,0}    & \bfb_{0,1} & \bfb_{0,2} & \cdots & \bfb_{0,\mB}\\ \hline\hline
    \bfb_{1,0} & \BBlk{1}{1} &  \BBlk{1}{2}
        & \cdots & \BBlk{1}{\mB} \\ \hline
    \bfb_{2,0} & \BBlk{2}{1} &  \BBlk{2}{2}
        & \cdots & \BBlk{2}{\mB} \\ \hline
    \vdots & \vdots &  \vdots & \vdots & \vdots \\ \hline
    \bfb_{\mA,0} & \BBlk{\mA}{1} &  \BBlk{\mA}{2}
        & \cdots & \BBlk{\mA}{\mB}\\
    \end{array}
    \right),
\end{equation}
where each of the boldfaced entries in the above matrix is a
block matrix of appropriate dimension. For example,
$\bfb_{1,1}$ in the above matrix is the following
$(\nA-1)\times(\nB-1)$ matrix
\begin{equation}\label{Eq:Dfn:BlockMatrix:b}
    \bfb_{1,1}\equiv\left(
    \begin{array}{cccc}
    \Coeff{1}{1}{1}{1} &  \Coeff{1}{1}{1}{2} & \cdots & \Coeff{1}{1}{1}{\nB} \\
    \Coeff{1}{2}{1}{1} &  \Coeff{1}{2}{1}{2} & \cdots & \Coeff{1}{2}{1}{\nB} \\
    \vdots & \vdots &  \vdots & \vdots \\
    \Coeff{\quad1}{\nA-1}{1}{\,1} &  \Coeff{\quad1}{\nA-1\,}{1}{2} &
    \cdots & \Coeff{\quad1}{\nA-1\,}{1}{\nB-1}\\
    \end{array}
    \right),
\end{equation}
whereas $\bfb_{1,0}$ and $\bfb_{0,1}$ are, respectively, column
vector and row vector of length $\nA-1$ and
$\nB-1$.\footnote{The length of these vectors as well as the
dimension of each block matrix can be traced back to the
discussion around Eq.~\eqref{Eq:Dfn:No-signaling}.
Consequently, for a two-outcome Bell inequality (e.g. the
Bell-CH inequality), or a Bell correlation inequality, we will
collapse each block matrix and write it as a single number.} It
is then expedient to write a Bell inequality explicitly as
\begin{equation}\label{Ineq:BI:Generic}
    I^{(k)}_{\bfm;\bfn}: \SLHV^{(k)}(\bfm;\bfn;\bfp)\le\bLHV{k},
\end{equation}
but compactly as
\begin{equation}\label{Ineq:BI:Generic2}
    I^{(k)}_{\bfm;\bfn}: b^{(k)}\le\bLHV{k},
\end{equation}
where $b^{(k)}$ is the corresponding matrix of coefficients,
Eq.~\eqref{Eq:Dfn:b} -- Eq.~\eqref{Eq:Dfn:BlockMatrix:b}, associated
with the specific Bell inequality.

As an example, let us look at the simplest nontrivial scenario
where $\bfm=(2,2)$ and $\bfn=(2,2)$. In this case, it is known
for a long time~\cite{A.Fine:PRL:1982,I.Pitowsky:Book:1989,
I.Pitowsky:PRA:2001,D.Collins:JPA:2004} that the only class of
nontrivial\footnote{The other tight Bell inequalities are
trivial in the sense that they either require probabilities to
be non-negative or not larger than unity.} tight Bell
inequalities are the Bell-CH inequalities listed
in~\eqref{Ineq:Prob:CH} and their equivalents. In this case, we
have\footnote{Hereafter, we will drop the arguments of $\SLHV$
for brevity of notation.}
\begin{equation}\label{Ineq:CH:Functional}
    \SLHVBI{CH}=\ProbTwJ{1}{\oA}{1}{\oB}+\ProbTwJ{1}{\oA}{2}{\oB}
    + \ProbTwJ{2}{\oA}{1}{\oB}-\ProbTwJ{2}{\oA}{2}{\oB}
    -\ProbTwMA{1}{\oA} -\ProbTwMB{1}{\oB}\le 0.
\end{equation}
Making use of the matrix representation introduced above, we
will write this class of inequalities
as~\cite{D.Collins:JPA:2004}
\begin{equation}\label{Ineq:CH:Matrix}
    I^{\mbox{\tiny (CH)}}_{(2,2);(2,2)}:~
    \left(
    \begin{array}{c||c|c}
    \cdot & -1 & \cdot\\ \hline\hline
    -1  &  1 & 1\\ \hline
     \cdot  & 1 & -1
    \end{array}
    \right)\le 0,
\end{equation}
where for ease of reading, we will always replace each null
entry in a matrix by a single dot.

A great advantage of this matrix representation is that a Bell
inequality that only differs from another in its label of {\em
measurement settings} can be obtained from (the matrix
representation of) the other by applying an appropriate
permutation to the rows and/or columns of blocks in the
associated matrix of coefficients, c.f. Eq.~\eqref{Eq:Dfn:b}.
Similarly, two Bell inequalities that differ from another only
in their label of {\em measurement outcomes} for a particular
local measurement setting can be obtained from one another by
applying appropriate permutation to the rows and/or columns
within the entire row/column of {\em blocks of} matrix of
coefficients, c.f. Eq.~\eqref{Eq:Dfn:BlockMatrix:b}. And
finally, two Bell inequalities that only differ in their label
of observers, e.g. ``Alice"$\leftrightarrow$``Bob", can be
obtained from one another by transposing their respective
matrix of coefficients (see Appendix~\ref{App:Sec:CGLMP} for
examples). With this compact notation, the stage is now set for
us to look into Bell inequalities that arise in the more
complicated experimental scenarios.

\section{The Zoo of Bell Inequalities} \label{Sec:Zoo:BI}

To date, a zoo of Bell inequalities is available in the
literature. In particular, a handful of these were constructed
in the 1980s~\cite{N.D.Mermin:PRD:1980,A.Garg:PRL:1982,
A.Garg:PRL:1982b, N.D.Mermin:FOP:1982,A.Garg:PRD:1983,
A.Garg:FP:1984,N.D.Mermin:1986} primarily to investigate if
Bell inequality violation would vanish in one of the plausible
classical limits. Some of these early attempts, however,
suffered by their rather {\em ad hoc} construction of
(non-tight) Bell inequalities. In what follows, we will review,
via the characterization of various classical correlation
polytopes $\polycl$, some of the more well-known (tight) Bell
inequalities beyond Bell-CH and Bell-CHSH that can be, or have
been constructed using the geometrical approach presented
above.

For bipartite Bell inequalities, that is, Bell inequalities
involving only two parties, our discussion will be carried out
primarily for Bell inequalities for probabilities, as this is
where most of the work was
done~\cite{I.Pitowsky:PRA:2001,D.Collins:JPA:2004,T.Ito:PRA:2006}.
On the contrary, most of the work for multipartite Bell
inequalities were carried out in the context of Bell
correlation inequalities, in particular those involving only
the full correlation functions.\footnote{A full correlation
function, as opposed to a restricted correlation function, for
an $N$-party Bell experiment is a correlation function that
takes the local observables at all the $N$ sites as
arguments~\cite{R.F.Werner:PRA:2001} (see the discussion at
pp.~\pageref{Eq:Dfn:MarginalCorrelationFn} for more details).}

\subsection{Other Bipartite Bell Inequalities for Probabilities}
\label{Sec:BipartiteBI}

\subsubsection{Two Outcomes $\bfn=(2,2)$}

Now, let us focus on bipartite Bell inequalities for
probabilities involving only dichotomic observables, i.e.,
$\bfn=(2,2)$. For scenarios involving more than two
measurements on one side, but not on the other, i.e.,
$\bfm=(2,m)$ or $\bfm=(m,2)$ with $m>2$, Collins and Gisin have
shown in Ref.~\cite{D.Collins:JPA:2004} that there are no new
tight Bell inequalities. In other words, all facets of the
correlation polytope $\PLHV{2}{2}{m}{2}$ (equivalently
$\PLHV{m}{2}{2}{2}$) either correspond to the trivial
requirement of probabilities being positive, or to a Bell-CH
type inequality involving only two out of the $m$ possible
measurements. An example of such an inequality would
be~\cite{D.Collins:JPA:2004}
\begin{equation}
    \ProbTwJ{1}{\oA}{1}{\oB}+\ProbTwJ{1}{\oA}{m}{\oB}
    + \ProbTwJ{2}{\oA}{1}{\oB}-\ProbTwJ{2}{\oA}{m}{\oB}
    -\ProbTwMA{1}{\oA} -\ProbTwMB{1}{\oB}\le 0,
\end{equation}
in which case only statistics of Bob's first and $\idx{m}$
local measurement are involved in the above inequality.

In the case when each party is allowed to perform three
alternative measurements, i.e., for the correlation polytope
$\PLHV{3}{2}{3}{2}$, a complete list of 684 facets was first
obtained by Pitowsky and Svozil in
Ref.~\cite{I.Pitowsky:PRA:2001}. Among these, Collins and
Gisin~\cite{D.Collins:JPA:2004} have found that there are 36
positive probability facets, 72 Bell-CH-type facets while the
remaining facets are associated with inequalities that are
equivalent to\footnote{The analogous analysis for Bell
correlation inequalities with $\bfm=(3,3)$ has also been
carried out independently by \'Sliwa~\cite{C.Sliwa:PLA:2003}
(see also~\cite{M.Froissart:NCB:1981}).}
\begin{equation}\label{Ineq:Prob:I3322}
    I^{(1)}_{(3,3);(2,2)}:~
    \left(
    \begin{array}{r||r|r|r}
    \cdot & -2 & -1 & \cdot\\ \hline\hline
    -1 &  1 &  1 & 1\\ \hline
     \cdot &  1 &  1 &-1\\ \hline
     \cdot &  1 & -1 & \cdot
    \end{array}
    \right)\le 0.
\end{equation}
Equivalently, in the notation of
Eq.~\eqref{Eq:Prob:Convention}, inequality
$I^{(1)}_{(3,3);(2,2)}$ can be written more explicitly as:
\begin{align}
    \SLHV^{(I_{3322})}=~&\ProbTwJ{1}{\oA}{1}{\oB} + \ProbTwJ{1}{\oA}{2}{\oB} + \ProbTwJ{1}{\oA}{3}{\oB}
    + \ProbTwJ{2}{\oA}{1}{\oB} + \ProbTwJ{2}{\oA}{2}{\oB} - \ProbTwJ{2}{\oA}{3}{\oB}\nonumber\\
    +~&\ProbTwJ{3}{\oA}{1}{\oB} - \ProbTwJ{3}{\oA}{2}{\oB} - \ProbTwMA{1}{\oA}
    - 2\ProbTwMB{1}{\oB} -\ProbTwMB{2}{\oB}\le0,\label{Ineq:Functional:I3322}
\end{align}
which is understood to hold true for arbitrary but fixed choice
of $\oA$ and $\oB$.

Here, we again see that a lower dimensional Bell inequality,
namely, the Bell-CH inequality occurring as a facet of a more
complicated correlation polytope. As was shown by
Pironio~\cite{S.Pironio:JMP:2005}, this is actually a generic
feature of tight Bell inequalities for
probabilities\footnote{See Avis\etal~\cite{D.Avis:JPA:2006} for
the analogous proof for Bell correlation inequalities.}, i.e.,
when lifted to a more complicated experimental scenario, say,
involving more local measurement settings and/or outcomes
and/or number of parties, the lower dimensional Bell inequality
will still serve as a tight Bell inequality in the higher
dimensional space. Since a direct enumeration of all tight Bell
inequalities is computationally intensive and may not be
feasible in practice, this property of tight Bell inequalities
will enable us to find out, at least, a partial list of facets
in the higher dimensional correlation
polytope~\cite{S.Pironio:JMP:2005}.

For example, for the correlation polytope $\PLHV{4}{2}{4}{2}$,
even though a complete characterization of tight Bell
inequalities for probabilities is not known, we do know from
Pironio's result~\cite{S.Pironio:JMP:2005} that all the tight
inequalities derived from, say, $\PLHV{3}{2}{4}{2}$ will also
serve as tight inequalities in the higher-dimensional space.
This lower dimensional case has been fully characterized in
Ref.~\cite{D.Collins:JPA:2004} and the correlation polytope
$\PLHV{3}{2}{4}{2}$ is known to made up of from five different
classes of facets.

For $\PLHV{4}{2}{4}{2}$, however, it is known that there are
also other classes of tight Bell inequalities. For example,
Collins and Gisin~\cite{D.Collins:JPA:2004} have shown that a
generalization of $I^{(1)}_{(3,3);(2,2)}$, namely,
\begin{equation}\label{Ineq:Prob:I4422}
    I^{(3)}_{(4,4);(2,2)}:~
    \left(
    \begin{array}{r||r|r|r|r}
    \cdot   & -3 & -2 &-1 & \cdot \\ \hline\hline
    -1 &  1 &  1 & 1 & 1 \\ \hline
     \cdot &  1 &  1 & 1 &-1 \\ \hline
     \cdot &  1 &  1 &-1 & \cdot \\ \hline
     \cdot &  1 & -1 & \cdot & \cdot
    \end{array}
    \right)\le 0,
\end{equation}
is also a tight Bell inequality. By brute force,
Ito\etal~\cite{T.Ito:PRA:2006} have found two other tight Bell
inequalities for this experimental scenario:
\begin{equation}\label{Ineq:Prob:I4422:Ito}
    I^{(1)}_{(4,4);(2,2)}:~
    \left(
    \begin{array}{r||r|r|r|r}
    \cdot   &  \cdot & -1 &-1 &-1 \\ \hline\hline
    -1 & -1 &  1 & \cdot & 2 \\ \hline
     \cdot &  \cdot &  1 &-1 &-1 \\ \hline
    -1 &  1 & -1 & 1 & 1 \\ \hline
    -1 & -1 &  1 & 2 &-1
    \end{array}
    \right)\le 0,\quad
    I^{(2)}_{(4,4);(2,2)}:~
    \left(
    \begin{array}{r||r|r|r|r}
    \cdot   &  \cdot &  \cdot &-1 &-1 \\ \hline\hline
    \cdot &  1 &  1 & 1 & \cdot \\ \hline
    -1 &  1 & -1 & \cdot & 1 \\ \hline
    -1 & -1 &  1 & 1 & 1 \\ \hline
    \cdot &  \cdot & -1 & 1 & \cdot
    \end{array}
    \right)\le 0,
\end{equation}
and by the method of {\em triangular
elimination}~\cite{D.Avis:JPA:2005}, they have also found at
least one other tight Bell inequality:
\begin{equation}\label{Ineq:Prob:A5}
    I^{(4)}_{(4,4);(2,2)}:~
    \left(
    \begin{array}{r||r|r|r|r}
    \cdot & \cdot & -1 &-1 &-1 \\ \hline\hline
    -2    & -1    &  1 & \cdot & 2 \\ \hline
    -1    & \cdot &  1 &-1 &-1 \\ \hline
    -1    &  1    & -1 & 1 & 1 \\ \hline
    \cdot & -1    &  1 & 2 &-1
    \end{array}
    \right)\le 0.
\end{equation}
which they have labeled as ``A5". Very recently, a partial list
of 26 inequivalent facet-inducing inequalities for
$\PLHV{4}{2}{4}{2}$ was presented in
Ref.~\cite{N.Brunner:quant-ph/0711.3362}.

Beyond this, a systematic characterization of all the tight
Bell inequalities with more local measurements seems
formidable. However, we do know that both
$I^{(1)}_{(3,3);(2,2)}$ and $I^{(3)}_{(4,4);(2,2)}$ are members
of a broader class of Bell inequalities, called $I_{mm22}$ by
Collins and Gisin~\cite{D.Collins:JPA:2004}. It is worth noting
that this class of inequalities is {\em asymmetric} with
respect to swapping Alice and Bob. In particular, for
$\bfm=(m,m)$, the inequality admits the following compact
representation~\cite{D.Collins:JPA:2004,D.Avis:JPA:2005}:\footnote{Note
that for the specific case of $m=3$ and $m=4$,
Eq.~\eqref{Ineq:Prob:Imm22} is related to, respectively,
Eq.~\eqref{Ineq:Prob:I3322} and Eq.~\eqref{Ineq:Prob:I4422} by
a transposition, which corresponds to swapping the label
``Alice"~$\leftrightarrow$~``Bob". The current form of
Eq.~\eqref{Ineq:Prob:Imm22}, as opposed to
Eq.~\eqref{Ineq:Prob:I3322} and Eq.~\eqref{Ineq:Prob:I4422},
looks closer to the original form presented in
Refs.~\cite{D.Collins:JPA:2004,D.Avis:JPA:2005}.}
\begin{equation}\label{Ineq:Prob:Imm22}
    I_{mm22}:~
    \left(
    \begin{array}{r||r|r|r|r|r|r|r|r}
     \cdot & -1 & \cdot & \cdot & \hdots &  \cdot & \cdot & \cdot & \cdot \\ \hline\hline
    -(m-1) &  1 &  1 & 1 & \hdots &  1 & 1 & 1 & 1\\  \hline
    -(m-2) &  1 &  1 & 1 & \hdots &  1 & 1 & 1 & -1 \\  \hline
    -(m-3) &  1 &  1 & 1 & \hdots &  1 & 1 &-1 & \cdot \\  \hline
    -(m-4) &  1 &  1 & 1 & \hdots &  1 & -1 & \cdot  & \cdot \\  \hline
    \vdots & \vdots &  \adots & \adots & \adots & \adots & \adots & \vdots & \vdots\\ \hline
    \vdots & \vdots & \adots & \adots & \adots & \adots & \vdots & \vdots & \vdots\\ \hline
        -1 &  1 &  1 &-1 & \cdot & \hdots & \cdot & \cdot & \cdot\\  \hline
         \cdot &  1 & -1 & \cdot & \cdot & \hdots & \cdot & \cdot & \cdot
    \end{array}
    \right)\le 0.
\end{equation}
For $m\le7$, Collins and Gisin computationally verified that
each $I_{mm22}$ is a tight Bell inequality, and for general
$m$, the tightness of these inequalities has also been proven
very recently by Avis and Ito~\cite{D.Avis:DAP:2007}. Apart
from this, Avis\etal~\cite{D.Avis:JPA:2005} have also obtained
a huge number of tight Bell inequalities by applying the method
of triangular elimination to a list of tight inequalities for
the so-called {\em cut polytope}. On top of inequality
\eqref{Ineq:Prob:A5}, the explicit form of some of these
inequalities with $\mA, \mB\le5$ can also be found in
Ref.~\cite{T.Ito:PRA:2006,T.Ito:url:BI}.\\

\subsubsection{More than Two Outcomes}
\label{Sec:BI:>2outcome}

The set of classical correlations involving greater number of
measurement outcomes is apparently not as well known. In
particular, investigation carried out by Collins and
Gisin~\cite{D.Collins:JPA:2004} suggests that for $2<n\le 5$,
all facets of the correlation polytope $\PLHV{2}{2}{2}{n}$
(equivalently $\PLHV{2}{n}{2}{2}$) are either of the
Bell-CH-type or the trivial type that requires non-negativity
of probabilities. For $\PLHV{2}{3}{2}{3}$, it was shown by
Masanes~\cite{Ll.Masanes:QIC:2003} that there is only one other
class of tight Bell inequalities, which was discovered
independently by Collins\etal~\cite{D.Collins:PRL:2002} and
Kaslikowski\etal~\cite{D.Kaszlikowski:PRA:2002}. Following
Ref.~\cite{D.Collins:JPA:2004}, we will write this inequality
as
\begin{equation}\label{Ineq:Prob:I2233}
    I^{(1)}_{(2,2);(3,3)}:~
    \left(
    \begin{array}{r||r|r}
      \cdot     & -\mathbf{1}_2\t & \zero_2\t\\ \hline\hline
    -\mathbf{1}_2 &  M_1 &  M_2\\ \hline
     \zero_2 &  M_2 &  -M_2\\
    \end{array}
    \right)\le 0,
\end{equation}
where $\mathbf{1}_2$ and $\zero_2$ are, respectively, column
vector of ones and zeros with length 2,
\begin{equation}
    M_1\equiv\left(\begin{array}{cc}
    1 & 1 \\
    1 & \cdot
    \end{array}\right),\quad
    M_2\equiv\left(\begin{array}{cc}
    \cdot & 1 \\
    1 & 1
    \end{array}\right).
\end{equation}

In Ref.~\cite{D.Collins:JPA:2004}, the inequality
\eqref{Ineq:Prob:I2233} was actually presented as a special
case of a class of inequalities --- which Collins and Gisin
labeled as $I_{22nn}$  --- that holds for arbitrary
$\bfn=(n,n)$. Specifically, for $n=4$, it takes the form of
\begin{equation}\label{Ineq:Prob:I2244}
    I^{(1)}_{(2,2);(4,4)}:~
    \left(
    \begin{array}{r||rrr|rrr}
       & -1 & -1 & -1 & \cdot & \cdot &  \cdot\\ \hline\hline
    -1 &  1 &  1 &  1 & \cdot & \cdot &  1\\
    -1 &  1 &  1 &  \cdot & \cdot & 1 &  1\\
    -1 &  1 &  \cdot &  \cdot & 1 & 1 &  1\\ \hline
     \cdot &  \cdot &  \cdot &  1 & \cdot & \cdot & -1\\
     \cdot &  \cdot &  1 &  1 & \cdot &-1 & -1\\
     \cdot &  1 &  1 &  1 &-1 &-1 & -1\\
    \end{array}
    \right)\le 0,
\end{equation}
where inequalities for higher values of $n$ involve the obvious
modifications on individual blocks. For general $n$, we can
write $I_{22nn}$ in the following functional form:
\begin{align}
    \SLHV^{(I_{22nn})}&=\sum_{\oA=1}^{n-1}\sum_{\oB=1}^{n-\oA}\ProbTwJ{1}{\oA}{1}{\oB}
    +\sum_{\oA=1}^{n-1}\sum_{\oB=n-\oA}^{n-1}\Big[\ProbTwJ{1}{\oA}{2}{\oB}
    +\ProbTwJ{2}{\oA}{1}{\oB}-\ProbTwJ{2}{\oA}{2}{\oB}\Big]\nonumber\\
    &-\sum_{\oA=1}^{n-1}\ProbTwMA{1}{\oA}-\sum_{\oB=1}^{n-1}\ProbTwMB{1}{\oB}\le 0.
    \label{Ineq:Prob:I22nn:Functional}
\end{align}

This class of inequalities is
believed~\cite{D.Collins:JPA:2004} to be equivalent to the more
well-known $n$-outcome Collins-Gisin-Linden-Massar-Popescu
(henceforth abbreviated as CGLMP)
inequality~\cite{D.Collins:PRL:2002}, which admits the
following functional form:\footnote{Here, we have swapped $B_1$
and $B_2$ (i.e., Bob's first and second measurement settings)
and followed Ref.~\cite{Ll.Masanes:QIC:2003} by grouping terms
for the same setting together. Moreover, we have also shifted
the constant ``2" to the {\em lhs} of the inequality.}
\begin{align}
    \SLHVBI{$I_{n}$}=&\sum_{k=0}^{\lfloor
    \frac{n}{2}-1\rfloor}\left(1-\frac{2k}{n-1}\right)
        \sum_{\oB=1}^n\Big[
        \ProbTwJ{1}{\oB-k}{1}{\,\oB}-\ProbTwJ{1}{\oB+k+1}{1}{\,\oB}
        +\ProbTwJ{1}{\oB+k}{2}{\,\oB}-\ProbTwJ{1}{\oB-k-1}{2}{\,\oB} \nonumber\\
        &
        +\ProbTwJ{2}{\oB+k}{1}{\,\oB}-\ProbTwJ{2}{\oB-k-1}{1}{\,\oB}
        +\ProbTwJ{2}{\oB-k-1}{2}{\,\oB}-\ProbTwJ{2}{\oB+k}{2}{\,\oB}
        \Big]\le 2,
        \label{Ineq:Prob:CGLMP:Functional}
\end{align}
where expression such as $\oB-k$ in the above inequality is
understood to be evaluated modulo $n$. A proof of their
equivalence is, however, not available in the literature. In
Appendix~\ref{App:Sec:CGLMP}, we have provided this missing
proof. The tightness of the CGLMP inequality, and hence
$I_{22nn}$ was proven by Masanes in
Ref.~\cite{Ll.Masanes:QIC:2003}. They therefore correspond to
facets of $\PLHV{2}{n}{2}{n}$ for arbitrary $n\ge2$.

Finally, we note that a family of (tight) Bell inequalities --
the $I_{mmnn}$ inequality --- involving more than two
measurements per site, and more than two outcomes per
measurement has also been presented in
Ref.~\cite{D.Collins:JPA:2004}. However, the $I_{mmnn}$
inequality is only known to correspond to facets of
$\PLHV{m}{n}{m}{n}$ with $m,n>2$ for some relatively small
values of $m$ and $n$.

\subsection{Other Bipartite Correlation Inequalities}
\label{Sec:BI:Correlations}

Thus far, we have focused on the analysis of correlation
polytopes living in the space of probability vectors $\Cp$ and
looked at the corresponding tight Bell inequalities bounding
these polytopes. Now, let us turn our attention to the space of
correlations defined in terms of correlation functions --
denoted by $\Cct$ --- for an experimental scenario involving
only two parties performing $\mA$ and $\mB$
dichotomic\footnote{Strictly, many of the subsequent discussion
will still hold true even if we have more outcomes in the
experiments, provided that all measurement outcomes are bounded
between ``$-1$" and ``1".} measurements, and whose measurement
outcomes are labeled by $\pm1$. Specifically, in this bipartite
scenario, $\Cct$ is a space of dimension $d_c=\mA\mB+\mA+\mB$,
which can be labeled by the following
coordinates~\cite{M.Froissart:NCB:1981,D.Avis:JPA:2006}
\begin{gather}
    \big\{E(A_1,B_1),\ldots,E(A_1,B_{\mB}),E(A_2,B_1),
    \ldots,E(A_{\mA},B_{\mB}),\nonumber\\
    E(A_1),\ldots,E(A_{\mA}),E(B_1),
    \ldots,E(B_{\mB})\big\},
\end{gather}
where the {\em restricted} correlation
functions~\cite{R.F.Werner:PRA:2001} are defined as
\begin{equation}\label{Eq:Dfn:MarginalCorrelationFn}
    E(A_{\sA})=\ProbTwMA{\sA}{+}-\ProbTwMA{\sA}{-},\quad
    E(B_{\sB})=\ProbTwMB{\sB}{+}-\ProbTwMB{\sB}{-}.
\end{equation}
More often than not, however, we are only interested in the
(sub)space of correlations that is defined solely in terms of
the {\em full} correlation functions $E(A_{\sA},B_{\sB})$. We
shall denote this subspace by $\Ccs$. Note that it is a
subspace of dimension $d_s=\mA\mB$. As with $\Cp$, the set of
classical correlations in $\Cct$ ($\Ccs$) is a convex polytope
which we shall denote by $\polyclt$ ($\polycls$).

A well-known fact in relation to these polytopes is that the
two polytopes $\polyclt$ and $\PLHV{\mA}{2}{\nA}{2}$ are
actually isomorphic (see for example~\cite{D.Avis:JPA:2006}).
Therefore, any tight Bell inequality defining
$\PLHV{\mA}{2}{\nA}{2}$ can also be mapped to a tight
correlation inequality defining $\polyclt$ via
Eq.~\eqref{Eq:pLHV->ELHV} and
Eq.~\eqref{Eq:Dfn:MarginalCorrelationFn}. Nevertheless, for the
purpose of performing this mapping, it is more convenient to
make use of an equivalent form of Eq.~\eqref{Eq:pLHV->ELHV},
\begin{gather}\label{Eq:pLHV->ELHV3}
    E(A_{\sA},B_{\sB})=1-2\ProbTwMA{\sA}{+}-2\ProbTwMB{\sB}{+}
    +4\ProbTwJ{\sA}{+}{\sB}{+}.
\end{gather}
For instance, in the simplest scenario of $\bfm=(2,2)$, one
obtains the Bell-CHSH inequality from the Bell-CH inequality
via Eq.~\eqref{Eq:pLHV->ELHV3}. Similarly, by applying
Eq.~\eqref{Eq:Dfn:MarginalCorrelationFn} and
Eq.~\eqref{Eq:pLHV->ELHV3} to $I^{(1)}_{(3,3);(2,2)}$,
Eq.~\eqref{Ineq:Functional:I3322}, one can obtain the following
correlation inequality~\cite{C.Sliwa:PLA:2003,D.Avis:JPA:2006}
\begin{gather}
    E(A_1,B_1)+E(A_1,B_2)+E(A_1,B_3)+E(A_2,B_1)
    +E(A_2,B_2)-E(A_2,B_3)\nonumber\\
    +E(A_3,B_1)-E(A_3,B_2)-E(A_1)-E(A_2)+E(B_1)+E(B_2)\le
    4.\label{Ineq:Cor:I3322:Explicit}
\end{gather}
Note, nonetheless, that as opposed to the Bell-CHSH inequality,
inequality \eqref{Ineq:Cor:I3322:Explicit} does not live in the
subspace of full correlations $\Cors{3}{2}{3}{2}$, i.e., it
{\em is not} a facet-inducing inequality for
$\PLHVs{3}{2}{3}{2}$. In fact, recent work by
Avis\etal~\cite{D.Avis:JPA:2006} has demonstrated that for
$\bfm=(\mA,\mB)$ with $\min\{\mA,\mB\}\le3 $, the Bell-CHSH
inequalities and the trivial inequalities\footnote{In relation
to Eq.~\eqref{Eq:pLHV->ELHV3} and
Eq.~\eqref{Eq:Dfn:MarginalCorrelationFn}, these trivial
inequalities are in one-to-one correspondence with the trivial
requirement of probabilities being non-negative and less than
or equal to one.}
\begin{equation}\label{Ineq:Corr:trivial}
    -1\le E(A_{\sA},B_{\sB})\le 1,
\end{equation}
for all $\sA=1,\ldots,\mA$ and all $\sB=1,\ldots,\mB$, are the
only tight correlation inequalities defining
$\PLHVs{3}{2}{3}{2}$.

On the contrary, when four alternative measurements are allowed
at each site, Gisin has constructed the following Bell
correlation inequalities~\cite{N.Gisin:0702021}
\begin{equation}\label{Ineq:Corr:AS4}
    AS_4:~\left(\begin{array}{rrrr}
        1 & 1 & 1 & 1\\
        1 & 1 & 1 &-1\\
        1 & 1 &-2 & \cdot\\
        1 &-1 & \cdot & \cdot
        \end{array}\right)\le 6,
\end{equation}
\begin{equation}\label{Ineq:Corr:D4}
    D_4:~\left(\begin{array}{rrrr}
        2 & 1 & 1 & 2\\
        1 & 1 & 2 &-2\\
        1 & 2 &-2 &-1\\
        2 &-2 &-1 &-1
        \end{array}\right)\le 10,
\end{equation}
where the $(\sA,\sB)$ entry in each matrix is the coefficient
associated with the full correlation function
$E(A_{\sA},B_{\sB})$. These inequalities are tight. Together
with the trivial inequalities \eqref{Ineq:Corr:trivial} and the
Bell-CHSH inequality, they form a complete set of tight
correlation inequalities defining
$\PLHVs{4}{2}{4}{2}$~\cite{D.Avis:JPA:2006}.

As a last remark, we note that the correlation inequality
$AS_4$ has been generalized to an arbitrary even number of
measurement settings. Moreover, they can also be seen as a
correlation inequality that is valid for arbitrary number of
measurement outcomes if instead of Eq.~\eqref{Eq:pLHV->ELHV} or
Eq.~\eqref{Eq:pLHV->ELHV3}, which are only for two-outcome Bell
experiments, the full correlation function
$E(A_{\sA},B_{\sB})$, Eq.~\eqref{Eq:Dfn:CorrelationFn:LHV3}, is
interpreted as the difference between the probability of
observing the same outcomes at the two sites and the
probability of observing different outcomes at the two sites,
as it was done in
Eq.~\eqref{Eq:pLHV->ELHV2}~\cite{N.Gisin:0702021}.

\subsection{Multipartite Bell Inequalities}
\label{Sec:MultipartiteBI}

In sharp contrast with the study of bipartite Bell inequalities
 --- where most developments were carried out in the context of
{\em probability vectors} --- the multipartite analog was
primarily developed in the context of {\em correlation
functions}, and in particular the full correlation functions.
The pioneering work in this regard was initiated by
Mermin~\cite{N.D.Mermin:PRL:1990} who, in turn, was inspired by
the results presented by Greenberger, Horne and Zeilinger
(henceforth abbreviated as GHZ) on a demonstration of
incompatibility between local causality and quantum mechanical
prediction without resorting to any
inequalities~\cite{GHZ,GHZ:0712.0921} .

In his seminal work, Mermin~\cite{N.D.Mermin:PRL:1990}
investigated a scenario involving $n$ parties and where each of
them can perform two dichotomic measurements. Starting from the
assumption of a general LHVM, Mermin constructed a Bell
correlation inequality which involves only  $n$-partite full
correlation functions; his inequality therefore defines a
closed halfspace in $\CcM$ where $\polycM$ resides (here, there
are $n$ indices in both the superscript\footnote{We are
generalizing the notation introduced in
Sec.~\ref{Sec:BipartiteBI} such that indices in the subscript
(sequentially) indicate the number of possible measurements at
each site and indices in the superscript indicate the number of
possible outcomes per measurement at each site.} and
subscript). This work was further developed by Roy and
Singh~\cite{S.M.Roy:PRL:1991},
Ardehali~\cite{M.Ardehali:PRA:1992}, and eventually by
Belinski\v{\i} and
Klyskho~\cite{A.V.Belinskii:UFN:1993,A.V.Belinskii:PU:1993}
whereby the current form of Mermin inequality (also commonly
known as Mermin-Ardehali-Belinski\v{\i}-Klyskho, or in short,
MABK inequality) was culminated.\footnote{The inequality
developed by Roy and Singh~\cite{S.M.Roy:PRL:1991} is actually
equivalent to the current form of Mermin's inequality developed
by Belinski\v{\i} and
Klyskho~\cite{A.V.Belinskii:UFN:1993,A.V.Belinskii:PU:1993}.}

An interesting feature of the present form of Mermin's
inequality is that all inequalities involving $n>2$ parties can
be obtained from the Bell-CHSH inequality in a recursive
manner. To see that, let us now denote by $\oN{j}{s_j}=\pm1$
the outcome of measurement when the $\idx{j}$ observer chooses
to measure the $\idx{{s_j}}$ dichotomic observable. As a
classical variable,\footnote{That is, a variable that can be
defined using local hidden-variable.} $o^{[j]}_{s_j}$ can be
defined independently for each $j$ and each $s_j$. Thus, in
each run of the experiment, the expression
\begin{equation}\label{Eq:Dfn:F2}
    F_2\equiv\half\left(o^{[1]}_1+o^{[1]}_2\right)o^{[2]}_1 +
    \half\left(o^{[1]}_1-o^{[1]}_2\right)o^{[2]}_1,
\end{equation}
must either end up as $1$ or $-1$, since either
$o^{[1]}_1=o^{[1]}_2$ or $o^{[1]}_1=-o^{[1]}_2$. Averaging this
expression over many runs of the experiment, we see that the
average value of $F_2$ must be less than or equal to 1, since
each term in the average is at most 1. This is essentially a
statement of the Bell-CHSH inequality given in
Eq.~\eqref{Ineq:CHSH:Convention}.

To obtain the $n$-partite Mermin's inequality, we now follow
Ref.~\cite{N.Gisin:PLA:1998} and define
\begin{equation}\label{Eq:Dfn:Fn}
    F_n\equiv\half\left(o^{[n]}_1+o^{[n]}_2\right)F_{n-1} +
    \half\left(o^{[n]}_1-o^{[n]}_2\right)F_{n-1}',
\end{equation}
where $F_{n-1}'$ is the same expression as $F_{n-1}$ except
that all the $o^{[j]}_1$ and $o^{[j]}_2$ are interchanged. By
going through the same reasoning as before, it is not difficult
to see that the average value of $F_n$ must be bounded above by
1, i.e.,
\begin{equation}\label{Ineq:Mermin}
    {\rm Exp}\left(F_n(o^{[1]}_1,\ldots,o^{[n]}_1,o^{[1]}_2,\ldots,o^{[n]}_2)\right)\le 1,
\end{equation}
where here, ${\rm Exp}(x)$ refers to the expectation value of
$x$. It is also not difficult to see from Eq.~\eqref{Eq:Dfn:Fn}
that $F_n$ is an expression that is linear in all the local
variable $o^{[j]}_{s_j}$, therefore by generalizing the
notation introduced in Eq.~\eqref{Eq:SLHV:Generic}, we can
write
\begin{equation}\label{Eq:Fn}
    F_n=\sum_{s_1,s_2,\ldots,s_n=1}^2 b_{s_1s_2\ldots s_n}\prod_{j=1}^n o^{[j]}_{s_j},
\end{equation}
for some specific $b_{s_1s_2\ldots s_n}$. Now, we can write the
entire class of Mermin inequalities in a form that is closer to
inequality~\eqref{Ineq:CHSH:Convention}, i.e.,
\begin{equation}\label{Ineq:MultipartiteCn:General}
    \sum_{s_1=1}^2\cdots\sum_{s_n=1}^2 b_{s_1s_2\ldots s_n}
    E\left(o^{[1]}_{s_1},o^{[2]}_{s_2},\ldots,o^{[n]}_{s_n}\right)\le 1,
\end{equation}
where $E(.)$ is the $n$-party correlation function defined
analogous to Eq.~\eqref{Eq:Dfn:CorrelationFn:LHV3}.

As is now well-known, Mermin inequalities are not the only
class of $n$-partite Bell correlation inequalities. In fact, a
{\em complete}\footnote{Complete, in the sense that a vector of
full correlation functions is classical if and only if it
satisfies all of these inequalities.} set of $2^{2^n}$ Bell
correlation inequalities involving only the full correlation
functions has been obtained independently by Werner \&
Wolf~\cite{R.F.Werner:PRA:2001} and \.Zukowski \&
Brukner~\cite{M.Zukowski:PRL:2002} (the complete set of
inequalities for $n=4$ was also obtained by Weinfurter and
\.Zukowski in Ref.~\cite{H.Weinfurter:PRA:2001}). All these
inequalities are uniquely characterized by the tensor
$b_{s_1s_2\ldots s_n}$, which can be written
as~\cite{R.F.Werner:PRA:2001}
\begin{equation}\label{Eq:Dfn:Multipartite:b}
    b_{s_1s_2\ldots s_n}=2^{-n}\sum_{r_1=0}^1\sum_{r_2=0}^1\cdots\sum_{r_n=0}^1
    f(r_1,r_2,\ldots,r_n)(-1)^{\sum_j r_j (s_j-1)}
\end{equation}
where $f(r_1,r_2,\ldots,r_n)\in\{+1,-1\}$ is a binary function
that takes an $n$-bit-vector $r$ (with components $r_i$) as
argument. There are altogether $2^{2^n}$ of such functions,
each of them gives rise to a unique tensor $b_{s_1s_2\ldots
s_n}$ which, in turn, defines a Bell correlation inequality via
Eq.~\eqref{Ineq:MultipartiteCn:General}. These inequalities are
tight~\cite{R.F.Werner:PRA:2001,M.Zukowski:PRL:2002}, and
therefore are facet inducing for the correlation polytope of
$n$-partite correlation functions $\polycM$. It happens that
$\polycM$ is actually a $2^n$-dimensional
hyperoctahedron~\cite{R.F.Werner:PRA:2001}, and hence the
complete set of $2^{2^n}$ inequalities is equivalent to a
single nonlinear
inequality~\cite{R.F.Werner:PRA:2001,M.Zukowski:PRL:2002}.

More recently, by generalizing the work of Wu and Zong on
$\PLHVsM{4;2;2;\cdots;2}{2;2;2;\cdots;2}$~\cite{XH.Wu:PRA:2003},
Laskowski and coworkers~\cite{W.Laskowski:PRL:2004} have come
up with a systematic way to generate a huge class of tight Bell
correlation inequalities for
$\PLHVsM{2^{n-1};2^{n-1};2^{n-2};2^{n-3};\ldots;2}
{2;2;2;2;\ldots\ldots\ldots\ldots;2}$.  In particular, explicit
forms of these facet-inducing inequalities for
$\PLHVsM{4;4;2}{2;2;2}$ and $\PLHVsM{8;8;4;2}{2;2;2;2}$ can be
found in Ref.~\cite{W.Laskowski:PRL:2004}. A first step towards
the complete characterization of facets for a more symmetrical
experimental scenario, namely,
$\PLHVsM{3;3;3;\cdots;3}{2;2;2;\cdots;2}$ was carried out in
Ref.~\cite{M.Zukowski:QIP:2006} by \.Zuwkoski. Apparently, a
complete characterization for this experimental scenario has
subsequently been achieved in
Ref.~\cite{M.Zukowski:quant-ph:0611086}. Based on these
findings,  the explicit form of a tight correlation inequality
for $\PLHVsM{3;3;3}{2;2;2}$  has very recently been derived and
presented in Ref.~\cite{M.Wiesniak:PRA:2007}.

Finally, we note that at present, only one facet-inducing
inequality for $\PLHV{2}{3}{2;2}{3;3}$ is known, and is
presented in the form of a coincidence Bell
inequality~\cite{A.Acin:PRL:2004}. Other multipartite Bell
inequalities, such as those involving restricted correlation
functions~\cite{V.Scarani:PRA:2005,K.Chen:PRA:2006,C.F.Wu:PRA:2007}
or in the form of probability
inequality~\cite{J.L.Chen:PRL:2004} can also be found in the
literature. Their tightness, however, is not well studied.

\section{Conclusion}

In this chapter, we have looked at the set of classical
correlations, i.e., correlations (either in the form of
probability vector or a vector of correlation functions) that
are describable within the framework of LHVTs and how it is
related to the zoo of Bell inequalities that one can find in
the literature. Equipped with a solid understanding of the set
of classical correlations, we will next investigate what
quantum mechanics has to offer, both in terms of classical
correlations and correlations that cannot be accounted for
using any locally causal theory.

\chapter{Quantum Correlations and Locally Causal Quantum States}
\label{Chap:Q.Cn:Classical}

In the last chapter, we have looked at the set of classical
correlations and the characterization of its boundaries in
terms of Bell inequalities. Now, in this chapter, we will move
on to study the set of quantum correlations and see how they
are related to the set of classical correlations. Some
well-known examples of quantum states admitting locally causal
description will also be reviewed.

\section{Introduction}

In a nutshell, {\em quantum correlations} are simply points in
the space of correlations, c.f.
Sec.~\ref{Sec:CorrelationSpace}, that are realizable by quantum
mechanics through some choice of quantum states and some local
measurement operators. Ironically, despite the statistical
nature of quantum predictions, there was no known study on this
specific aspect of quantum predictions prior to the seminal
work by Bell in 1964~\cite{J.S.Bell:1964}.

After that, it seems to have taken another 16 years before the
first quantitative study on the set of quantum correlations was
carried out by Tsirelson\footnote{Incidentally, in response to
a question raised by A.~M.~Vershik (see pp.~884 of
Ref.~\cite{L.A.Khalfin:FOP:1992}).}~\cite{B.S.Cirelson:LMP:1980}.
In his work~\cite{B.S.Cirelson:LMP:1980}, Tsirelson showed that
the set of quantum correlations in $\Cors{2}{2}{2}{2}$ is also
bounded by some very similar linear inequalities like its
classical partner. However, these linear inequalities (often
known as the Tsirelson inequalities) are, in general, not
sufficient to distinguish a correlation that is realizable by
quantum mechanics from one that is not. In fact, it took a few
more years before Tsirelson came up with a set of necessary and
sufficient conditions --- in terms of inequalities that are
non-linear in the correlation functions --- for the
realizability of a point in $\Cors{2}{2}{2}{2}$ using quantum
mechanics~\cite{B.S.Tsirelson:LOMI:1985,
B.S.Tsirelson:JSM:1987}.

Meanwhile, a general study on the structure of the set of
quantum correlations beyond the simplest scenario of
$\mA=\mB=\nA=\nB=2$ was taken up by
Pitowsky~\cite{I.Pitowsky:JMP:1986}. In fact, it was in
Ref.~\cite{I.Pitowsky:JMP:1986} that the convexity of this set
and its relationship with the set of classical correlations
were, for the first time, formally established (see also
Ref.~\cite{I.Pitowsky:Book:1989}).

From Bell's theorem~\cite{J.S.Bell:1964}, we have learned that
there are quantum correlations that fall outside the classical
correlation polytope. A characterization of quantum states that
can give rise to such nonclassical correlations is,
nevertheless, still lacking. The seminal work by
Werner~\cite{R.F.Werner:PRA:1989} has established that {\em
entanglement} between spatially separated subsystems is a
necessary condition to establish nonclassical correlation.
Nonetheless, in the same article~\cite{R.F.Werner:PRA:1989},
Werner also provided an example of an entangled state which
does not violate any Bell inequalities if the source is
directly subjected to local, projective measurements without
any preprocessing. In fact, there are now a few known examples
of entangled quantum states which admit an explicit
LHVM~\cite{R.F.Werner:PRA:1989,J.Barrett:PRA:2002,
G.Toth:PRA:2006,M.L.Almeida:PRL:2007}.

In this chapter, we will start off, in Sec.~\ref{Sec:QC}, by
reviewing some well-known facts about the set of quantum
correlations. In the same section, we will also specify what we
mean by a {\em standard Bell experiment}, a key notion that is
used in this, as well as the subsequent chapters. After that,
in Sec.~\ref{Sec:LocalQS}, we will review some of the
well-known examples of quantum states admitting either a
partial, or a full LHVM for projective or generalized
measurements given by positive-operator-valued measures (POVM).

\section{Quantum Correlations}\label{Sec:QC}

Consider again the set of two-party correlations that respects
the relativistic causality condition, $\Cp$. In analogy with
the idea of a classical probability vector introduced in
Chapter~\ref{Chap:ClassicalCn}, we will now define a {\em
quantum probability vector} as follows.\footnote{The definition
is given for probability vectors considered in a bipartite
correlation experiment and where correlations are expressed in
terms of probability vectors. Nonetheless, it should be clear
as to how this definition can be generalized to the
multipartite scenario, or the space of correlations defined in
terms of correlation functions.}
\begin{dfn}\label{Dfn:QuantumCn}
A probability vector $\pQM$ in $\Cp$ is said to be a quantum
probability vector if there exists a bipartite quantum state
$\rho$ acting on $\HA\ten\HB$, i.e.,
$\rho\in\B\left(\HA\ten\HB\right)$, and some (local) POVM
elements $A_{\sA}^{\oA}\in\B(\HA)$, $B_{\sB}^{\oB}\in\B(\HB)$,
i.e., operators satisfying {\em
\begin{subequations}\label{Eq:POVM}
    \begin{gather}
        \sum_{\oA=1}^{\nA}\POVMAg=\unit_{\dA}\quad
        \text{and}\quad\sum_{\oB=1}^{\nB}\POVMBg=\unit_{\dB}
        \quad\forall\quad \sA, \sB,\label{Eq:POVM:Normalization}\\
         \POVMAg\ge 0,\quad \POVMBg\ge
        0\quad\forall\quad \sA, \sB,\oA,\oB,\label{Eq:POVM:PSD}
    \end{gather}
\end{subequations}}
such that the components of the probability vector satisfy{\em
    \begin{subequations}\label{Eq:QM:probabilities}
        \begin{gather}
            \ProbTwGJ= \tr\left(\rho\,\POVMAg\ten
            \POVMBg\right)\label{Eq:QM:probabilities:joint}\\
            \ProbTwGMA=\tr\left(\rho\,\POVMAg\ten \unit_{\dB}\right),\quad
            \ProbTwGMB= \tr\left(\rho\,\unit_{\dA}\ten
            \POVMBg\right),\label{Eq:QM:probabilities:marginal}
        \end{gather}
    \end{subequations}}
    where $\dA=\dim(\HA)$ and $\dB=\dim(\HB)$.
\end{dfn}
\noindent Note that in the above definition of a quantum
probability vector $\pQM$, the dimension of the Hilbert spaces
is not fixed {\em a priori}. In other words, the dimension of
the Hilbert spaces involved may vary depending on the given
probability vector. As with a classical probability vector, we
shall also refer to a quantum probability vector, loosely, as a
{\em quantum correlation}. Moreover, the set of quantum
correlations will be denoted by $\Qp$.

Physically, a quantum probability vector $\pQM\in\Qp$ is one
whose components can be realized via what we shall call a {\em
standard Bell experiment}.\footnote{In the literature, the term
{\em standard Bell experiment} has been used in various
different contexts. In particular, it is commonly used to refer
to a Bell experiment that involves measurements of
two-dichotomic observables per site (i.e.,
$\mA=\mB=\nA=\nB=2$). When there are only two parties involved
in the experiment, this reduces to an experiment that tests
against the Bell-CHSH/ Bell-CH inequality. Here, we are using
this term in the same sense as that used in
Ref.~\cite{M.Zukowski:PRA:1998}, which distinguishes it from
nonstandard Bell experiment that typically involves (either
active or passive) preprocessing prior to an actual Bell test.}
\begin{dfn}\label{Dfn:StandardBellExperiment}
    A standard Bell experiment (in relation to $\Cp$) on a source
    characterized by some {\em quantum} state $\rho$ is one whereby
    the source distributes pairs of physical systems to Alice and
    Bob, and where each of them can perform (on {\em each} physical
    system that they receive), respectively, $\mA$ and $\mB$
    alternative measurements  that would each generate $\nA$ and
    $\nB$ distinct outcomes.
\end{dfn}
\noindent Here, we have implicitly assumed that at the
receiving ends, the composite systems that Alice and Bob
receive are still well characterized by the same physical state
$\rho$ and this is the assumption that we will make whenever we
deal with a standard Bell experiment. With this assumption,
then via a standard Bell experiment, the sets of local POVM
elements $\{\{A_{\sA}^{\oA}\}_{\oA=1}^{\nA}\}_{\sA=1}^{\mA}$,
$\{\{B_{\sB}^{\oB}\}_{\oB=1}^{\nB}\}_{\sB=1}^{\mB}$ and the
(bipartite) quantum state $\rho$  give rise to a quantum
correlation $\pQM\in\Qp$ via Eq.~\eqref{Eq:QM:probabilities}.
As such, we will also say that $\rho$, together with these POVM
elements form a {\em quantum strategy} that realizes $\pQM$.

Of course, at a more general level, one can also imagine a
scenario where Alice and Bob choose to perform local
measurements on $N>1$ copies of the quantum systems at a time;
this is the scenario of performing a standard Bell experiment
on $\rho^{\ten N}$. Alternatively, one could also imagine that
while the source is well characterized by $\rho$, Alice and Bob
may choose to perform some local preprocessing on $\rho$ which
effectively transforms it to some other state $\rho'$ prior to
a standard Bell experiment. Loosely, we shall say that these
are nonstandard Bell experiments on $\rho$, since the source is
still well characterized by $\rho$. However, these and other
scenarios which do not fit within the framework of a standard
Bell experiment on $\rho$ will be the topics of future
discussion in Chapter~\ref{Chap:BellViolation} and
Chapter~\ref{Chap:Hidden.Nonlocality}.

\subsection{General Structure of the Set of Quantum Correlations}

Now, let us take a closer look at the structure of $\Qp$, and
in particular its relationship with $\Cp$. To begin with, we
note that for any two quantum probability vectors $\pQM$,
$\pQM'\in\Qp$, an arbitrary convex combination of them
\begin{equation}
    \bfp''=q~\pQM + (1-q)~\pQM',
\end{equation}
where $0\le q\le 1$ also gives rise to another quantum
probability vector $\bfp''\in\Qp$. To see this, let us denote
by $\rho$ and
$\{\{A_{\sA}^{\oA}\}_{\oA=1}^{\nA}\}_{\sA=1}^{\mA}$,
$\{\{B_{\sB}^{\oB}\}_{\oB=1}^{\nB}\}_{\sB=1}^{\mB}$,
respectively, a quantum state and some local POVM which
together form a quantum strategy for $\pQM$; likewise, $\rho'$,
$\{\{A_{\sA}^{'\oA}\}_{\oA=1}^{\nA}\}_{\sA=1}^{\mA}$ and
$\{\{B_{\sB}^{'\oB}\}_{\oB=1}^{\nB}\}_{\sB=1}^{\mB}$ which
together realize the quantum probability vector $\pQM'$. Then
it is easy to see that the quantum state
\begin{equation}
    \rho''\equiv q~\rho \oplus (1-q)~\rho',
\end{equation}
and the local POVM (elements) defined by
\begin{equation}
    A_{\sA}^{''\oA}\equiv A_{\sA}^{\oA} \oplus A_{\sA}^{'\oA},\quad
    B_{\sB}^{''\oB}\equiv B_{\sB}^{\oB} \oplus B_{\sB}^{'\oB},
\end{equation}
for all $\oA$, $\oB$, $\sA$ and $\sB$ do realize the
probability vector $\bfp''$ in the sense of
Eq.~\eqref{Eq:QM:probabilities}. Hence, as with the set of
classical correlations, $\Qp$ is convex. However, in sharp
contrast with $\polycl$, the set of quantum correlations is
{\em not} a convex polytope~\cite{I.Pitowsky:Book:1989}.
Nonetheless, for the simplest scenario where
$\mA=\mB=\nA=\nB=2$, the boundary of the set of quantum
correlations, or more precisely $\Qs{2}{2}$ has already been
characterized in
Refs.~\cite{B.S.Tsirelson:LOMI:1985,B.S.Tsirelson:JSM:1987}.

How is $\polycl$ related to $\Qp$? Intuitively, one would
expect $\polycl$ to be a subset of $\Qp$. To see that this is
indeed the case, it suffices to show that all extreme points of
$\polycl$ are contained in $\Qp$, that is, all extremal
classical probability vectors can be realized by some quantum
strategy. For definiteness, let us consider the extreme point
${}^{\bfa,\bfb}\bfB_{\A\B}$ whose components are given by
Eq.~\eqref{Eq:VertPLHV:Boolean}. A particular trivial way to
realize this classical probability vector is to pick any
(normalized) quantum state $\rho$ acting on $\HA\ten\HB$ and
the following local POVM elements
\begin{equation}\label{Eq:QM->CL}
    A_{\sA}^{\oA}=\delta_{\oA \vartheta^{[1]}_{\sA}}\unit_{\dA},\quad
    B_{\sB}^{\oB}=\delta_{\oB \vartheta^{[2]}_{\sB}}\unit_{\dB},
\end{equation}
for all $\oA$, $\oB$, $\sA$ and $\sB$. Then, from
Eq.~\eqref{Eq:QM:probabilities}, it is straightforward to see
that this quantum strategy does realize the classical
probabilities given in Eq.~\eqref{Eq:VertPLHV:Boolean}.
Therefore, $\ver{\polycl}\subset\Qp$ and by convexity of
$\polycl$ and $\Qp$, it follows that $\polycl\subseteq\Qp$,
i.e., the set of classical correlations is contained in the set
of quantum correlations.

On the other hand, as we recall from Bell's theorem
(Theorem~\ref{Thm:Bell}), there are quantum correlations which
violate a Bell inequality and hence fall outside the set of
classical correlations (hereafter we will also refer to a
probability vector $\bfp$ which is in $\Qp$ but not in $\Cp$ as
a {\em nonclassical correlation}. Therefore, the set of quantum
correlations $\Qp$ is a strict superset of the set of classical
correlations $\polycl$, i.e., $\polycl\subset\Qp$, for at least
some choices of $\mA$, $\mB$, $\nA$ and $\nB$. Meanwhile, it
has also been known for some time that the set of quantum
correlations $\Qp$ is a strict subset of the set of
correlations satisfying the no-signaling condition, i.e.,
$\Qp\subset\Cp$~\cite{S.Popescu:FP:1994}. In fact, some
quantitative understanding on the volume of these three sets,
namely, $\polycl$, $\Qp$ and $\Cp$ has recently been
established for the simplest scenario of
$\mA=\mB=\nA=\nB=2$~\cite{A.Cabello:PRA:2005a}.

\subsection{Quantum Correlation and Bell Inequality Violation}

Although all quantum states are capable of generating classical
correlations, only some quantum states are capable of
generating correlations outside the classical correlation
polytope. Necessarily, in this case, the nonclassical
correlation $\pQM\in\Qp$ must gives rise to a violation of some
Bell inequality. As we shall see in the later chapters, the
kind of correlation that a quantum state $\rho$ can offer
depends very much on whether it is a standard or a nonstandard
Bell experiment that is carried out on $\rho$. However, even if
we restrict our attention to standard Bell experiments, c.f.
Definition~\ref{Dfn:StandardBellExperiment}, whether a given
quantum state can offer nonclassical correlation may still
depend on the actual number of possible measurements --- $\mA$
and $\mB$ --- as well as the actual number of possible outcomes
for each measurement --- $\nA$ and $\nB$ (see
Chapter~\ref{Chap:QuantumBounds} and \ref{Chap:BellViolation}
for examples).

In this regard, let us now introduce the following definition
for a Bell inequality violation by a given state $\rho$ with
respect to some specific choice of the parameters
$\bfm\equiv(\mA,\mB)$ and $\bfn\equiv(\nA,\nB)$.
\begin{dfn}\label{Dfn:BIViolation}
    A quantum state $\rho\in\B(\H_\A\ten\H_\B) $ is said to violate
    a Bell inequality $I^{(k)}_{\bfm;\bfn}$,
    Eq.~\eqref{Ineq:Halfspace} -- Eq.~\eqref{Ineq:Halfspace:LHV},
    via a standard Bell experiment if and only if $\exists$ local
    measurement operators
    $\{\{A_{\sA}^{\oA}\}_{\oA=1}^{\nA}\}_{\sA=1}^{\mA}\subset\B(\HA)$,
    $\{\{B_{\sB}^{\oB}\}_{\oB=1}^{\nB}\}_{\sB=1}^{\mB}\subset\B(\HB)$
    such that the resulting quantum probability vector $\pQM$
    obtained via Eq.~\eqref{Eq:QM:probabilities} violates
    $I^{(k)}_{\bfm;\bfn}$.
\end{dfn}

Hereafter, unless otherwise stated, Bell inequality violation
for a given state $\rho$ will always be used in relation to a
standard Bell experiment and with respect to some specific Bell
inequality $I^{(k)}_{\bfm;\bfn}$. As we shall see later in
Sec.~\ref{Sec:LocalQS}, it is possible that $\rho$ {\em does
not violate} any Bell inequalities or is {\em known to satisfy}
a large class of Bell inequalities for all choices of local
measurements. At this stage, it is worth noting that, as with
the set of quantum states, the set of quantum states that do
not violate a given Bell inequality is {\em convex} (see
Appendix~\ref{App:Sec:Convexity:LocalStates} for a proof).

Nevertheless, as long as $\rho$ {\em does violate} a Bell
inequality with some choice of local measurements, we will say
that $\rho$ is Bell-inequality-violating:
\begin{dfn}\label{Dfn:BIV&NBIV}
    A quantum state $\rho$ is said to be Bell-inequality-violating
    (henceforth abbreviated as BIV) if and only if for some $\bfm$
    and $\bfn$, $\rho$ violates a Bell inequality
    $I^{(k)}_{\bfm;\bfn}$ for some $k$ in the sense defined in
    Definition~\ref{Dfn:BIViolation}. Similarly, a quantum state
    $\rho$ is said to be non-Bell-inequality-violating (henceforth
    abbreviated as NBIV) if and only if for all $\bfm$ and $\bfn$,
    $\rho$ does not violate any Bell inequality in the sense
    defined in Definition~\ref{Dfn:BIViolation}.\footnote{A quantum
    state that is BIV is commonly known in the literature as a {\em
    nonlocal} state; likewise, a quantum state that is NBIV is
    commonly known in the literature as a {\em local} state. This
    convention, however, is not unanimously accepted (see, for
    example, Ref.~\cite{M.Zukowski:JMO:2003}).}
\end{dfn}
Clearly, since the set of quantum states not violating a
specific Bell inequality is convex, so is the set of quantum
states that are NBIV. Let us denote this set by $\NVg$. As far
as a standard Bell experiment is concerned, the behavior of
NBIV quantum states is entirely classical, since any
experimental statistics generated from these states can be
mimicked by some LHVM. In the next section, we will review some
well-known examples of quantum states which are NBIV as well as
quantum states which are known to satisfy a large class of Bell
inequalities.

\section{Locally Causal Quantum States}\label{Sec:LocalQS}

Historically, Bell inequality violation has served as one of
the first means, both theoretical and experimental, to
demonstrate stronger than classical correlations. Nevertheless,
as is now well-known, not all quantum states are capable of
demonstrating such nonclassical correlations. In fact, some
quantum states are {\em only} capable of generating classical
correlations in {\em any} (standard) Bell experiments.

\subsection{Separable States}\label{Sec:SeparableStates}

An obvious example of a quantum state that is only capable of
producing classical correlations is a {\em separable state}
({\em aka} a {\em classically correlated
state}~\cite{R.F.Werner:PRA:1989}). In its simplest form, an
$n$-partite separable pure state $\ket{\Psi_\text{Sep}}$ is
just the tensor product of $n$ pure states, i.e.,
\begin{equation}\label{Eq:PureProductState}
    \ket{\Psi_{\mbox{\tiny Sep}}}=\ket{\phi^{[1]}}
    \ten\ket{\phi^{[2]}}\ten\cdots\ket{\phi^{[n]}},
\end{equation}
where $\ket{\phi^{[i]}}\in\H^{[i]}$. Due to its form, these
separable states are also known as {\em product states}, or
sometimes {\em uncorrelated states}. It is easy to see that
measurement statistics on a single particle, say that described
by $\ket{\phi^{[i]}}$, can be modeled in a purely classical
manner. In particular, the state vector $\ket{\phi^{[i]}}$, or
more generally $\ket{\Psi_{\mbox{\tiny Sep}}}$, serves as a
perfectly legitimate LHVM that reproduces the quantum
mechanical predictions. This can be seen, for example, by
noting that the joint probability of observing some local
measurement outcomes always factorizes into the $n$ marginal
probabilities and thus the correlation generated is always
classical, c.f. Eq.~\eqref{Eq:Dfn:ClassicalProbVec}.

Of course, separable states can also be correlated. The most
general separable state involves one that can be decomposed as
a convex combination of product states, i.e.,
\begin{subequations}\label{Eq:SeparableState}
    \begin{gather}
        \rho_{\mbox{\tiny Sep}}=\sum_kp_k~\rho_k^{[1]}\ten\rho_k^{[2]}\ten
        \cdots\ten\rho_k^{[n]},\\
        p_k\ge0,\quad\sum_kp_k=1,
    \end{gather}
\end{subequations}
where $\rho_k^{[i]}\equiv\ketbra{\phi_k^{[i]}}$. Operationally,
these are states that can be prepared from classical
correlations using only local quantum operations assisted by
classical communication (henceforth abbreviated as
LOCC\,\footnote{This is also commonly known in the literature
as LQCC (see, for example Ref.~\cite{D.Jonathan:PRL:1999} and
references therein).})~\cite{R.F.Werner:PRA:1989}. Since each
term in the sum, i.e., $\ten_{i=1}^n\rho_k^{[i]}$ is NBIV, and
hence can be modeled classically, so is their convex
combination. An immediate consequence of this is thus the
following Lemma~\cite{R.F.Werner:PRA:1989}.
\begin{lemma}\label{Thm:BIV->Entangled}
    A quantum state describing a composite system is BIV only if it
    is entangled, i.e., non-separable\footnote{A non-separable
    state was originally called an EPR correlated state in
    Ref.~\cite{R.F.Werner:PRA:1989}.} across its
    subsystems~\cite{R.F.Werner:PRA:1989}. Hence, a BIV state
    cannot be written in the form of Eq.~\eqref{Eq:SeparableState}.
\end{lemma}

\subsection{Quantum States Admitting General LHVM}
\label{Sec:States:LHVM:General}

Naively, it seems plausible that the converse of
Lemma~\ref{Thm:BIV->Entangled}, i.e., ``all entangled states
are BIV", is true. In other words, for any given entangled
state, there exist appropriate measurements such that the
corresponding correlation vector obtained from
Eq.~\eqref{Eq:QM:probabilities} lies outside the classical
correlation polytope. However, it turns out that there are
entangled quantum states whose measurement statistics in a
standard Bell experiment can be reproduced entirely using some
LHVM. In this section, we will look at some specific examples
of quantum states whereby a general LHVM can be constructed to
reproduce the quantum mechanical prediction for projective and/
or POVM measurements on the respective quantum states.

\subsubsection{$U\ten{U}$ Invariant States --- Werner States}
\label{Sec:WernerStates}

The first counterexample to the commonly held intuition that
``entanglement $\Rightarrow$ Bell inequality violation" was
given by Werner~\cite{R.F.Werner:PRA:1989} who considered
bipartite quantum states $\rw\in\B(\Cdg\ten\Cdg)$ that are
invariant under $U\ten{U}$, i.e.,
\begin{equation}\label{Eq:Dfn:WernerState}
    \rw=U\ten{U}~\rw~U^\dag\ten{U}^\dag,
\end{equation}
where $U$ is an arbitrary unitary operator acting on $\Cdg$. It
can be shown that $\rw$, now known as the Werner state, admits
the following compact form\footnote{Note that Werner has used,
instead, the following parametrization in
Ref.~\cite{R.F.Werner:PRA:1989}:
\begin{equation}\label{Eq:WernerState:Original}
    \rw(\Phi)=\frac{1}{d^3-d}\left[(d-\Phi)\unit_{d^2}+(d\Phi-1)V\right],
    \quad -1\le\Phi\le 1,
\end{equation}
where $V\in\B(\HA\ten\HB)$ is the flip operator such that
$V\ket{\alpha}_\A\ket{\beta}_\B=\ket{\beta}_\A\ket{\alpha}_\B$
and $\Phi=1-2q$.}
\begin{equation}\label{Eq:WernerState:Explicit}
    \rw(q)=(1-q)\frac{\PiS}{\tr(\PiS)}+q\frac{\PiA}{\tr(\PiA)},\quad
    0\le q\le 1,
\end{equation}
where $\PiS$ and $\PiA$ are, respectively, the projector onto
the symmetric and antisymmetric subspace of $\Cdg\ten\Cdg$.
Using the identities $\PiS+\PiA=\unit_{d}\ten\unit_d$,
$\tr(\Pi_{\pm})=d(d\pm1)/2$, Werner states can also be written
as an affine combination of the antisymmetric projector and the
$d\times d$-dimensional maximally mixed state, i.e.,
\begin{equation}\label{Eq:WernerState:NoisyProjector}
    \rw(p)=p\frac{2~\PiA}{d(d-1)}+(1-p)\frac{\unit_{d}\ten\unit_d}{d^2},\quad
    1-\frac{2d}{d+1}\le p\le1,
\end{equation}
where $p=1-\frac{2d}{d+1}(1-q)$.\footnote{Note that in this
case, the weight $p$ could also take on negative values.} The
separability of Werner states has been fully characterized in
Ref.~\cite{R.F.Werner:PRA:1989}: a Werner state is separable if
and only if $p\le\pS{W$_d$}\equiv1/(d+1)$ or equivalently,
$q\le\half$.

\begin{figure}[h!]
    \centering\rule{0pt}{4pt}\par
    \scalebox{1}{
    \includegraphics[width=11.58cm,height=9.0cm]
    {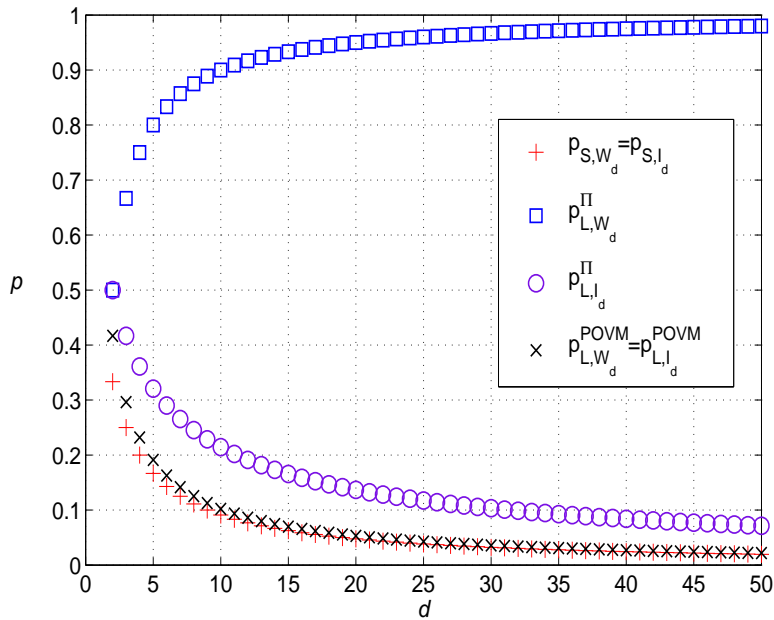}}
    \caption{\label{Fig:Thresholdp:WernerStates}
Plot of the various threshold weights $\pS{W$_d$}$,
$\pLPi{W$_d$}$, $\pLPOVM{W$_d$}$ for Werner states $\rw(p)$,
and $\pS{I$_d$}$, $\pLPi{I$_d$}$, $\pLPOVM{I$_d$}$ for
isotropic states $\rI(p)$ as a function of $d$. Notice that for
these two classes of states, the threshold weights for
separability, i.e., $\pS{W$_d$}$ and $\pS{I$_d$}$ are
identical; likewise for the threshold weights whereby a LHVM
for POVM measurements is known to exist, i.e., $\pLPOVM{W$_d$}$
and $\pLPOVM{I$_d$}$. For each $d$, the vertical line joining
these two threshold weights, which are, respectively, marked by
a red $+$ and a black $\times$, correspond to weights $p$ of
$\rw(p)$ and $\rI(p)$ whereby the states are entangled but do
not violate any Bell inequalities. Similarly, for each $d$, the
vertical line joining a blue $\square$ and a red $+$
corresponds to $\rw(p)$ which are entangled but are NBIV with
projective measurements whereas the vertical line joining a
purple circle and a red $+$  corresponds to $\rI(p)$ that are
entangled but are NBIV with projective measurements.
    }
\end{figure}

For any von Neumann (projective) measurement on $\rw(p)$ with
\begin{equation}
    p=\pLPi{W$_d$}\equiv1-\frac{1}{d},
\end{equation}
Werner~\cite{R.F.Werner:PRA:1989} has constructed an LHVM that
reproduces the corresponding quantum mechanical
prediction.\footnote{It is worth noting that alternative
derivations of Werner's LHVM for the $d=2$ case could also be
found in Refs.~\cite{J.Degorre:PRA:2005,G.Toth:PRA:2006}.}
Recall that the set of quantum states not violating a given
Bell inequality is convex (c.f.
Appendix~\ref{App:Sec:Convexity:LocalStates}), therefore for
$d\ge2$, $\rw(p)$ with $\pS{W$_d$}<p\le \pLPi{W$_d$}$ is
entangled but does not violate any Bell inequality with
projective measurements. Since $\pLPi{W$_d$}$ is an increasing
function of $d$, an interesting feature of Werner's model is
that it covers a greater range of entangled Werner states as
$d$ increases (Figure~\ref{Fig:Thresholdp:WernerStates}).

Given that this is not the most general measurement that one
can perform on $\rw(p)$, one may be tempted to conjecture that
all entangled Werner states could produce nonclassical
correlations with generalized measurements given by POVMs. In
2002, Barrett showed that this line of thought is
untenable~\cite{J.Barrett:PRA:2002}. In particular, he
constructed a LHVM for Werner states with
\begin{equation}
    p=\pLPOVM{W$_d$}\equiv\frac{3d-1}{d^2-1}\left(1-\frac{1}{d}\right)^d,
\end{equation}
for any POVM measurement. It is not difficult to show that
$\pS{W$_d$}\le\pLPOVM{W$_d$}$ for any $d\ge 2$, thus from the
convexity of NBIV states, it follows that any Werner state with
$\pS{W$_d$}\le p\le\pLPOVM{W$_d$}$ is entangled but does not
violate {\em any} Bell inequality. In contrast with Werner's
model~\cite{R.F.Werner:PRA:1989}, $\pLPOVM{W$_d$}$ decreases as
$d$ increases, hence the applicability of Barrett's LHVM
shrinks as $d$ increases
(Figure~\ref{Fig:Thresholdp:WernerStates}). In fact, it can be
easily checked that in the asymptotic limit of $d\to\infty$,
Barrett's model is barely applicable to any entangled Werner
states.

For the specific case of $d=2$, since the antisymmetric
projector $\PiA$ is none other than the projector onto the Bell
singlet state $\ket{\Psi^-}$,  a two-qubit\footnote{A qubit is
a two-level quantum system, which can be physically realized,
for example, by the polarization of a photon, the spin of an
electron etc. (see, for example,
Ref.~\cite{M.A.Nielsen:Book:2000}). A two-qubit state, in this
context, is the state of a two-party system, where each
subsystem can be represented by a qubit.} Werner state is
essentially a noisy Bell singlet state. Building on earlier
work by Tsirelson~\cite{B.S.Tsirelson:JSM:1987}, Ac\'in {\em et
al.}~\cite{A.Acin:PRA:2006} showed that in this case, i.e.,
$d=2$. there exists an LHVM for projective measurements on
Werner states with $p\lesssim 0.659\,50$. This is done by first
showing that the threshold $p$ whereby $\rwd{2}(p)$ becomes
NBIV with projective measurements, denoted by $\pcLPi{W$_2$}$,
is related to the Grothendieck's constant of order three, i.e.,
$K_{\mbox{\tiny G}}(3)$ by $\pcLPi{W$_2$}=1/K_{\mbox{\tiny
G}}(3)$. Then, by using an upper bound\footnote{The exact value
of $K_{\mbox{\tiny G}}(3)$ is not known.} on $K_{\mbox{\tiny
G}}(3)$ due to Krivine~\cite{J.L.Krivine:AIM:1979}, the above
lower bound on $\pcLPi{W$_2$}$ follows immediately. Moreover,
if any one of the observers has his/ her projective
measurements restricted to a plane in the Bloch sphere, then
there is a LHVM for $\rwd{2}(p)$ if and only if $p\le
1/\sqrt{2}$~\cite{A.Acin:PRA:2006,J.L.Krivine:AIM:1979}.

\subsubsection{$U\ten{\overline{U}}$ Invariant States --- Isotropic
States} \label{Sec:IsotropicStates}

Recently, a similar construction of an LHVM was also obtained
by Almeida {\em et~al.} for another class of bipartite mixed
states with a high degree of
symmetry~\cite{M.L.Almeida:PRL:2007}. Isotropic states, as they
are now known, were first introduced in
Ref.~\cite{M.Horodecki:PRA:1999} and have the nice property of
being invariant under $U\ten\overline{U}$, i.e.,
\begin{equation}\label{Eq:Dfn:IsotropicState}
    \rI(p)=U\ten\overline{U}~\rI~U^\dag\ten\overline{U}^\dag,
\end{equation}
where $\overline{U}$ denotes the complex conjugate of an
arbitrary $d\times{d}$ unitary matrix $U$. As for Werner
states, the isotropic states $\rI\in\B(\Cdg\ten\Cdg)$ admit the
explicit form~\cite{M.Horodecki:PRA:1999}
\begin{equation}\label{Eq:IsotropicState:Explicit}
    \rI(p)=p~\ProjMEd+(1-p)\frac{\unit_{d}\ten\unit_d}{d^2},
\end{equation}
which, for $0\le p\le1$, can be interpreted as a convex mixture
of the maximally entangled state $\MEd\in\Cdg\ten\Cdg$,
\begin{equation}\label{Eq:Dfn:State:ME-d}
    \MEd\equiv\frac{1}{\sqrt{d}}\sum_{i=1}^d\ket{i}_{\A}
    \ten\ket{i}_{\B},
\end{equation}
and the maximally mixed state,  where $\{\ket{i}_{\A}\}$ and
$\{\ket{i}_{\B}\}$ are, respectively, local orthonormal bases
of $\H_\A=\Cdg$ and $\H_\B=\Cdg$.\footnote{Due to the explicit
form given in Eq.~\eqref{Eq:Dfn:State:ME-d}, some authors also
refer to the isotropic state $\rI(p)$ as the (generalized)
Werner states (see, for example,
Ref.~\cite{V.M.Kendon:PRA:2002}).} If, and only if
$p\le\pS{I$_d$}\equiv1/(d+1)$, the mixture represents a
separable state~\cite{M.Horodecki:PRA:1999}. It is worth
nothing that in this case, partial transposition of $\rI(p)$
gives a legitimate and separable $\rw(p')$ with
$p'=(1-d)p$~\cite{K.G.H.Vollbrecht:PRA:2001}.

In the same spirit as Werner's and Barrett's construction,
Almeida {\em et~al.} constructed an explicit LHVM for
projective measurements as well as an LHVM for generalized
measurements on $\rI(p)$. Specifically, their models work for
mixtures with weight $p$ given by
\begin{equation}\label{Eq:CriticalP:IsotropicState}
    \pLPi{I$_d$}\equiv\frac{1}{d-1}\sum_{k=2}^d\frac{1}{k}\quad
    \text{and}\quad\pLPOVM{I$_d$}\equiv\frac{3d-1}{d^2-1}
    \left(1-\frac{1}{d}\right)^d=\pLPOVM{W$_d$}
\end{equation}
respectively.\footnote{It is interesting to note that, from
here, Almeida\etal~\cite{M.L.Almeida:PRL:2007} have also found,
using existing results from Ref.~\cite{M.A.Nielsen:PRL:1999}, a
lower bound on $p$ whereby an arbitrary convex mixture of a
bipartite pure state and the maximally mixed state would admit
a LHVM for both projective and the generalized measurements.}
Just like $\pLPOVM{W$_d$}$, $\pLPi{I$_d$}$is monotonically
decreasing with $d$. Nevertheless, for any $d\ge1$, it can
again be shown that the latter is always greater than or equal
to $\pS{I$_d$}$ (Figure~\ref{Fig:Thresholdp:WernerStates}).
Hence, experimental statistics obtained from projective
measurements on isotropic states with $\pS{I$_d$}<
p\le\pLPi{I$_d$}$  cannot violate any Bell inequalities. More
generally, isotropic states with $\pS{I$_d$}<
p\le\pLPOVM{I$_d$}$ are entangled but are NBIV.

Notice that when $d=2$, the isotropic state is local unitarily
equivalent to a Werner state. Therefore, all the bounds
obtained by Ac\'in {\em et al.}~\cite{A.Acin:PRA:2006} for
2-dimensional Werner state are also applicable to the
2-dimensional isotropic state. In particular, this means that
$\pcLPi{W$_d$}=\pcLPi{I$_d$}$. On the other hand, if only
traceless observables are measured on the isotropic states, it
was also shown in Ref.~\cite{A.Acin:PRA:2006} that a LHVM for
the experimental statistics exists for $p\le1/K_{\mbox{\tiny
G}}(d^2-1)$, where $K_{\mbox{\tiny G}}(n)$ is the Grothendieck
constant of order $n$.

\subsubsection{$U\ten{U}\ten{U}$ Invariant States}
\label{Sec:UUUStates}

In the multipartite scenario, Werner's LHVM has also been
extended to cover some tripartite states with $U\ten{U}\ten{U}$
symmetry~\cite{T.Eggeling:PRA:2001}. In
Ref.~\cite{G.Toth:PRA:2006}, T\'oth and Ac\'in gave an
alternative derivation of Werner's LHVM, which allows them to
generalize straightforwardly to tripartite states of the form
\begin{equation}\label{Eq:Dfn:rho3c}
    \rTA(p)=\frac{1}{8}\unit_2\ten\unit_2\ten
    +\frac{1}{24}\sum_{k=x,y,z}\unit_2\ten\sigma_k\ten\sigma_k
    -\frac{p}{16}(\sigma_k\ten\unit_2\ten\sigma_k+\sigma_k\ten\sigma_k\ten\unit_2),
\end{equation}
where $\{\sigma_i\}_{i=1}^3$ are Pauli matrices introduced in
Eq.~\eqref{Eq:Dfn:PauliMatrices}. Their model works for
projective measurements and $p\le 1$ whereas $\rTA(p)$ with $
p>\pS{TA}\equiv\frac{1}{3}(\sqrt{13}-1)$ are states that cannot
be written either in the form of Eq.~\eqref{Eq:SeparableState}
or the form
\begin{gather}
    \sum_k p_k^{[\A\B]}\rho_k^{[AB]}\ten\rho_k^{[\C]}+
    p_k^{[\A\C]}\rho_k^{[\A\C]}\ten\rho_k^{[\B]}+
    p_k^{[\B\C]}\rho_k^{[\B\C]}\ten\rho_k^{[\A]},\nonumber\\
    p_k^{[ij]}\ge 0\quad\forall~i,j\in\{\A,\B,\C\},\quad
    \sum_k p_k^{[\A\B]}+p_k^{[\A\C]}+p_k^{[\B\C]}=1,
    \label{Eq:States:TwoEntangled}
\end{gather}
where $\rho_k^{[i]}\in\B(\H_i)$, and
$\rho_k^{[ij]}\in\B(\H_i\ten\H_j)$ for all
$i,j\in\{\A,\B,\C\}$. Tripartite states that can be written in
the form of Eq.~\eqref{Eq:States:TwoEntangled} are
biseparable~\cite{A.Acin:PRL:2001} and can be prepared by
mixing pure states on one side and an entangled two-party state
at the remaining sites. Hence, $\rTA(p)$ with $\pS{TA}< p\le 1$
contains genuine tripartite entanglement but admits LHVM for
projective measurements.

\subsection{Quantum States Satisfying Some Bell Inequalities}
\label{Sec:States:LHVM:Partial}

The LHVMs that have been constructed for $\rw(p)$, $\rI(p)$ and
$\rTA(p)$ are very general in that they can reproduce exactly
the quantum mechanical prediction for projective and/or POVM
measurements on the respective states. As a result, these
states do not violate {\em any} Bell inequalities via
measurements where the models are applicable. However,
construction of these general LHVMs are by no means trivial,
and could only be done, so far, for states with a high-degree
of symmetry. In this section, we will look at some examples of
entangled states that are known to satisfy a large class of,
rather than {\em all} Bell inequalities. For the examples
presented in Sec.~\ref{Sec:States:PPT}, no explicit LHVM is
constructed, but the states are known to satisfy a large class
of Bell inequalities whereas for the examples presented in
Sec.~\ref{Sec:States:Extension}, an LHVM is constructed for
Bell inequalities with a specific number of measurement
settings per site.

\subsubsection{PPT Entangled States}\label{Sec:States:PPT}

Historically, positive-partial-transposed (henceforth
abbreviated as PPT) entangled states referred to bipartite
entangled states that remain positive semidefinite (henceforth
abbreviated as PSD) after partial transposition with respect to
one of its subsystems~\cite{P.Horodecki:PLA:1997}. By virtue of
this property, entanglement of PPT states cannot be decided
using the Peres-Horodecki criterion ({\em aka} PPT criterion)
for
separability~\cite{A.Peres:PRL:1996,MPR.Horodecki:PLA:1996}.
The very first example of a PPT entangled state in the
literature is the following 1-parameter family of
two-qutrit\footnote{A qutrit is a three-level quantum system.}
mixed states~\cite{P.Horodecki:PLA:1997}:
\begin{subequations}\label{Eq:State:H3}
\begin{equation}
    \rH(p)=\frac{8p}{8p+1}\rho_\text{Ent}+\frac{1}{8p+1}\ketbra{\Psi_p},\quad
    0< p <1,
\end{equation}
where
\begin{align}
    \rho_\text{Ent}&=\frac{1}{8}\sum_{i,j=0,i\neq j}^2\ketbra{i}\ten\ketbra{j}
    -\frac{1}{8}\ketbra{2}\ten\ketbra{0}+\frac{3}{8}\ProjME{3},\\
    \ket{\Psi_p}&=\ket{2}\ten\left(\sqrt{\frac{1+p}{2}}\ket{0}
    +\sqrt{\frac{1-p}{2}}\ket{2}\right),
\end{align}\end{subequations}
and $\ME{3}$ is the maximally entangled state for $d=3$, c.f.
Eq.~\eqref{Eq:Dfn:State:ME-d}.

As was first demonstrated by
Horodecki\etal~\cite{MPR.Horodecki:PRL:1998}, a bipartite PPT
state cannot be distilled~\cite{C.H.Bennett:PRL:1996} to a Bell
singlet state $\ket{\Psi^-}$ using LOCC. Hence, PPT entangled
states are also known as {\em bound entangled states}. The
entanglement contained in a bound entangled state is rather
weak and often has to be used in conjunction with other
entangled states to demonstrate its nonclassical features. In
fact, it was even conjectured by Peres~\cite{A.Peres:FP:1999}
that no PPT entangled states violate any Bell inequalities. The
first result that was in favor of this conjecture was given by
Werner and Wolf~\cite{R.F.Werner:PRA:2000} where they showed,
using the variance inequality,\footnote{That is, the variance
of a random variable is non-negative.} that an $n$-partite
(entangled) state that is PPT with respect to all combinations
of its subsystems cannot violate any of the $n$-partite Mermin
inequalities, Eq.~\eqref{Eq:Dfn:F2} --
Eq.~\eqref{Ineq:MultipartiteCn:General}. Since the Mermin
inequality reduces to the Bell-CHSH inequality when $n=2$, an
immediate corollary of Werner and Wolf's result is that no
bipartite PPT entangled states can violate the Bell-CHSH
inequality.

It is still possible, however, to see a Bell inequality
violation coming from an $n$-partite entangled state
$\rho\in\B\left(\H^{[1]}\ten\H^{[2]}\ten\ldots\H^{[n]}\right)$
that is PSD with respect to transposition of each individual
subsystem, i.e.,
\begin{equation}
    \rho^{\mbox{\tiny $T_k$}}\ge 0\quad\forall\quad k\in\{1,2,\ldots,n\},
\end{equation}
where $(.)^{\mbox{\tiny $T_k$}}$ denotes the partial
transposition with respect to subsystem $k$. In particular,
D\"ur~\cite{W.Dur:PRL:2001} showed that the $n$-partite Mermin
inequality (with $n\ge8$) is violated by an $n$-partite
entangled state that is of this sort. Specifically, the
multipartite mixed entangled state that D\"ur considered
reads:\footnote{For $n\ge 4$, $\rho_D$ has positive partial
transposition with respect to each of the subsystem $k$, where
$k\in\{1,2,\ldots,n\}$ but the state is not PSD if a partial
transposition is carried out with respect to
$\H^{[i]}\ten\H^{[j]}$ for any $i\neq j$. Hence, by the PPT
criterion for separability~\cite{A.Peres:PRL:1996,
MPR.Horodecki:PLA:1996}, $\rho_D$ for $n\ge4$ cannot be fully
separable, i.e., cannot be written in the form of
Eq.~\eqref{Eq:SeparableState}.}
\begin{equation}
    \rho_D=\frac{1}{n+1}\left(\ketbra{\Psi_{\mbox{\tiny GHZ}}}
    +\half\sum_{k=1}^n\left(\ketbra{\Phi_{k,0}}+\ketbra{\Phi_{k,1}}
    \right)\right),
\end{equation}
where $\GHZ\in\H^{[1]}\ten\H^{[2]}\ten\cdots\ten\H^{[n]}
=\Cd{2}\ten\Cd{2}\ten\cdots\ten\Cd{2}$ is the $n$-partite
generalized GHZ state~\cite{GHZ:AJP:1990},
\begin{equation}\label{Eq:Dfn:States:GHZ}
    \GHZ\equiv\frac{1}{\sqrt{2}}\left(\ket{0}^{\otimes n}+
    \exp{\ii\alpha_n}\ket{1}^{\otimes n}\right),
\end{equation}
$\alpha_n$ is an arbitrary phase factor,
$\{\ketbra{\Phi_{k,0}},\ketbra{\Phi_{k,1}}\}$ are product
states defined by
\begin{gather*}
    \ket{\Phi_{k,0}}\equiv\ket{0^{[1]}}\ten\ket{0^{[2]}}\ten\cdots\ket{0^{[k-1]}}
    \ten\ket{1^{[k]}}\ten\ket{0^{[k+1]}}\ten\cdots\ket{0^{[n]}},\\
    \ket{\Phi_{k,1}}\equiv\ket{1^{[1]}}\ten\ket{1^{[2]}}\ten\cdots\ket{1^{[k-1]}}
    \ten\ket{0^{[k]}}\ten\ket{1^{[k+1]}}\ten\cdots\ket{1^{[n]}},
\end{gather*}
and $\{\ket{0^{[j]}},\ket{1^{[j]}}\}$ are local orthonormal
basis vectors for $\H^{[j]}$. This is, nevertheless, not in
contradiction with the result given by Werner and
Wolf~\cite{R.F.Werner:PRA:2000}. In fact, follow up work by
Ac\'in~\cite{A.Acin:PRL:2002} showed that for all these states
violating the Mermin inequality, there is at least one
bipartite splitting of the system such that the state becomes
distillable.\footnote{See also Ref.~\cite{A.Acin:PRA:2002} for
a more thorough discussion between distillability and violation
of $n$-partite Bell inequality.}

Of course, as reviewed earlier in
Sec.~\ref{Sec:MultipartiteBI}, Mermin inequality is not the
only class of tight Bell correlation inequalities for the
$n$-partite correlation polytope $\polycM$, thus a natural
question that follows is whether Werner and Wolf's
result~\cite{R.F.Werner:PRA:2000} generalizes to all the
$2^{2^n}$ tight Bell correlation inequalities with two
dichotomic observables per site, c.f.
Eq.~\eqref{Ineq:MultipartiteCn:General} and
Eq.~\eqref{Eq:Dfn:Multipartite:b}. In 2001, Werner and
Wolf~\cite{R.F.Werner:PRA:2001} provided a positive answer to
this question --- $n$-partite states that are PPT with respect
to all combinations of its subsystems do not violate any of the
$n$-partite Bell correlation inequalities with two dichotomic
observables per site. By far, this is the strongest result in
support of Peres' conjecture. Although a counterexample to this
conjecture is not known in the literature, the same goes for a
proof, despite the wide range of supporting evidence. In what
follows, we will review some other examples of bipartite PPT
entangled states which are known to satisfy a large class of
Bell inequalities.

\subsubsection{Entangled States with Symmetric Quasiextension}
\label{Sec:States:Extension}

Given that it is nontrivial to come up with a general LHVM, a
natural question that follows is whether there is any
systematic way to generate, perhaps not as general, LHVM for
arbitrary quantum states. In 2003, an important breakthrough
along this line came about following Terhal\etal's
consideration of symmetric quasiextension for multipartite
quantum states~\cite{B.M.Terhal:PRL:2003}. To appreciate that,
let us recall the following definition from
Ref.~\cite{B.M.Terhal:PRL:2003}:
\begin{dfn}
    \label{Dfn:SymmetricQuasiExtension}
    Let $\pi:\H^{\ten{s}}\to\H^{\ten{s}}$ be a permutation of Hilbert
    spaces $\H$ in $\H^{\ten{s}}$ and let
    \begin{equation}
        \Sym_{\H^{\ten s}}(\rho)\equiv\frac{1}{s!}
        \sum_\pi\pi\,\rho\,\pi^\dag,
    \end{equation}
    then $\rho$ acting on $\H_\A\ten\H_\B$ has a
    $(s_a,s_b)$-symmetric quasiextension when there exists a
    multipartite entanglement witness\footnote{An $n$-partite entanglement
    witness $W$ is a Hermitian matrix that satisfies
    \begin{equation*}
        \tr\left(W\rho_{\mbox{\tiny Sep}}\right)\ge 0
    \end{equation*}
    for all $n$-partite separable states $\rho_{\mbox{\tiny Sep}}$,
    c.f. Eq.~\eqref{Eq:SeparableState}.}
    $W_\rho\in\B(\H_\A^{\ten{s_a}}\ten\H_\B^{\ten{s_b}})$ such that
    {\em
    $\tr_{\H_\A^{\ten{(s_a-1)}}\ten\H_\B^{\ten{(s_b-1)}}}W_\rho=\rho$}
    and
    $W_\rho=\Sym_{\H_\A^{\ten{s_a}}}\ten\Sym_{\H_\B^{\ten{s_b}}}(W_\rho)$.
\end{dfn}

With this definition of symmetric quasiextension,
Terhal\etal~\cite{B.M.Terhal:PRL:2003} then went on to show
that if $\rho$ has an $(s_a,s_b)$-symmetric quasiextension,
then an LHVM can be constructed for $\rho$ for all Bell
experiments with $\bfm=(s_a,s_b)$; hence, $\rho$ does not
violate any Bell inequality with $\bfm=(s_a,s_b)$ settings. In
fact, the following strengthened version of the theorem was
also proven in the same paper~\cite{B.M.Terhal:PRL:2003}.
\begin{theorem}\label{Thm:Extension}
    If $\rho$ has a $(1,s_b)$-symmetric quasiextension, then $\rho$ does
    not violate a Bell inequality with $s_b$ settings for Bob and any
    number of settings for Alice. Similarly, if $\rho$ has a
    $(s_a,1)$-symmetric quasiextension, then $\rho$ does not violate a
    Bell inequality with $s_a$ settings for Alice and any number of
    settings for Bob.
\end{theorem}

From Definition~\ref{Dfn:SymmetricQuasiExtension}, it follows
that if a given state has an $(s_a,s_b)$-symmetric
quasiextension, it must necessarily have a $(1,s_b)$-symmetric
quasiextension and an $(s_a,1)$-symmetric quasiextension. This,
together with Theorem~\ref{Thm:Extension}, implies that if
$\rho$ has an $(s_a,s_b)$-symmetric quasiextension, it cannot
violate any Bell inequalities with $\bfm=(s_{a'},s_{b'})$
settings where $\min\{s_{a'},s_{b'}\}\le\max\{s_{a},s_{b}\}$.
As a first application of their technique,
Terhal\etal~constructed a $(2,2)$-symmetric extension for any
bipartite bound entangled state based on a real {\em
unextendible product basis}~\cite{C.H.Bennett:PRL:1999}.
Therefore, if any of such states is to violate a Bell
inequality, it must involve more than 2 measurement settings on
at least one of the sites.

The construction of a symmetric (quasi)extension of a
given quantum state $\rho$, if it exists, can be done, to some
extent, numerically. In particular, the search for an
$(s_a,s_b)$-symmetric quasiextension with
non-negative\footnote{In this case, the corresponding
entanglement witness $W_\rho$ is a trivial one and it actually
corresponds to what is called a symmetric extension of
$\rho$~\cite{A.C.Doherty:PRL:2002,A.C.Doherty:PRA:2004}.} or
decomposable $W_\rho$ is a semidefinite programming feasibility
problem (Appendix~\ref{App:Sec:SDP:EW}), which can be
efficiently solved on a computer. In some cases, these
semidefinite programs (henceforth abbreviated as SDP) can even
be solved analytically. For example, in the case of Werner
states, c.f. Eq.~\eqref{Eq:WernerState:NoisyProjector}, it was
established in
Ref.~\cite{Terhal:eprintquant-ph:2002,B.M.Terhal:PRL:2003} that
all $\rw(p)$ have symmetric extensions as long as
\begin{equation}\label{Eq:ExtensionBound:WernerState}
    s_a+s_b\le d.
\end{equation}
In the case of $d=2$, a bound better than
Eq.~\eqref{Eq:ExtensionBound:WernerState} was also derived in
Ref.~\cite{Terhal:eprintquant-ph:2002}, namely, a
$(2,2)$-symmetric extension and hence an LHVM with 2 settings
or less can be constructed for $\rwd{2}(p)$ with
$-\frac{1}{3}\le p\le\frac{2}{3}.$\footnote{It should be
emphasized that these bounds were obtained by considering a
symmetric quasiextension derived from either a non-negative
entanglement witness or a decomposable entanglement witness. It
could very well be that the state of interest has a symmetric
quasiextension that is not of either of these two forms.} Note
that, in comparison with the work presented by
Ac\'in\etal~\cite{A.Acin:PRA:2006}, the LHVM derived in this
manner is actually applicable to more entangled $\rwd{2}(p)$
even for POVM measurements. The tradeoff, however, is that it
is only applicable to scenarios where $\min\{\sA,\sB\}\le 2$.

Numerically, Terhal\etal's construction has also been applied
to the following one-parameter family of two-qutrit mixed
state~\cite{PMR.Horodecki:PRL:1999}:
\begin{subequations}\label{Eq:State:Choi-H3}
    \begin{equation}
        \rCH(\alpha)=\frac{2}{7}\ketbra{\Phi_3^+}+\frac{\alpha}{7}\sigma_+
        +\frac{5-\alpha}{7}\sigma_-, \quad 2 < \alpha <5,
    \end{equation}
    where
    \begin{align}
        \sigma_\pm=\frac{1}{3}\sum_{j=0}^2\ketbra{j}\ten\ketbra{j\pm1{\rm~mod}~3},
\end{align}
\end{subequations}
which is known to be separable for $2\le\alpha\le3$, bound
entangled for $3<\alpha\le4$ and having negative partial
transposition for $4<\alpha\le5$. In particular, entangled
$\rCH(\alpha)$ was found to possess a $(2,2)$-symmetric
quasiextension and a $(3,3)$-symmetric quasiextension derived
from a decomposable entanglement witness for
$\alpha\in[3,4.84]$ and $\alpha\in[3,4]$ respectively.
Therefore any potential Bell inequality violation of the bound
entangled $\rCH(\alpha)$ must involve at least four alternative
measurements on one of the sites.

\section{Conclusion}

In this chapter, we have formally defined what we mean by
quantum correlations, and the closely related concept of a
standard Bell experiment. We have also looked at some of the
basic structure of the set of quantum correlations and its
relationship with the classical correlation polytope. In
addition, we have also reviewed some well-known examples of
quantum states admitting either a partial, or a full LHVM for
projective/ POVM measurements. The stage is finally set for us
to look into genuine quantum correlations which cannot be
accounted for by any LHVM.

\chapter{Bounds on Quantum Correlations in Standard Bell Experiments}
\label{Chap:QuantumBounds}

As we have seen in the previous chapter, correlations generated
by quantum systems can sometimes be described in a purely
classical manner via a local hidden variable model. By Bell's
theorem, of course, we know that some entangled quantum states
can also offer correlations that are not describable within the
classical framework. In this and the next chapter, we will look
at such nonclassical behavior displayed by entangled quantum
systems in standard Bell experiments.

\section{Introduction}

Before pursuing any in-depth study on the nonclassical
correlations offered by quantum systems, it seems natural to
first determine if a given entangled state is
Bell-inequality-violating (BIV) and hence capable of
demonstrating nonclassical correlations in a standard Bell
experiment. In the terminologies that we have introduced
earlier in Sec.~\ref{Sec:Convexity:ClassicalCn}, this amounts
to determining if a given quantum state, with a judicious
choice of local measurements, can give rise to correlations
that lie outside the classical correlation polytope. Typically,
this is done by varying over the local measurements that each
observer may perform and checking if the resulting statistics
can violate any Bell inequalities.

Surprisingly, relatively little is known in terms of which
quantum states are BIV. For bipartite quantum systems, the
strongest results that we know in this regard are due to Gisin
and Peres~\cite{N.Gisin:PLA:1992}, who showed that all
bipartite pure entangled states violate the Bell-CHSH
inequality (a weaker version of Gisin and Peres's result was
first presented by Capasso\etal~\cite{V.Capasso:IJTP:1973} and
later rediscovered by Gisin~\cite{N.Gisin:PLA:1991}). In other
words, a bipartite pure quantum system is capable of
demonstrating nonclassical correlations {\em if and only if} it
is entangled.

The corresponding situation for multipartite quantum systems is
a lot more complicated and it is still not known if all
multipartite pure entangled states are BIV. To begin with,
Scarani and Gisin~\cite{V.Scarani:JPA:2001} noticed that some
generalized GHZ states, despite being entangled, do not violate
any of the Mermin inequalities, Eq.~\eqref{Eq:Dfn:F2} --
Eq.~\eqref{Ineq:MultipartiteCn:General}. Although some of these
states were later found to violate some, among the complete set
of $2^{2^n}$ $n$-partite correlation inequalities,
Eq.~\eqref{Ineq:MultipartiteCn:General} --
Eq.~\eqref{Eq:Dfn:Multipartite:b}, the rest were proved to
satisfy this set of inequalities
also~\cite{M.Zukowski:PRL:2002b}. A twist came about when
Chen\etal~\cite{J.L.Chen:PRL:2004} constructed a tripartite
Bell inequality for probabilities and proved that all the
above-mentioned generalized GHZ states, as well as any
2-entangled pure tripartite states\footnote{These are
tripartite states of the form
\begin{equation}
    \ket{\Psi_{\A\B\C}}=\ket{\Psi_{\A\B}}\ten\ket{\Psi_\C},
\end{equation}
where $\ket{\Psi_{\A\B}}$ is a bipartite pure entangled state.}
violate the constructed Bell inequality. In addition, they have
presented some numerical evidence that this inequality is also
violated by other kinds of tripartite entangled pure states.
For $N>3$ parties, some further investigations exist (see
Refs.~\cite{V.Scarani:PRA:2005,C.F.Wu:PRA:2006,C.F.Wu:PRA:2007}
and references therein) but nothing as strong as the results
presented by Gisin and Peres in Ref.~\cite{N.Gisin:PLA:1992} is
known yet.

As for mixed quantum states, Horodecki {\em et~al.} have also
provided an analytic criterion~\cite{RPM.Horodecki:PLA:1995} to
determine if a two-qubit state violates the Bell-CHSH
inequality. This criterion is, unfortunately, also the only
analytic criterion that we have in determining if a broad class
of quantum states, namely two-qubit states, can be simulated by
some LHVM in a standard Bell experiment. Nonetheless, for
specific quantum states, such as those that we have looked at
in Chapter~\ref{Chap:Q.Cn:Classical}, the existence of LHVMs
for these states will exclude the possibility of them violating
a Bell inequality (via measurements where the models are
applicable).

In general, to determine if a quantum state violates a Bell
inequality is a high-dimensional variational problem, which
requires a nontrivial optimization of a Hermitian operator
$\Bell$ (now known as the {\em Bell}
operator~\cite{S.L.Braunstein:PRL:1992}) over the various
possible measurement settings that each observer may perform.
This optimization does not appear to be convex and is possibly
NP-hard~\cite{M.M.Deza:Book:1997}. In fact, a closely related
problem, namely to determine if a given probability vector is a
member of the set of classical correlations is known to be
NP-complete~\cite{I.Pitowsky:Book:1989}.

Except for the simplest scenario  where one deals with the
Bell-CHSH inequality, in conjunction with a two-qubit
state~\cite{RPM.Horodecki:PLA:1995}, or a (bipartite) maximally
entangled pure
state~\cite{N.Gisin:PLA:1992,S.Popescu:PLA:1992b}, and its
mixture with the maximally mixed state~\cite{T.Ito:PRA:2006},
very few analytic results for the optimal measurements are
known. As such, for the purpose of characterizing quantum
states that are incompatible with locally causal description,
efficient algorithms to perform this state-dependent
optimization are very desirable.

On the other hand, state-independent bounds of quantum
correlations have also been investigated since the early 1980s.
In particular, Tsirelson~\cite{B.S.Cirelson:LMP:1980} has
demonstrated, using what is now known as Tsirelson's vector
construction, that in a Bell-CHSH setup, bipartite quantum
systems of arbitrary dimensions cannot exhibit correlations
stronger than $2\sqrt{2}$ - a value now known as Tsirelson's
bound. Recently, analogous bounds for more complicated Bell
inequalities have also been investigated by Filipp and
Svozil~\cite{S.Filipp:PRL:2004}, Buhrman and
Massar~\cite{H.Buhrman:PRA:2005},
Wehner~\cite{S.Wehner:PRA:2006},
Toner~\cite{B.F.Toner:quant-ph:0601172}, Avis {\em
et~al.}~\cite{D.Avis:JPA:2006} and Navascu\'es {\em
et~al.}~\cite{M.Navascues:PRL:2007}. On a related note, bounds
on quantum correlations for given local measurements, rather
than given quantum state, have also been investigated by
Cabello~\cite{A.Cabello:PRL:2004} and Bovino {\em
et~al.}~\cite{F.A.Bovino:PRL:2004}.

The main purpose of this chapter is to look into the
algorithmic aspect of determining if a quantum state can
violate a given Bell inequality. In particular, we will
present, respectively, in Sec.~\ref{Sec:LB} and
Sec.~\ref{Sec:UB}, two algorithms that were developed to
provide a lower bound and an upper bound on the maximal
expectation value of a Bell operator for a given quantum state.
The second algorithm is another instance where a nonlinear
optimization problem is approximated by a hierarchy of
semidefinite programs, each giving a better bound of the
original optimization problem~\cite{M.Navascues:PRL:2007,
P.A.Parrilo:MP:2003,J.B.Lasserre:SJO:2001,A.C.Doherty:PRA:2004,J.Eisert:PRA:2004}.
In its simplest form, it provides a bound that is apparently
state-independent.

In Sec.~\ref{Sec:Bell-CHSH:Analytic}, we will derive, based on
the second algorithm, a necessary condition for a class of
two-qudit states\footnote{A two-qudit state is a bipartite
quantum state describing two $d$-level quantum systems. Some
authors refer to them, instead, as a two-qunit state for two
$n$-level quantum systems.} to violate the Bell-CHSH
inequality. Next, in Sec.~\ref{Sec:Bell-CHSH:Horodecki}, we
will illustrate how the lower bound algorithm can be used to
derive the Horodecki criterion~\cite{RPM.Horodecki:PLA:1995}
for two-qubit states. After that, we will demonstrate how the
two algorithms can be used in tandem to determine if some
quantum states violate a given Bell inequality. Some
limitations of these algorithms will then be discussed. We will
conclude with a summary of results and some possibilities for
future research.

\section{Bounds on Quantum Correlations} \label{Sec:QuantumBounds}

\subsection{Preliminaries}\label{Sec:QuantumBounds:Preliminaries}

In the earlier chapter, we have learned that a particular Bell
inequality deals with a specific experimental setup, say
involving two experimenters Alice and Bob,\footnote{For
definiteness, we will restrict our attention to bipartite
setups and point out, when relevant, how the arguments can be
extended to the multipartite scenario.}  where each of them can
perform, respectively, $\mA$ and $\mB$ alternative measurements
that would each generate $\nA$ and $\nB$ distinct outcomes. For
each of these setups, a Bell inequality places a bound on the
experimental statistics obtained from the corresponding Bell
experiments. In particular, we recall from
Sec.~\ref{Sec:Polytope&BI} that a (linear) Bell inequality
takes the form :
\begin{equation}\label{Eq:Scl:dfn}
    \SLHV\le \betaLHV,
\end{equation}
where $\betaLHV$ is a real number and $\SLHV$ involves a
specific {\em linear} combination of correlation functions or
joint and marginal probabilities of experimental outcomes.

To determine if a quantum state violates a given Bell
inequality with some choice of measurements, we need to
evaluate these correlation functions, or probabilities
according to the quantum mechanical rules [see
Eq.~\eqref{Eq:Dfn:CorrelationFn:Singlet} for an example]. The
bounds on $\SLHV$ then translate into corresponding bounds
$\betaLHV$ on the expectation value of some Hermitian
observable that describes the (standard) Bell inequality
experiment, this observable is known as the {\it Bell operator}
$\Bell$~\cite{S.L.Braunstein:PRL:1992}. The restriction that
the given Bell inequality is satisfied in the experiment is
then
\begin{equation}
    \Sqm(\rho,\Bell) = \tr\left(\rho~\Bell \right)\le \betaLHV.
    \label{Eq:S:dfn}
\end{equation}
The Bell operator depends on the choice of measurements at each
of the sites (polarizer angles for example). These measurements
will be described by a set of Hermitian operators $\{O_m\}$.
For correlation inequalities these are simply the measured
observables at each stage of the Bell measurement, while for
general probability inequalities the $O_m$ are POVM elements
that describe the measurements at each site. We will denote
this expectation value by $\Sqm(\rho, \{O_m\})$ when we want to
emphasize its dependence on the choice of local Hermitian
observables $O_m$. Ideally the choice of measurement should
give the maximal expectation value of the Bell operator, for
which we will give the notation
\begin{equation}\label{Eq:Dfn:Sqm}
    \Sqm(\rho)\equiv\max_{\{O_m\}}\Sqm(\rho,\{O_m\}).
\end{equation}
Moreover, we will explicitly include a superscript in
$\Sqm(\rho)$, e.g. $\SqmBI{$k$}(\rho)$, when we want to make
reference to a specific Bell inequality labeled by ``$k$". It
is this implicitly-defined function that will give us
information about which states violate a given Bell inequality.

As an example, let us recall the Bell-CHSH inequality,
Eq.~\eqref{Ineq:CHSH:Convention}, which is reproduced here for
ease of reference
\begin{equation}\label{Ineq:Bell-CHSH}
    \SLHV^{\mbox{\tiny (CHSH)}}=E(A_1,B_1)+E(A_1,B_2)+E(A_2,B_1)-E(A_2,B_2)\le 2.
\end{equation}
In quantum mechanics, each of these correlation functions
$E(A_{\sA},B_{\sB})$ is computed using
\begin{equation}
    \Eqm(A_{\sA},B_{\sB})=\tr\left(\rho~A_{\sA}\ten B_{\sB}\right).
\end{equation}
Substituting this into Eq.~\eqref{Ineq:Bell-CHSH} and comparing with
Eq.~\eqref{Eq:S:dfn}, one finds that the corresponding Bell operator
reads
\begin{equation}\label{Eq:Bell-CHSH:operator}
    \Bell_\text{CHSH}=A_1\ten(B_1+B_2)+A_2\ten(B_1-B_2).
\end{equation}

To determine the maximal Bell-inequality violation for a given
$\rho$, $\Sqm(\rho)$,  requires a maximization by varying over
all possible choices of $\{O_m\}$, i.e., $A_{\sA}$ and
$B_{\sB}$ in the case of Eq.~\eqref{Eq:Bell-CHSH:operator}.
Whether we are interested in correlation inequalities or in
Bell inequalities for probabilities the (bipartite) Bell
operator has the general structure
\begin{equation}\label{Eq:BellOperator:General}
    \Bell =\sum_{K,L}b_{KL} A_K\ten B_L,
\end{equation}
which essentially follows from the linearity of Bell inequality
as well as the linearity of expectation values in quantum
mechanics. In the case of a Bell inequality for probabilities
the indices $K$, $L$ are collective indices, c.f.
Eq.~\eqref{Eq:SLHV:Generic}, describing both a particular
measurement setting and a particular outcome for each observer;
the $A_K$ and $B_L$ are then POVM elements corresponding to
specific outcomes in the Bell experiment. For correlation
inequalities, the indices $K$, $L$ refer simply to the
measurement settings as in the Bell-CHSH case described in
detail above.

In what follows, we will present two algorithms which we have
developed specifically to perform the maximization over the
choice of measurements. The first, which we will abbreviate as
LB, provides a {\em lower bound} on the maximal expectation
value and can be implemented for any Bell inequality. This
bound makes use of the fact that the objective function
$\Sqm(\rho,\{O_m\})$ is bilinear in the observables $O_m$, that
is it is linear in the $A_K$ for fixed $B_L$ and likewise
linear in the $B_L$ for fixed $A_K$. The second bound, which we
will abbreviate as UB, provides an {\em upper bound} on
$\Sqm(\rho)$ by regarding $\Sqm(\rho,\{O_m\})$ as a polynomial
function of the variables that define the various $O_m$ and
applying general techniques for finding such bounds on
polynomials~\cite{P.A.Parrilo:MP:2003,J.B.Lasserre:SJO:2001}.
Both of these make use of convex optimization techniques in the
form of a semidefinite program (SDP).  An SDP is a linear
optimization over positive semidefinite (PSD) matrices which
are subjected to affine constraints. Readers who are unfamiliar
with semidefinite programming are referred to
Appendix~\ref{App:Sec:SDP}.

\subsection{Algorithm to Determine a Lower Bound on $\Sqm(\rho)$}
\label{Sec:LB}

The key idea behind the LB algorithm is to realize that when
measurements for all but one party are {\em fixed}, the optimal
measurements for the remaining party can be obtained
efficiently using convex optimization techniques, in particular
an SDP. Thus we can fix Bob's measurements and find Alice's
optimal choice, at least numerically; with these optimized
measurements for Alice, we can further find the optimal
measurements for Bob (for this choice of Alice's settings), and
then Alice again and so on and so forth until
$\Sqm(\rho,\{O_m\})$ converges within the desired numerical
precision.\footnote{That such an iterative algorithm using SDP
can lead to a local maximum of $\Sqm(\rho,\POVMAg,\POVMBg)$ was
also discovered independently by Ito {\em et
al.}~\cite{T.Ito:PRA:2006}.}

Back in 2001, Werner and Wolf~\cite{R.F.Werner:QIC:2001} presented a
similar iterative algorithm, by the name of {\em See-Saw iteration}, to
maximize the expectation value of the Bell operator for a Bell
correlation inequality involving only dichotomic
observables.\footnote{A dichotomic observable is a Hermitian observable
with only two distinct eigenvalues.} As a result we will focus here on
the (straightforward) generalization to the widest possible class of
Bell inequalities. In the work of Werner and
Wolf~\cite{R.F.Werner:QIC:2001} it turned out that once the dichotomic
observables for one party are fixed, optimization of the other party's
observables can be carried out explicitly. This turns out to be true
for any dichotomic Bell inequality and we will return to this question
in Sec.~\ref{Sec:two-outcome}.

\subsubsection{General Settings}

Let us now consider a Bell inequality for probabilities for
$\polycl$.\footnote{For a Bell correlation inequality, we can
apply LB by first rewriting the corresponding Bell operator,
Eq.~\eqref{Eq:BellOperator:Correlation1}, in terms of POVM
elements that form the measurement outcomes of each $O_m$.} We
will denote the POVM element associated with the $\idx{\oA}$
outcome of Alice's $\idx{\sA}$ measurement by $\POVMAg$  while
$\POVMBg$ is the POVM element associated with the $\idx{\oB}$
outcome of Bob's $\idx{\sB}$ measurement. Moreover, let $\dA$
and $\dB$, respectively, be the dimension of the state space
that each of the $\POVMAg$ and $\POVMBg$ acts on. Then it
follows from Born's rule that
\begin{subequations}
    \begin{gather}
        \ProbTwGJ= \tr\left(\rho\,\POVMAg\ten
        \POVMBg\right)\tag{\ref{Eq:QM:probabilities:joint}}\\
        \ProbTwGMA=\tr\left(\rho\,\POVMAg\ten \unit_{\dB}\right),\quad
        \ProbTwGMB= \tr\left(\rho\,\unit_{\dA}\ten
        \POVMBg\right),\tag{\ref{Eq:QM:probabilities:marginal}}
    \end{gather}
\end{subequations}
where, as defined in Sec.~\ref{Sec:Bell-CH}, $\ProbTwGJ$ refers
to the joint probability that the $\idx{\oA}$ experimental
outcome is observed at Alice's site and the $\idx{\oB}$ outcome
at Bob's, given that Alice performs the $\idx{\sA}$ measurement
and Bob performs the $\idx{\sB}$ measurement; likewise for the
marginal probabilities $\ProbTwGMA$ and $\ProbTwGMB$. A general
Bell operator for probabilities can then be expressed as
\begin{equation}\label{Eq:BellOperator:Probabilities}
    \Bell =\sum_{\sA=1}^{\mA}\sum_{\oA=1}^{\nA}\sum_{\sB=1}^{\mB}
    \sum_{\oB=1}^{\nB} \CoeffG \POVMAg\ten
    \POVMBg,
\end{equation}
where $\CoeffG$ are determined from the given Bell inequality,
c.f. Eq.~\eqref{Eq:SLHV:Generic}. Again, one is reminded that
the sets of POVM elements
$\left\{\POVMAg\right\}_{\oA=1}^{\nA}$ and
$\left\{\POVMBg\right\}_{\oB=1}^{\nB}$ satisfy
\begin{subequations}
    \begin{gather}
        \sum_{\oA=1}^{\nA}\POVMAg=\unit_{\dA}\quad
        \text{and}\quad\sum_{\oB=1}^{\nB}\POVMBg=\unit_{\dB}
        \quad\forall\quad \sA, \sB,\tag{\ref{Eq:POVM:Normalization}}\\
         \POVMAg\ge 0,\quad \POVMBg\ge
        0\quad\forall\quad \sA,
        \sB,\oA,\oB.\tag{\ref{Eq:POVM:PSD}}
    \end{gather}
\end{subequations}

\subsubsection{Iterative Semidefinite Programming Algorithm}

To see how to develop a lower bound on $\Sqm(\rho)$ by fixing
the observables at one site and optimizing  the other, we
observe that upon substituting
Eq.~\eqref{Eq:BellOperator:Probabilities} into
Eq.~\eqref{Eq:S:dfn}, the {\em lhs} of the inequality can be
rewritten as
\begin{align}
    \Sqm(\rho,\POVMAg,\POVMBg)&=\sum_{\sB,\oB} \tr
    \left(\rho_{\POVMBg}\,\POVMBg\right),\label{Eq:S2}
\end{align}
where
\begin{equation}\label{Eq:Dfn:rhoAB}
    \rho_{\POVMBg}\equiv\sum_{\sA,\oA}\CoeffG~
    \tr_\A\left[\rho\left(\POVMAg\ten\unit_{\dB}\right)\right],
\end{equation}
and $\tr_\A~\cdot$ is the partial trace over subsystem $\A$.

Notice that if all $\rho_{\POVMBg}$ are held constant by fixing
all of Alice's measurement settings  (given by the set of
$\POVMAg$) then $\rho_{\POVMBg}$ is a constant matrix
independent of the $\POVMBg$. Thus the objective function is
linear in these variables. The constraints that
$\{\POVMBg\}_{\oB=1}^{\nB}$ form a POVM for each value of $\sB$
is a combination of affine and matrix nonnegativity
constraints. As a result it is fairly clear that the following
problem is an SDP in {\em standard form},
Eq.~\eqref{Eq:SDP:Matrix},
\begin{subequations}\label{Eq:sdpiter}
\begin{eqnarray}
    &{\rm maximize}_{\{\POVMBg\}}&\Sqm(\rho,\POVMAg,\POVMBg)\label{Eq:SDPIter:obj}\\
    &{\rm subject \ to} \quad &
    \sum_{\oB=1}^{\nB}\POVMBg=\unit_{\dB}
    \quad\forall\quad \sB, \label{Eq:SDPIter:eq}\\
     && \POVMBg\ge
    0\quad\forall\quad \sB, \oB. \label{Eq:SDPIter:ineq}
    \end{eqnarray}
\end{subequations}
The detailed formulation of this optimization problem in terms
of an SDP in standard form can be found in
Appendix~\ref{App:Sec:SDP:LB}.

Exactly the same analysis follows if we fix Bob's measurement
settings, and optimize over Alice's POVM elements instead. To
arrive at a local maximum of $\Sqm(\rho,\POVMAg,\POVMBg)$, it
therefore suffices to start with some random measurement
settings for Alice (or Bob), and optimize over the two parties'
settings iteratively.  A (nontrivial) lower bound on
$\Sqm(\rho)$ can then be obtained by optimizing the measurement
settings starting from a set of randomly generated initial
guesses.

It is worth noting that in any implementation of this
algorithm, physical observables $\{\POVMAg, \POVMBg\}$
achieving the lower bound are constructed when the
corresponding SDP is solved. In the event that the lower bound
is greater than the classical threshold $\betaLHV$, then these
observables can, in principle, be measured in the laboratory to
demonstrate a  Bell-inequality violation of the given quantum
state.

We have implemented this algorithm in MATLAB\footnote{MATLAB is
a trademark of  The Math Works, Inc., Natick, MA.} to search
for a lower bound on $\Sqm(\rho)$ in the case of Bell-CH,
$I_{3322}$, $I_{4422}$, $I_{2233}$ and $I_{2244}$ inequalities
(Sec.~\ref{Sec:BipartiteBI}), and with the local dimension
$d=\dA=\dB$ up to 32. Typically, with no more than 50
iterations, the algorithm already converges to a point that is
different from a local maximum by no more than $10^{-9}$. To
test the effectiveness of finding $\Sqm(\rho)$ using LB, we
have randomly generated 200 Bell-CH violating two-qubit states
and found that on average, it takes about 6 {\em random initial
guesses} before the algorithm gives $\Sqm(\rho,\{O_m\})$ that
is close to the actual maximum, computed using Horodecki's
criterion~\cite{RPM.Horodecki:PLA:1995}, to within $10^{-5}$.
Specific examples of the implementation of this algorithm will
be discussed in Sec.~\ref{Sec:examples} and
Chapter~\ref{Chap:BellViolation}.

Two other remarks concerning this algorithm should now be made.
Firstly,  the algorithm is readily generalized to multipartite
Bell inequalities for probabilities: one again starts with some
random measurement settings for all but one party, and
optimizes over each party iteratively. Also, it is worth noting
that this algorithm is not only useful as a numerical tool, but
for specific cases, it can also provide a useful analytic
criterion. In particular, when applied to the Bell-CH
inequality for two-qubit states, the LB algorithm may lead us
to the Horodecki criterion~\cite{RPM.Horodecki:PLA:1995},
i.e.,~the necessary and sufficient condition for two-qubit
states to violate the Bell-CH/ Bell-CHSH inequality (see
Sec.~\ref{Sec:Bell-CHSH:Horodecki} and
Appendix~\ref{App:Sec:HorodeckiCriterion} for more details).

\subsubsection{Two-outcome Bell Experiment}
\label{Sec:two-outcome}

We will show that, just as in the case of Bell correlation
inequalities~\cite{R.F.Werner:QIC:2001}, the local optimization
can be solved analytically for two-outcome measurements. If we
denote by ``$\pm$" the two outcomes of the experiments, it
follows from Eq.~\eqref{Eq:POVM} that the POVM element
$B^-_{\sB}$ can be expressed as a function of the complementary
POVM element $B^+_{\sB}$,
i.e.,~$B^-_{\sB}=\unit_{\dB}-B^+_{\sB}$, subjected to $0\le
B^+_{\sB} \le \unit_{\dB}$. We then have
\begin{align*}
    \sum_{\oB=\pm}\tr \left(\rho_{\POVMBg}\,\POVMBg\right)&=
    \tr \left[\left(\rho_{B^+_{\sB}}-\rho_{B^-_{\sB}}\right)B^+_{\sB}\right] +\tr
    \left(\rho_{B^-_{\sB}}\right).
\end{align*}
The above expression can be maximized by setting the PSD
operator $B^+_{\sB}$ to be the projector onto the positive
eigenspace of $\rho_{B^+_{\sB}}-\rho_{B^-_{\sB}}$. In a similar
manner, we can also write
\begin{align*}
    \sum_{\oB=\pm}\tr
    \left(\rho_{\POVMBg}\,\POVMBg\right)=\tr
    \left[\left(\rho_{B^-_{\sB}}-\rho_{B^+_{\sB}}\right)B^-_{\sB}\right]+\tr
    \left(\rho_{B^+_{\sB}}\right),
\end{align*}
which can be maximized by setting  $B^-_{\sB}$ to be the
projector onto the non-positive eigenspace of
$\rho_{B^+_{\sB}}-\rho_{B^-_{\sB}}$. Notice that this choice is
consistent  with our earlier choice of $B\,^+_{\sB}$ for the
``$+$" outcome POVM element in that they form a valid POVM.
Since there can be no difference in these maxima, we may write
the maximum as their average, i.e.,~
\begin{align*}
    \sum_{\oB=\pm}\tr
    \left(\rho_{\POVMBg}\,\POVMBg\right)&=\half\left|\left|
    \rho_{B^+_{\sB}}-\rho_{B^-_{\sB}}\right|\right|
    +\half\sum_{\oB=\pm}\tr\left(\rho_{\POVMBg}\right),
\end{align*}
where $||O||$ is the trace norm of the Hermitian operator
$O$~\cite{R.A.Horn:Book:1990,G.Strang:Book:1988}. Carrying out
the optimization for each of the $\mB$ settings, the optimized
$\Sqm(\rho,\POVMAg,\POVMBg)$, as an implicit function  of
Alice's POVM $\left\{\POVMAg\right\}$, is given by
\begin{align}
    \Sqm(\rho,\POVMAg)&=\half\sum_{\sB}\left|
    \left|\rho_{B^+_{\sB}}-\rho_{B^-_{\sB}}\right|\right|
    +\half\sum_{\sB}\sum_{\oB=\pm}\tr
    \left(\rho_{\POVMBg}\right).\label{Eq:S:implicitA}
\end{align}
Notice that this calculation is essentially the same as that
which shows that the Helstrom
measurement~\cite{C.W.Helstrom:Book:1976} is optimal for
distinguishing two quantum states.

An immediate corollary of the above result is that for the
optimization of a two-outcome Bell operator for probabilities,
it is unnecessary for any of the two observers to perform
generalized measurements described by a POVM; von Neumann
projective measurements are sufficient.\footnote{As was pointed
out in  Ref.~\cite{T.Ito:PRA:2006}, this sufficiency also
follows from Theorem 5.4 in Ref.~\cite{R.Cleve:Pr.IEEE:2004}.}
In practice, this simplifies any analytic treatment of the
optimization problem as a generic parametrization of a POVM is
a lot more difficult to deal with, thereby supporting the
simplification adopted in Ref.~\cite{S.Filipp:PRL:2004}.

Nevertheless, it may still be advantageous to consider generic
POVMs as our initial measurement settings when implementing the
algorithm numerically. This is because the local maximum of
$\Sqm(\rho,\{O_m\})$ obtained using the iterative procedure is
a function of the initial guess. In particular, it was found
that the set of local maxima attainable could change
significantly if the ranks of the initial measurement
projectors are altered. As such, it seems necessary to step
through various ranks of the starting projectors to obtain a
good lower bound on $\Sqm(\rho)$. Even then, we have also found
examples where this does not give a lower bound on $\Sqm(\rho)$
that is as good as when generic POVMs are used as the initial
measurement operators.

\subsection{Algorithm to Determine an Upper Bound on $\Sqm(\rho)$}
\label{Sec:UB}

A major drawback of the above algorithm, or the analogous
algorithm developed by Werner and
Wolf~\cite{R.F.Werner:QIC:2001} for Bell correlation
inequalities is that, except in some special cases, it is
generally impossible to tell if the maximal
$\Sqm(\rho,\{O_m\})$ obtained through this optimization
procedure corresponds to the global maximum $\Sqm(\rho)$.

Nontrivial upper bounds on $\Sqm(\rho)$, nevertheless, can be
obtained by considering {\em relaxations} of the global
optimization  problem given by Eq.~\eqref{Eq:Dfn:Sqm}. In a
relaxation, a (possibly nonconvex) maximization problem is
modified in some way so as to yield a more tractable
optimization that bounds the optimization of interest. One
example of a variational upper bound that exists for any
optimization problem is the {\em Lagrange dual} optimization
that arises in the method of Lagrange
multipliers~\cite{S.Boyd:Book:2004}.

To see how to apply existing studies in the optimization
literature to find upper bounds on $\Sqm(\rho)$, let us first
remark that the global objective function $\Sqm(\rho,\{O_m\})$
can be mapped to a polynomial function in real variables, for
instance, by expanding all the local observables $\{O_m\}$ and
the density matrix $\rho$ in terms of Hermitian basis
operators.  In the same manner, matrix equality constraints,
such as that given in Eq.~\eqref{Eq:POVM:Normalization} can
also mapped to a set of polynomial equalities by requiring that
the matrix equality holds component wise. Now, it is known from
the work of Lasserre~\cite{J.B.Lasserre:SJO:2001} and
Parrilo~\cite{P.A.Parrilo:MP:2003} that a hierarchy of global
bounds of a polynomial function, subjected to polynomial
equalities and inequalities, can be achieved by solving
suitable SDPs. Essentially, this is achieved by approximating
the original nonconvex optimization problem by a series of
convex ones in the form of a SDP, each giving a better bound of
the original polynomial objective function.

At the bottom of this hierarchy is the lowest order relaxation
provided by the {\em Lagrange dual} of the original nonconvex
problem. By considering {\em Lagrange multipliers} that depend
on the original optimization variables, higher order
relaxations to the original problem can be constructed to give
tighter upper bounds on $\Sqm(\rho)$ (see
Appendix~\ref{Sec:semidefinite relaxation} for more details).

In the following, we will focus our discussion on a general
two-outcome Bell (correlation) inequality, where the
observables $\{O_m\}$ are only subjected to matrix equalities.
In particular, we will show that the global optimization
problem for these Bell inequalities is a {\em
quadratically-constrained quadratic-program} (QCQP), i.e., one
whereby the objective function and the constraints are both
{\em quadratic} in the optimization variables. Then, we will
demonstrate explicitly how the Lagrange dual of this QCQP,
which is known to be an SDP, can be constructed. The analogous
analytic treatment is apparently formidable for higher order
relaxations. Nonetheless, there exists third-party MATLAB
toolbox known as the SOSTOOLS which is tailored specifically
for this kind of optimization
problem~\cite{SOSTOOLS,S.Prajna:LNCIS:2005}.

Numerically, we have implemented the algorithm for several
two-outcome correlation inequalities and will discuss the
results in greater detail in Sec.~\ref{Sec:examples}. For a
general Bell inequality where each $O_m$ is also subjected to a
linear matrix inequality (henceforth abbreviated as LMI) like
Eq.~\eqref{Eq:POVM:PSD}, the algorithm can still be
implemented, for instance, by requiring that all the principle
submatrices of $O_m$ have non-negative
determinants~\cite{R.A.Horn:Book:1990,G.Strang:Book:1988}. This
then translates into a set of polynomial inequalities which fit
into the framework of a general polynomial optimization problem
(see Appendix~\ref{Sec:semidefinite relaxation}). However a
more effective approach would retain the structure of linear
matrix inequalities constraining a polynomial optimization
problem; we leave the investigation of these bounds to further
work.

\subsubsection{Global Optimization Problem}
\label{Sec:GlobalOptimizationProblem}

Now, let us consider a dichotomic Bell correlation inequality
where Alice an Bob can respectively perform $\mA$ and $\mB$
alternative measurements. A general Bell correlation operator
for such an experimental setup can be written
as\footnote{Strictly, a general Bell correlation operator may
also contain marginal terms like $O_m\ten\unit_{\dB}$ and
$\unit_{\dA}\ten O_m$, that result from {\em restricted
correlation function},
Eq.~\eqref{Eq:Dfn:MarginalCorrelationFn}, defining the
correlation inequality. For brevity, we will not consider such
Bell operators in the following discussion. It should,
nevertheless, be clear to the readers that the following
arguments are readily generalized to include Bell operators of
this more general kind.}
\begin{equation}
    \Bell =\sum_{\sA=1}^{\mA}\sum_{\sB=1}^{\mB} \CoeffCnG O_{\sA}\ten
    O_{\sB+\mA},\label{Eq:BellOperator:Correlation1}
\end{equation}
where $\CoeffCnG$ are determined from the given Bell
correlation inequality, $O_{\sA}$ for $\sA=1,\ldots,\mA$ refers
to the $\idx{\sA}$ Hermitian observable measured by Alice, and
$O_{\sB+\mA}$ for $\sB=1,\ldots,\mB$ refers to the $\idx{\sB}$
Hermitian observable measured by Bob. Furthermore, these
dichotomic observables are usually chosen to have eigenvalues
$\pm1$ and thus
\begin{gather}
    O_m^\dag O_m=(O_m)^2=\unit_{d}
    \label{Eq:Constraint:Equality}
\end{gather}
for all $m=1,2,\ldots,\mA+\mB$, where we have assumed for
simplicity that all the local observables $O_m$ act on a state
space of dimension $d$.\footnote{In general, the composite
system may consist of subsystems of different dimensions. All
the following arguments can be readily generalized to this more
general scenario.}

The global optimization problem derived from a dichotomic Bell
correlation inequality thus takes the form of
\begin{subequations}\label{Eq:GlobalOptimization:Correlation}
    \begin{align}
        &\text{maximize \ \ \ } \tr\left(\rho~\Bell\right),
        \label{Eq:GlobalOptimization:Correlation:Obj}\\
        &\text{subject to \ \ } O_m^2=\unit_d
        \label{Eq:GlobalOptimization:Correlation:Constraint}
    \end{align}
\end{subequations}
for all $m=1,2,\ldots,\mA+\mB$. For any $m\times n$ complex
matrices, we will now define \vvec{$A$} to be the $m\cdot n$
dimensional vector obtained by stacking all columns of $A$ on
top of one another. By collecting all the vectorized
observables together
\begin{equation*}
    \bfw^\dag\equiv[\vvec{O_1}^\dag\,\,\vvec{O_2}^\dag\,
    \ldots\,\,\vvec{O_{\mA+\mB}}^\dag],
\end{equation*}
and using the identity
\begin{equation}
    \tr(\rho~O_{\sA}\ten
    O_{\sB+\mA})=\vvec{O_{\sA}}^\dag(V\rho)
    ^{\mbox{\tiny $T_\A$}}\vvec{O_{\sB+\mA}},
\end{equation}
with $V$ being the flip operator introduced in
Eq.~\eqref{Eq:WernerState:Original} and $(.)^{\mbox{\tiny
$T_\A$}}$ being the partial transposition with respect to
subsystem $\A$, we can write the objective function more
explicitly as
\begin{equation}\label{Eq:eBell:Correlation:Poly}
    \Sqm(\rho,\{O_m\})=\tr(\rho~\Bell)=-\bfw^\dag\Omega_0\bfw
\end{equation}
where
\begin{equation}\label{Eq:Dfn:Omega0}
    \Omega_0\equiv\half\left(
    \begin{array}{cc}
    \zero & -b\ten R \\
    -b^T\ten R^\dag & \zero\\
    \end{array}\right),
\end{equation}
$b$ is a $\mA\times \mB$ matrix with $[b]_{\sA,\sB}=\CoeffCnG$,
c.f. Eq.~\eqref{Eq:BellOperator:Correlation1}, and $R\equiv
(V\rho)^{\mbox{\tiny $T_A$}}$. In this form, it is explicit
that the objective function is {\em quadratic} in $\vvec{O_m}$.
Similarly, by requiring that the matrix equality,
Eq.~\eqref{Eq:Constraint:Equality}, holds component-wise, we
can get a set of equality constraints, which are each {\em
quadratic} in $\vvec{O_m}$. The global optimization problem
given by Eq.~\eqref{Eq:GlobalOptimization:Correlation} is thus
an instance of a QCQP.

On a related note, for any Bell inequality experiments where
measurements are restricted to the projective type, the global
optimization problem is also a QCQP. To see this, we first note that
the global objective function for the general case, as follows from
Eq.~\eqref{Eq:S:dfn} and Eq.~\eqref{Eq:BellOperator:General}, is always
quadratic in the local Hermitian observables $\{A_K,B_L\}$. The
requirement that these measurement operators are projectors amounts to
requiring
\begin{align}
    A_K^2=A_K,\quad B_L^2=B_L,\quad \forall~K,L,
\end{align}
which are quadratic constraints on the local Hermitian
observables. Since we have shown in Sec.~\ref{Sec:two-outcome}
that for any two-outcome Bell inequality for probabilities, it
suffices to consider projective measurements in optimizing
$\Sqm(\rho,\{O_m\})$, it follows that the global optimization
problem for these Bell inequalities is always a QCQP.

\subsubsection{State-independent Bound}
\label{Sec:State-independent Bound}

As mentioned above, the lowest order relaxation to the global
optimization problem --- given by
Eq.~\eqref{Eq:GlobalOptimization:Correlation} --- is simply the
Lagrange dual of the original QCQP. This can be obtained via
the {\em Lagrangian}~\cite{S.Boyd:Book:2004} of the global
optimization problem, i.e.,
\begin{equation}
    \mathcal{L}(\{O_m\},\Lambda_m)=\Sqm(\rho,\{O_m\})-\sum_{m=1}^{\mA+\mB}
    \tr\left[\Lambda_m\left(O_m^2-\unit_d\right)\right],
    \label{Eq:Lagrangian}
\end{equation}
where $\Lambda_m$ is a matrix of Lagrange multipliers
associated with the $m^{\rm th}$ matrix equality constraint.
With no loss of generality, we can assume that the
$\Lambda_m$'s are Hermitian.

Notice that for all values of $\{O_m\}$ that satisfy the
constraints, the Lagrangian
\begin{equation*}
    \mathcal{L}(\rho, \{O_m\},\Lambda_m)= \Sqm(\rho,\{O_m\}).
\end{equation*}
As a result, if we maximize the Lagrangian without regard to
the constraints we obtain an upper bound on the maximal
expectation value of the Bell operator
\begin{equation}
  \max_{\{O_m\}} \mathcal{L} (\rho, \{O_m\},\Lambda_m) \geq \Sqm(\rho).
\end{equation}
The Lagrange dual optimization simply looks for the best such
upper bound.

In order to maximize the Lagrangian we rewrite the Lagrangian with
Eq.~\eqref{Eq:eBell:Correlation:Poly} and the identity
\begin{equation}
    \tr\left(\Lambda_mO_mO_m^\dag\right)=\vvec{O_m}^\dag
    \left(\unit_{d}\ten\Lambda_m\right)\vvec{O_m},
\end{equation}
to obtain
\begin{equation}
    \mathcal{L}(\bfw,
    \Lambda_m)=-\bfw^\dag\Omega\bfw+\sum_{m=1}^{\mA+\mB}\tr~\Lambda_m,
    \label{Eq:Lagrangian:Explicit}
\end{equation}
where
\begin{equation}\label{Eq:Dfn:Omega}
    \Omega\equiv\Omega_0+\bigoplus_{m=1}^{\mA+\mB}\unit_{d}\ten\Lambda_m.
\end{equation}
Note that each of the diagonal blocks $\unit_{d}\ten \Lambda_m$
is of the same size as the matrix $R$.

To obtain the dual optimization problem, we maximize the
Lagrangian over $\bfw$ to obtain the {\em Lagrange dual
function}
\begin{equation}\label{Eq:Lagrange Dual}
    g(\Lambda_m)\equiv \sup_{\bfw}\mathcal{L}(\bfw,\Lambda_m).
\end{equation}
As noted above $g(\Lambda_m)\ge \Sqm(\rho)$ for all choices of
$\Lambda_m$. Moreover, this supremum  over $\bfw$ is unbounded
above unless $\Omega \ge 0$, in which case the supremum is
attained by setting $\bfw=0$ in
Eq.~\eqref{Eq:Lagrangian:Explicit}. Hence, the {\em Lagrange
dual} optimization, which seeks for the best upper bound of
Eq.~\eqref{Eq:GlobalOptimization:Correlation} by minimizing
Eq~\eqref{Eq:Lagrange Dual} over the Lagrange multipliers,
reads
\begin{gather}
    \text{minimize \ \ } \sum_{m=1}^{\mA+\mB}\tr~\Lambda_m,\nonumber\\
    \text{subject to \ \ \ }\quad\Omega \geq
    0.\qquad\label{Eq:SDP-relaxation:deg2}
\end{gather}
By expanding $\Lambda_m$ in terms of Hermitian basis operators
satisfying Eq.~\eqref{Eq:basis:Gell-Mann},
\begin{equation}\label{Eq:Lambda:Expansion}
    \Lambda_m=\sum_{n=0}^{d^2-1}\lambda_{mn}\sigma_n,
\end{equation}
the optimization problem given by
Eq.~\eqref{Eq:SDP-relaxation:deg2} is readily seen to be an SDP
in the {\em inequality form}, Eq.~\eqref{Eq:SDP:Vector}.

For Bell-CHSH inequality and the correlation equivalent of
$I_{3322}$ inequality given by
Eq.~\eqref{Ineq:Cor:I3322:Explicit}, it was observed
numerically that the upper bound obtained via the SDP
\eqref{Eq:SDP-relaxation:deg2} is always {\em
state-independent}. For 1000 randomly generated two-qubit
states, and 1000 randomly generated two-qutrit states, the
Bell-CHSH upper bound of $\Sqm(\rho)$ obtained through
\eqref{Eq:SDP-relaxation:deg2} was never found to differ from
the Tsirelson bound~\cite{B.S.Cirelson:LMP:1980} by more than
$10^{-7}$. In fact by finding an explicit feasible solution to
the optimization problem dual to
Eq.~\eqref{Eq:SDP-relaxation:deg2}, Wehner has shown that the
upper bound obtained here can be no better than that obtained
by Tsirelson's vector construction for correlation
inequalities\footnote{S.~Wehner (private communication). See
also Ref.~\cite{S.Wehner:PRA:2006}.}.

In a similar manner, we have also investigated the upper bound
of $\Sqm(\rho)$ for some dichotomic Bell probability
inequalities using the lowest order relaxation to the
corresponding global optimization problem. Interestingly, the
numerical upper bounds obtained from the analog of
Eq.~\eqref{Eq:SDP-relaxation:deg2} for these inequalities --
namely the Bell-CH inequality, Eq.~\eqref{Ineq:CH:Functional},
the $I_{3322}$ inequality, Eq.~\eqref{Ineq:Functional:I3322},
and the $I_{4422}$ inequality, Eq.~\eqref{Ineq:Prob:I4422} ---
are also found to be state-independent and are given by
0.207~106~7, 0.375 and 0.669~346~1 respectively.

\subsubsection{State-dependent Bound}
\label{Sec:State-dependent Bound}

Although the state-independent upper bounds obtained above are
interesting in their own right, our main interest here is to find an
upper bound  on $\Sqm(\rho)$ that does depend on the given quantum
state $\rho$. This can be obtained, with not much extra cost, from
the Lagrange dual to a more-refined version of the original
optimization problem.

To appreciate that, let us first recall that each dichotomic
Hermitian observable $O_m$ can only have eigenvalues $\pm1$. It
follows that their trace\begin{subequations}\label{Eq:z}
\begin{equation}
    z_m\equiv\tr(O_m),\label{Eq:z_m:Defn}
\end{equation}
can only take on the following values
\begin{equation}
    z_m=-d, -d+2,\ldots,d-2, d.\label{Eq:Range:z}
\end{equation}\end{subequations}
In particular, if $z_{m}=\pm d$ for any $m$, then
$O_m=\pm\unit_d$ and it is known that the Bell-CHSH inequality
cannot be violated for this choice of
observable~\cite{S.L.Braunstein:PRL:1992} (see also
Appendix~\ref{App:Sec:FullRank}).

Better Lagrange dual bounds arise from taking these additional
constraints \eqref{Eq:z} explicitly into account. For that
matter, we found it most convenient to express the original
optimization problem in terms of real variables given by the
expansion coefficients of $O_m$ in terms of a basis for
Hermitian matrices that includes the (traceless) Gell-Mann
matrices and the identity matrix, c.f.
Eq.~\eqref{Eq:basis:Gell-Mann}. The resulting calculation is
very similar to what we have done in the previous section (for
details see Appendix~\ref{App:Sec:Fixed-rank}). Here, we will
just note that the result is a set of SDPs, one for each of the
various choices of $z_m$. The lowest order upper bound on
$\Sqm(\rho)$ can then be obtained by stepping through the
various choices of $z_m$ given in Eq.~\eqref{Eq:Range:z},
solving each of the corresponding SDPs, and taking their
maximum. The results of this approach will be discussed later,
for now it suffices to note that tighter bounds can be obtained
that are explicitly state dependent.

\subsubsection{Higher Order Relaxations}
The higher order relaxations simply arise from allowing the
Lagrange multipliers $\lambda$ to be polynomial functions of
$\{O_m\}$ rather than constants. In this case, it is no longer
possible to optimize over the primal variables in the
Lagrangian analytically  but let us consider the following
optimization
\begin{align}
    &{\rm minimize\ \ \ }\qquad\qquad  \gamma \nonumber \\
    &{\rm subject \ to}\quad  \gamma - \Sqm(\rho, \bfx) = \mu(\bfx)
    +\sum_{i} \lambda_i(\bfx) f_{\text{eq},i}(\bfx),
    \label{Eq:HigherOrder QCQP}
\end{align}
where each $\lambda_i(\bfx)$ is a polynomial function of $\bfx$
and $\mu(\bfx)$ is a sum of squares (SOS) polynomial and
therefore non-negative. That is $\mu(\bfx) = \sum_j
[h_j(\bfx)]^2\geq 0$ for some set of real polynomials
$h_j(\bfx)$. The variables of the optimizations are $\gamma$
and the coefficients that define the polynomials $\mu(\bfx)$
and $\lambda_i(\bfx)$. Notice that we have $\gamma \geq
\Sqm(\rho,\bfx) $ whenever the constraints are satisfied so
that once again we have a global upper bound on
$\Sqm(\rho,\bfx)$. This optimization can be implemented
numerically by restricting $\mu(\bfx)$ and  $\lambda_i(\bfx)$
to be of some fixed degree. The Lagrange dual
optimization~\eqref{Eq:SDP-relaxation:deg2} arises from
choosing the degree of $\lambda_i(\bfx)$ to be zero. It is
known that for any fixed degree this optimization is an
SDP~\cite{P.A.Parrilo:MP:2003,J.B.Lasserre:SJO:2001} and we
have implemented up to degree four using
SOSTOOLS~\cite{SOSTOOLS,S.Prajna:LNCIS:2005}. Schm\"{u}dgen's
theorem~\cite{K.Schmudgen:MA:1991} guarantees that by
increasing the degree of the polynomials in the relaxation we
obtain bounds approaching the true maximum $\Sqm(\rho)$. This
is a special case of the general procedure described
in~\cite{P.A.Parrilo:MP:2003,
S.Prajna:LNCIS:2005,K.Schmudgen:MA:1991} which is also able to
handle inequality constraints. For more details see
Appendix~\ref{Sec:semidefinite relaxation}.

\section{Applications \& Limitations of the Two Algorithms}
\label{Sec:examples}

In this section, we will look at some concrete examples of how
the two algorithms can be used to determine if some quantum
states violate a Bell inequality. Specifically, we begin by
looking at how the second algorithm can be used to determine,
both numerically and analytically, if some bipartite qudit
state violates the Bell-CHSH inequality. Then in
Sec.~\ref{Sec:Bell-CHSH:Horodecki}, we will illustrate how LB
can used to recover the Horodecki criterion. After that, in
Sec.~\ref{Sec:Bell-3322}, we demonstrate how the two algorithms
can be used in tandem to determine if a class of two-qubit
states violate the $I_{3322}$ inequality,
Eq.~\eqref{Ineq:Functional:I3322}. We will conclude this
section by pointing out some limitations of the UB algorithm
that we have observed.

\subsection{Bell-CHSH violation for Two-Qudit
States}\label{Sec:Bell-CHSH:Analytic}

The Bell-CHSH inequality, as given by
Eq.~\eqref{Ineq:Bell-CHSH}, is one that amounts to choosing
[c.f. Eq.~\eqref{Eq:BellOperator:Correlation1}]
\begin{equation}
    b=\left(\begin{array}{cc}
    1 & 1\\
    1 & -1
    \end{array}\right).\label{Eq:b-CHSH:dfn}
\end{equation}
For low-dimensional quantum systems, an upper bound on
$\SqmCHSH(\rho)$ can be efficiently computed in MATLAB
following the procedures described in
Sec.~\ref{Sec:State-dependent Bound}. However, for
high-dimensional quantum systems, intensive computational
resources are required to compute this upper bound, which may
render the computation infeasible in practice. In this regard,
it is worth noting that for each choice of the Lagrange
multipliers in the Lagrange dual function~\eqref{Eq:Lagrange
Dual}, there is a corresponding upper bound on
$\SqmCHSH(\rho)$. In fact, for a specific class of two-qudit
states, namely those whose coherence vectors\footnote{These are
generalization of the Bloch vectors representation for higher
dimensional quantum systems. See also
Refs.~\cite{M.S.Byrd:PRA:2003,G.Kimura:PLA:2003}.} vanish, and
using some choice of the Lagrange multipliers, it can be shown
(Appendix~\ref{App:Sec:UpperBound:Analytic}) that
$\SqmCHSH(\rho)$ cannot exceed
\begin{equation}\label{Eq:CHSHViolation:UB}
    \max_{z_1,z_2,z_3,z_4} 2\sqrt{2}s_1d\sqrt{\prod_{i=1}^2
    \frac{2d^2-z_{2i-1}^2-z_{2i}^2}{2d^2}}
    +\sum_{\sA,\sB=1}^2b_{\sA\sB}\frac{z_{\sA}z_{\sB+2}}{d^2},
\end{equation}
where $s_1$ is the largest singular value of the matrix $R'$
defined in Eq.~\eqref{Eq:R':Defn}, and $z_m$ is the trace of
the dichotomic observable $O_m$ given in Eq.~\eqref{Eq:z}.
Since this bound is derived by considering a specific choice of
the Lagrange multipliers, it is generally not as tight as the
upper bound obtained numerically using the procedures described
in Sec.~\ref{Sec:State-dependent Bound}.

To violate the Bell-CHSH inequality, we must have $\Sqm(\rho)>2$,
hence for this class of quantum states, the Bell-CHSH inequality
cannot be violated if
\begin{equation}
    \max_{z_1,z_2,z_3,z_4}\sqrt{2}s_1d
    \sqrt{\prod_{i=1}^2\frac{2d^2-z_{2i-1}^2-z_{2i}^2}{2d^2}}
    +\sum_{\sA,\sB=1}^2b_{\sA\sB}\frac{z_{\sA}z_{\sB+2}}{2d^2}\le
    1.\label{Eq:CHSHViolation:Criterion}
\end{equation}
In addition, since the Bell-CHSH inequalities are the only
class of nontrivial facet-inducing inequalities for
$\PLHVc{2}{2}{2}{2}$, Eq.~\eqref{Eq:CHSHViolation:Criterion}
guarantees the existence of an LHVM for the experimental setup
defined by $\mA=\mB=\nA=\nB=2$~\cite{A.Fine:PRL:1982}.

\begin{table}[!h]
    \centering\rule{0pt}{4pt}\par
    \caption{\label{tbl:IsotropicStates:Bell-CHSH}
    The various threshold values for isotropic states $\rI(p)$. The
    first column of the table is the dimension of the local
    subsystem $d$. From the second column to the seventh column, we
    have, respectively, the value of $p$ below which the state is
    separable $\pS{I$_d$}$; the value of $p$ below which
    Eq.~\eqref{Eq:CHSHViolation:Criterion} is satisfied
    $p_\text{UB-semianalytic}$, and hence the state does not
    violate the Bell-CHSH inequality; the value of $p$ below which
    the upper bound obtained from lowest order relaxation is
    compatible with Bell-CHSH inequality; the value of $p$ below
    which the state cannot violate any Bell inequality via
    projective measurements; the value of $p$ below which the state
    cannot violate any Bell inequality
    (Sec.~\ref{Sec:IsotropicStates}); and the value of $p$ above
    which a Bell-CHSH violation has been observed using the LB
    algorithm.
    }
    \begin{tabular}{r|cccccc}
    $d$ & $\pS{I$_d$}$ & $p_\text{UB-semianalytic}$ &
    $p_\text{UB-numerical}$ & $\pLPi{I$_d$}$ & $\pLPOVM{I$_d$}$ & $p_\text{LB}$
    \\ \hline
      2 &  0.33333 & 0.70711 & 0.70711 & 0.50000 & 0.41667 & 0.70711\\
      3 &  0.25000 & 0.70711 & 0.70711 & 0.41667 & 0.29630 & 0.76297\\
      4 &  0.20000 & 0.65465 & 0.65465 & 0.36111 & 0.23203 & 0.70711\\
      5 &  0.16667 & 0.63246 & 0.63246 & 0.32083 & 0.19115 & 0.74340\\
     10 &  0.09091 & 0.51450 & -       & 0.21433 & 0.10214 & 0.70711 \\
     25 &  0.03846 & 0.36490 & -       & 0.11733 & 0.04274 & 0.71516 \\
     50 &  0.01961 & 0.26963 & -       & 0.07141 & 0.02171 & 0.70711 \\
    \end{tabular}
\end{table}

As an example, consider the $d$-dimensional isotropic state
$\rI(p)$ introduced in Eq.~\eqref{Eq:IsotropicState:Explicit}.
Recall from Sec.~\ref{Sec:IsotropicStates} that this class of
states is entangled if and only if $p>\pS{I$_d$}=1/(d+1)$.
Using the procedures outlined in Sec.~\ref{Sec:State-dependent
Bound}, we can numerically compute, up till $d=5$, the
threshold value of $p$ below which there can be no violation of
the Bell-CHSH inequality; these critical values, denoted by
$p_\text{UB-numerical}$ can be found in column 4 of
Table~\ref{tbl:IsotropicStates:Bell-CHSH}. Similarly, we can
numerically compute the corresponding threshold values given by
Eq.~\eqref{Eq:CHSHViolation:Criterion}, denoted by
$p_\text{UB-semianalytic}$. It is worth noting that these
threshold values, as can be seen from column 3 and 4 of
Table~\ref{tbl:IsotropicStates:Bell-CHSH}, agree exceptionally
well, thereby suggesting that the computationally feasible
criterion given by Eq.~\eqref{Eq:CHSHViolation:Criterion} may
be exact for the isotropic states.

\subsection{Bell-CH violation for Two-Qubit States}
\label{Sec:Bell-CHSH:Horodecki}

The semianalytic criterion presented in
Eq.~\eqref{Eq:CHSHViolation:Criterion} is general enough that
it can be applied to any two-qudit states with vanishing
coherence vectors. The price of such generality, however, is
that the bound is often not tight. In particular, for $d=2$,
the exact value of $\SqmCHSH(\rho)$ for any two-qubit state
$\rho$ is known~\cite{RPM.Horodecki:PLA:1995} and is often
below the upper bound given by Eq.~\eqref{Eq:CHSHViolation:UB},
i.e.,  $4\sqrt{2}~s_1$.

Nevertheless, in this case, it turns out that we can use LB to
obtain analytically $\SqmCH(\rho)$ and hence $\SqmCHSH(\rho)$
via
\begin{equation}\label{Eq:CH-CHSH}
    \SqmCHSH(\rho) =4\left(\SqmCH(\rho)+\half\right).
\end{equation}
The Horodecki criterion~\cite{RPM.Horodecki:PLA:1995} can then
be recovered by imposing the condition $\SqmCH(\rho)>0$, or
equivalently $\SqmCHSH(\rho)>2$. To see this, let us first note
that we can write a general two-qubit state $\rho$ in the
so-called Hilbert-Schmidt
form~\cite{RPM.Horodecki:PLA:1995}\footnote{We can easily
obtain this particular representation from the coherence vector
representation, Eq.~\eqref{Eq:rho:CoherenceRepn}, by defining
the rescaled basis matrices as $\sigma_0=\unit_2/\sqrt{2}$,
$\sigma_1=\sigma_x/\sqrt{2}$, $\sigma_2=\sigma_y/\sqrt{2}$,
$\sigma_3=\sigma_z/\sqrt{2}$ and rescaling the various
coefficients in Eq.~\eqref{Eq:rho:CoherenceRepn} as
$\bfr_\A\to\bfr_\A/2$, $\bfr_\B\to\bfr_\B/2$, $R'\to T\equiv
R'/2$.}
\begin{equation}\label{Eq:TwoQubit:HS}
    \rho=\frac{1}{4}\left(\unit_{2}\ten\unit_2+\bfr_\A\cdot\vec{\sigma}
    \ten\unit_2+\unit_2\ten\bfr_\B\cdot\vec{\sigma}
    \sum_{i,j=x,y,z}[T]_{ij}\sigma_i\ten\sigma_j\right),
\end{equation}
where $\vec{\sigma}$ is defined in Eq.~\eqref{Eq:Dfn:vecSigma},
$\{\sigma_i\}_{i=x,y,z}$ are the Pauli matrices introduced in
Eq.~\eqref{Eq:Dfn:PauliMatrices} and
\begin{equation}\label{Eq:Dfn:T}
    [T]_{ij}\equiv\tr\left(\rho~\sigma_i\ten\sigma_j\right).
\end{equation}

For ease of reference, we will now reproduce the functional
form of the Bell-CH inequality as follows:
\begin{equation}
    \SLHVBI{CH}=\ProbTwJ{1}{\oA}{1}{\oB}+\ProbTwJ{1}{\oA}{2}{\oB}
    + \ProbTwJ{2}{\oA}{1}{\oB}-\ProbTwJ{2}{\oA}{2}{\oB}
    -\ProbTwMA{1}{\oA} -\ProbTwMB{1}{\oB}\le 0.
    \tag{\ref{Ineq:CH:Functional}}
\end{equation}
where $\oA$, $\oB=``\pm"$ are the two possible local
measurement outcomes in a two-outcome Bell-CH experiment.
Substituting Eq.~\eqref{Eq:QM:probabilities} into
Eq.~\eqref{Ineq:CH:Functional} and comparing the resulting
expression with Eq.~\eqref{Eq:S:dfn}, one finds that the Bell
operator for this Bell inequality with $\oA=\oB=``+"$ can be
written as
\begin{align}
    \Bell_{\rm CH}
    &=A_1^+\ten(B_1^++B_2^+)+A_2^+\ten(B_1^+-B_2^+)-A_1^+\ten
    \unit_{\dB}-\unit_{\dA}\ten{B_1^+},\nonumber\\
    &=A_1^+\ten(B_2^+-B_1^-)-A_2^-\ten B_1^+-A_2^+\ten B_2^+
    \label{Eq:BellOperator:CH}
\end{align}
where we have also made used of
Eq.~\eqref{Eq:POVM:Normalization} to arrive at the final
form.\footnote{It should be clear that there is no unique way
of writing the Bell operator derived from a given Bell
inequality. The function that is of our interest, $\Sqm(\rho)$,
however is in no way affected by this degeneracy.}

Now, recall that it suffices to consider projective
measurements (Sec.~\ref{Sec:two-outcome}) for a two-outcome
Bell inequality and that the Bell-CH inequality cannot be
violated when any of the POVM elements considered are of full
rank (Appendix~\ref{App:Sec:FullRank}). Therefore, without loss
of generality, we can restrict our attention to the following
rank one projectors:
\begin{equation}\label{Eq:TwoOutomceProjectors}
    A^{\pm}_{\sA}=\half\left(\unit_2 \pm\hat{a}_{\sA}
    \cdot\vec{\sigma}\right),
\end{equation}
where $\hat{a}_{\sA}\in\mathbb{R}^3$ for $\sA=1,2$ are unit
vectors.

Next, we would like to optimize over Bob's measurements for
this choice of Alice's measurement using Helstrom-like
optimization~\cite{C.W.Helstrom:Book:1976} which has been
discussed in Sec.~\ref{Sec:two-outcome}. This allows us to
obtain $\SqmCH(\rho,{\POVMAg})$ which can further be optimized
to obtain $\SqmCH(\rho)$ using simple variational techniques.
Substituting Eq.~\eqref{Eq:TwoQubit:HS} and
Eq.~\eqref{Eq:TwoOutomceProjectors} into
Eq.~\eqref{Eq:Dfn:rhoAB} and Eq.~\eqref{Eq:S:implicitA}, and
after some computation
(Appendix~\ref{App:Sec:HorodeckiCriterion}), it can be shown
that for a general two-qubit state, Eq.~\eqref{Eq:TwoQubit:HS},
\begin{equation}\label{Eq:SqmCH:TwoQubit}
    \SqmCH(\rho)=\max\left\{0, \half\left(
    \sqrt{\varsigma_1^2+\varsigma_2^2}-1\right)\right\},
\end{equation}
where $\varsigma_1$ and $\varsigma_2$ are the two largest
singular values of $T$. Since a Bell-CH violation for $\rho$
occurs if and only if $\SqmCH(\rho)>0$, the necessary and
sufficient condition for a two-qubit state $\rho$ to violate
the Bell-CH inequality, and hence the Bell-CHSH inequality
[c.f. Eq.~\eqref{Eq:CH-CHSH}] is
\begin{equation}\label{Eq:HorodeckiCriterion}
    \varsigma_1^2+\varsigma_2^2>1,
\end{equation}
which is just the Horodecki
criterion~\cite{RPM.Horodecki:PLA:1995}.\footnote{It is worth
noting that yet another alternative derivation of
Eq.~\eqref{Eq:HorodeckiCriterion} has also been given in
Ref.~\cite{F.Verstraete:PRL:2002}.}

\subsection{$I_{3322}$-violation for a Class of Two-Qubit
States}\label{Sec:Bell-3322}

Next, we look at how the two algorithms can be used in tandem
to determine if some two-qubit states violates the $I_{3322}$
inequality introduced in Eq.~\eqref{Ineq:Functional:I3322}.
This Bell inequality is interesting in that there are quantum
states that violate this new inequality but not the
Bell-CH/Bell-CHSH inequality. The analogue of Horodecki's
criterion for this inequality is thus very desirable.

To the best of our knowledge, such an analytic criterion is yet
to be found. However, by combining the two algorithms presented
above, we can often offer a definitive, yet nontrivial,
conclusion about the compatibility of a quantum state with a
locally causal description. For ease of reference, we will also
reproduce the functional form of $I_{3322}$ inequality as
follows:
\begin{align}
    \SLHV^{(I_{3322})}=~&\ProbTwJ{1}{\oA}{1}{\oB} + \ProbTwJ{1}{\oA}{2}{\oB} + \ProbTwJ{1}{\oA}{3}{\oB}
    + \ProbTwJ{2}{\oA}{1}{\oB} + \ProbTwJ{2}{\oA}{2}{\oB} - \ProbTwJ{2}{\oA}{3}{\oB}\nonumber\\
    +~&\ProbTwJ{3}{\oA}{1}{\oB} - \ProbTwJ{3}{\oA}{2}{\oB} - \ProbTwMA{1}{\oA}
    - 2\ProbTwMB{1}{\oB} -\ProbTwMB{2}{\oB}\le0,
    \tag{\ref{Ineq:Functional:I3322}}
\end{align}
where the outcomes $\oA$ and $\oB$ are labeled as ``$\pm$".
Without loss of generality, we can restrict our attention to
$\oA=\oB=``+"$. Then, from Eq.~\eqref{Eq:S:dfn},
Eq.~\eqref{Eq:QM:probabilities} and
Eq.~\eqref{Eq:POVM:Normalization}, it can be shown that the
Bell operator corresponding to this Bell inequality reads:
\begin{align}
    \Bell _{I_{3322}}
    &=A^+_1\ten (B^-_1-B^+_2+B^-_3) -A^+_2\ten B^-_3
    -A^-_2\ten (B^-_1+B^-_2)
        - A^+_3\ten B^-_2 - A^-_3 \ten B^-_1.
    \label{Eq:Bell-3322:Operator}
\end{align}
For convenience, we will adopt the notation that
$O^{\pm}_m\equiv A^{\pm}_m$ for $m=1,2,3$ and $O^{\pm}_m\equiv
B^{\pm}_{m-3}$ for $m=4,5,6$. In these notations, the global
optimization problem for this Bell inequality can be written as
\begin{subequations}\label{Eq:GlobalOptimization:3322}
\begin{gather}
    \text{maximize \ \ } \tr(\rho~\Bell_{I_{3322}}) \label{Eq:eBell-3322}\\
    \text{subject to \ \ }
    \left(O^\pm_m\right)^2=O_m
    \label{Eq:Constraints:Projector}
\end{gather}\end{subequations}
for $m=1,2,\ldots,6$, which is a QCQP. The lowest order
relaxation to this problem can thus be obtained by following
similar procedures as that described in Sec.~\ref{Sec:UB}.

To obtain a {\em  state-dependent} upper bound on $\Sqm(\rho)$ for
this inequality, we have to impose the analogue of
Eq.~\eqref{Eq:Range:z}, i.e.,
\begin{equation}
    z^{\pm}_m=\tr(O^{\pm}_m)=0,1,\ldots,d,\label{Eq:Range:xpm0}
\end{equation}
for each of the POVM elements. For small $d$, numerical upper
bounds on $\Sqm(\rho)$ can then be solved for using SOSTOOLS.
As an example, let's now look at how this upper bound, together
with the LB algorithm, has enabled us to determine if a class
of mixed two-qubit states violates the $I_{3322}$ inequality.

The mixed two-qubit state
\begin{gather}
    \rCG(p)=p\,\ketbra{\Psi_{2:1}}+(1-p)\,\AProj{0}\ten\BProj{1},\quad 0\le
    p\le 1,\label{Eq:State:CG}
\end{gather}
can be understood as a mixture of the pure product state
$\ket{0}_\A\ket{1}_\B$ and the non-maximally entangled
two-qubit state
$\ket{\Psi_{2:1}}=\frac{1}{\sqrt{5}}(2\ket{0}_\A\ket{0}_\B+\ket{1}_\A\ket{1}_\B)$.
As can be easily verified using the PPT
criterion~\cite{A.Peres:PRL:1996,MPR.Horodecki:PLA:1996}, this
state is entangled for $0<p\le 1$. In particular, the mixture
with $p=0.85$ was first presented in
Ref.~\cite{D.Collins:JPA:2004} as an example of a two-qubit
state that violates the $I_{3322}$ inequality but not the
Bell-CH/ Bell-CHSH inequality.

\begin{figure}[h!]
    \centering\rule{0pt}{4pt}\par
    \includegraphics[width=8cm,height=2.29cm]{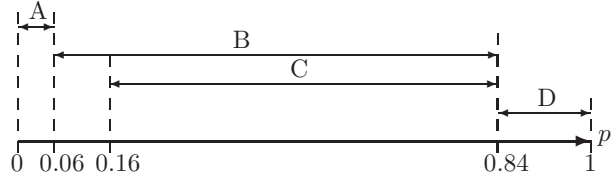}
    \caption{\label{Fig:S:CG2:Bounds}
    Domains of $p$ where the compatibility between a locally causal
    description and quantum mechanical prediction given by
    $\rCG(p)$ was studied via the LB and UB algorithms in
    conjunction with the $I_{3322}$ inequality. From right to left
    are respectively the domain of $p$ whereby $\rCG(p)$ is: (D)
    found to violate the $I_{3322}$ inequality; (C) found to give a
    lowest order upper bound that is compatible with the $I_{3322}$
    inequality; (B) found to give a higher order upper bound that
    is compatible with the $I_{3322}$ inequality; (A) not known if
    it violates the $I_{3322}$ inequality.
    }
\end{figure}

Given the above observation, a natural question that one can
ask is, at what values of $p$ does $\rCG(p)$ violate the
$I_{3322}$ inequality? Using the LB algorithm, we have found
that for\footnote{Throughout, we will use $p\gtrsim p'$ and
$p\lesssim p'$ to denote $p'$ as, respectively, a numerical
(approximate) lower bound and upper bound for $p$.} $p\gtrsim
0.83782$ (domain D in Figure~\ref{Fig:S:CG2:Bounds}), $\rCG(p)$
violates the $I_{3322}$ inequality. As we have pointed out in
Sec.~\ref{Sec:LB}, observables that lead to the observed level
of $I_{3322}$-violation can be readily read off from the output
of the SDP.

On the other hand, through the UB algorithm, we have also found
that, with the lowest order relaxation, the states do not
violate this 3-setting inequality for $0.16023 \lesssim p
\lesssim 0.83625$ (domain C in Figure~\ref{Fig:S:CG2:Bounds});
with a higher order relaxation, this range expands to $0.06291
\lesssim p \lesssim 0.83782$ (domain B in
Figure~\ref{Fig:S:CG2:Bounds}). Notice that at the presented
accuracy, the upper bound of $p$ where there can be no
violation of the $I_{3322}$ inequality now agrees with the
lower bound of $p$ where an $I_{3322}$ violation was found.

The algorithms alone therefore leave a tiny gap at $0 < p
\lesssim 0.06291$ (domain A in Figure~\ref{Fig:S:CG2:Bounds})
where we could not conclude if $\rCG(p)$ violates the
$I_{3322}$ inequality. Nevertheless, if we recall that the set
of quantum states not violating a given Bell inequality is
convex and that $\rCG(0)$, being a pure product state, cannot
violate any Bell inequality, we can immediately conclude that
$\rCG(p)$ with $0\le p \lesssim 0.83782$ cannot violate the
$I_{3322}$ inequality. As such, together with convexity
arguments, the two algorithms allow us to fully characterize
the state $\rCG(p)$ compatible with LHVTs, when each observer
is only allowed to perform three different dichotomic
measurements.

\subsection{Limitations of the UB algorithm}
\label{Sec:Limitation:UB}

As can be seen in the above examples, the UB algorithm does not
always provide a very good upper bound for $\Sqm(\rho)$. In
fact, it has been observed that for pure product states, the
algorithm with lowest order relaxation always returns a
state-independent bound (the Tsirelson bound in the case of
Bell-CHSH inequality). As such, for mixed states that can be
decomposed as a high-weight mixture of pure product state and
some other entangled state, the upper bound given by UB is
typically bad. To illustrate this, let us consider the
1-parameter family of PPT bound entangled state
$\rH(p)$~\cite{P.Horodecki:PLA:1997,MPR.Horodecki:PRL:1998},
Eq.~\eqref{Eq:State:H3}, and recall from
Sec.~\ref{Sec:States:PPT} that a bipartite PPT entangled state
cannot violate the Bell-CH or the Bell-CHSH
inequality~\cite{R.F.Werner:PRA:2000}.

\begin{figure}[h!]
    \centering\rule{0pt}{4pt}\par
    \includegraphics[width=12cm,height=9.32cm]
    {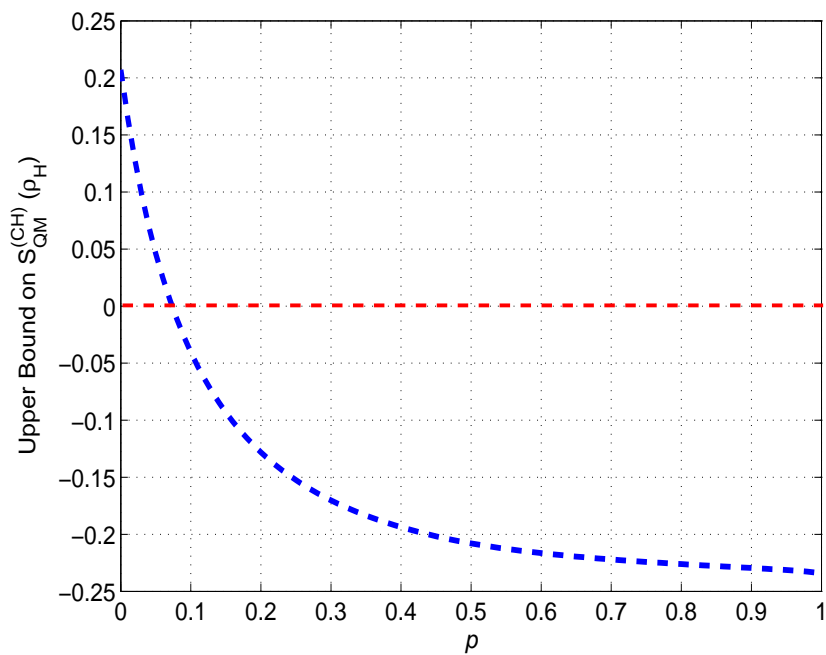}
    \caption{\label{Fig:HBE}
    Numerical upper bound on $\SqmCH(\rH)$ obtained from the UB
    algorithm using lowest order relaxation and
    Eq.~\eqref{Eq:Range:xpm0}. The dashed horizontal line is the
    threshold above which no locally causal description is
    possible.
    }
\end{figure}

However, when tested with the UB algorithm using the lowest
order relaxation, it turned out that some of these upper bounds
are actually above the threshold of Bell-CH violation (see
Figure~\ref{Fig:HBE}). In fact, the upper bound obtained for
the pure product state, $\rH(0)=\ketbra{\Psi_p}$ is actually
the maximal achievable Bell-CH violation given by a quantum
state~\cite{B.S.Cirelson:LMP:1980}. Nonetheless, as with the
example presented in Sec.~\ref{Sec:Bell-3322}, we can exclude
the possibility of $\rH(p)$ violating the Bell-CH inequality by
combining the upper bound on $\SqmCH(\rH)$ and the convexity of
NBIV states.

\section{Conclusion}

In this chapter, we have looked specifically into the problem
of determining if a given (entangled) quantum state is
Bell-inequality-violating (BIV) for some fixed but arbitrary
Bell inequality. For that purpose, we have presented two
algorithms which can be used to determine, respectively, a
lower bound (LB) and an upper bound (UB) on the maximal
expectation value of a Bell operator for a given quantum state,
i.e., $\Sqm(\rho)$.

In particular, we have demonstrated how one can make use of the
upper bound to derive a necessary condition for two-qudit
states with vanishing coherence vectors to violate the
Bell-CHSH inequality. When $d=2$, we have also illustrated how
the LB algorithm can be used to rederive Horodecki's criterion
for two-qubit states. For more complicated Bell inequalities
where analytic treatment seems formidable, we have demonstrated
how one can make use of the two algorithms in tandem to
determine, numerically, if the quantum mechanical prediction is
compatible with a locally causal description. In
Chapter~\ref{Chap:BellViolation}, we will also see how these
algorithms have been applied to the search of
maximal-Bell-inequality-violation in the context of collective
measurements without postselection.

As with many other numerical optimization algorithms, the LB
algorithm can only guarantee the convergence to a local maximum
of $\Sqm\left(\rho,\{O_m\}\right)$. The UB algorithm, on the
other hand, provides an (often loose)  upper bound on
$\Sqm(\rho)$. In the event that these bounds agree (up to
reasonable numerical precision), we know that optimization of
the corresponding Bell operator using LB has been achieved.
This ideal scenario, however, is not as common as we would like
it to be. In particular, the UB algorithm with lowest order
relaxation has been observed to give rather bad bounds for
states with a high-weight mixture of pure product states
(although we can often rule out the possibility of a violation
in this situation by convexity arguments as in
Sec.~\ref{Sec:Bell-3322} and Sec.~\ref{Sec:Limitation:UB}). A
possibility to improve these bounds, as suggested  by the work
of Nie {\em et~al.}~\cite{J.Nie:MP:2006}, is to incorporate the
{\em Karush-Kuhn-Tucker} optimality condition as an additional
constraint to the problem. We have done some preliminary
studies on this but have not so far found any improvement in
the bounds obtained but this deserves further study.

As of now, we have only implemented the UB algorithm to
determine upper bounds on $\Sqm(\rho)$ for dichotomic Bell
inequalities. For Bell inequalities with more outcomes, the
local Hermitian observables are generally also subjected to
constraints in the form of a LMI. Although the UB algorithm can
still be implemented for these Bell inequalities by first
mapping the LMI to a series of polynomial inequalities, this
approach seems blatantly inefficient. Future work to remedy
this difficulty is certainly desirable.

Finally, despite the numerical and analytic evidence at hand,
it is still unclear why the lowest order relaxation to the
global optimization problem, as described in
Sec.~\ref{Sec:State-independent Bound}, seems always gives rise
to a bound that is state-independent and how generally this is
true. Some further investigation on this may be useful,
particularly to determine whether the lowest order relaxation
is always state-independent even for inequalities that are not
correlation inequalities. If so this could complement the
methods of
Refs.~\cite{S.Wehner:PRA:2006,D.Avis:JPA:2006,M.Navascues:PRL:2007}
for finding state-independent bounds on Bell inequalities. In
fact, recently, very similar techniques were found to give
provably state-independent bounds on maximal Bell inequality
violation~\cite{M.Navascues:PRL:2007,A.C.Doherty:QMP:2007}.

\chapter{Bell-Inequality Violations by Quantum States}
\label{Chap:BellViolation}

In this chapter, we will make use of the toolkits developed in
Chapter~\ref{Chap:QuantumBounds} to analyze the extent to which
specific quantum states can violate a given Bell inequality.
Geometrically, the degree of violation of a given
facet-inducing Bell inequality $I^{(k)}$ provides a measure of
the distance of the Bell-inequality-violating quantum
correlation from the boundary of the convex set of classical
correlations corresponding to $I^{(k)}$. We will consider this
problem both in the typical scenario where a quantum system is
measured one copy at a time, and the other scenario where
multiple copies of the same quantum system are measured
collectively.

\section{Introduction}

Pioneering investigation on the extent to which a given quantum
state can violate a given Bell inequality can be traced back to
as early as 1980s. At that time, Mermin and
Garg~\cite{N.D.Mermin:PRD:1980,
A.Garg:PRL:1982,A.Garg:PRL:1982b, A.Garg:PRD:1983} were mainly
interested to know if this nonclassical feature displayed by
two entangled spin-$j$ quantum systems could survive in the
``classical limit" of $j\to\infty$. Their initial
attempt~\cite{N.D.Mermin:PRD:1980} seemed to have suggested
that this nonclassical feature does indeed diminish with
increasing quantum numbers, in agreement with the mentality
that the classical world arises in the $j\to\infty$ limit. That
this observation is an artefact of their analysis was almost
immediately confirmed by their follow up
work~\cite{A.Garg:PRL:1982, A.Garg:PRD:1983}, in which they
showed that the spin-$j$ singlet state for any $j$ could indeed
contradict predictions given by any LHVT.

A quantitative study of the strength of Bell-CHSH-violation for
arbitrary spin-$j$ singlet states was nonetheless not available
until Peres revisited the problem almost a decade
later~\cite{A.Peres:PRA:1992}. The measurements that Peres
considered in Ref.~\cite{A.Peres:PRA:1992} are, however, not
optimal and only lead to a Bell-CHSH-violation of 2.481 in the
asymptotic limit of $j\to\infty$. This result was soon
strengthened by Gisin and Peres~\cite{N.Gisin:PLA:1992}, who
showed that for the spin-$j$ singlet state, i.e., the
$(2j+1)$-dimensional maximally entangled state $\ME{2j+1}$, the
corresponding Bell-CHSH-violation is just $2\sqrt{2}$ (the
Tsirelson bound) when $j$ is a half integer, and tends towards
the same bound as $j\to\infty$ if $j$ is an integer.

The strength of a Bell inequality violation is also relevant
from an experimental point of view. Given that in a realistic
experimental scenario, pure entangled states are hard, if not
impossible, to prepare, a natural question that follows is the
robustness of nonclassical correlations against the mixture of
noise. How is the robustness of nonclassical correlations
against noise related to the strength of violation? Crudely
speaking, in the presence of noise, the strength of violation
decreases, therefore the stronger an entangled state violates a
given Bell inequality, the more robust are the corresponding
nonclassical correlations against the mixture of noise. Along
this line of investigation, Kaszlikowski and
coauthors~\cite{D.Kaszlikowski:PRL:2000} made an interesting
discovery that, as opposed to the mentality of $j\to\infty$
being the classical limit, the inconsistency between LHVT and
quantum mechanical prediction for $\ME{2j+1}$ actually gets
more robust against the mixture of noise as $j$
increases.\footnote{The noise is modeled by the incoherent
mixture of the state in question with a maximally mixed state
(see, for example, the discussion on $\rw(p)$ and $\rI(p)$ in
Sec.~\ref{Sec:WernerStates} and
Sec.~\ref{Sec:IsotropicStates}).}

Indeed, using the $d$-outcome CGLMP inequality that they
derived, Collins\etal~\cite{D.Collins:PRL:2002} showed that as
$d$, the dimension of the local Hilbert space increases, the
maximal violation found for $\MEd$ against this class of
inequalities also increases (see also
Ref.~\cite{D.Kaszlikowski:PRA:2002}). This finding is, of
course, consistent with the above intuition, and the discovery
presented in Ref.~\cite{D.Kaszlikowski:PRL:2000} that as $d$
increases, the nonclassical correlations derived from $\MEd$
are more robust against the mixture of (white) noise. In this
regard, it is also worth noting that, somewhat surprisingly,
for a given $d$, $\MEd$ is not the quantum state whose Bell
inequality violation is most robust against the mixture of
noise~\cite{A.Acin:PRA:2002b,L.B.Fu:PRA:2003,
Li-Bin.Fu:PRA:2004}.

On the other hand, experiments to test Bell inequalities
usually involve making many measurements on individual copies
of the quantum system with the system being prepared in the
same way for each measurement. In this chapter, we will also
consider a somewhat different scenario and ask if {\em quantum
nonlocality}\,\footnote{The term ``quantum nonlocality" is used
here merely as a widely, but not universally accepted synonym
for the violation of a Bell inequality (see e.g.
Ref.~\cite{S.Popescu:PLA:1992} and
Ref.~\cite{M.Zukowski:JMO:2003} for opposing views).} can be
enhanced by making joint local measurements on multiple copies
of the entangled state. We will use the maximal Bell inequality
violation of a quantum state $\rho$ as our measure of
nonlocality. Our interest is to determine if $\rho^{\otimes
N}$, when compared with $\rho$, can give rise to a higher Bell
inequality violation for some $N>1$.

A very similar problem was introduced by
Peres~\cite{A.Peres:PRA:1996} who considered Bell inequality
violations under collective measurements but allowed the
experimenters to make an auxiliary measurement on their systems
and postselect on both getting a specific outcome of their
measurement. Numerically, Peres showed that with collective
measurements and postselection, a large class of two-qubit
states give rise to better Bell inequality violation. However,
note that the postselection in Peres' scheme is stronger than
that in realistic Bell inequality experiments where detector
inefficiencies require a postselection on events where both
detectors fired. In such a case the failure probability is
independent of the quantum state.

As with Peres' examples, existing results in the literature on
nonlocality enhancement always involve some kind of
postselection, it is thus of interest to investigate the power
of collective measurements, without postselection, in terms of
increasing Bell inequality violation. Indeed, it is one of the
main purposes of this chapter to show that postselection is not
necessary to improve Bell inequality violation. In order to
find such examples for mixed states we have resorted to various
numerical approaches that are described in
Sec.~\ref{Sec:QuantumBounds} to provide upper (UB) and lower
bounds (LB) on the optimal violation of a given Bell inequality
by a given quantum state. Unless otherwise stated, Bell
inequality violations presented hereafter refer to the best
violation that we could find either analytically, or
numerically using the LB algorithm. For ease of reference,
upper bounds obtained via UB are marked where they appear with
$^\dag$. In the event that a violation presented is known to be
maximal (such as those computable using the Horodecki's
criterion~\cite{RPM.Horodecki:PLA:1995}), an * will be
attached.

This chapter is organized as follows. In
Sec.~\ref{Sec:Bell-CH:Pure2Qudit:Measurements}, we present a
measurement scheme which we will use to determine the Bell-CH
inequality violation for any bipartite pure state. These
measurements led to the largest violation that we were able to
find and may even be maximal. This is then followed by a review
of what is known about the best $I_{22nn}$-violation for some
two-qudit states in Sec.~\ref{Sec:I22nnViolation}. Then, in
Sec.~\ref{Sec:pure-states}, we show that for bipartite pure
entangled states, collective measurement can lead to a greater
violation of the Bell-CH inequality. The corresponding scenario
for mixed entangled states is analyzed in
Sec.~\ref{Sec:mixed-states}. We then conclude with a summary of
results and some future avenues of research.

\section{Single Copy Bell Inequality Violation}

\subsection{Bell-CH-violation for Pure Two-Qudit States}
\label{Sec:Bell-CH:Pure2Qudit:Measurements}

In this section, we present a measurement scheme which gives
rise to the largest Bell-CH inequality violation that we have
found for arbitrary pure two-qudit states. We find using this
inequality for probabilities rather than correlations to be
convenient for our purposes. From Eq.~\eqref{Eq:CH-CHSH}, we
know that if the conjectured measurement scheme is optimal for
the Bell-CH inequality, it will also give rise to the maximal
Bell-CHSH inequality violation for any pure two-qudit state.

For ease of reference, let us again reproduce the functional form of
Bell-CH inequality here:
\begin{equation}
    \SLHVBI{CH}=\ProbTwJ{1}{\oA}{1}{\oB}+\ProbTwJ{1}{\oA}{2}{\oB}+
    \ProbTwJ{2}{\oA}{1}{\oB}-\ProbTwJ{2}{\oA}{2}{\oB}
    -\ProbTwMA{1}{\oA} -\ProbTwMB{1}{\oB}\le 0,\tag{\ref{Ineq:CH:Functional}}
\end{equation}
where in quantum mechanics, the relevant joint and marginal
probabilities are calculated according to
Eq.~\eqref{Eq:QM:probabilities}. Without loss of generality, in
the following discussion, we will focus on the above inequality
with $\oA=\oB=``+"$.

The maximal Bell inequality violation for a quantum state is
invariant under a local unitary transformation. As such, the
maximal Bell inequality violation for any bipartite pure
quantum state is identical to its maximal violation when
written in the Schmidt
basis~\cite{E.Schmidt:MA:1906,S.M.Barnett:PLA:1991}. In this
basis, an arbitrary pure two-qudit state, i.e.,
$\ket{\Phi_d}\in\Cdg\ten\Cdg$ takes the form
\begin{equation}\label{Eq:TwoQudit:SchmidtForm}
    \ket{\Phi_d}=\sum_{i=1}^d c_i\ket{i}_{\A}\ket{i}_{\B},
\end{equation}
where $\{\ket{i}_{\A}\}$ and $\{\ket{i}_{\B}\}$ are local
orthonormal bases of subsystem possessed by observer $\A$ and
$\B$ respectively, and $\{c_i\}_{i=1}^d$ are the Schmidt
coefficients of $\ket{\Phi_d}$. Without loss of generality, we
may also assume that $c_1\ge c_2\ge\ldots\ge c_d> 0$. Then
$\ket{\Phi_d}$ is entangled if and only if $d>1$. Now, let us
consider the following measurement settings for Alice, which
were first adopted in
Ref.~\cite{N.Gisin:PLA:1992},\footnote{\label{fn:odd-d}Here, as
well as Eq.~\eqref{Eq:DiffAlicePOVM} and
Eq.~\eqref{Eq:rhoB:intermediate}, we will adopt the convention
that when $d$ is odd, the end product of the direct sum is
appended with zero entries to make the dimension of the
resulting matrix $d\times d$.}
\begin{gather}
    A_1^{\pm}=\half\left[\unit_d\pm Z\right],\quad
    A_2^{\pm}=\half\left[\unit_d\pm X\right],\nonumber\\
    Z\equiv \bigoplus_{i=1}^{\lfloor d/2\rfloor}\sigma_z+\Xi,\quad
    X\equiv
    \bigoplus_{i=1}^{\lfloor d/2\rfloor}\sigma_x+\Xi,\nonumber\\
    \label{Eq:Alice-POVM}\left[\Xi\right]_{ij}=0\quad\forall\quad
    i,j\neq d,\qquad \left[\Xi\right]_{dd}=d\mod 2,
\end{gather}
where $\sigma_x$ and $\sigma_z$ are respectively the Pauli $x$
and $z$ matrices introduced in
Eq.~\eqref{Eq:Dfn:PauliMatrices}.

Notice, however, that the
$\left\{B_{\sB}^\pm\right\}_{\sB=1}^2$ given in
Ref.~\cite{N.Gisin:PLA:1992} are not optimal for a general pure
two-qudit state. In fact, as we have seen in
Sec.~\ref{Sec:two-outcome}, given the measurements for Alice in
Eq.~\eqref{Eq:Alice-POVM}, the optimization of Bob's
measurement settings can be carried out explicitly. Using the
resulting analytic expression for Bob's optimal POVM
(Appendix~\ref{App:Sec:Bell-CH:Pure2Qudit:Measurements}), the
optimal expectation value of the Bell-CH operator,
Eq.~\eqref{Eq:BellOperator:CH}, for $\ket{\Phi_d}$ can be
computed and we find
\begin{equation}\label{Eq:Sqm:CH:pure}
    \eBellCH_{\ket{\Phi_d}}=\half\sum_{n=1}^{\lfloor  d/2\rfloor}
    \sqrt{(c_{2n-1}^2+c_{2n}^2)^2
    +4c_{2n}^2 c_{2n-1}^2}+\frac{\xi}{2}c_d^2-\half,
\end{equation}
where $\xi\equiv d\mod 2$. From here, it is easy to see that
for any entangled $\ket{\Phi_d}$, i.e., $d>1$,
\begin{equation}
    \eBellCH_{\ket{\Phi_d}}>\half\sum_{n=1}^{\lfloor  d/2\rfloor}
    \sqrt{(c_{2n-1}^2+c_{2n}^2)^2}+\frac{\xi}{2}c_d^2-\half=0,
\end{equation}
where we have made use of the normalization condition
$\sum_{i=1}^d c_i^2=1$. Therefore, as was first shown by Gisin
and Peres~\cite{N.Gisin:PLA:1992}, a pure two-qudit state
violate the Bell-CH, or equivalently the Bell-CHSH inequality
if and only if it is entangled.

Effectively, the measurement scheme presented above corresponds
to first ordering each party's local basis vectors
$\{\ket{i}\}_{i=1}^d$ according to their Schmidt coefficients,
and grouping them pairwise in descending order from the Schmidt
vector with the largest Schmidt coefficient. Physically, this
can be achieved by Alice and Bob each performing an appropriate
local unitary transformation. Each of their Hilbert spaces can
then be represented as a direct sum of 2-dimensional subspaces,
which can be regarded as a one-qubit space, plus a
1-dimensional subspace if $d$ is odd. The final step of the
measurement consists of performing the optimal measurement
(\cite{RPM.Horodecki:PLA:1995}, see also
Appendix~\ref{App:Sec:HorodeckiCriterion}) in each of these
two-qubit spaces as if the other spaces did not exist.

From here, it is easy to see that if we have a $d$-dimensional
maximally entangled  state $\MEd$,
Eq.~\eqref{Eq:Dfn:State:ME-d}, then Eq.~\eqref{Eq:Sqm:CH:pure}
gives\footnote{Although Bob's measurements
$\left\{B_{\sB}^\pm\right\}_{\sB=1}^2$ given in
Ref.~\cite{N.Gisin:PLA:1992} are generally not optimal when
Alice's measurements are given by Eq.~\eqref{Eq:Alice-POVM},
the measurement settings given in Ref.~\cite{N.Gisin:PLA:1992}
do give rise to the same $\eBellCH_{\MEd}$ as we have got here
for maximally entangled state.}
\begin{equation}\label{Eq:S:ME}
    \eBellCH_{\MEd}=\left\{
    \begin{array}{r@{\quad\quad}}
        \frac{1}{\sqrt{2}}-\half^*:d\,\rm{even} \\
        \frac{\sqrt{2}(d-1)+1}{2d}-\half: d~\rm{odd} \\
    \end{array}\right. .
\end{equation}
With this measurement scheme, the  Bell-CH inequality violation
for a maximally entangled state with even $d$ is thus the
maximum allowed by Tsirelson's
bound~\cite{B.S.Cirelson:LMP:1980} whereas that of maximally
entangled state with odd $d$ is not. This may seem surprising
at first glance, but as was pointed out by Popescu and Rohrlich
in Ref.~\cite{S.Popescu:PLA:1992b}, the Tsirelson bound can
{\em never} be attained by any $\ket{\Phi_d}$ with odd $d$.

How good is the measurement scheme given by
Eq.~\eqref{Eq:Alice-POVM} and Eq.~\eqref{Eq:Bob-POVM}? It is
constructed so that for pure two-qubit states, i.e. when $d=2$,
Eq.~\eqref{Eq:Sqm:CH:pure} gives the same violation found in
Refs.~\cite{N.Gisin:PLA:1991,N.Gisin:PLA:1992}, and is the
maximal violation determined by
Horodecki\etal~\cite{RPM.Horodecki:PLA:1995}
(Appendix~\ref{App:Sec:HorodeckiCriterion}). The measurement
given by Eq.~\eqref{Eq:Alice-POVM} is hence optimal for any
two-qubit state $\ket{\Phi_2}$. Moreover, for the 3-dimensional
isotropic state $\rId{3}(p)$, c.f.
Eq.~\eqref{Eq:IsotropicState:Explicit},
\begin{equation}\label{Eq:IsotropicState:d3}
    \rId{3}(p)=p~\ProjME{3}+(1-p)\frac{\unit_3\ten\unit_3}{9},
\end{equation}
the measurement scheme given by Eq.~\eqref{Eq:Alice-POVM} and
Eq.~\eqref{Eq:Bob-POVM} gives rise to
\begin{equation}
    \SqmCH(\rId{3})=\max\left\{\left(\frac{1+3\sqrt{2}}{9}\right)p
    -\frac{4}{9},0\right\},
\end{equation}
which is exactly the maximum Bell-CH violation of $\rId{3}(p)$
as determined by Ito\etal~\cite{T.Ito:PRA:2006}. In other
words, the measurement operators given by
Eq.~\eqref{Eq:Alice-POVM} and Eq.~\eqref{Eq:Bob-POVM} are also
optimal for $\ME{3}$ and its mixture with the maximally mixed
state.

In general, for higher dimensional quantum systems, we have
looked at randomly generated  pure two-qudit states
($d=3,\ldots,10$) with their (unnormalized) Schmidt
coefficients uniformly chosen at random from the interval
$(0,1)$. For all the 20,000 states generated for each $d$, we
found that with Eq.~\eqref{Eq:Alice-POVM} as the initial
measurement setting, the (iterative) LB algorithm never gives a
$\eBellCH_{\ket{\Phi_d}}$ that is different from
Eq.~\eqref{Eq:Sqm:CH:pure} by more than $10^{-15}$, thus
indicating that Eq.~\eqref{Eq:Sqm:CH:pure} is, at least, a
local maximum of the optimization problem.

Furthermore, for another 8,000 randomly generated pure
two-qudit states, 1,000 each for $d=3,\ldots,10$, an extensive
numerical search using more than $4.6\times 10^6$ random
initial measurement guesses have not led to a single instance
where $\eBellCH_{\ket{\Phi_d}}$ is higher than that given in
Eq.~\eqref{Eq:Sqm:CH:pure}\footnote{It is worth noting that
among the 1,000  random pure states generated for each $d$,
there are always some  whose best  Bell-CH inequality violation
found differs from Eq.~\eqref{Eq:Sqm:CH:pure} by no more than
$10^{-10}$.}. These numerical results suggest that the
measurement scheme given by Eq.~\eqref{Eq:Alice-POVM} and
Eq.~\eqref{Eq:Bob-POVM} may be the optimal measurement that
maximizes the Bell-CH inequality violation for arbitrary pure
two-qudit states.

\subsection{CGLMP and $I_{22nn}$-violation for Some Two-Qudit States}
\label{Sec:I22nnViolation}

Apart from the Bell-CH/ Bell-CHSH inequalities, the other class
of bipartite Bell inequalities whose quantum violations are
most well-studied in the literature is probably the CGLMP
inequality, Eq.~\eqref{Ineq:Prob:CGLMP:Functional}, which is
equivalent to the $I_{22nn}$ inequality,
Eq.~\eqref{Ineq:Prob:I22nn:Functional}. For any quantum state
$\rho$, its violations of these two inequalities are shown in
Appendix~\ref{App:Sec:CGLMP} to be related linearly as follows:
\begin{equation}\label{Eq:In-I22nn}
    \tr\left(\rho~\Bell_{I_n}\right)=
    \frac{2n}{n-1}\tr\left(\rho~\Bell_{I_{22nn}}\right)+2,
\end{equation}
where $\Bell_{I_n}$ is the Bell operator derived from the
$n$-outcome CGLMP inequality,
Eq.~\eqref{Ineq:Prob:CGLMP:Functional}. In
Eq.~\eqref{Eq:In-I22nn}, $\Bell_{I_{22nn}}$ is the Bell
operator associated with the $I_{22nn}$ inequality, which can
be written explicitly as
\begin{align}
\Bell_{I_{22nn}}&=\sum_{\oA=1}^{n-1}\sum_{\oB=1}^{n-\oA}\POVMA{1}{\oA}\ten\POVMB{1}{\oB}
    +\sum_{\oA=1}^{n-1}\sum_{\oB=n-\oA}^{n-1}\Big(\POVMA{1}{\oA}\ten\POVMB{2}{\oB}
    +\POVMA{2}{\oA}\ten\POVMB{1}{\oB}-\POVMA{2}{\oA}\ten\POVMB{2}{\oB}\Big)\nonumber\\
    &-\sum_{\oA=1}^{n-1}\POVMA{1}{\oA}\ten\unit_{\dB}
    -\sum_{\oB=1}^{n-1}\unit_{\dA}\ten\POVMB{1}{\oB},
    \label{Eq:BellOperator:I22nn}
\end{align}
where $\dA$ and $\dB$ are, respectively, the dimension of
Alice's and Bob's Hilbert spaces.

In this section, we will give a brief review of the best
CGLMP-violation and hence --- via Eq.~\eqref{Eq:In-I22nn}
--- the best $I_{22nn}$-violation known for the isotropic state
$\rI(p)$,
\begin{equation}
    \rI(p)=p~\ProjMEd+(1-p)\frac{\unit_{d}\ten\unit_d}{d^2},
    \tag{\ref{Eq:IsotropicState:Explicit}}
\end{equation}
where $p$ is the weight of the $d$-dimensional maximally
entangled state $\ME{d}$ in the mixture. In what follows, we
shall thus be contented with the scenario where $\dA=\dB=n=d$.
Interestingly, it turned out that the best known
$I_{22nn}$-violation for $\rI(p)$ is achieved with rank-one
projective measurements. By linearity of expectation value, it
therefore suffices to determine the maximal
$I_{22dd}$-violation for $\MEd$; the best $I_{22dd}$-violation
for $\rI(p)$ will follow immediately. These best known
violations will come in handy when we need to compare the best
$I_{22dd}$-violation that we have found against what is known
in the literature.

Now, let us recall the best known $I_{22dd}$-violation for
$\MEd$. From the pioneering result of
Collins\etal~(\cite{D.Collins:PRL:2002}, see also
Ref.~\cite{D.Kaszlikowski:PRA:2002}), it follows that with
rank-one projective measurements, the $d$-dimensional maximally
entangled state $\MEd$ can violate the $I_{22dd}$ inequality by
as much as
\begin{equation}\label{Eq:Sqm:MEd:I22nn}
    \eBell{I_{22dd}}{\MEd}=\frac{d-1}{2d}\left[4d\sum_{k=0}
    ^{\lfloor \frac{d}{2}\rfloor -1}(q_k-q_{-(k+1)})-2\right],
\end{equation}
where $q_k\equiv
\frac{1}{2d^3\sin^2\left[\pi\left(k+\quar\right)d\right]}$. In
particular, in the asymptotic limit of $d\to\infty$, this best
$I_{22dd}$-violation by $\MEd$ converges to
\begin{equation}\label{Eq:Sqm:I22nn:ifty}
    \lim_{d\to\infty}\eBell{I_{22dd}}{\MEd}
    =\frac{1}{\pi^2}\sum_{k=0}^\infty\frac{1}{(k+1/4)^2}-\frac{1}{(k+3/4)^2}
    =\frac{16}{\pi^2}\times\text{Catalan}-1
    \approx 0.484~91
\end{equation}
where Catalan$\approx 0.915~97$ is the Catalan constant.
Explicit values for some of these best known violations can be
found in column 4 of Table~\ref{tbl:MaximalI22nnViolation}.
From column 2 and 3 of the same table, it can also be seen that
the best known violation of this inequality is apparently not
attained by the maximally entangled state $\MEd$ --- an
interesting phenomenon that was first discovered by
Ac\'in\etal~\cite{A.Acin:PRA:2002b}.

\begin{table}[!h]
\centering\rule{0pt}{4pt}\par
\caption{\label{tbl:MaximalI22nnViolation} Best known
CGLMP-violation and $I_{22dd}$-violation for the maximally
entangled two-qudit state $\MEd$. The first column of the table
gives the dimension of the local subsystem $d$. The second
column gives the largest possible quantum violation of the
CGLMP inequality for $d\le 8$, first obtained in
Ref.~\cite{A.Acin:PRA:2002b}, and subsequently verified in
Ref.~\cite{M.Navascues:PRL:2007}; these maximal violations also
set an upper bound on the maximal violation attainable by
$\MEd$ for each $d$. The third column of the table gives the
best known $d$-outcome CGLMP-violation for $\MEd$ whereas the
fourth column gives the corresponding best known
$I_{22dd}$-violation obtained from Eq.~\eqref{Eq:In-I22nn}.
Also included in the fifth column of the table is the threshold
weight $p_d$ below which no violation of either inequality by
isotropic state $\rI(p)$ is known.}
    \begin{tabular}{r|cccc}
    $d$ & $\SqmBI{CGLMP}(\rho)$ & $\eBell{\mbox{\tiny CGLMP}}{\MEd}$
        & $\eBell{I_{22dd}}{\MEd}$ & $p_d$
    \\ \hline\hline
      2 & 2.8284  & 2.8284 & 0.20711 & 0.70711\\
      3 & 2.9149  & 2.8729 & 0.29098 & 0.69615\\
      4 & 2.9727  & 2.8962 & 0.33609 & 0.69055\\
      5 & 3.0157  & 2.9105 & 0.36422 & 0.68716\\
      8 & 3.1013  & 2.9324 & 0.40793 & 0.68203\\
     10 & -       & 2.9398 & 0.42291 & 0.68032\\
     100 & -      & 2.9668 & 0.47856 & 0.67413\\
     1000 & -     & 2.9695 & 0.48427 & 0.67351\\
     $\infty$ & - & 2.9698 & 0.48491 & 0.67349\\
    \end{tabular}
\end{table}

Now, it is not difficult to see from
Eq.~\eqref{Eq:BellOperator:I22nn} that when restricted to
rank-one projective measurements, the expectation value of
$\Bell_{I_{22nn}}$ with respect to the $d\times d$-dimensional
maximally mixed state $\rho_{d\times d}$ reads:
\begin{equation}
    \tr\left(\rho_{d\times d}\,\Bell_{I_{22dd}}\right)=-1+\frac{1}{d}.
\end{equation}
Therefore, from the linearity of expectation value and
Eq.~\eqref{Eq:Sqm:MEd:I22nn}, it follows that the best known
$I_{22dd}$-violation for the isotropic states is:
\begin{equation}\label{Eq:SqmI22dd:Isotropic}
    \tr\Big[\rI(p)\,\Bell_{I_{22dd}}\Big]
    =p\times\frac{d-1}{2d}\left[4d\sum_{k=0}
    ^{\lfloor \frac{d}{2}\rfloor -1}(q_k-q_{-(k+1)})-2\right]
    +(1-p)\left(-1+\frac{1}{d}\right),
\end{equation}

On the other hand, given that this best known violation
increases linearly with $p$, it is also easy to see that there
exists a threshold weight $p=p_d$ (sometimes called the
visibility parameter) below which $\rI(p)$ is not known to
violate the $I_{22dd}$ inequality. Explicit values for some of
these threshold weights can be found in column 6 of
Table~\ref{tbl:MaximalI22nnViolation}. In principle, it is of
course possible that $\rI(p)$ with $1/(d+1)\le p<p_d$ violates
$I_{22dd}$ and/or other Bell inequalities for
$\PLHV{2}{d}{2}{d}$ with some other choice of measurements.
However, preceding results due to
Kaslikowski\etal~\cite{D.Kaszlikowski:PRL:2000} suggest that
$p_d$ could very well be the threshold $p$ below which $\rI(p)$
does not violate any Bell inequalities for $\PLHV{2}{d}{2}{d}$
(see also Refs.~\cite{J.L.Chen:PRA:2001,L.B.Fu:PRA:2003,
Li-Bin.Fu:PRA:2004} in this regard). In other words,
Eq.~\eqref{Eq:SqmI22dd:Isotropic} may very well give the
maximal $I_{22dd}$-violation for the isotropic states.

\section{Better Bell-inequality Violation by Collective Measurements}
\label{Sec:NonlocalityEnhancement}

\subsection{Multiple Copies of Pure States}
\label{Sec:pure-states}

Let us now look into the problem of whether stronger
nonclassical correlations can be derived by performing
collective measurements on $N>1$ copies of an entangled quantum
state\footnote{Notice that the maximal Bell inequality
violation for $N>M$ copies of a quantum system is never less
than that involving only $M$ copies. This follows from the fact
that the maximal  $M$-copy violation can always be recovered in
the $N$-copy scenario by performing the
$M$-copy-optimal-measurement on $M$ of the $N$ copies, while
leaving the remaining $N-M$ copies untouched.}. As our first
example of nonlocality  enhancement, consider again those pure
maximally entangled two-qudit states residing in Hilbert space
with odd $d$. As remarked earlier, it is well-known that their
maximal Bell-CH/ Bell-CHSH inequality violation cannot saturate
Tsirelson's bound~\cite{S.Popescu:PLA:1992b}. In fact, their
best known Bell-CH inequality violation~\cite{N.Gisin:PLA:1992}
is that given in Eq.~\eqref{Eq:S:ME}. By combining $N$ copies
of these quantum states, it is readily seen that we effectively
end up with another maximally entangled state of
$d^N$-dimension. It then follows from Eq.~\eqref{Eq:S:ME} that
their Bell-CH violation under collective measurements increases
monotonically with the number of copies $N$ (see also
Table~\ref{tbl:MaximalViolation}, column 3 and 7). In fact, it
can be easily shown that this violation approaches
asymptotically the Tsirelson bound~\cite{B.S.Cirelson:LMP:1980}
in the limit of large $N$. Therefore, if the maximal violation
of these quantum states is given by Eq.~\eqref{Eq:S:ME}, which
is the case for $d=3$~\cite{T.Ito:PRA:2006}, collective
measurements can already give better Bell-CH violation with
$N=2$. Even if the maximal violation is not given by
Eq.~\eqref{Eq:S:ME}, it can be seen (by comparing the upper
bound of the single-copy violation from the UB algorithm and
the lower bound of the $N$-copy violation) from
Table~\ref{tbl:MaximalViolation} that for $d=5$, a Bell-CH
violation better than the maximal single-copy violation can
always be obtained when $N$ is sufficiently large.

Such an enhancement is even more pronounced in the case of
non-maximally entangled states. In particular, for $N$ copies
of a (non-maximally entangled) two-qubit state written in the
Schmidt basis,
\begin{equation}\label{Eq:NME-2-qubit}
    \ket{\Phi_2}^{\otimes N}=\left(\cos\phi\ket{0}_{\A}\ket{0}_\B
    +\sin\phi\ket{1}_\A\ket{1}_\B\right)^{\otimes  N},
\end{equation}
where $0<\phi\le\frac{\pi}{4}$~\footnote{For
$\frac{\pi}{4}<\phi<\frac{\pi}{2}$, we just have to redefine
$\phi$ as $\frac{\pi}{2}-\phi$ and all the subsequent results
follow.}, the Bell-CH violation given by
Eq.~\eqref{Eq:Sqm:CH:pure} is
\begin{equation}\label{eqn:S:NME}
    \eBellCH_{\ket{\Phi_2}}=\frac{p}{\sqrt{2}}+
    \frac{1-p}{2}\sqrt{1+\sin^22\phi}-\half,
\end{equation}
where
\begin{equation*}
    p=1-\half\cos^{2(N-1)}\phi\sum_{m=0}^{N-1}\tan^{2m}\phi
    \left[1-(-1)^{\frac{(N-1)!}{m!(N-1-m)!}}\right],
\end{equation*}
is the total probability of finding $\ket{\Phi_2}^{\ten N}$  in
one of the {\em perfectly correlated 2-dimensional subspaces}
(i.e., a subspace with $c_{2n-1}=c_{2n}$) upon reordering  of
the Schmidt coefficients in descending order.

\begin{figure}[h!]
    \centering\rule{0pt}{4pt}\par
    \includegraphics[width=12.5cm,height=9.7cm]{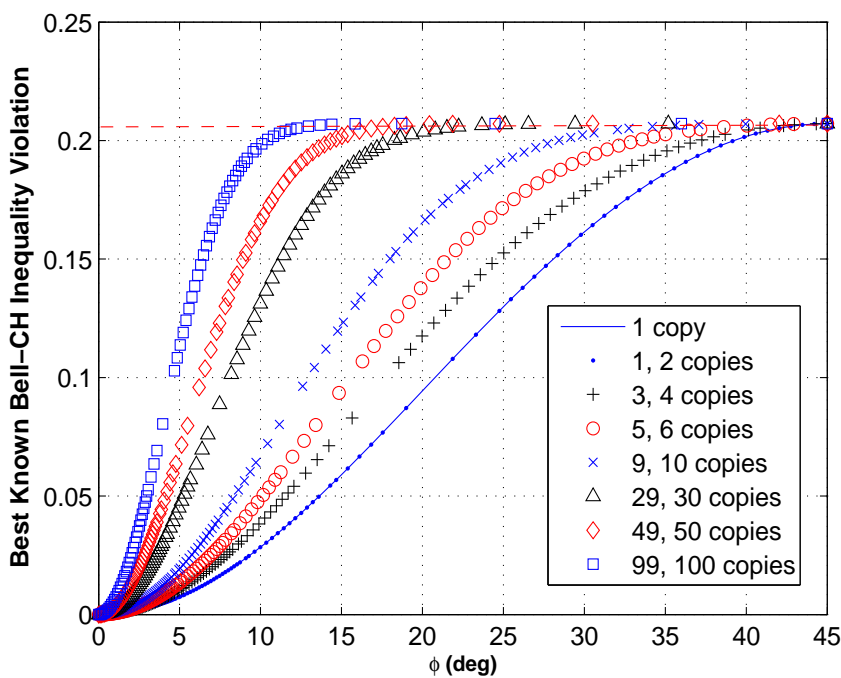}
    \caption{\label{Fig:S:NME:Qubit}
    Best known Bell-CH inequality violation of pure two-qubit
    states obtained from Eq.~\eqref{Eq:Sqm:CH:pure}, plotted as  a
    function of $\phi$, which gives a primitive measure of
    entanglement; $\phi=0$ for bipartite pure product state and
    $\phi=45^o$ for bipartite maximally entangled state.The curves
    from right to left represent increasing numbers of copies. The
    dotted horizontal line at $\frac{1}{\sqrt{2}}-\half$ is the
    maximal possible violation of Bell-CH inequality; correlations
    allowed by locally causal theories have values less than or
    equal to zero. The solid line is the maximal Bell-CH inequality
    violation of $\ket{\Phi_2}$ determined using the Horodecki
    criterion, c.f. Appendix~\ref{App:Sec:HorodeckiCriterion}.
    }
\end{figure}

It is interesting to note that for these two-qubit states,
their Bell-CH inequality violation for $N=2k-1$ copies, and
$N=2k$ copies are identical\footnote{This can be rigorously
shown using combinatoric arguments (private communication,
Henry Haselgrove).} for all $k\ge 1$, as illustrated in column
2 of Table~\ref{tbl:MaximalViolation} and in
Figure~\ref{Fig:S:NME:Qubit}. This feature, however, does not
seem to generalize to higher dimensional quantum states.

Like the odd-dimensional maximally entangled state, the
violation of the Bell-CH inequality for {\em any} pure
two-qubit entangled states, as given by
Eq.~\eqref{Eq:Sqm:CH:pure}, increases asymptotically towards
the Tsirelson bound~\cite{B.S.Cirelson:LMP:1980} with the
number of copies $N$, as can be seen in
Figure~\ref{Fig:S:NME:Qubit}.

\begin{table}[!h]
\centering\rule{0pt}{4pt}\par
\caption{\label{tbl:MaximalViolation} Best known Bell-CH
inequality violation for some bipartite pure entangled states,
obtained from  Eq.~\eqref{Eq:Alice-POVM} and
Eq.~\eqref{Eq:Bob-POVM} with and without collective
measurements. Also included below is the upper bound on
$\SqmCH(\ketbra{\Phi})$ obtained from the UB algorithm. Each of
these upper bounds is marked with a $^\dag$. The first column
of the table gives the number of copies $N$ involved in the
measurements. Each quantum state is labeled by its non-zero
Schmidt coefficients, which are separated by : in the
subscripts attached to the ket vectors; e.g.,
$\ket{\Phi}_{3:3:2:1}$  is the state with unnormalized Schmidt
coefficients $\{c_i\}_{i=1}^4=\{3,3,2,1\}$. For each quantum
state there is a box around the entry corresponding to the
smallest $N$ such that the lower bound on
$\SqmCH(\ketbra{\Phi})$ exceeds the single-copy upper bound
(coming from the UB algorithm or otherwise). A violation that
is known to be maximal is marked with a $^*$.}
    \begin{tabular}{r|cccccc}
    $N$ & $\ket{\Phi_{2:1}}$ & $\ket{\Phi_{1:1:1}}$
        & $\ket{\Phi_{3:2:1}}$ & $\ket{\Phi_{4:3:2:1}}$ &
        $\ket{\Phi_{3:3:2:1}}$ & $\ket{\Phi_{1:1:1:1:1}}$ \\ \hline
      &Lower & Bound& & & & \\
    \hline
      1 & 0.14031* & 0.13807* & 0.16756 & 0.18431 & 0.19259 & 0.16569\\
      2 & 0.14031 & \boxed{0.18409} & 0.18307 & 0.19624 & 0.20516 & 0.19882\\
      3 & \boxed{0.16169} & 0.19944 & 0.19451 & 0.20275 & 0.20685 & 0.20545\\
      4 & 0.16169 & 0.20455 & \boxed{0.19642} & 0.20388 & 0.20706 & \boxed{0.20678}\\
      5 & 0.17964 & 0.20625 & 0.20254 & 0.20596 & 0.20710 & 0.20704\\
     10 & 0.19590 & 0.20710 & 0.20643 & 0.20704 & 0.20711 & 0.20711 \\
     \hline
      &Upper & Bound& & & & \\
    \hline
      1 & 0.14031* & 0.13807* & 0.19624$^\dag$
      & 0.20711$^\dag$ & 0.20711$^\dag$ & 0.20569$^\dag$ \\
    \end{tabular}
\end{table}

Similarly, if we consider $N$ copies of pure two-qutrit
entangled states written in the Schmidt form,
\begin{equation}
    \ket{\Phi_3}^{\otimes N}= \left(\cos\phi\ket{0}_\A\ket{0}_\B
    +\sin\phi\cos\theta\ket{1}_\A\ket{1}_\B+
    \sin\phi\sin\theta\ket{2}_\A\ket{2}_\B\right)^{\otimes N},
\end{equation}
where $0<\phi\le\frac{\pi}{4}$, $0<\theta\le\frac{\pi}{4}$, it
can be verified that their Bell-CH inequality violation, as
given by  Eq.~\eqref{Eq:Sqm:CH:pure}, also increases steadily
with the number of copies. Thus, if  Eq.~\eqref{Eq:Sqm:CH:pure}
gives the maximal Bell-CH violation for pure two-qutrit states,
better Bell-inequality violation can also be attained by
collective measurements using two copies of these quantum
states. The explicit value of the violation can be found in
column 3 and 4 of Table~\ref{tbl:MaximalViolation} for two
specific two-qutrit states. As above, even if the maximal
Bell-CH violation is not given by  Eq.~\eqref{Eq:Sqm:CH:pure},
collective measurements with  Eq.~\eqref{Eq:Alice-POVM} can
definitely give a violation that is better than the
maximal-single-copy ones as a result of the bound coming from
the UB algorithm for a single copy (see
Table~\ref{tbl:MaximalViolation}). Corresponding examples for
pure bipartite 4-dimensional and 5-dimensional quantum states
can also be found in the table.

Some intuition for the way in which better Bell-CH inequality
violation may be obtained with collective measurements and the
measurement scheme given by Eq.~\eqref{Eq:Alice-POVM} and
Eq.~\eqref{Eq:Bob-POVM} is that the reordering of subspaces
prior to these measurements generally increases the total
probability of finding 2-dimensional subspaces with
$c_{2n}=c_{2n-1}$, while ensuring that the remaining
2-dimensional subspaces are at least as correlated as any of
the corresponding single-copy 2-dimensional subspaces. The
measurement then effectively projects onto each of these
subspaces (with Alice and Bob being guaranteed to obtain the
same result) and then performs the optimal measurement on the
resulting shared two-qubit state. Since the optimal
measurements in each of these perfectly correlated
2-dimensional subspaces gives the maximal Bell-CH inequality
violation, while the same measurements in the remaining
2-dimensional subspaces give as much violation as the
single-copy violation, the multiple-copy violation is thus
generally greater than that of a single copy.

As one may have noticed, our measurement protocol bears some
resemblance with the entanglement concentration protocol
developed by Bennett\etal~\cite{C.H.Bennett:PRA:1996}. In
entanglement concentration, Alice and Bob make slightly
different projections onto subspaces that are spanned by all
those ket vectors sharing the same Schmidt coefficients thus
obtaining a maximally entangled state in a bipartite system of
some dimension. One can also obtain improved Bell inequality
violations by adopting their protocol and first projecting
Alice's Hilbert space into one of the perfectly correlated
subspaces and performing the best known measurements for a Bell
inequality violation in each of these (not necessary
2-dimensional) subspaces. We have compared the Bell-CH
inequality violation of an arbitrary pure two-qubit state
derived from each of these protocols and found that the
violation obtained using our protocol always outperforms the
one based on entanglement concentration. The difference,
nevertheless, diminishes as $N\to\infty$. This observation
provides another consistency check of the optimality of
Eq.~\eqref{Eq:Sqm:CH:pure}.

\subsection{Multiple Copies of Mixed States}
\label{Sec:mixed-states}

The impressive enhancement in a pure state Bell-CH inequality
violation naturally leads us  to ask if the same conclusion can
be drawn for mixed entangled states. The possibility of
obtaining better Bell inequality violation with collective
measurements, however, does not seem to generalize to all
entangled states.

\begin{figure}[h!]
    \centering\rule{0pt}{4pt}\par
    \includegraphics[width=12cm,height=9.32cm]
    {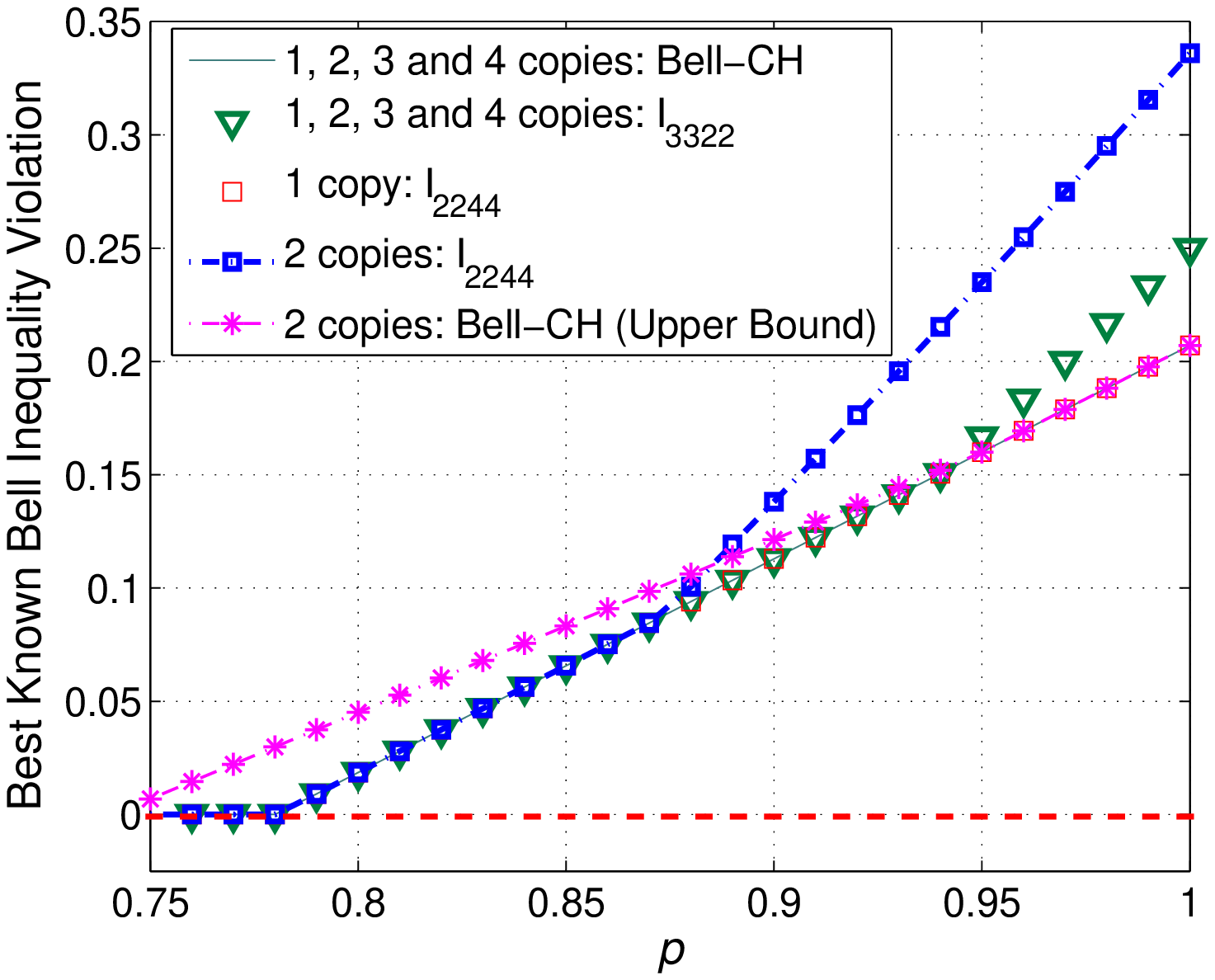}
    \caption{\label{Fig:Werner}
    Best known expectation value of the Bell operator coming from
    the Bell-CH inequality [$\Bell_{\rm CH}$,
    Eq.~\eqref{Eq:BellOperator:CH}], $I_{3322}$ inequality
    [$\Bell_{I_{3322}}$, Eq.~\eqref{Eq:Bell-3322:Operator}], and
    the $I_{2244}$ inequality [$\Bell_{I_{2244}}$,
    Eq.~\eqref{Eq:BellOperator:I22nn}] with respect to the
    2-dimensional Werner states $\rwd{2}(p)$; $p$ represents the
    weight of the spin-$\half$ singlet state in the mixture. Note
    that the best $I_{2244}$-violation found for
    $\ket{\Psi^-}^{\otimes2}$ agrees with the best known violation
    presented in Table~\ref{tbl:MaximalI22nnViolation}. Also
    included is the upper bound on $\SqmBI{CH}(\rwd{2}^{\ten 2})$
    obtained from the UB algorithm.
    }
\end{figure}

Our first counterexample comes from the 2-dimensional Werner
state, Eq.~\eqref{Eq:WernerState:NoisyProjector}, which can
seen as a mixture of the spin-$\half$ singlet state and the
maximally mixed state,
\begin{equation}
    \rwd{2}(p)=p\,\ketbra{\Psi^-}(1-p)\frac{\unit_2\ten\unit_2}{4},
\end{equation}
where $p$ is the weight of $\ket{\Psi^-}$ in the mixture. This
state is entangled for $p>1/3$ (c.f.
Sec.~\ref{Sec:WernerStates}) and from the Horodecki criterion
(Appendix~\ref{App:Sec:HorodeckiCriterion}) one can easily show
that it violates the Bell-CH inequality if and only
if~\cite{RPM.Horodecki:PLA:1995}
\begin{equation}
    p>p_w\equiv\frac{1}{\sqrt{2}}\simeq 0.707~11
\end{equation}

\begin{figure}[h!]
    \centering\rule{0pt}{4pt}\par
    \includegraphics[width=12cm,height=9.32cm]{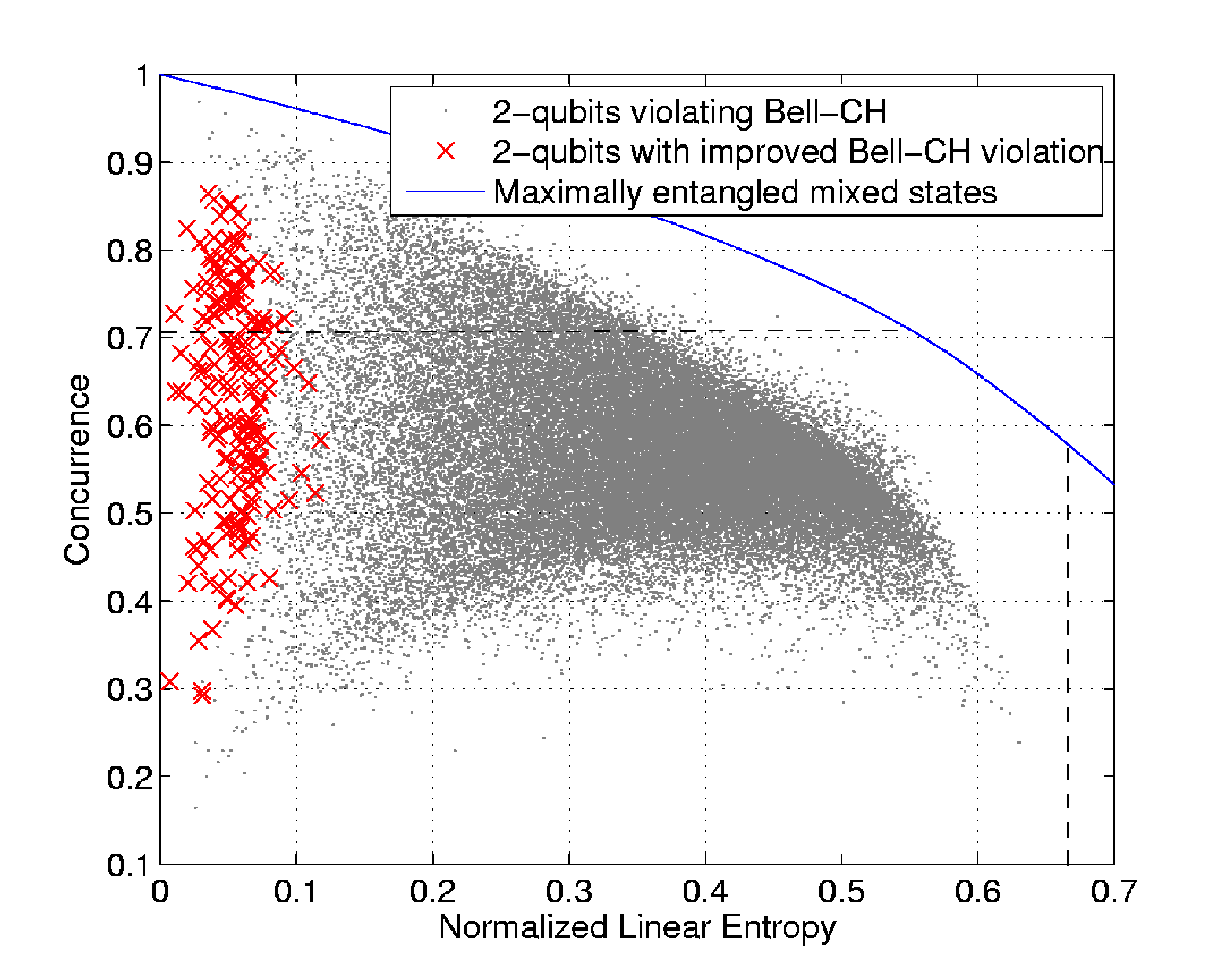}
    \caption{\label{Fig:CS-CH-Enhancement}
    Distribution of two-qubit states sampled for better Bell-CH
    violation by collective measurements. The maximally entangled
    mixed states (MEMS), which demarcate the boundary of the set of
    density matrices on this concurrence-entropy
    plane~\cite{W.J.Munro:PRA:2001,T-C.Wei:PRA:2003}, are
    represented by the solid line. Note that as a result of the
    chosen distribution over mixed states this region is not well
    sampled. The region bounded by the solid line and the
    horizontal dashed line (with concurrence equal to $1/\sqrt{2}$)
    only contain two-qubit states that {\em violate} the Bell-CH
    inequality~\cite{L.Derkacz:PRA:2005}; the region bounded by the
    solid line and the vertical dashed line (with normalized linear
    entropy equal to $2/3$) only contain states that {\em do not
    violate} the Bell-CH inequality~\cite{E.Santos:PRA:2004,
    E.Santos:PRA:2004b,L.Derkacz:PRA:2005}. Two-qubit states found
    to give better 3-copy Bell-CH violation are marked with red
    crosses.
    }
\end{figure}

Using the LB algorithm, we have searched for the maximal
violation of $\rwd{2}(p)$ with $p>p_w$ for $N\le 4$ copies but
no increase in the maximal violation of Bell-CH inequality has
ever been observed  (see Figure~\ref{Fig:Werner}). In fact, by
using the UB algorithm, we find that for two copies of some
Bell-CH violating Werner states, their maximal Bell-CH
inequality violation are identical to the corresponding
single-copy violation within a numerical precision of
$10^{-12}$. This strongly suggests  that for some Werner states
the maximal Bell-CH inequality violation does not depend on the
number of copies $N$.

There are, nevertheless, some two-qubit states whose maximal
Bell-CH inequality violation for $N=3$ is higher than the
corresponding single-copy violation. In contrast to the pure
state scenario, the set of mixed two-qubit states seems to be
dominated by those whose 3-copy Bell-CH inequality violation is
not enhanced.  In fact, among 50,000 randomly generated Bell-CH
violating two-qubit states\footnote{We follow the algorithm
presented in Ref.~\cite{K.Zyczkowski:PRA:1998} to generate
random two-qubit states. In particular, the eigenvalues
$\{\lambda_i\}_{i=1}^4$ of the quantum states were chosen from
a uniform distribution on the 4-simplex defined by $\sum_i
\lambda_i=1$.},  only about $0.38\%$ of them were found to have
their 3-copy Bell-CH inequality violation greater than their
maximal single-copy violation. Moreover, as can be seen in
Figure~\ref{Fig:CS-CH-Enhancement}, they are all clustered at
regions with relatively low linear entropy.

\begin{figure}[h!]
    \centering\rule{0pt}{4pt}\par
    \includegraphics[width=12cm,height=9.32cm]{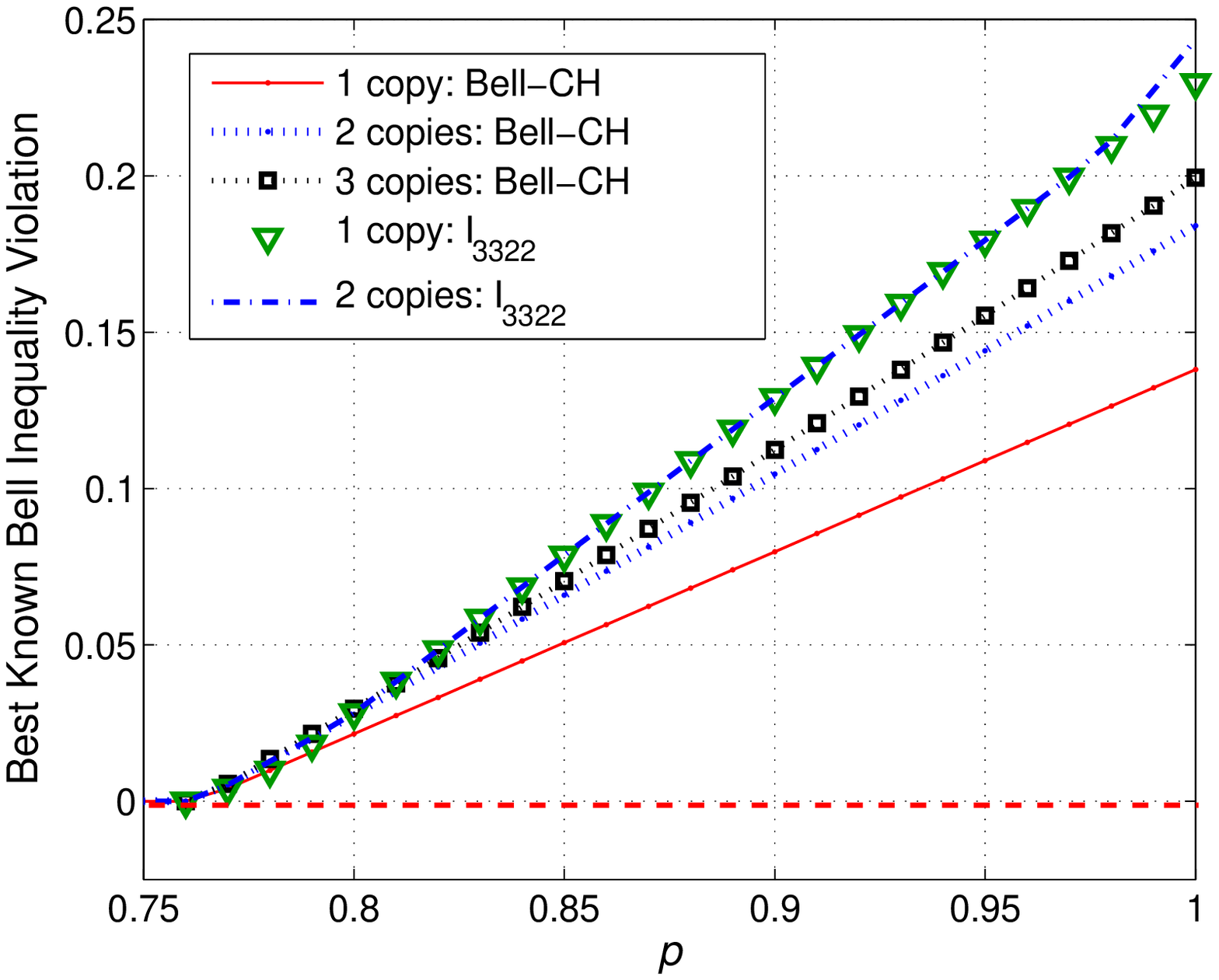}
    \caption{\label{Fig:MEQt}
    Best known expectation value of the Bell operator coming from
    the Bell-CH inequality [$\Bell_{\rm CH}$,
    Eq.~\eqref{Eq:BellOperator:CH}] and the $I_{3322}$ inequality
    [$\Bell_{I_{3322}}$, Eq.~\eqref{Eq:Bell-3322:Operator}], with
    respect to the 3-dimensional isotropic states, $\rId{3}(p)$;
    $p$ is the weight of maximally entangled two-qutrit state in
    the mixture. The single copy Bell-CH inequality violation found
    here through LB is identical with the maximal violation,
    $\SqmBI{CH}(\rId{3})$, found by Ito\etal~in
    Ref.~\cite{T.Ito:PRA:2006}.
    }
\end{figure}

As with the pure state scenario, an enhancement of nonclassical
correlations in the Bell-CH setting seems to be more prevalent
in higher dimensional quantum systems. In particular, for all
the 3-dimensional isotropic states
[Eq.~\eqref{Eq:IsotropicState:d3}] that violate the Bell-CH
inequality, numerical results obtained from the LB algorithm
suggest that the maximal violation increases steadily with the
number of copies. The results are summarized in
Figure~\ref{Fig:MEQt}.

Yet another question that one can ask is how much does the
enhancement of nonclassical correlations depend on the choice
of Bell inequality. To address this question, we have also
studied the enhancement of nonclassical correlations with
respect to other Bell inequalities for probabilities, in
particular the $I_{3322}$ inequality given in
Eq.~\eqref{Ineq:Functional:I3322}, the $I_{2233}$ inequality
given in Eq.~\eqref{Ineq:Prob:I2233} and the $I_{2244}$
inequality given in Eq.~\eqref{Ineq:Prob:I2244}. For these Bell
inequalities, we find that the possibility of enhancing
nonclassical correlations does seem to depend on both the
number of alternative settings and the number of possible
outcomes involved in a Bell experiment. The dependence on the
number of outcomes is particularly prominent in the case of
2-dimensional Werner states, where a large range of
$I_{2244}$-inequality-violating $\rwd{2}(p)$ seem to achieve a
higher two-copy violation, even though their maximal Bell-CH
inequality violation apparently remains unchanged up to $N=4$
(Figure~\ref{Fig:Werner}).

The dependence on the number of alternative settings can be
seen in the best known violation of $\rId{3}(p)$ with respect
to the Bell-CH inequality and the $I_{3322}$ inequality
(Figure~\ref{Fig:MEQt}). In particular, when the number of
alternative settings is increased from 2 (in the case of
Bell-CH inequality) to 3 (in the case of $I_{3322}$
inequality), the range of states whereby collective
measurements were found to improve the Bell inequality
violation is drastically {\em reduced}.\\

\section{Conclusion}

In this chapter, we have focused on bipartite entangled systems
and analyzed the extent to which a given entangled state can
violate a given Bell inequality. For the Bell-CH inequality,
the measurement scheme that we have presented in
Sec.~\ref{Sec:Bell-CH:Pure2Qudit:Measurements} has allowed us
to obtain the best known violation of any pure two-qudit states
for this inequality. A general proof that the measurement is
indeed optimal seems formidable. However, the resulting
violation does reproduce known (optimal) results in various
special cases, including the maximal Bell-CH violation for
3-dimensional isotropic states $\rId{3}(p)$. In addition,
intensive numerical studies have not provided a single instance
where the presented measurement is outperformed. In
Sec.~\ref{Sec:I22nnViolation}, we have also briefly reviewed
the best known $I_{22dd}$-violation for the $d$-dimensional
isotropic states, $\rI(p)$.

Next, we considered the enhancement of nonclassical
correlations by collective measurements without postselection.
This amounts to allowing an experiment in which a local unitary
is applied to a number of copies of the state $\rho$ prior to
the Bell inequality experiment. We find that the Bell-CH
inequality violation of all bipartite pure entangled states,
can be enhanced by allowing collective measurements even
without postselection. For mixed entangled states, however,
explicit examples (Werner states) have been presented to
demonstrate that there may be entangled states whose
nonclassical correlations cannot be enhanced in any Bell-CH
experiments. In fact, the set of mixed two-qubit states whose
Bell-CH violation can be increased with collective measurements
seems to be relatively small.

We have also done some preliminary studies on how the
usefulness of collective measurements depends on the choice of
Bell inequality and on the dimension of the subsystem. Our data
at the moment are consistent with the hypothesis that the
usefulness of collective measurements in Bell inequality
experiments increases with the Hilbert space dimension and with
the number of measurement outcomes allowed by Bell inequality.
On the other hand as the number of measurement settings allowed
by the Bell inequality increases the advantage provided by
collective measurements seems to diminish. However, note that
we have not really performed the systematic study required to
establish such trends, if they exist, due to the great
numerical effort that would be required. Given these
observations, it does seem that postselection is a lot more
powerful than collective measurements on their own in
increasing Bell-inequality violation.

An immediate question that follows from the present work is
what is the class of quantum states whereby collective
measurements can increase their Bell inequality violation? One
motivation for studying our problem is to understand better the
set of quantum states that can lead to a Bell inequality
violation and are thus inconsistent with a locally causal
description. It has been known for a long time that this set is
a strict subset of the entangled states if
projective~\cite{R.F.Werner:PRA:1989} or even generalized
measurements~\cite{J.Barrett:PRA:2002} on single copies of a
system are permitted. One might wonder whether collective
measurements without postselection allow us to violate Bell
inequalities for a larger set of states. However we do not know
of examples where a state that does not violate a given Bell
inequality becomes violating under collective measurements when
no postselection is allowed~\cite{O.Krueger:0504166}. Moreover,
for mixed states, the set of states whose violations increase
when collective measurements are allowed appears to be rather
restricted. This is consistent with the recent work by
Masanes~\cite{Ll.Masanes:PRL:2006} which suggests that the set
of states that violates a given Bell inequality under
collective measurements without postselection is a subset of
all distillable states.

Finally, the analysis that we have presented in this chapter
only concerns bipartite quantum systems. Given that
multipartite entanglement is fundamentally richer than the
bipartite analogue, it should also be interesting to
investigate the possibility of enhancing nonclassical
correlations by collective measurements in the multipartite
setting.

\chapter{Nonstandard Bell Experiments and Hidden Nonlocality}
\label{Chap:Hidden.Nonlocality}

As we have seen in Chapter~\ref{Chap:Q.Cn:Classical}, some
quantum states, despite being entangled, {\em cannot} violate
any Bell inequalities via a standard Bell experiment.
Nonetheless, it is now well-known that nonclassical
correlations can be derived from many of these entangled states
if we consider more sophisticated Bell experiments which also
allow appropriate local preprocessing --- deriving nonclassical
correlations from all entangled states via such nonstandard
Bell experiments will be the subject of discussion in this
chapter.

\section{Introduction}

Clearly, entanglement, being one of the most striking features
offered by quantum mechanics, is in some way responsible for
the generation of nonclassical correlations and hence the
bizarre phenomenon of Bell inequality violation. Operationally,
entanglement is defined in terms of the physical resources
needed for the {\em preparation} of the state (c.f.
Sec.~\ref{Sec:SeparableStates}): a multipartite state is said
to be entangled if it cannot be prepared from classical
correlations using local quantum operations assisted by
classical communication (LOCC)~\cite{R.F.Werner:PRA:1989}. This
definition, however, does not tell us anything about the
``behavior'' of such a state. For example, is an entangled
state useful in some quantum information processing task such
as teleportation,\footnote{For bipartite systems, this question
has been answered in Ref.~\cite{Ll.Masanes:PRL:2006b}.} or does
the state violate a Bell inequality? We have learned in
Sec.~\ref{Sec:States:LHVM:General} that with a standard Bell
experiment, not all entangled states can violate a Bell
inequality. But some of these states do violate Bell
inequalities if, prior to the measurement, the state is
subjected to appropriate local preprocessing. This phenomenon
has been termed {\it hidden
nonlocality}~\cite{S.Popescu:PRL:1995,N.Gisin:PLA:1996}.

Thus far, all existing protocols that demonstrate hidden
nonlocality in a nonstandard Bell experiment involve some kind
of {\it local filtering operations}. These are local
measurements that if successful are followed by a standard Bell
inequality experiment, but if unsuccessful result in the state
being discarded. Moreover, by allowing joint measurements on
several copies of the state in conjunction with local filtering
operations, Peres~\cite{A.Peres:PRA:1996} has shown that an
even larger set of two-qubit entangled states could be detected
through their violation of a Bell inequality. However, the
question of whether all entangled states might display some
kind of (hidden) nonlocality has remained open.

A possible generalization of Peres' idea would be to perform
local filtering operations and collective measurements on
arbitrarily many copies of a quantum state, and subject the
resulting state to a standard Bell inequality test. If the
resulting correlation violates a Bell inequality, the original
state is said to {\em violate this inequality
asymptotically}~\cite{Ll.Masanes:PRL:2006}. In
Ref.~\cite{Ll.Masanes:PRL:2006} it was shown that a bipartite
state violates the Bell-CHSH inequality asymptotically if, and
only if, it is distillable. This result suggests that
undistillable entangled states may admit a locally causal
description even when experiments are performed on an
arbitrarily large number of copies of the state.

As a result, it does seem necessary to consider even more
general protocols to derive nonclassical correlations that may
be hidden in an arbitrary entangled state. One natural
possibility is to allow joint processing with auxiliary states
(that do not themselves violate the Bell inequality) rather
than just with more copies of the state in question. In this
chapter, we will show that this kind of protocol is indeed
useful to derive nonclassical correlations from all entangled
states. This gives a conclusive answer to the long-standing
question of whether or not all entangled states can lead to
observable nonlocality~\cite{S.Popescu:PRL:1995,
N.Gisin:PLA:1996, J.Barrett:PRA:2002,N.Gisin:0702021}.

The structure of this chapter is as follows. In
Sec.~\ref{Sec:NonstandardBE:SingleCopy}, we will start off by
reviewing some of the nonstandard Bell tests where the system
of interest is measured one copy at a time. This is then
followed by a more general scenario whereby collective
measurements on multiple copies of the quantum system are
allowed in the nonstandard Bell experiment. After that, in
Sec.~\ref{Sec:Nonlocality:All}, we will provide a protocol
involving shared ancilla states to demonstrate the nonlocality
associated with all bipartite entangled states. Finally, we
will conclude with some possible avenues for future research.

\section{Single Copy Nonstandard Bell Experiments}
\label{Sec:NonstandardBE:SingleCopy}

\subsection{Nonstandard Bell Experiments on Pure Entangled
States}\label{Sec:NonstandardBE:SingleCopy:PureState}

The very first (implicit) proposal on a nonstandard Bell
experiment could be traced back to the influential work by
Gisin~\cite{N.Gisin:PLA:1991}. There, he considered a general,
entangled pure two-qudit state $\ket{\Phi_d}\in\HA\ten\HB$ with
$d\ge2$,
\begin{equation}
    \ket{\Phi_d}=\sum_{i=1}^d c_i\ket{i}_{\A}\ket{i}_{\B},
    \tag{\ref{Eq:TwoQudit:SchmidtForm}}
\end{equation}
but where local measurements are performed only on an entangled
two-qubit subspace.\footnote{From
Eq.~\eqref{Eq:TwoQudit:SchmidtForm}, it is evident that any
pair of correlated local bases $\{\ket{i}_\A,\ket{j}_\A\}$,
$\{\ket{i}_\B,\ket{j}_\B\}$ would define an entangled pure
two-qubit subspace for $\ket{\Phi_d}$.} By showing that any
entangled pure two-qubit state can violate the Bell-CHSH
inequality, Gisin has essentially also demonstrated that any
entangled pure two-qudit state can lead to a Bell-CHSH
violation by first performing the following {\em projections}
on the local subsystems
\begin{equation}
    \HA\to\Pi^{(2)}_\A\HA,\quad \HB\to\Pi^{(2)}_\B\HB,
\end{equation}
where
\begin{equation}\label{Eq:Dfn:2dProjections}
    \Pi^{(2)}_\A\equiv\AProj{i}+\AProj{j},\quad
    \Pi^{(2)}_\B\equiv\BProj{i}+\BProj{j},
\end{equation}
$\ket{i}_\A$, $\ket{j}_\A\in\HA$ are any pair of orthogonal
basis vectors defined in Eq.~\eqref{Eq:TwoQudit:SchmidtForm}
and $\ket{i}_\B$, $\ket{j}_\B$ are the corresponding correlated
basis states in $\HB$.

In effect, these local projections bring the pure two-qudit
state $\ket{\Phi_d}$ into a pure two-qubit state $\ket{\Phi_2}$
\begin{equation}\label{Eq:SLOCC:d->2}
    \ket{\Phi_d}\to\ket{\Phi_2}\propto\Pi^{(2)}_\A\ten\Pi^{(2)}_\B\ket{\Phi_d},
\end{equation}
with some probability of success. Clearly, such local
transformation does not always succeed. In the event that it
fails, the resulting state is discarded\footnote{A modification
to this scheme, proposed by Popescu and
Rohrlich~\cite{S.Popescu:PLA:1992}, would bypass postselection
but, instead, perform trivial local measurements $\unit_{d-2}$
whenever the received subsystem falls outside the qubit
subspace. In this case, they showed that such measurement
scheme could also lead to a (non-optimal) Bell-CHSH violation
for any entangled pure two-qudit state.} but whenever the
transformation succeeds, the resulting two-qubit state
$\ket{\Phi_2}$ is further subjected to a standard Bell-CHSH
experiment to unveil its nonclassical feature. Of course, as
Gisin and Peres~\cite{N.Gisin:PLA:1992} subsequently
demonstrated, nonclassical correlations can also be derived
directly from any pure two-qudit entangled states via a
standard Bell experiment (see
Sec.~\ref{Sec:Bell-CH:Pure2Qudit:Measurements}).

Whether the same can be said for multipartite entangled states
still remains unclear at present. When the number of parties
(denoted by $n$) is 3, Chen\etal~\cite{J.L.Chen:PRL:2004} have
presented strong evidence that all tripartite pure entangled
states violate a Bell inequality that they have derived. Beyond
this, it is still not known if a general $n$-partite pure
entangled state can violate some Bell inequality via a standard
Bell experiment. Nonetheless, as Popescu and Rohrlich showed in
Ref.~\cite{S.Popescu:PLA:1992}, all $n$-partite pure entangled
states do lead to a Bell inequality violation after appropriate
local filtering operations. The key idea behind their proof is
to realize that by suitable choice of local projection on $n-2$
out of $n$ subsystems, a local transformation that brings an
$n$-partite pure entangled state to a bipartite pure entangled
state is always achievable with some nonzero probability. Then,
conditioned on the success of this local transformation, the
desired bipartite entangled state can further be subjected to,
say, the above-mentioned scheme proposed by
Gisin~\cite{N.Gisin:PLA:1991}, or to the measurement scheme
described in Sec.~\ref{Sec:Bell-CH:Pure2Qudit:Measurements},
which will lead to a Bell-CHSH violation coming from any
$n$-partite pure entangled state.

\subsection{Nonstandard Bell Experiments on Mixed Entangled
States}

Let us now turn our attention to mixed states. Clearly, since
some mixed entangled states, e.g. the Werner states $\rw(p)$
with $p\le\pLPOVM{W}$, admit a general LHVM description, we
cannot hope to find a Bell inequality violation of such states
via a standard Bell experiment. Nonetheless, as we will see in
this section, nonstandard Bell experiments --- in the form of
standard Bell experiments preceded with appropriate local
filtering operations --- can also help to demonstrate the
nonlocality that is apparently {\em hidden} in some of these
entangled quantum states.

\subsubsection{Nonlocality Hidden in Werner States}
\label{Sec:Nonlocality:Popescu}

At first glance, Werner's LHVM for $\rw(p)$ with
$p\le\pLPi{W}$, c.f. Sec.~\ref{Sec:WernerStates}, seems to have
suggested the impossibility of deriving nonclassical
correlations from such mixed entangled states. However, another
nonclassical feature displayed by {\em all\,} entangled,
2-dimensional Werner states --- namely, all entangled
$\rwd{2}(p)$ were found to be useful for
teleportation~\cite{S.Popescu:PRL:1994}
--- has led Popescu to think that there may be other less
straightforward way to derive nonclassical correlations from
these quantum states.

Indeed, via a nonstandard Bell experiment of the kind described
in Sec.~\ref{Sec:NonstandardBE:SingleCopy:PureState}, Popescu
\cite{S.Popescu:PRL:1995} has managed to show that for $d\ge5$,
Werner states admitting explicit LHVM can also lead to a
Bell-CHSH violation. Specifically, Popescu has considered the
Werner state, Eq.~\eqref{Eq:WernerState:NoisyProjector}, with
$p=\pLPi{W}$, i.e., the entangled Werner state whereby an
explicit LHVM for projective measurements is known. This
mixture can be written more explicitly as
\begin{equation}\label{Eq:WernerState@pLPi}
    \rw\left(\pLPi{W}\right)=\rw\left(1-\frac{1}{d}\right)
    =\frac{1}{d^2}\left(2\,\PiA+\frac{1}{d}\unit_{d}\ten\unit_d\right).
\end{equation}
By locally projecting each subsystems onto a 2-dimensional
subspace via Eq.~\eqref{Eq:Dfn:2dProjections}, i.e.,
\begin{equation}
    \rw\left(\pLPi{W}\right)\to
    \Pi^{(2)}_\A\ten\Pi^{(2)}_\B~\rw\left(\pLPi{W}\right)
    ~\Pi^{(2)}_\A\ten\Pi^{(2)}_\B,
\end{equation}
and after renormalization,\footnote{Note that this local
transformation only succeeds with probability
$\frac{2d+4}{d^3}$.} one obtains a 2-dimensional Werner state
with $p=p\,'\equiv d/(d+2)$, i.e.,
\begin{equation}
    \rwd{2}\left(p\,'\right)=\frac{d}{d+2}
    \left(\ketbra{\Psi^-}+\frac{1}{2\,d}\unit_{2}\ten\unit_2\right).
\end{equation}
Now, if this 2-dimensional state is further subjected to local
measurements that give maximal Bell-CHSH
violation\footnote{This can be obtained by applying appropriate
local unitary transformation to the measurement described in
Eq.~\eqref{Eq:Alice-POVM} and Eq.~\eqref{Eq:Bob-POVM}.} for the
singlet state $\ket{\Psi^-}$, one finds that\footnote{It can be
easily shown that these local measurements give zero
expectation value for the maximally mixed state
$\frac{\unit_2\ten\unit_2}{4}$.}
\begin{equation}
    \Sqm\left(\rwd{2}(p\,')\right)=\frac{d}{d+2}\times2\sqrt{2},
\end{equation}
which is greater than 2 for all $d\ge 5$. Therefore, for all
Werner states $\rw(p)$ with $p=\pLPi{W}$ and $d\ge 5$, even
though there exists an explicit LHVM which reproduces their
quantum mechanical probabilities (for projective measurements),
nonclassical correlations can be still derived from them by
first projecting the states locally, each onto an appropriate
2-dimensional subspace. This is an illustration of what is now
commonly called {\em hidden nonlocality}, where the
nonclassical correlations hidden in an entangled state only
shows up in a more sophisticated, nonstandard Bell
experiment.

\subsubsection{Nonlocality Hidden in Standard Bell-CHSH Experiment}
\label{Sec:Nonlocality:Gisin}

As opposed to Popescu's example, the term, ``hidden
nonlocality" has also been used in a looser sense where
non-Bell-CHSH-inequality-violating quantum states become
Bell-CHSH-inequality-violating in a nonstandard Bell
experiment~\cite{N.Gisin:PLA:1996}. In
Ref.~\cite{N.Gisin:PLA:1996}, Gisin has considered a class of
two-qubit states that is local unitarily equivalent to
\begin{equation}\label{Eq:Dfn:State:Gisin}
    \rG(p,\theta)\equiv p~\ketbra{\Phi_2}+\half(1-p)
    \Big(\AProj{0}\ten\BProj{1}+\AProj{1}\ten\BProj{0}\Big),
\end{equation}
where $p_0\equiv 1/(2-\sin2\theta)<p<1$. This state can be
interpreted as a mixture of the non-maximally entangled
pure-two qubit state $\ket{\Phi_2}$, c.f.
Eq.~\eqref{Eq:NME-2-qubit}, and the pure product states
$\ket{0}_\A\ket{1}_\B$, $\ket{1}_\A\ket{0}_\B$. Using the PPT
criterion for
separability~\cite{A.Peres:PRL:1996,MPR.Horodecki:PLA:1996}, it
can be easily shown that this mixture represents an entangled
state whenever $p>p_E\equiv 1/(1+\sin2\theta)$.

Moreover, from the Horodecki criterion
(Appendix~\ref{App:Sec:HorodeckiCriterion}), it is also not
difficult to show that despite being entangled, $\rG(p,\theta)$
with
\begin{equation}\label{Eq:LocalBeforeLocalFiltering}
    p_E< p\le p_L\equiv \frac{4}{4+\sin^22\theta},
\end{equation}
does not violate the standard Bell-CHSH inequality with {\em
any} choice of dichotomic measurements. However, in practice,
even if the source emits physical systems that are well
described by $\rG(p,\theta)$, it is not inconceivable that the
{\em end users} Alice and Bob may receive states that are
better described by a different density matrix
$\rG'(p,\theta)$, which could well lead to a Bell inequality
violation.

In particular, if $\rG(p,\theta)$ describes the polarization
state of photon pairs emitted from some source and where each
pair of photons are distributed, respectively, to Alice and Bob
via channels that both perform the following local filtering
operation
\begin{equation}\label{Eq:LocalFilters:Gisin}
    F_\A=F_\B=\left(
    \begin{array}{cc}
        \sqrt{\tan\theta} & 0\\
        0                 & 1\\
    \end{array}\right),
\end{equation}
then, at the end of the channels, Alice and Bob will receive a
state that is actually better described by
\begin{equation*}
    \rG'(p,\theta)\propto F_\A\ten F_\B~\rG(p,\theta)~F_\A^\dag\ten F_\B^\dag.
\end{equation*}
More explicitly, after normalization, the locally filtered
state reads
\begin{equation*}
    \rG'(p,\theta)
    =\frac{\tan\theta}{ p_{\mbox{\tiny suc.}} }
    \left[p\,\sin2\theta\,\ketbra{\Phi^+_2}+\half(1-p)
    \Big(\AProj{0}\ten\BProj{1}+\AProj{1}\ten\BProj{0}\Big)\right],
\end{equation*}
where $p_{\mbox{\tiny suc.}}=
\tan\theta\left[1-p\left(1-\sin2\theta\right)\right]$ is the
probability that they both receive a photon at their end. Note
that in contrast with the original state given by
Eq.~\eqref{Eq:Dfn:State:Gisin}, the resulting state
$\rG'(p,\theta)$ can now be described as a mixture of the
maximally entangled pure two-qubit state $\ket{\Phi^+_2}$ and
the same set of pure product states involved in
Eq.~\eqref{Eq:Dfn:State:Gisin}.

\begin{figure}[h!]
    \centering\rule{0pt}{4pt}\par
    \scalebox{0.8}{
    \includegraphics[width=14cm,height=10.88cm]
    {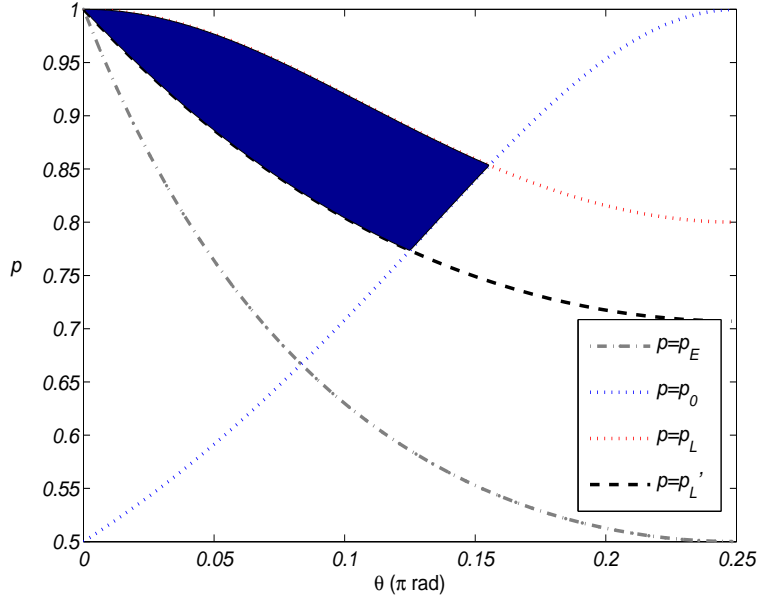}}
    \caption{\label{Fig:LocalFiltering:Gisin}
    The relevant parameter space for $\rG(p,\theta)$. The set of
    states that do not violate the Bell-CHSH inequality but which
    do after the local filtering operations given by
    Eq.~\eqref{Eq:LocalFilters:Gisin} is the shaded region bounded
    by the black dashed line ($p=p_L'$), the blue dotted line
    ($p=p_0$) and the red dotted line ($p=p_L$).
    }
\end{figure}

Again, from the Horodecki criterion, it can be shown that the
locally filtered state $\rG'(p,\theta)$ violates the Bell-CHSH
inequality if and only if
\begin{equation}\label{Eq:NonlocalAfterLocalFiltering}
    p>p_L'\equiv\frac{1}{1+(\sqrt{2}-1)\sin2\theta}.
\end{equation}
Now, if the intersection of the sets satisfying $p>p_0$,
Eq.~\eqref{Eq:LocalBeforeLocalFiltering} and
Eq.~\eqref{Eq:NonlocalAfterLocalFiltering} is not empty, one
will have found example(s) of two-qubit state not violating the
Bell-CHSH inequality but which does after the local filtering
operations given by Eq.~\eqref{Eq:LocalFilters:Gisin}. Indeed,
as can be seen from Figure~\ref{Fig:LocalFiltering:Gisin}, a
substantial subset of the class of states $\rG(p,\theta)$ do
satisfy the conjunction of all the above requirements. Hence,
as was first shown by Gisin~\cite{N.Gisin:PLA:1996}, there are
two-qubit states whose nonclassical correlations cannot be
observed directly in a standard Bell-CHSH experiment but if the
experiment is preceded with appropriate local filtering
operations, their hidden nonlocality do lead to observable
nonclassical correlations. It is worth noting that an
experimental demonstration of a very similar example has been
carried out and presented in Ref.~\cite{P.G.Kwiat:Nature:2001}.

\subsection{Justification of Single Copy Nonstandard Bell
Experiment}\label{Sec:HiddenNonlocality:Justification}

As we have seen in the examples given above, even if a
bipartite entangled quantum state $\rho\in\B(\HA\ten\HB)$ is
not known to violate a Bell inequality (or in some cases is
known to be NBIV), it may still be possible to observe a Bell
inequality violation coming from $\rho$, if, prior to the
standard Bell experiment, appropriate local filtering
operations are carried out. In effect, this transforms the
state $\rho$ locally to another quantum state $\Omega(\rho)$
via:
\begin{equation}\label{Eq:Dfn:LocalFiltering}
    \rho\to\Omega(\rho)=\sum_i F_{\A,i}\ten F_{\B,i}~\rho~
    F_{\A,i}^\dag\ten F_{\B,i}^\dag,
\end{equation}
where $F_{\A,i}$ and $F_{\B,i}$ are, respectively, local
filtering operators ({\em aka} Kraus operators\footnote{After
Kraus' work on completely positive
maps~\cite{K.Kraus:AP:1971,K.Kraus:Book:1983}.
Eq.~\eqref{Eq:Dfn:LocalFiltering} is also commonly known in the
literature as the Kraus decomposition of $\Omega(\rho)$.})
acting on subsystem $\A$ and $\B$. Up to some constant,
$\Omega(\rho)$ is also known as a {\em separable map} acting on
$\rho$. Evidently, since these local transformations do not
always succeed, some form of postselection, and hence
(classical) communication is involved when transforming the
state locally from $\rho$ to $\Omega(\rho)$. Indeed, this
nondeterministic nature of the local operations also result in
them being more commonly known in the literature as {\em
stochastic local quantum operations assisted by classical
communication} (henceforth abbreviated as SLOCC,
Appendix~\ref{App:Sec:SeparableMap})~\cite{W.Dur:PRA:2000}.

Naturally, the postselection involved in such SLOCC prior to a
standard Bell experiment reminds one of the {\em detection
loophole} discussed in a standard Bell test. An important
distinction between the two, as was first pointed out by
Popescu~\cite{S.Popescu:PRL:1995}, and subsequently by
\.Zukowski\etal~\cite{M.Zukowski:PRA:1998}, is that the
postselection is carried out {\em prior to}\footnote{In the
relativistic sense.} the standard Bell experiment. Therefore,
{\em a priori}, the postselection involved does not causally
depend on the choice of measurements made subsequently. In
addition, one should note that local filtering operation on any
quantum state $\rho$ {\em cannot} create nonclassical
correlations in the resulting state $\rho'$ --- local
operations assisted by classical communication cannot create
entanglement. As such, any nonclassical correlations derivable
from the resulting state $\rho'$ must have inherited from the
original state $\rho$. For a more rigorous version of this
argument, see the proof presented by \.Zukowski and coauthors
in Ref.~\cite{M.Zukowski:PRA:1998}.

\section{Nonstandard Bell Experiments on Multiple Copies}
\label{Sec:NonstandardBE:MultipleCopies}

The single-copy nonstandard Bell experiments that we have
considered in the previous section has certainly shed some
light on what can be done to reveal the nonclassical
correlations associated with an entangled quantum system. A
natural question that follows is whether this aspect of
nonclassicality can be demonstrated for arbitrary entangled
states. To this end, a negative answer was provided by
Verstraete and Wolf~\cite{F.Verstraete:PRL:2002} who showed
that a large class of two-qubit entangled states, including
some of the entangled Werner states, do not violate the
Bell-CHSH inequality even after an arbitrary local filtering
operation.

Of course, as with the complication involved in a standard Bell
experiment, it is still possible, at least in principle, that
some of these states actually violate some other more
complicated Bell inequalities after appropriate SLOCC.
Nevertheless, given that not much is known in this regard ---
even in the simpler scenario of a standard Bell experiment ---
it seems natural to consider other alternatives. In particular,
one could consider running a standard Bell experiment using
collective measurements on multiple copies of a quantum system.
The idea is that perhaps, one can find a quantum state $\rho$
not known to violate {\em any} Bell inequality when measured
one copy at a time but for $N$ large enough, one finds that
$\rho^{\ten N}$ does violates some Bell inequality. However, as
we have discussed in Chapter~\ref{Chap:BellViolation} (see
Sec.~\ref{Sec:mixed-states} in particular) no such example has
ever been found.

\subsection{Nonstandard Bell Experiments with Collective Measurements}
\label{Sec:NonstandardBE:ManyCopies:CollectiveMeasurements}

In the same vein as the single-copy scenario, why not consider
a standard Bell experiment that is preceded with SLOCC on
multiple copies of a quantum system? More precisely, even when
$\rho$, as well as $\rho^{\ten N}$ is not found to violate any
Bell inequality, it could still be that the following local
filtering operations prior to a standard Bell experiment is
useful in deriving nonclassical correlations from $\rho^{\ten
N}$:
\begin{equation}\label{Eq:Dfn:LocalFiltering:MoreCopies}
    \rho^{\ten N}\to\rho'\propto \sum_i F_{\A,i}\ten F_{\B,i}~\rho^{\ten N}~
    F_{\A,i}^\dag\ten F_{\B,i}^\dag,
\end{equation}
where here, it is worth noting that the tensor product between
$F_{\A,i}$ and $F_{\B,i}$ acts differently from the tensor
product involved in $\rho^{\ten N}$.

Indeed, this is exactly what Peres has contemplated to
demonstrate the nonlocality hidden in 2-dimensional Werner
states~\cite{A.Peres:PRA:1996}. More specifically, Peres has
considered a scenario where $N$ copies of $\rwd{2}(p)$ are
collected and further subjected to some local unitary
transformation acting on all the $N$ copies of the local
subsystems. After that, for both Alice and Bob, projective
measurements are carried out in the $Z$ basis for all but one
of the $N$ particles.\footnote{If there is a need to perform
measurement in any other basis, one can achieve that by first
performing additional unitary transformation on the particle in
question prior to a measurement on the $Z$
basis~\cite{A.Peres:PRA:1996}.} If all the $2(N-1)$ measurement
results are ``$\spup$", the remaining 2 particles are subjected
to a standard Bell-CHSH experiment, otherwise they are
discarded and the experiment is restarted.

With this protocol, Peres has shown that many $\rwd{2}(p)$ not
known to violate any Bell inequality do violate the Bell-CHSH
inequality after the described postselection. In particular,
with $N=5$ copies, Peres has found that, despite the explicit
LHVM constructed by Werner (see Sec.~\ref{Sec:WernerStates}),
the Werner state $\rwd{2}(1/2)$ does lead to a Bell-CHSH
inequality violation of 2.0087 via the above-mentioned
nonstandard Bell experiment. Moreover, due to the
distillability~\cite{C.H.Bennett:PRL:1996} of all 2-dimensional
entangled states~\cite{MPR.Horodecki:PRL:1997}, it is expected
that for sufficiently large $N$, all entangled $\rwd{2}(p)$
will lead to a Bell-CHSH inequality violation in this manner.

\subsection{Nonstandard Bell Experiments and Distillability}
\label{Sec:NonstandardBE:ManyCopies:Distillability}

As we have just seen, a nonstandard Bell experiment that allows
collective measurement on many copies of a quantum system and
postselection on some desired outcome is clearly more powerful
than all the other Bell experiments that we have described so
far. In particular, if we allow $N$ --- the number of copies --
to be arbitrarily large, it seems like we can go through these
procedures to derive nonclassical correlations out of a large
set of entangled states. The immediate question that follows is
whether this is a {\em strict subset} of the set of entangled
states. Evidently, if a state $\rho$ is distillable, one can
extract a spin-1/2 singlet state from $\rho^{\ten N}$ via some
local filtering operations, c.f.
Eq.~\eqref{Eq:Dfn:LocalFiltering:MoreCopies}, and therefore
$\rho$ violates a standard Bell-CHSH experiment that is
preceded with some SLOCC.

What about the converse? Must {\em undistillable entangled}
states ({\em aka} {\em bound entangled} states)  satisfy Bell
inequalities even if the Bell experiment is preceded with
arbitrary SLOCC? To answer this question, Masanes has
introduced the following definition in
Ref.~\cite{Ll.Masanes:PRL:2006}.
\begin{dfn}\label{Dfn:AsymptoticViolation}
A quantum state $\rho$ is said to violate a Bell inequality
asymptotically if $\rho^{\ten N}$ for an arbitrarily large $N$
violates the Bell inequality after some stochastic local
quantum operations without communication (SLO).
\end{dfn}
Notice that no communication is allowed in the above
definition. However, as it turns out, allowing classical
communication (i.e., with SLOCC instead of SLO) does not allow
more states to violate a Bell inequality in this
manner~\cite{Ll.Masanes:PRL:2006}.\footnote{Intuitively, one
can see that this is true by noting that the role of classical
communication, if any, in a nonstandard Bell experiment is
primarily to facilitate any postselection involved.} A partial
result to the above question is then provided by Masanes in the
following theorem~\cite{Ll.Masanes:PRL:2006}.
\begin{theorem}\label{Thm:AsymptoticViolation}
    A bipartite state $\rho$ is distillable if and only if it
    violates the Bell-CHSH inequality asymptotically. In other
    words $\rho$ is distillable if and only if there
    exists an $N\in\Zp$ and some SLO, denoted by $\Omega$ such that
    $\Omega\left(\rho^{\ten N}\right)$ violates the Bell-CHSH
    inequality.
\end{theorem}

Again, it is still logically possible that undistillable states
can violate some other Bell inequalities asymptotically.
Nonetheless, this theorem due to Masanes has clearly suggested
that one should also look for other alternatives to derive
nonclassical correlations, if any, associated with arbitrary
bipartite entangled states, especially the bound entangled
states.

\section{Observable Nonlocality for All Bipartite Entangled States}
\label{Sec:Nonlocality:All}

In this section, we will go beyond the typical nonstandard Bell
experiment and consider one that also involves shared ancilla
states. In particular, we will prove that via a local filtering
protocol that involves a specific ancilla state (which by
itself does not violate the Bell-CHSH inequality), one can
always observe a Bell-CHSH violation coming from a single copy
of any bipartite entangled state.

\subsection{Bipartite States with No Bell-CHSH Violation after SLOCC}
\label{Sec:CSLOCC}

To this end, let us first introduce the following definition
regarding the set of bipartite states that {\em do not violate}
the Bell-CHSH inequality even after arbitrary local filtering
operations. The nonstandard Bell experiment that we are going
to consider will involve an ancilla state which is a member of
this set.

\begin{dfn}
    Denote by $\CSLOCC{CHSH}$ the set of bipartite states that do
    not violate the Bell-CHSH inequality, even after SLO on a
    single copy of the state of interest.
\end{dfn}

As for Theorem~\ref{Thm:AsymptoticViolation}, it follows from
the results presented in Ref.~\cite{Ll.Masanes:PRL:2006} that
states in $\CSLOCC{CHSH}$ also do not violate the Bell-CHSH
inequality even after SLOCC --- hence the notation
$\CSLOCC{CHSH}$. The exact nature of the local operations
allowed in the definition of $\CSLOCC{CHSH}$ is thus not
important. Clearly, states that do not violate the Bell-CHSH
inequality asymptotically, c.f.
Definition~\ref{Dfn:AsymptoticViolation}, are in
$\CSLOCC{CHSH}$. Therefore, it follows from
Theorem~\ref{Thm:AsymptoticViolation} that $\CSLOCC{CHSH}$
contains all undistillable states~\cite{Ll.Masanes:PRL:2006}
(which include the set of bound entangled states as a subset).
As remarked earlier, there are no undistillable two-qubit
entangled states~\cite{MPR.Horodecki:PRL:1997}. However, from
the results presented in Ref.~\cite{F.Verstraete:PRL:2002}, we
know that there are also two-qubit entangled states that are in
$\CSLOCC{CHSH}$.

In what follows, we will describe a set of necessary and
sufficient conditions for a general bipartite state $\rho$ to
be in $\CSLOCC{CHSH}$. To begin with, we note that
$\CSLOCC{CHSH}$ is a {\em convex set}\footnote{The proof is
similar to the one presented in
Appendix~\ref{App:Sec:Convexity:LocalStates}.} and thus it can
be characterized via hyperplanes that separate this set from
any point outside the set. In particular, for any state $\rho$
that is not in $\CSLOCC{CHSH}$, a hyperplane that separates
$\rho$ from $\CSLOCC{CHSH}$ can be constructed; this hyperplane
therefore serves as a kind of witness operator that detects
Bell-CHSH violation of $\rho$ after some SLOCC.

\begin{lemma}\label{Lem:CharacterizationCSLOCC}
A bipartite state $\rho$ acting on $\H_{\A}\otimes\H_{\B}$
belongs to $\CSLOCC{CHSH}$ if, and only if, it satisfies
\begin{equation}\label{C}
    \tr \left[\rho~(F_\A\otimes F_\B)^\dag~  H_\theta~
    (F_\A\otimes F_\B)\right]
    \geq 0,
\end{equation}
for all matrices of the form $F_\A\!
:\H_{\A}\rightarrow\Cd{2}$, $F_\B\! :\H_{\B}\rightarrow\Cd{2}$
and all $\theta \in [0,\pi/4]$, where
\begin{equation} \label{Eq:Dfn:Htheta}
    H_\theta\equiv \unit_2\otimes\unit_2-\cos\theta\,
    \sigma_x\otimes\sigma_x - \sin\theta\,
    \sigma_z\otimes \sigma_z,
\end{equation}
$\unit_2$ being the $2\times 2$ identity matrix and
$\{\sigma_i\}_{i=x,y,z}$ are the Pauli matrices introduced in
Eq.~\eqref{Eq:Dfn:PauliMatrices}.
\end{lemma}

\begin{proof}
We shall prove this Lemma in two stages. Firstly, we will prove
a criterion analogous to inequality \eqref{C} for the scenario
where no local filtering operation is involved and when $\rho$
is a two-qubit state. Then, we will provide a proof for the
general scenario by incorporating existing results in
Ref.~\cite{Ll.Masanes:PRL:2006}.

Now, let us start with the special case of a two-qubit state
and where no local filtering operation is involved. Recall from
Sec.~\ref{Sec:Bell-CHSH:Horodecki} that in a standard Bell
experiment --- a Bell experiment without local preprocessing
--- a two-qubit state $\varrho$ violates the Bell-CH/ Bell-CHSH
inequality if and only
if~\cite{RPM.Horodecki:PLA:1995,F.Verstraete:PRL:2002}
\begin{equation}
    \varsigma_1^2+\varsigma_2^2>1,\tag{\ref{Eq:HorodeckiCriterion}}
\end{equation}
where $\varsigma_k$ is the $\idx{k}$ largest singular value of
the $3\times 3$ real matrix $T$ defined in
Eq.~\eqref{Eq:Dfn:T}. Equivalently, $(\varsigma_1,\varsigma_2)$
derived from $\varrho$ must lie outside the unit circle
$\varsigma_1^2+\varsigma_2^2=1$, which is true if and only if
there exists $\theta\in [0,2\pi]$ such that
\begin{equation}\label{Eq:Linear:CHSH}
    \varsigma_1\cos\theta+\varsigma_2\sin\theta >1.
\end{equation}
Now, it is also well-known that by appropriate local unitary
transformations $U$, $V$, it is always possible to arrive at a
local basis such that $T$ is diagonal\footnote{See for example
pp.~2227 of Ref.~\cite{B.G.Englert:JMO:2000}.} with
$\varsigma_1=T_{xx}$ and $\varsigma_2=T_{zz}$. From the
definition of $T$ it then follows that
\begin{equation}
  \varsigma_1\cos\theta = \tr \left[(U \otimes V)~
    \varrho~ (U \otimes V)^\dag~ (\cos\theta~
    \sigma_x\otimes\sigma_x)
    \right],
\end{equation}
with the expression for $\varsigma_2 \sin \theta$ involving
obvious modifications.  Since singular values are non-negative,
it thus follows that if $\varrho$ violates the Bell-CHSH
inequality then there exist $U,V\in\rm SU(2)$, $\theta\in
\left[0,\frac{\pi}{4}\right]$ such that
\begin{equation}\label{Cqubits}
    \tr \left[\varrho~(U \otimes V)^\dag~ H_\theta~
    (U \otimes V) \right]< 0.
\end{equation}

Conversely, suppose that there exists some $U,V\in\rm SU(2)$,
$\left[0,\frac{\pi}{4}\right]$ satisfying inequality
\eqref{Cqubits}, then it follows that
\begin{equation}\label{Eq:Txx+Tzz}
    T_{xx}\cos \theta +T_{zz}\sin \theta > 1.
\end{equation}
Since $\varsigma_1 \geq \varsigma_2$ by definition, the
inequalities
\begin{equation}\label{Eq:Tii:Bound}
    |T_{ii}|\leq \varsigma_1 \leq 1,\quad i\in\{x,y,z\},
\end{equation}
follow from the definition of singular
values~\cite{R.Bhatia:Book:1996} and the well-known fact that
all singular values of $T$ are less than or equal to
one.\footnote{See, for example, pp. 1840 of
Ref.~\cite{R.Horodecki:PRA:1996}.} In addition, since $0 \le
\theta \le \frac{\pi}{4}$, we must also have
\begin{equation*}
    \cos \theta \ge \sin \theta\ge 0.
\end{equation*}
These inequalities, together with Eq.~\eqref{Eq:Txx+Tzz} and
Eq.~\eqref{Eq:Tii:Bound}, imply that both $T_{xx}$ and $T_{zz}$
must be non-negative. Moreover, we may assume without loss of
generality that $T_{xx}\ge T_{zz}$. This is because if it
happens that $T_{xx}=\min\{T_{xx},T_{zz}\}$, then since
\begin{equation*}
    T_{zz}\cos\theta+T_{xx}\sin\theta\ge
    T_{xx}\cos\theta+T_{zz}\sin\theta>1,
\end{equation*}
we may also take the larger of $\{T_{xx},T_{zz}\}$ as the
coefficient of $\cos\theta$. Finally, note that the singular
values of $T$ obey the inequality $|T_{xx}+T_{zz}|\leq
\varsigma_1+\varsigma_2$ (pp. 76,
Ref.~\cite{R.Bhatia:Book:1996}). As a result, we find
\begin{align*}
    \varsigma_1\cos\theta+\varsigma_2\sin\theta&=
    \varsigma_1(\cos\theta-\sin\theta)+(\varsigma_1+\varsigma_2)\sin\theta,\\
    &\ge\varsigma_1(\cos\theta-\sin\theta)+(T_{xx}+T_{zz})\sin\theta,\\
    &\ge T_{xx}\cos\theta+T_{zz}\sin\theta>1,
\end{align*}
so $\varrho$ violates the Bell-CHSH inequality. Thus, a
two-qubit state {\em $\varrho$ violates the Bell-CHSH
inequality if and only if inequality \eqref{Cqubits} holds}.
This completes our proof for the scenario where no local
filtering operation is involved and when $\rho$ is a two-qubit
state.

Let us now come back to the question of Bell-CHSH violation
after local filtering operations. Assume that $\rho$ violates
Bell-CHSH inequality after SLO. Let us show that it must
violate inequality \eqref{C} for some $(F_\A,F_\B,\theta)$. In
Ref.~\cite{Ll.Masanes:PRL:2006}, it was proven that, if a state
violates the Bell-CHSH inequality, then it can be transformed
by SLO into a two-qubit state which also violates the Bell-CHSH
inequality. Therefore, there must exist a separable map
$\Omega$ with two-qubit output, such that the resulting state
$\varrho=\Omega(\rho)$ satisfies inequality \eqref{Cqubits} for
some $(U,V,\theta)$ which we shall denote by
$(U_0,V_0,\theta_0)$, i.e.,
\begin{align*}
    &\tr \left[\Omega(\rho)~(U_0 \otimes V_0)^\dag~ H_{\theta_0}~
    (U_0 \otimes V_0) \right]<0,
\end{align*}
Clearly, if this is true, it also follows from the Kraus
decomposition of $\Omega(\rho)$,
Eq.~\eqref{Eq:Dfn:LocalFiltering}, such that
\begin{align*}
    \tr \left[\left(F_{\A,i}\ten F_{\B,i}~\rho~
    F_{\A,i}^\dag\ten F_{\B,i}^\dag\right)
    ~(U_0 \otimes V_0)^\dag~ H_{\theta_0}~
    (U_0 \otimes V_0) \right]< 0,
\end{align*}
for some $i$. This implies that $\rho$ violates inequality
\eqref{C} for $F_\A= U_0\,F_{\A,i} $, $F_\B= V_0\,F_{\B,i}$ and
$\theta=\theta_0$. This proves one direction of the lemma, we
shall next show the other.

Assume that $\rho$ violates inequality \eqref{C} for
$(F_{\A,0},F_{\B,0},\theta_0)$. It is straightforward to see
that $\rho$ violates the Bell-CHSH inequality after SLOCC.
Consider the operation that transforms $\rho$ into the
two-qubit state $\varrho\propto(F_{\A,0}\otimes
F_{\B,0})~\rho~(F_{\A,0}\otimes F_{\B,0})^\dag$. By assumption,
the final state $\varrho$ satisfies inequality \eqref{Cqubits}
with $U=V=\unit_2$ and $\theta=\theta_0$, which implies that it
violates the Bell-CHSH inequality. This completes our proof of
the Lemma.
\end{proof}

\subsection{Nonstandard Bell Experiment with Shared Ancillary State}
\label{Sec:Nonlocality:All:details}

With the characterization given above, we are now ready to
state and prove the main result of this section, namely:
\begin{theorem}\label{Thm:ObservableNonlocality}
    A bipartite state $\sigma$ is entangled if, and only if, there
    exists a state $\rho \in \CSLOCC{CHSH}$ such that
    $\rho\otimes\sigma$ is not in $\CSLOCC{CHSH}$.
\end{theorem}

Let us first try to clarify the physical significance behind
this theorem. If $\rho$ belongs to $\CSLOCC{CHSH}$, no matter
how much additional classical correlation (which can always be
represented by a separable state $\eta_{\mbox{\scriptsize
sep}}$) we supply to it, the result $\rho \otimes
\eta_{\mbox{\scriptsize sep}}$ is still in $\CSLOCC{CHSH}$. On
the contrary, for every entangled state $\sigma$, we can always
find a $\rho\in\CSLOCC{CHSH}$ such that the combined state
$\rho \otimes \sigma$ is not in $\CSLOCC{CHSH}$, and hence
violates the Bell-CHSH inequality after appropriate SLOCC. {\em
This is true, remarkably, even if both $\rho$ and $\sigma$ are
in $\CSLOCC{CHSH}$}.

Here, the violation of Bell-CHSH inequality manifests the
qualitatively different behavior between $\rho \otimes \sigma$
and $\rho \otimes \eta_{\mbox{\scriptsize sep}}$, where
$\eta_{\mbox{\scriptsize sep}}$ is any separable state, and
$\sigma$ is any entangled state. In other words,
Theorem~\ref{Thm:ObservableNonlocality} says that for each
entangled state $\sigma$ there exists a protocol (which also
involves the ancilla state $\rho$ associated with the theorem)
in which $\sigma$ cannot be substituted by an arbitrarily large
amount of classical correlations without changing the
experimental statistics.\footnote{On the contrary, recall from
the discussion in Sec.~\ref{Sec:LHVM:Example} that the
existence of an LHVM for some experimental data ensures that
the latter can be replaced by classical correlations which do
preserve the experimental statistics given by all the joint and
marginal probabilities.} Consequently, yet another way of
putting the theorem would be: {\em bipartite entangled states
are the ones that cannot always be simulated by classical
correlations}.

The proof of the above theorem relies on an explicit
characterization of the set $\CSLOCC{CHSH}$, which we have
already obtained in Sec.~\ref{Sec:CSLOCC}. We can then make use
of convexity arguments similar to those given in
Ref.~\cite{Ll.Masanes:PRL:2006b} to prove by contradiction that
there exists some $\rho\in \CSLOCC{CHSH}$ such that one of
those witness-like operators may be constructed for
$\rho\otimes \sigma$ whenever $\sigma$ is entangled. To carry
this argument through we also require a characterization of the
separable completely positive maps between Bell diagonal
states, which we have included in Appendix~\ref{App:SLOCC}.
With these characterizations in hand, we may now proceed to the
actual proof of the theorem.

\begin{proof}
Firstly, we note that if $\sigma$ is separable, then for all
$\rho \in\CSLOCC{CHSH}$ we must have $\rho\otimes\sigma \in
\CSLOCC{CHSH}$. Intuitively, one can see that this is so
because $\sigma$ can only generate classical correlations which
will not lead to any Bell inequality violation. In fact,
starting from $\rho$, one can prepare $\rho\ten\sigma$ for any
separable $\sigma$ during the LOCC preprocessing of $\rho$.
Therefore, if $\rho\ten\sigma$ for any separable $\sigma$ were
to violate Bell-CHSH inequality after SLOCC, so would $\rho$,
which contradicts our assumption that $\rho\in\CSLOCC{CHSH}$.
Hence, {\em if there exists a bipartite state $\rho
\in\CSLOCC{CHSH}$ such that $\rho\otimes\sigma \not\in
\CSLOCC{CHSH}$, we know that $\sigma$ has to be entangled}.
Next, we will proceed to prove the other direction of the
theorem, namely
\begin{center}
    $\sigma$ is entangled $\Rightarrow\exists\quad\rho\in\CSLOCC{CHSH}$
    such that $\rho\otimes\sigma\in\CSLOCC{CHSH}$
\end{center}

Denote by  $\H=\H_{\A} \otimes\H_{\B}$ the state space that
$\sigma$ acts on and by $d_\A$, $d_\B$, respectively, the
dimension of the local subsystem $\H_{\A}$ and $\H_{\B}$. From
now onwards, we will assume that $\sigma$ is entangled across
$\H_{\A}$ and $\H_{\B}$. Our goal is to show for every
$\sigma$, there always exists an ancilla state
$\rho\in\CSLOCC{CHSH}$ such that $\rho\otimes\sigma \not\in
\CSLOCC{CHSH}$. To achieve that, we will consider ancilla state
$\rho$ that acts on the bipartite Hilbert space
$\left[\H_{\A'}\otimes\H_{\A''}\right] \otimes
\left[\H_{\B'}\otimes\H_{\B''}\right]$, where $\H_{\A'}=\H_\A$,
$\H_{\B'}=\H_{\B}$ and $\H_{\A''} =\H_{\B''} =\Cd{2}$ (see
Figure~\ref{Fig:Protocol}). To prove the above theorem, we then
need to show that the state $\rho\otimes\sigma$ violates
inequality \eqref{C} for some choice of $F_\A$, $F_\B$, and
$\theta$.

\begin{figure}[h!]
    \centering\rule{0pt}{4pt}\par
    \scalebox{1.2}{\includegraphics{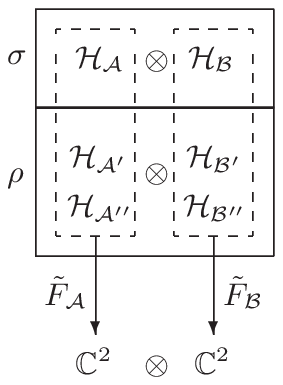}}
    \caption{\label{Fig:Protocol}
    Schematic diagram illustrating the local filtering operations
    $\tilde{F}_\A$ and $\tilde{F}_\B$ involved in our protocol. The
    solid box on top is a schematic representation of the state
    $\sigma$ whereas that on the bottom is for the ancilla state
    $\rho$. Left and right dashed boxes, respectively, enclose the
    subsystems possessed by the two experimenters $\A$ and $\B$.
    }
\end{figure}

In particular, let
\begin{equation}\label{Eq:LocalFilteringProtocol}
  \tilde{F}_\A = \bra{\Phi_{\A \A'}}\otimes
  \unit_{\A''}\ ,\quad
  \tilde{F}_\B = \bra{\Phi_{\B \B'}}\otimes
  \unit_{\B''}\ , \quad \theta = \frac{\pi}{4},
\end{equation}
where $\ket{\Phi_{\A \A'}}=\sqrt{d_\A}\ket{\Phi^+_{d_\A}}$,
c.f. Eq.~\eqref{Eq:Dfn:State:ME-d}, is the (unnormalized)
maximally entangled state between the spaces $\H_{\A}$ and
$\H_{\A'}$ (which have the same dimension), and $\unit_{\A''}$
is the identity matrix acting on $\Cd{2}$ (analogously for
Bob). After some simple calculations, it can be shown that for
any $\rho$ acting on $\left[\H_{\A'}\otimes\H_{\A''}\right]
\otimes \left[\H_{\B'}\otimes\H_{\B''}\right]$
\begin{equation*}\label{eq}
    \tr \left[\rho\otimes\sigma\,
    (\tilde{F}_\A\otimes\tilde{F}_\B)^\dag~ H_{\frac{\pi}{4}}~
    (\tilde{F}_\A\otimes \tilde{F}_\B)\right]
    = \tr \left[\rho\, (\sigma\t\otimes H_{\frac{\pi}{4}})\right],
\end{equation*}
where $\sigma\t$ stands for the transpose of
$\sigma$.\footnote{More generally, for any $\rho$ acting on
$\left[\H_{\A'}\ten\H_{\A''}\right]
\ten\left[\H_{\B'}\ten\H_{\B''}\right]$, any $\sigma$ acting on
$\left[\H_{\A}\right]\ten\left[\H_{\B}\right]$, and any $M$
acting on $\left[\H_{\A''}\right]\ten\left[\H_{\B''}\right]$,
it can be shown that
\begin{equation*}
    \tr \left[\rho\otimes\sigma\,
    (\tilde{F}_\A\otimes\tilde{F}_\B)^\dag~ M~
    (\tilde{F}_\A\otimes \tilde{F}_\B)\right]
    = \tr \left[\rho\, (\sigma\t\otimes M)\right].
\end{equation*}
} Hence, the requirement that inequality \eqref{C} is violated
(i.e., $\rho\ten\sigma\not\in\CSLOCC{CHSH}$) with
$\theta=\pi/4$, $F_\A=\tilde{F}_\A$, $F_\B=\tilde{F}_\B$
becomes
\begin{equation}\label{cond}
    \tr\left[\rho \left(\sigma\t
    \otimes H_{\frac{\pi}{4}}\right) \right]<0.
\end{equation}
What remains is to show that there exists some physical state
$\rho\in\CSLOCC{CHSH}$ such that the above inequality holds
true.

For convenience, in the rest of the proof we allow $\rho$ to be
unnormalized. The only constraints on the matrices $\rho \in
\CSLOCC{CHSH}$ are then positive semidefiniteness ($\rho \in
\s$), and satisfiability of all the inequalities \eqref{C} in
Lemma~\ref{Lem:CharacterizationCSLOCC}. $\CSLOCC{CHSH}$ is now
a convex cone, and its dual cone is defined as
\begin{equation}\label{Eq:Dfn:DualCone}
    \C_{\mbox{\tiny SLOCC}}^{\mbox{\tiny (CHSH)}~*}
    =\{X : \tr \left(\rho\, X\right)\geq 0,\
    \forall~\rho\in\CSLOCC{CHSH}\},
\end{equation}
where $X$ are Hermitian matrices. An important point to note
now is that Farkas' Lemma \cite{B.D.Craven:SJMA:1977} states
that all matrices in $\C_{\mbox{\tiny SLOCC}}^{\mbox{\tiny
(CHSH)}~*}$ can be written as non-negative linear combinations
of matrices $P \in \s$ and matrices of the form
\begin{equation}\label{Eq:Matrices:FAB}
    (F_\A\otimes F_B)^\dag~ H_\theta~ (F_\A\otimes F_\B)
\end{equation}
with $F_\A:\H_{\A'}\otimes\H_{\A''}\rightarrow \Cd{2}$ and
$F_\B:\H_{\B'} \otimes\H_{\B''}\rightarrow \Cd{2}$.

We now show that there always exists $\rho\in\CSLOCC{CHSH}$
satisfying inequality \eqref{cond} by supposing otherwise and
arriving at a contradiction. Suppose that {\em for all} $\rho
\in \CSLOCC{CHSH}$, the converse inequality of Eq.~\eqref{cond}
holds true, i.e.,
\begin{equation}
    \tr \left[\rho\,(\sigma\t\otimes H_{\frac{\pi}{4}})\right]\ge 0.
\end{equation}
It then follows from the definition of $\C_{\mbox{\tiny
SLOCC}}^{\mbox{\tiny (CHSH)}~*}$, Eq.~\eqref{Eq:Dfn:DualCone},
that the matrix $\sigma\t\!\otimes H_{\frac{\pi}{4}}$ belongs
to $\C_{\mbox{\tiny SLOCC}}^{\mbox{\tiny (CHSH)}~*}$. Applying
Farkas' Lemma \cite{B.D.Craven:SJMA:1977} we can write
\begin{equation*}
    \sigma\t\! \otimes H_{\frac{\pi}{4}}=
    \int\dd{x}\, (F_{\A,x}\otimes F_{\B,x})^\dag~ H_{\theta_x}
    (F_{\A,x}\otimes F_{\B,x}) + \int\dd{y}~P_y,
\end{equation*}
where $x$ is a label for matrices of the form given by
Eq.~\eqref{Eq:Matrices:FAB} and $y$ is a label for element in
$\s$. It is easy to see that the above matrix equality is
equivalent to the matrix inequality
\begin{equation}\label{principal0}
    \sigma\t\! \otimes H_{\frac{\pi}{4}} - \int\dd{x}
    \ \Omega_x\!\left( H_{\theta\!_x} \right) \geq 0,
\end{equation}
where each $\Omega_x$ is a separable map, c.f.
Eq.~\eqref{Eq:Dfn:LocalFiltering}, that takes matrices acting
on $[\Cd{2}] \otimes [\Cd{2}]$ to matrices acting on
$[\H_{\A'}\otimes\H_{\A''}]\ten[\H_{\B'} \otimes\H_{\B''}]$.
The following Lemma, however, requires that this is true only
if $\sigma$ is separable (see
Appendix~\ref{App:Sec:ProofOfLemma} for details).
\begin{lemma}\label{Lem:LMI->Separable}
    Let $\Omega_x: [\Cd{2}] \otimes [\Cd{2}] \rightarrow
    [\H_{\A}\otimes \Cd{2}]\otimes[\H_{\B} \otimes \Cd{2}]$ be a family of
    maps, separable with respect to the partition denoted by the brackets.
    Let $\mu$ be a unit-trace, PSD matrix acting on
    $[\H_{\A}]\otimes[\H_{\B}]$ such that
    \begin{equation}
    \label{principal}
        \mu\t\! \otimes H_{\frac{\pi}{4}} - \int\dd{x}
        \ \Omega_x\!\left( H_{\theta_x} \right)
        \geq 0,
    \end{equation}
    where $H_\theta$ is defined in Eq.~\eqref{Eq:Dfn:Htheta}, then
    $\mu$ has to be separable.
\end{lemma}

Hence, if all $\rho\in\CSLOCC{CHSH}$ are such that none of them
can give rise to a Bell-CHSH violation for $\rho\ten\sigma$ via
the protocol given in Eq.~\eqref{Eq:LocalFilteringProtocol}
(see also Figure~\ref{Fig:Protocol}), it must be the case that
$\sigma$ is a separable state. As a result, the corresponding
contrapositive positive statement reads: {\em for every
entangled $\sigma$, there exists $\rho\in\CSLOCC{CHSH}$ such
that $\rho\ten\sigma$ violates the Bell-CHSH inequality via the
protocol given by Eq.~\eqref{Eq:LocalFilteringProtocol}}. This
completes our proof of Theorem~\ref{Thm:ObservableNonlocality}.

\end{proof}

At this stage, it is worth making a few other remarks
concerning the nonstandard Bell experiments that we have just
described. To fix ideas, we will restrict ourselves to the
nontrivial case that both $\rho$ and $\sigma$ are members of
$\CSLOCC{CHSH}$ and where $\sigma$ is entangled. Then for
$\rho\ten\sigma$ to violate the Bell-CHSH inequality via our
protocol, it must also be that (1) $\rho$ is entangled and (2)
at least one of $\rho$ and $\sigma$ has negative partial
transposition. That $\rho$ is entangled can be easily seen by
following the argument given in the proof of
Theorem~\ref{Thm:ObservableNonlocality}, but with the role of
$\rho$ and $\sigma$ reversed
(pp.~\pageref{Thm:ObservableNonlocality}). On the other hand,
it is also not difficult to see that if both $\rho$ and
$\sigma$ were to have positive partial transposition, then
after the local filtering operation given by
Eq.~\eqref{Eq:LocalFilteringProtocol}, the resulting two-qubit
state would still be PPT~\cite{MPR.Horodecki:PRL:1998} and
hence separable~\cite{MPR.Horodecki:PLA:1996}. Since no
separable state can violate the Bell-CHSH inequality, at least
one of $\rho$ and $\sigma$ must have negative partial
transposition.

Meanwhile, we have only required that the ancilla state $\rho$
does not violate the Bell-CHSH inequality, and therefore it may
violate other Bell inequalities, like anyone among the zoo of
inequalities presented in Sec.~\ref{Sec:BipartiteBI} and
Sec.~\ref{Sec:BI:Correlations}. However, even if $\rho$ does
violate another Bell inequality, we know by definition that
$\rho$, and thus $\rho \otimes \eta_{\mbox{\scriptsize sep}}$
(with $\eta_{\mbox{\scriptsize sep}}$ being any separable
state) does not violate the Bell-CHSH inequality. Hence, in the
Bell-CHSH experiment that we are considering, $\sigma$ cannot
be replaced by any classical correlations or separable state
$\eta_{\mbox{\scriptsize sep}}$.

\section{Conclusion}

In this chapter, we have reviewed the phenomenon of {\em hidden
nonlocality} associated with entangled states, and the various
kinds of nonstandard Bell experiments that have been proposed
to derive nonclassical correlations from them. To date, it is
still not known if all entangled states can violate some Bell
inequalities via a nonstandard Bell experiment that only
involves the state in question. Given this state of affair, we
have looked into the possibility of deriving nonclassical
correlations from all bipartite entangled states by considering
nonstandard Bell experiments that also involve shared auxiliary
states. Evidently, the choice of such an ancilla state cannot
be arbitrary. In particular, the protocol that we have
considered involves an ancilla state $\rho$ which by itself
{\em does not violate} the Bell-CHSH inequality even after
arbitrary local filtering operations. In the notation that we
have developed, we say that $\rho\in\CSLOCC{CHSH}$. Then, by
considering a specific local filtering protocol, we have shown
that for every entangled state $\sigma$, there exists an
ancilla state $\rho\in\CSLOCC{CHSH}$ such that the combined
state $\rho\ten\sigma$ {\em does violate} the Bell-CHSH
inequality after the prescribed local filtering operations.

This provides us with a new way to interpret (bipartite)
entanglement in terms of the behavior of the states, in
contrast with the usual definition in terms of the preparation
of the states. Entangled states are, by definition, the ones
that cannot be generated from classical correlations using
LOCC. We have shown that in the bipartite case, one can
equivalently define entangled states as the ones that cannot be
simulated by classical correlations alone.\footnote{However,
some nonclassical correlations can be simulated by classical
correlations when supplemented with only one bit of classical
communication (see, for example Ref.~\cite{B.F.Toner:PRL:2003}
and references therein).} In addition, this also gives a
conclusive answer to the long-standing question of whether all
(bipartite) entangled states can display some hidden
nonlocality~\cite{N.Gisin:0702021,S.Popescu:PRL:1995,
N.Gisin:PLA:1996, J.Barrett:PRA:2002}.

Despite that, it is worth reminding that our proof of the key
result is a non-constructive one. Therefore, even though we
know that there exists some ancilla state $\rho$ such that
$\rho\ten\sigma$ can lead to observable nonlocality for any
entangled $\sigma$, we do not know much about the property of
the ancilla state. A natural task that follows from our
findings is thus to obtain an explicit expression for the
ancilla state $\rho$ for some given $\sigma$. From an
experimental point of view, a better understanding of this
ancilla state $\rho$ is also relevant, since distillation
protocol involving many copies of the same quantum system is
hard to implement. Therefore, a protocol to demonstrate
nonclassical correlations involving only a single copy of
$\rho$ and $\sigma$ may be preferable over those other which
involve, say, 10 copies of $\sigma$ or $\rho$.

On the other hand, as with the bipartite scenario, there are
also mixed multipartite entangled states that admit explicit
LHVM for projective measurement~\cite{G.Toth:PRA:2006} (see
Sec.~\ref{Sec:UUUStates}). An interesting question that follows
from the present work is therefore to determine if the current
proof of observable nonlocality also generalizes to this more
complicated scenario, and hence establishes some kind of
equivalence between entanglement and states that cannot always
be simulated by classical correlations.

\chapter{Conclusion} \label{Chap:Conclusion}

It is one of the most phenomenal discoveries that quantum
mechanical predictions on entangled, spatially separated
systems cannot always be given a locally causal description.
By now, it is well-known that entanglement is necessary, but
may not always be sufficient to demonstrate this fact through a
Bell inequality violation in a standard Bell experiment. A
nonstandard Bell experiment, which involves local preprocessing
and some kind of postselection, may however unveil the
nonclassical correlations hidden in some entangled quantum
states. Given this state of affairs, this thesis aims to
clarify further the relationships between the notions of {\em
correlations}, {\em Bell inequality violation} and {\em quantum
entanglement} in discrete variable quantum systems. In this
chapter, we will summarize our key findings and outline some
possible avenues for future research.

Our study began in Chapter~\ref{Chap:QuantumBounds}, where we
looked into the problem of determining if a given quantum state
$\rho$ can violate some fixed but arbitrary Bell inequality in
a standard Bell experiment. This is a high-dimensional
variational problem where, in general, nontrivial optimization
over the choice of local observables is required. To this end,
we have derived two algorithms which can be used to determine,
respectively, a lower bound (LB) and an upper bound (UB) on the
{\em strength} of correlation that a quantum state $\rho$ can
offer in a given Bell experiment corresponding to some Bell
inequality $I_k$ --- a quantity which we have given the
notation $\SqmBI{$I_k$}(\rho)$. Both of these algorithms make
use of convex optimization techniques in the form of a
semidefinite program (SDP), which is readily solved on a
computer. The LB algorithm requires one to solve a series of
SDPs iteratively, whereas the UB algorithm provides a hierarchy
of SDPs, with each giving a better upper bound on
$\SqmBI{$I_k$}(\rho)$. These algorithms can also be implemented
analytically. In fact, we have made use of the UB algorithm to
derive a necessary condition for bipartite qudit states with
vanishing coherence vectors to violate the Bell-CHSH
inequality; a simple implementation of the LB algorithm has
also enabled us to rederive the Horodecki criterion for
two-qubit states. Since the bounds derived from these algorithm
are usually not tight, these algorithms  often need to be used
in tandem to determine if $\rho$ can violate some Bell
inequality $I_k$.

Next, in Chapter~\ref{Chap:BellViolation}, we looked at some of
the best known Bell inequalities violations by bipartite
quantum states. In particular, using the LB algorithm derived
in Chapter~\ref{Chap:QuantumBounds}, we have obtained the local
measurements giving the best known Bell-CH, and hence Bell-CHSH
inequality violation for arbitrary pure two-qudit states. Then,
by establishing a formal equivalence between the $n$-outcome
CGLMP inequality and the $I_{22nn}$ inequality, we have also
obtained the best known $I_{22dd}$ violation for the
$d$-dimensional isotropic states $\rI(p)$. Together with the UB
algorithm derived in Chapter~\ref{Chap:QuantumBounds}, these
best known violations were then used  to show that for
(arbitrary) bipartite pure two-qudit entangled state $\rho$, a
better Bell-CH inequality violation can be obtained via
collective measurements on $\rho^{\otimes N}$, i.e., $N$ copies
of $\rho$ for $N>2$. The same, however, cannot be said for
mixed entangled states. In fact, we have strong numerical
evidence suggesting that the maximal Bell-CH inequality
violation for some entangled states may not depend on the
number of copies $N$. Further numerical evidence even indicates
that the set of mixed two-qubit states is dominated by those
whose maximal Bell-CH inequality violation remains unchanged
even when $N\ge3$.

After that, in Chapter~\ref{Chap:Hidden.Nonlocality}, we
studied the possibility of deriving nonclassical correlations
from all entangled states via a nonstandard Bell experiment. In
other words, we wanted to know if it is actually possible to
demonstrate some kind of observable nonlocality for all
entangled states. To this end, we have explicitly characterized
the set of bipartite quantum states which do not violate the
Bell-CHSH inequality even after arbitrary local filtering
operations --- a set which we have given the notation
$\CSLOCC{CHSH}$. Then, by considering a specific type of local
filtering operation, we have (non-constructively) shown that
for every bipartite entangled state $\sigma$, there exists an
ancilla state $\rho\in\CSLOCC{CHSH}$ such that
$\rho\ten\sigma\not\in\CSLOCC{CHSH}$. Interestingly, this means
that even if both $\rho$ and $\sigma$ can be simulated,
individually, by classical correlations in the most general
single-copy nonstandard Bell-CHSH experiment, the combined
state $\rho\ten\sigma$ cannot be described by classical
correlations in some single-copy nonstandard Bell-CHSH
experiment. Consequently, we can now define a bipartite
entangled state $\sigma$ as precisely that which cannot be
simulated by classical correlations when one consider all
possible experiments that may be performed on $\sigma$ in
conjunction with non-Bell-CHSH-violating states.

Let us now make some remarks regarding future research. To
begin with, we note that possible follow-up projects in
relation to the work presented in each chapters have already
been presented in some details at the end of the corresponding
chapters. As such, we will not try to repeat all of them here,
but to merely remind the readers of some of the key ones.
Firstly, as one may have noticed, our analysis in this thesis
has been carried out exclusively for discrete variable quantum
systems and to a large extent, only for bipartite quantum
systems. There are, of course, many interesting problems that
are associated with Bell inequality violation in multipartite
and continuous variable quantum systems. For example, it is
still not known if all discrete multipartite pure entangled
states can violate a Bell inequality in a standard Bell
experiment. In continuous variable quantum systems, it is not
even known if all bipartite pure entangled states can violate a
Bell inequality. As a result, preliminary investigations on the
adaptability of the tools that we have developed here to this
latter scenario could be of some use.

Results that we have obtained in
Chapter~\ref{Chap:QuantumBounds}, as well as those presented in
Ref.~\cite{M.Navascues:PRL:2007} have indicated that upper
bound techniques similar to those that we have developed in
this thesis do allow us to investigate the extent to which
quantum mechanics can violate a fixed but arbitrary Bell
inequality. Further work on this is clearly desirable as it
will help us to learn something about the extreme points of the
set of quantum correlations. This is work in
progress~\cite{A.C.Doherty:QMP:2007}.

Given that we have only got a nonconstructive proof for the
nonclassical correlations hidden in an arbitrary entangled
state $\sigma$, it would be great if an explicit construction
of the ancilla state $\rho$ used in our protocol can be
obtained. A general construction of the ancilla state may be
formidable, but it would be helpful to at least solve this for
some simple cases like Werner states, or more desirably, some
bound entangled states. Finally, the arguably most important
problem that is left opened from the present work is whether it
is also possible to derive some observable, nonclassical
correlations from {\em all} multipartite entangled quantum
states, be it discrete or continuous. Any progress in this
regard would certainly help us to improve our understanding of
the quantum world, which is always full of surprises.



\appendix

\chapter{Bell-diagonal Preserving Separable Maps}
\label{App:SLOCC}

In this Appendix we classify the four-qubit states that commute
with $U\ten{U}\ten{V}\ten{V}$, where $U$ and $V$ are arbitrary
members of the Pauli group. We characterize the set of
separable states for this class, in terms of a finite number of
entanglement witnesses. Equivalently, we characterize the set
of two-qubit, Bell-diagonal-preserving, completely positive
maps (henceforth abbreviated as CPM) that are separable. These
separable CPMs correspond to protocols that can be implemented
with stochastic local quantum operations assisted by classical
communication (SLOCC). Explicit characterization of these CPMs
is an essential ingredient of the proof of
Lemma~\ref{Lem:LMI->Separable}.

\section{Four-qubit Separable States with $U\ten{U}\ten{V}\ten{V}$
Symmetry}\label{App:Sec:SLOCC:SeparableStates}

In this section, we will characterize the set of separable
states commuting with $U\ten{U}\ten{V}\ten{V}$, where $U$ and
$V$ are arbitrary members of the Pauli group. Let us begin by
reminding the reader about an important property of two-qubit
states which commute with all unitaries of the form $U\ten{U}$,
where $U$ is an arbitrary member of the Pauli group. The Pauli
group is generated by the Pauli matrices
$\{\sigma_i\}_{i=x,y,z}$, Eq,~\eqref{Eq:Dfn:PauliMatrices}, and
has 16 elements. The representation $U\ten{U}$ comprises four
1-dimensional irreducible representations, each acting on the
subspace spanned by one vector of the Bell basis\footnote{These
states are more conventionally denoted by
$\ket{\Phi_1}=\ket{\Phi^+}$, $\ket{\Phi_2}=\ket{\Phi^-}$,
$\ket{\Phi_3}=\ket{\Psi^+}$, $\ket{\Phi_1}=\ket{\Psi^-}$.}
\begin{subequations}\label{Eq:BellBases}
    \begin{align}
      \ket{\Phi_{^1_2}} &\equiv \frac{1}{\sqrt{2}}
      \left( \ket{0}\ket{0} \pm \ket{1}\ket{1} \right),
      \label{Eq:BellBases12}\\
      \ket{\Phi_{^3_4}} &\equiv \frac{1}{\sqrt{2}}
      \left( \ket{0}\ket{1} \pm \ket{1}\ket{0} \right).
      \label{Eq:BellBases34}
    \end{align}
\end{subequations}
This implies that~\cite{K.G.H.Vollbrecht:PRA:2001} any
two-qubit state which commutes with $U\ten{U}$ can be written
as $\rho=\sum_{k=1}^4 [r]_{k} \Pi_k$, where
$\Pi_k\equiv\ketbra{\Phi_k}$ is the $\idx{k}$ Bell projector.
With this information in mind, we are now ready to discuss the
case that is of our interest.

We would like to characterize the set of four-qubit states
which commute with all unitaries $U\ten{U}\ten{V}\ten{V}$,
where $U$ and $V$ are members of the Pauli group. Let us denote
this set of states by $\mathcal{R}$ and the state space of
$\rho\in\mathcal{R}$ as $\H\simeq
\H_{\A'}\ten\H_{\B'}\ten\H_{\A''}\ten\H_{\B''}$, where
$\H_{\A'}$, $\H_{\B'}$ etc. are Hilbert spaces of the
constituent qubits. In this notation, both the subsystems
associated with $\H_{\A'}\ten\H_{\B'}$ and that with
$\H_{\A''}\ten\H_{\B''}$ have $U\ten{U}$ symmetry and hence are
linear combinations of Bell-diagonal
projectors~\cite{K.G.H.Vollbrecht:PRA:2001}.

\begin{figure}[h!btp]
    \centering\rule{0pt}{4pt}\par
    \includegraphics[scale=1.25]{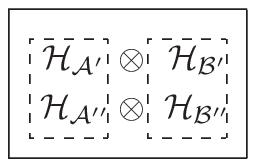}
    \caption{\label{Fig:StateSpace}
    A schematic diagram for the subsystems constituting $\rho$.
    Subsystems that are arranged in the same row in the diagram
    have $U\ten{U}$ symmetry and hence are represented by
    Bell-diagonal states~\cite{K.G.H.Vollbrecht:PRA:2001} (see text
    for details). In this Appendix, we are interested in states
    that are separable between subsystems enclosed in the two
    dashed boxes.
    }
\end{figure}

Our aim in this section is to provide a full characterization
of the set of $\rho$ that are separable between
$\H_\A\equiv\H_{\A'}\ten\H_{\A''}$ and
$\H_\B\equiv\H_{\B'}\ten\H_{\B''}$ (see
Figure~\ref{Fig:StateSpace}). Throughout this section, a state
is said to be {\em separable} if and only if it is separable
between $\H_\A$ and $\H_\B$.

The symmetry of $\rho$ allows one to write it as a {\em
non-negative} combination of (tensored-) Bell projectors:
\begin{gather}\label{Eq:Rep}
    \rho=\sum_{i=1}^4\sum_{j=1}^4 [r]_{i,j}\Pi_i\ten\Pi_j,
\end{gather}
where the Bell projector before and after the tensor product,
respectively, acts on $\H_{\A'}\ten\H_{\B'}$ and
$\H_{\A''}\ten\H_{\B''}$ (Figure~\ref{Fig:StateSpace}). Thus,
any state $\rho\in\mathcal{R}$  can be represented in a compact
manner, via the corresponding $4\times4$ matrix $r$. More
generally, any operator $\mu$ acting on the Hilbert space $\H$
and having the $U\ten{U}\ten{V}\ten{V}$ symmetry admits a
$4\times4$ matrix representation $M$ via:
\begin{gather}\label{Eq:Rep:General}
    \mu=\sum_{i=1}^4\sum_{j=1}^4 [M]_{i,j}\Pi_i\ten\Pi_j,
\end{gather}
where $[M]_{i,j}$ is now not necessarily non-negative. When
there is no risk of confusion, we will also refer to $r$ and
$M$, respectively, as a state and an operator having this
symmetry.

Evidently, in this representation, an operator $\mu$ is
non-negative if and only if all entries in the corresponding
$4\times4$ matrix $M$ are non-negative. Notice also that by
appropriate local unitary transformation, one can swap any
$\Pi_i$ with any other $\Pi_j$, $j\neq i$ while keeping all the
other $\Pi_k$, $k\neq i,j$ unaffected. Here, the term {\em
local} is used with respect to the $\A$ and $\B$ partitioning.
Specifically, via the local unitary transformation
\begin{equation}
    V_{ij}\equiv\left\{ \begin{array}{c@{\quad:\quad}l}
    \half(\unit_2+\ii\sigma_z)\ten(\unit_2+\ii\sigma_z) & i=1,j=2,\\
    \half(\sigma_x+\sigma_z)\ten(\sigma_x+\sigma_z) & i=2,j=3,\\
    \half(\unit_2+\ii\sigma_z)\ten(\unit_2-\ii\sigma_z) & i=3,j=4,
    \end{array} \right.
\end{equation}
one can swap $\Pi_i$ and $\Pi_j$ while leaving all the other
Bell projectors unaffected. In terms of the corresponding
$4\times 4$ matrix representation, the effect of such local
unitaries on $\mu$ amounts to permutation of the rows and/or
columns of $M$. For brevity, in what follows, we will say that
two matrices $M$ and $M'$ are local-unitarily equivalent if we
can obtain $M$ by simply permuting the rows and/or columns of
$M'$ and {\em vice versa}. A direct consequence of this
observation is that if $r$ represents a separable state, so is
any other $r'$ that is obtained from $r$ by independently
permuting any of its rows and/or columns.

Before we state the main result of this section, let us
introduce one more definition.
\begin{dfn}\label{dfn:rhos}
    Let $\P_s\subset\mathcal{R}$ be the convex hull of the states
    \begin{equation}\label{Eq:D0&G0}
        D_0\equiv \frac{1}{4}\left(
        \begin{array}{cccc}
        1 & \cdot & \cdot & \cdot \\
        \cdot & 1 & \cdot & \cdot \\
        \cdot & \cdot & 1 & \cdot \\
        \cdot & \cdot & \cdot & 1 \\
        \end{array}
        \right),\quad G_0\equiv \frac{1}{4}\left(
        \begin{array}{cccc}
        1 & 1 & \cdot & \cdot \\
        1 & 1 & \cdot & \cdot \\
        \cdot & \cdot & \cdot & \cdot \\
        \cdot & \cdot & \cdot & \cdot \\
        \end{array}
        \right),
    \end{equation}
    and the states that are local-unitarily equivalent to these
    two.
\end{dfn}
Simple calculations show that with respect to the $\A$ and $\B$
partitioning, $D_0$, $G_0$ are separable. In particular, when
written in the product basis of $\HA\ten\HB$, it can be shown
that $D_0$ admits the following convex decomposition in terms
of separable states:
\begin{align}
    \frac{1}{8}\Big[&\Big(\ket{0}\ket{0}\ten\ket{0}\ket{0}+\ket{1}\ket{1}\ten\ket{1}\ket{1}\Big)
    \Big(\bra{0}\bra{0}\ten\bra{0}\bra{0}+\bra{1}\bra{1}\ten\bra{1}\bra{1}\Big)\nonumber\\
    +&\Big(\ket{0}\ket{0}\ten\ket{1}\ket{1}+\ket{1}\ket{1}\ten\ket{0}\ket{0}\Big)
    \Big(\bra{0}\bra{0}\ten\bra{1}\bra{1}+\bra{1}\bra{1}\ten\bra{0}\bra{0}\Big)\nonumber\\
    +&\Big(\ket{0}\ket{1}\ten\ket{0}\ket{1}+\ket{1}\ket{0}\ten\ket{1}\ket{0}\Big)
    \Big(\bra{0}\bra{1}\ten\bra{0}\bra{1}+\bra{1}\bra{0}\ten\bra{1}\bra{0}\Big)\nonumber\\
    +&\Big(\ket{0}\ket{1}\ten\ket{1}\ket{0}+\ket{1}\ket{0}\ten\ket{0}\ket{1}\Big)
    \Big(\bra{0}\bra{1}\ten\bra{1}\bra{0}+\bra{1}\bra{0}\ten\bra{0}\bra{1}\Big)\Big].
\end{align}
Likewise, it can be shown that $G_0$ admits the following
convex decomposition in terms of product states:
\begin{align}
    \quar\Big(&\ket{0}\ket{0}\bra{0}\bra{0}\ten\ket{0}\ket{0}\bra{0}\bra{0}
    + \ket{0}\ket{1}\bra{0}\bra{1}\ten\ket{0}\ket{1}\bra{0}\bra{1}\nonumber\\
    + &\ket{1}\ket{0}\bra{1}\bra{0}\ten\ket{1}\ket{0}\bra{1}\bra{0}
    + \ket{1}\ket{1}\bra{1}\bra{1}\ten\ket{1}\ket{1}\bra{1}\bra{1}\Big).
\end{align}

Hence, $\P_s$ is a separable subset of $\mathcal{R}$. The main
result of this section consists of showing the converse, and
hence the following theorem.

\begin{theorem}\label{Thm:SeparableUUVV}
    $\P_s$ is the set of states in $\mathcal{R}$ that are separable with
    respect to the $\A$ and $\B$ partitioning.
\end{theorem}
Now, we note that $\P_s$ is a convex polytope. Its boundary is
therefore described by a finite number of
facets~\cite{B.Grunbaum:Book:2003}. Hence, to prove the above
theorem, it suffices to show that all these facets correspond
to valid entanglement witnesses. Denoting the set of facets by
$\mathcal{W}=\{W_i\}$. Then, using the software
PORTA,\footnote{This software package, which stands for
POlyhedron Representation Transformation Algorithm, is
available at
\texttt{http://www.zib.de/Optimization/Software/Porta/}} the
{\em nontrivial} facets were found to be equivalent under local
unitaries to one of the following:
\begin{gather}\label{Eq:W}
    W_1\equiv \left(
    \begin{array}{rrrr}
    1 & 1 & 1 &-1 \\
    1 & 1 & 1 &-1 \\
    1 & 1 & 1 &-1 \\
    -1 &-1 &-1 & 1 \\
    \end{array}
    \right),~W_2\equiv \left(
    \begin{array}{cccc}
    1 & 1 & \cdot & -1\\
    \cdot & \cdot & 1 & \cdot \\
    \cdot & \cdot & 1 & \cdot \\
    \cdot & \cdot & 1 & \cdot \\
    \end{array}
    \right),\nonumber\\
    W_3\equiv \left(
    \begin{array}{rrrr}
    3 & 3 & 1 &-1 \\
    3 &-1 & 1 & 3 \\
    1 & 1 & 3 & 1 \\
    -1 &-1 & 1 &-1 \\
    \end{array}
    \right),~W_4\equiv \left(
    \begin{array}{rrrr}
    3 & 3 & 1 &-1 \\
    3 &-1 & 1 & 3 \\
    3 &-1 & 1 &-1 \\
    1 & 1 &-1 & 1 \\
    \end{array}
    \right).
\end{gather}
Apart from these, there is also a facet $W_0$ whose only
nonzero entry is $[W_0]_{11}=1$. $W_0$ and the operators
local-unitarily equivalent to it give rise to  positive
definite matrices [c.f. Eq.~\eqref{Eq:ZW}], and thus correspond
to trivial entanglement witnesses. On the other hand, it is
also not difficult to verify that $W_1$ (and operators
equivalent under local unitaries) are decomposable and
therefore demand that $\rho_s$ remains positive semidefinite
after partial transposition. These are all the entanglement
witnesses that arise from the positive partial transposition
requirement~\cite{A.Peres:PRL:1996,MPR.Horodecki:PLA:1996} for
separable states.

To complete the proof of Theorem~\ref{Thm:SeparableUUVV}, it remains
to show that $W_2$, $W_3$, $W_4$ give rise to
Hermitian matrices
\begin{gather}\label{Eq:ZW}
    Z_{w,k}=\sum_{i=1}^4\sum_{j=1}^4
    [W_k]_{i,j}~\left(\Pi_i\ten\Pi_j\right)
\end{gather}
that are valid entanglement witnesses, i.e.,
$\tr(\rho_s\,Z_{w,k})\ge 0$ for any separable
$\rho_s\in\mathcal{R}$. It turns out that this can be proved
with the help of the following lemma from
Ref.~\cite{A.C.Doherty:PRL:2002,A.C.Doherty:PRA:2004}.

\begin{lemma}\label{Lem:ProofOfWitnesses}
    For a given Hermitian matrix $Z_w$ acting on $\H_\A\ten\H_\B$, with
    $dim(\H_\A)=d_\A$ and $dim(\H_\B)=d_\B$, if there exists
    $m,n\in\mathbb{Z}^+$,  positive semidefinite $\Z$ acting on
    $\H_\A^{\ten m}\ten\H_\B^{\ten n}$ and a subset $s$ of the $m+n$
    tensor factors such that
    \begin{equation}\label{Eq:Z:Dfn}
        \Pi_\A\ten\Pi_\B~\left(\unit_{d_\A}^{\otimes  m-1}\ten
        Z_w\ten\unit_{d_\B}^{\ten n-1}\right)~\Pi_\A\ten\Pi_\B =
        \Pi_\A\ten\Pi_\B~\left(\Z^{\mbox{\tiny $T_s$}}\right)~\Pi_\A\ten\Pi_\B,
    \end{equation}
    where $\Pi_\A$ is the projector onto the symmetric subspace of
    $\H_\A^{\ten m}$ (likewise for $\Pi_\B$) and $(.)^{\mbox{\tiny $T_s$}}$
    refers to partial transposition with respect to the subsystem $s$,
    then $Z_w$ is a valid entanglement witness across $\HA$ and $\HB$,
    i.e., $\tr(\rho_\text{sep}\,Z_w)\ge 0$ for any  state
    $\rho_\text{sep}$ that is separable with respect to the $\A$ and
    $\B$ partitioning.
\end{lemma}
\begin{proof}
Denote by $\A^{[k]}$ the subsystem associated with the $k$-th
copy of $\H_\A$ in $\H_\A^{\ten m}$; likewise for $\B^{[l]}$.
To prove the above lemma, let $\ket{\alpha}\in\H_\A$ and
$\ket{\beta}\in\H_\B$ be (unit) vectors, and for definiteness,
let $s=\B^{[n]}$ then it follows that
\begin{align*}
    &\bra{\alpha}\bra{\beta}~Z_w~\ket{\alpha}\ket{\beta}\\
    =&\bra{\alpha}^{\ten m}\bra{\beta}^{\ten n} \left(\unit_{d_\A}^{\ten
    m-1}\ten Z_w\ten\unit_{d_\B}^{\ten n-1}\right)\ket{\alpha}^{\ten
    m}\ket{\beta}^{\ten n}\\
    =&\bra{\alpha}^{\ten m}\bra{\beta}^{\ten n}\left[\Pi_\A\ten
      \Pi_\B\left(\Z^{\mbox{\tiny $T_s$}}\right)\Pi_\A\ten\Pi_\B \right]\ket{\alpha}^{\ten
      m}\ket{\beta}^{\ten n}\\
    =&\bra{\alpha}^{\ten m}\bra{\beta}^{\ten n}\left(\Z^{\mbox{\tiny $T_{\B^{[n]}}$}}
    \right)\ket{\alpha}^{\ten
      m}\ket{\beta}^{\ten n}\\
    =&\left(\bra{\alpha}^{\ten m} \bra{\beta}^{\ten n-1}
    \ten\bra{\beta^*}\right)~\Z~\left(\ket{\alpha}^{\ten
      m}\ket{\beta}^{\ten n-1}\ten\ket{\beta^*}\right)\\
    \ge &0,
\end{align*}
where $\ket{\beta^*}$ is the complex conjugate of
$\ket{\beta}$. We have made use of the identity
$\Pi_\A\ket{\alpha}^{\ten m}=\ket{\alpha}^{\ten m}$ (likewise
for $\Pi_\B$) in the second and third equality,
Eq.~\eqref{Eq:Z:Dfn} in the second equality, and the positive
semidefiniteness of $\Z$. To cater for general $s$, we just
have to modify the second to last line of the above computation
accordingly (i.e., to perform complex conjugation on all the
states in the set $s$) and the proof will proceed as before.
\end{proof}

More generally, let us remark that instead of having one $\Z$
on the {\em rhs} of Eq.~\eqref{Eq:Z:Dfn}, one can also have a
sum of different $\Z$'s, with each of them partial transposed
with respect to different subsystems $s$. Clearly, if the given
$Z_w$ admits such a decomposition, it is also an entanglement
witness~\cite{A.C.Doherty:PRL:2002,A.C.Doherty:PRA:2004}. For
our purposes these more complicated decompositions do not offer
any advantage over the simple decomposition given in
Eq.~\eqref{Eq:Z:Dfn}.

By solving some appropriate SDPs
(Appendix~\ref{App:Sec:SDP:EW}), we have found that when $m=3$,
$n=2$ and $s=\B^{[2]}$, there exist some $\Z_k\ge0$, such that
Eq.~\eqref{Eq:Z:Dfn} holds true for each $k\in\{1,2,3,4\}$. Due
to space limitations, the analytic expression for these
$\Z_k$'s will not be reproduced here but are made available
online at Ref.~\cite{url:z}. For $W_2$, the fact that the
corresponding $Z_{w,2}$ is a witness can even be verified by
considering $m=2$, $n=1$ and $s=\A^{[1]}$. In this case,
$d_\A=d_\B=4$. If we label the local basis vectors by
$\{\ket{i}\}_{i=0}^3$, the corresponding $\Z$ reads
\begin{gather*}
    \Z_2=\frac{1}{2}\sum_{i=1}^4\ketbra{z_i},\\
    \ket{z_1}=\ket{01,0}-\ket{02,3}+\ket{11,1}+\ket{13,3}+\ket{22,1}+\ket{23,0},\\
    \ket{z_2}=\ket{10,3}+\ket{11,2}+\ket{20,0}+\ket{22,2}-\ket{31,0}+\ket{32,3},\\
    \ket{z_3}=\ket{00,0}+\ket{02,2}+\ket{10,1}-\ket{13,2}+\ket{32,1}+\ket{33,0},\\
    \ket{z_4}=\ket{00,3}+\ket{01,2}-\ket{20,1}+\ket{23,2}+\ket{31,1}+\ket{33,3},
\end{gather*}
where we have separated $\A$'s degree of freedom from $\B$'s
ones by comma.\footnote{Note that to verify $\Z_2$ against
Eq.~\eqref{Eq:Z:Dfn}, one should also rewrite $Z_{w,2}$
obtained in Eq.~\eqref{Eq:ZW} in the appropriate tensor-product
basis such that $Z_{w,2}$ acts on
$\H_{\A'}\ten\H_{\A''}\ten\H_{\B'}\ten\H_{\B''}$.} This
completes the proof for Theorem~\ref{Thm:SeparableUUVV}.

An immediate corollary of the above characterization is that we
now know exactly the set of Bell-diagonal preserving
transformations that can be performed locally on a
Bell-diagonal state. In what follows, we will make use of the
Choi-Jamio\l{k}owski isomorphism~\cite{A.Jamiolkowski:RMP:1972,
M.D.Choi:LAA:1975,V.P.Belavkin:RMP:1986}, i.e., the one-to-one
correspondence between CPM and quantum state, to make these
SLOCC transformations explicit.

\section{Separable Maps and SLOCC}\label{App:Sec:SeparableMap}

Now, let us recall some well-established facts about CPM. To
begin with, a separable CPM, denoted by $\E_s$ takes the
following form~\cite{E.M.Rains:9707002,V.Vedral:PRA:1998}
\begin{equation}\label{Eq:SeparableMap}
    \E_s:\rho\to\sum_{i=1}^n(A_i\ten B_i)~\rho~(A_i^\dag\ten B_i^\dag),
\end{equation}
where $\rho$ acts on $\H_{\Ai}\ten\H_{\Bi}$, $A_i$ acts on
$\H_{\Ai}$, $B_i$ acts on $\H_{\Bi}$.\footnote{Following Kraus'
work on CPM~\cite{K.Kraus:AP:1971,K.Kraus:Book:1983}, this
specific form of the CPM is also known as a Kraus decomposition
of the CPM, with each $A_i\ten B_i$ in the sum conventionally
called the Kraus operator associated with the CPM.}

If, moreover,
\begin{equation}\label{Eq:TracePreserving}
    \sum_i \left(A_i\ten{B_i}\right)^\dag
    \left(A_i\ten{B_i}\right)=\unit,
\end{equation}
the map is trace-preserving, i.e., if $\rho$ is normalized, so
is the output of the map $\E_s(\rho)$. Equivalently, the
trace-preserving condition demands that the transformation from
$\rho$ to $\E_s(\rho)$ can always be achieved with certainty.
It is well-known that all LOCC transformations are of the form
Eq.~\eqref{Eq:SeparableMap} but the converse is not
true~\cite{C.H.Bennett:PRA:1999}.

However, if we allow the map $\rho\to\E_s(\rho)$ to fail with
some probability $p<1$, the transformation from $\rho$ to
$\E_s(\rho)$ can always be implemented probabilistically via
LOCC. In other words, if we do not impose
Eq.~\eqref{Eq:TracePreserving}, then
Eq.~\eqref{Eq:SeparableMap} represents, up to some
normalization constant, the most general LOCC possible on a
bipartite quantum system. These are the SLOCC
transformations~\cite{W.Dur:PRA:2000}.

To see that Eq.~\eqref{Eq:SeparableMap} can always be realized
with some non-zero probability of success, we first note that
each of the terms in the decomposition can always be
implemented with some probability of success. For instance, if
they wish to implement the $\idx{k}$ term in
Eq.~\eqref{Eq:SeparableMap}, i.e., $(A_k\ten
B_k)~\rho~(A_k^\dag\ten B_k^\dag)$ --- which by itself
represents uncorrelated local quantum operations on the
individual subsystems, they can do that by just by applying
some local unitary transformation and/or measurement on their
local subsystem. With the help of classical communication, they
can then postselect on the desired outcomes to achieve the
transformation $(A_k\ten B_k)~\rho~(A_k^\dag\ten B_k^\dag)$.

With that in mind, it is then easy to see that
implementation of the separable map can be carried out
by probabilistically selecting the term to implement in
the separable map given by Eq.~\eqref{Eq:SeparableMap}. Party $\A$ can
first toss a coin to decide on the term in the
decomposition [c.f. Eq.~\eqref{Eq:SeparableMap}] that she would like to
implement for that run of the experiment and
communicate this outcome to Bob. They then both perform
appropriate local operations and postselection to
achieve the desired transformation with some
probability. Clearly, since each term in Eq.~\eqref{Eq:SeparableMap} can
be implemented with some non-zero probability of
success, so can the separable map given by Eq.~\eqref{Eq:SeparableMap}.

Now, to make a connection between the set of SLOCC
transformations and the set of states that we have
characterized in Sec.~\ref{App:Sec:SLOCC:SeparableStates}, let
us also recall the Choi-Jamio{\l}kowski
isomorphism~\cite{A.Jamiolkowski:RMP:1972,
M.D.Choi:LAA:1975,V.P.Belavkin:RMP:1986} between CPM and
quantum states: for every (not necessarily separable) CPM
$\E:\H_{\Ai}\ten\H_{\Bi}\to\H_{\Ao}\ten\H_{\Bo}$ there is a
unique --- again, up to some positive constant $\alpha$ ---
quantum state $\rho_\E$ corresponding to $\E$:
\begin{equation}\label{Eq:JamiolkowskiState}
    \rho_\E=\alpha~\E_{\rm in}\otimes \mathcal{I}_{\rm
    out}\left(\ket{\Phi^+}_{\A\A}\bra{\Phi^+}\ten
    \ket{\Phi^+}_{\B\B}\bra{\Phi^+}\right),
\end{equation}
where $\ket{\Phi^+}_{\A}\equiv\sum_{i=1}^{d_{\Ai}}\ket{i}_{\rm
in}\ten\ket{i}_{\rm out}$ is the unnormalized maximally
entangled state of dimension $d_{\Ai}$ (likewise for
$\ket{\Phi^+}_{\B}$). In Eq.~\eqref{Eq:JamiolkowskiState}, it
is understood that $\E_{\rm in}$ only acts on the ``in" space
of $\ket{\Phi^+}_{\A}$ and $\ket{\Phi^+}_{\B}$. Clearly, the
state $\rho_\E$ acts on a Hilbert space of dimension
$d_{\Ai}\times d_{\Ao}\times d_{\Bi}\times d_{\Bo}$, where
$d_{\Ao}\times d_{\Bo}$ is the dimension of
$\H_{\Ao}\ten\H_{\Bo}$.

Conversely, given a state $\rho_\E$ acting on
$\H_{\Ao}\ten\H_{\Bo}\ten\H_{\Ai}\ten\H_{\Bi}$, the
corresponding action of the CPM $\E$ on some $\rho$ acting on
$\H_{\Ai}\ten\H_{\Bi}$ reads:
\begin{equation}\label{Eq:State->CPM}
    \E(\rho)=\frac{1}{\alpha}\tr_{\Ai\Bi}
    \left[\rho_\E\left(\unit_{\Ao\Bo}\ten\rho\t\right)\right],
\end{equation}
where $\rho\t$ denotes transposition of $\rho$ in some local
bases of $\H_{\Ai}\ten\H_{\Bi}$. For a trace-preserving CPM, it
then follows that we must have
$\tr_{\Ao\Bo}(\rho_\E)=\alpha\unit_{\Ai\Bi}$. A point that
should be emphasized now is that $\E$ is a separable map,
Eq.~\eqref{Eq:SeparableMap}, if and only if the corresponding
$\rho_\E$ given by Eq.~\eqref{Eq:JamiolkowskiState} is
separable across $\H_{\Ai}\ten\H_{\Ao}$ and
$\H_{\Bi}\ten\H_{\Bo}$~\cite{J.I.Cirac:PRL:2001}. Moreover, at
the risk of repeating ourselves, the map $\rho\to\E(\rho)$
derived from a separable $\rho_\E$ can always be implemented
locally, although it may only succeed with some (nonzero)
probability. Hence, if we are only interested in
transformations that can be performed locally, and not the
probability of success in mapping $\rho\to\E(\rho)$, the
normalization constant $\alpha$ as well as the normalization of
$\rho_\E$ becomes irrelevant. This is the convention that we
will adopt for the rest of this Appendix.

\section{Bell-diagonal Preserving SLOCC Transformations}
\label{App:Sec:BellMaps}

We shall now apply the isomorphism to the class of states
$\mathcal{R}$ that we have characterized in
Sec.~\ref{App:Sec:SLOCC:SeparableStates}. In particular, if we
identify $\Ai$, $\Ao$, $\Bi$ and $\Bo$ with, respectively,
$\A''$, $\A'$, $\B''$ and $\B'$, it  follows from
Eq.~\eqref{Eq:Rep} and Eq.~\eqref{Eq:State->CPM} that for any
two-qubit state $\rho_{\rm in}$, the action of the CPM derived
from $\rho\in\mathcal{R}$ reads:
\begin{equation}
    \E:\rho_{\rm in}\to\rho_{\rm out}\propto \sum_{i,j}[r]_{i,j}
    \tr\left(\rho\t_{\rm in}\Pi_j\right)\Pi_i.
\end{equation}
Hence, under the action of $\E$, any $\rho_{\rm in}$ is
transformed to another two-qubit state that is diagonal in the
Bell basis, i.e., a Bell-diagonal state. In particular, for a
Bell-diagonal $\rho_{\rm in}$, i.e.,
\begin{gather}
    \rho_{\rm in}=\sum_k[\beta]_k\Pi_k,\nonumber\\
    [\beta]_k\ge0,\quad \sum_k[\beta]_k=1,
\end{gather}
the map outputs another Bell-diagonal state
\begin{equation}
    \rho_{\rm out}=\E(\rho_{\rm in})\propto\sum_{i,j}
    [\beta]_j[r]_{i,j}\Pi_i.
\end{equation}
It is worth noting that for a general $\rho_\E\in\mathcal{R}$,
$\tr_{\A'\B'}\rho_\E$ is not proportional to the identity
matrix, therefore some of the CPMs derived from
$\rho\in\mathcal{R}$ are intrinsically
non-trace-preserving.\footnote{The $\rho_\E$ derived from $G_0$
in Eq.~\eqref{Eq:D0&G0} is an example of this sort. In fact, in
this case, if the input state has no support on $\Pi_1$ nor
$\Pi_2$, the map always outputs the zero matrix.}

By considering the convex cone\footnote{Since the mapping from
any $\rho\in\P_s$ to a separable CPM via
Eq.~\eqref{Eq:State->CPM} is only defined up to a positive
constant, for the subsequent discussion, we might as well
consider the cone generated by $\P_s$.} of separable states
$\P_s$ that we have characterized in
Sec.~\ref{App:Sec:SLOCC:SeparableStates}, we therefore obtain
the entire set of Bell-diagonal preserving SLOCC
transformations. Among them, we note that the extremal maps,
i.e., those derived from Eq.~\eqref{Eq:D0&G0}, admit simple
physical interpretations and implementations. In particular,
the extremal separable map for $D_0$, and the maps that are
related to it by local unitaries, correspond to permutation of
the input Bell projectors $\Pi_i$ --- which can be implemented
by performing appropriate local unitary transformations. The
other kind of extremal separable map, derived from $G_0$,
corresponds to making a measurement that determines if the
initial state is in a subspace spanned by a given pair of Bell
states and if successful discarding the input state and
replacing it by an equal but incoherent mixture of two of the
Bell states. This operation can be implemented locally since
the equally weighted mixture of two Bell states is a separable
state and hence  both the measurement step and the state
preparation step can be implemented locally.

\chapter{Some Miscellaneous Calculations}\label{App:MISC}

\section{Classical Correlations and Bell's Theorems}
\label{App:Sec:BellThm}

\subsection{Equivalence between the CGLMP and $I_{22nn}$ inequality}
\label{App:Sec:CGLMP}

In this section, we will provide a proof that the CGLMP
inequality for $\nA=\nB=n$ outcomes,
Eq.~\eqref{Ineq:Prob:CGLMP:Functional}, is equivalent to the
$I_{22nn}$ inequality, Eq.~\eqref{Ineq:Prob:I22nn:Functional}.
For the purpose of this proof, we will rewrite the CGLMP
inequality, Eq.~\eqref{Ineq:Prob:CGLMP:Functional}, by shifting
the constant ``2" to the {\em lhs} of the inequality, namely,
\begin{align}
    \SLHVBI{$I_{n}$}=&\sum_{k=0}^{\lfloor
    \frac{n}{2}-1\rfloor}\left(1-\frac{2k}{n-1}\right)
        \sum_{\oB=1}^n\Big[
        \ProbTwJ{1}{\oB-k}{1}{\,\oB}-\ProbTwJ{1}{\oB+k+1}{1}{\,\oB}
        +\ProbTwJ{1}{\oB+k}{2}{\,\oB}-\ProbTwJ{1}{\oB-k-1}{2}{\,\oB} \nonumber\\
        &
        +\ProbTwJ{2}{\oB+k}{1}{\,\oB}-\ProbTwJ{2}{\oB-k-1}{1}{\,\oB}
        +\ProbTwJ{2}{\oB-k-1}{2}{\,\oB}-\ProbTwJ{2}{\oB+k}{2}{\,\oB}
        \Big]-2\le 0,
        \label{Ineq:Prob:CGLMP:Functional2}
\end{align}
where we remind the reader that expression such as $\oB-k$ in
the above inequality is understood to be evaluated modulo $n$.
For ease of reference, we will also reproduce the $I_{22nn}$
inequality as follow:
\begin{align}
    \SLHV^{(I_{22nn})}&=\sum_{\oA=1}^{n-1}\sum_{\oB=1}^{n-\oA}\ProbTwJ{1}{\oA}{1}{\oB}
    +\sum_{\oA=1}^{n-1}\sum_{\oB=n-\oA}^{n-1}\Big[\ProbTwJ{1}{\oA}{2}{\oB}
    +\ProbTwJ{2}{\oA}{1}{\oB}-\ProbTwJ{2}{\oA}{2}{\oB}\Big]\nonumber\\
    &-\sum_{\oA=1}^{n-1}\ProbTwMA{1}{\oA}-\sum_{\oB=1}^{n-1}\ProbTwMB{1}{\oB}\le 0.
    \tag{\ref{Ineq:Prob:I22nn:Functional}}
\end{align}
Moreover, we shall make use of the matrix representation of a
Bell inequality for probabilities introduced in
Eq.~\eqref{Eq:SLHV:Generic} --
Eq.~\eqref{Eq:Dfn:BlockMatrix:b}.

Let us begin by showing the equivalence explicitly for $n=3$.
In this case, the {\em lhs} of
inequality~\eqref{Ineq:Prob:CGLMP:Functional2} can be
represented by the following matrix of
coefficients\footnote{Notice that here, we are writing each
block matrix in full dimension, i.e., each block is of
dimension $\nA\times\nB$.}, c.f. Eq.~\eqref{Eq:Dfn:b} and
Eq.~\eqref{Eq:Dfn:BlockMatrix:b},
\begin{subequations}
\begin{align}\label{Eq:b:I3}
    b^{(I_3)}:&\sim
    \left(
    \begin{array}{c||c|c}
    b_{0,0}    & \bfb_{0,1} & \bfb_{0,2} \\ \hline\hline
    \bfb_{1,0} & \BBlk{1}{1} &  \BBlk{1}{2}\\ \hline
    \bfb_{2,0} & \BBlk{2}{1} &  \BBlk{2}{1}
    \end{array}
    \right)
    =\left(
    \begin{array}{r||rrr|rrr}
            -2 &  \cdot &  \cdot &  \cdot & \cdot & \cdot &  \cdot\\ \hline\hline
            \cdot &  1 &  \cdot & -1 & 1 &-1 & \cdot\\
            \cdot & -1 &  1 &  \cdot & \cdot & 1 & -1\\
            \cdot &  \cdot & -1 &  1 &-1 & \cdot &  1\\ \hline
            \cdot &  1 & -1 &  \cdot &-1 & 1 &  \cdot\\
            \cdot &  \cdot &  1 & -1 & \cdot &-1 &  1\\
            \cdot & -1 &  \cdot &  1 & 1 & \cdot & -1\\
    \end{array}   \right)
\end{align}
where we recall that coefficients associated with Alice's
(Bob's) local measurement setting are separated by a single
horizontal (vertical) line; coefficients associated with
marginal probabilities are separated from the others via double
horizontal (vertical) lines.

Now, let us make use of the no-signaling condition,
Eq.~\eqref{Eq:Dfn:No-signaling} to express the joint
probabilities associated with Alice's third measurement outcome
$\ProbTwJ{\sA}{3}{\sB}{\oB}$ in terms of marginal probabilities
$\ProbTwGMB$ and the other joint probabilities $\ProbTwGJ$ for
$\oA\neq3$. For example, doing this for $\sA=\sB=1$, $\oB=2$
amounts to subtracting every entry in the second column of the
block matrix $\bfb_{1,1}$ by the entry $[\bfb_{1,1}]_{3,2}$ and
adding this specific entry to the marginal entry
$[\bfb_{0,1}]_2$ directly above it. Repeating this for all
combinations of $\sA$, $\sB$ and $\oB$ gives rise to an
equivalent inequality with matrix of coefficients given by
\begin{align}
    b^{'(I_3)}:&\sim\left(
    \begin{array}{r||rrr|rrr}
            -2 & -1 & -1 &  2 & \cdot & \cdot &  \cdot\\ \hline\hline
            \cdot &  1 &  1 & -2 & 2 &-1 & -1\\
            \cdot & -1 &  2 & -1 & 1 & 1 & -2\\
            \cdot &  \cdot &  \cdot &  \cdot & \cdot & \cdot &  \cdot\\ \hline
            \cdot &  2 & -1 & -1 &-2 & 1 &  1\\
            \cdot &  1 &  1 & -2 &-1 &-1 &  2\\
            \cdot &  \cdot &  \cdot &  \cdot & \cdot & \cdot &  \cdot\\
    \end{array}\right).
    \label{Eq:I3:Intermediate:NoA3}
\end{align}
Next, let us also make use of the no-signaling condition to
express the joint probabilities associated with Bob's third
measurement outcome $\ProbTwJ{\sA}{\oA}{\sB}{3}$ in terms of
marginal probabilities $\ProbTwGMA$ and the other joint
probabilities $\ProbTwGJ$ for $\oB\neq3$. In particular, for
$\sA=1$,$\sB=\oB=2$, this amounts to subtracting every entry in
the second row of the block matrix $\bfb'_{1,2}$ by the entry
$[\bfb'_{1,2}]_{2,3}$ and adding this specific entry to the
marginal entry $[\bfb'_{1,0}]_2$ that is on the same row. Doing
this for all combinations of $\sA$, $\sB$ and $\oA$ gives rise
to another equivalent inequality with matrix of coefficients
given by
\begin{align}
    b^{''(I_3)}&\sim\left(
    \begin{array}{r||rrr|rrr}
            -2 & -1 & -1 &  2 & \cdot & \cdot &  \cdot\\ \hline\hline
            -3 &  3 &  3 &  \cdot & 3 & \cdot &  \cdot\\
            -3 &  \cdot &  3 &  \cdot & 3 & 3 &  \cdot\\
            \cdot &  \cdot &  \cdot &  \cdot & \cdot & \cdot &  \cdot\\ \hline
            \cdot &  3 &  \cdot &  \cdot &-3 & \cdot &  \cdot\\
            \cdot &  3 &  3 &  \cdot &-3 &-3 &  \cdot\\
            \cdot &  \cdot &  \cdot &  \cdot & \cdot & \cdot &  \cdot\\
    \end{array}\right).
    \label{Eq:I3:Intermediate:NoB3}
\end{align}
Then, by using the normalization condition, c.f.
Eq.~\eqref{Eq:Probabilities:General:Normalization},
\begin{equation}\label{Eq:MProbalities:Normalization}
    \sum_{\oB=1}^n\ProbTwGMB=1,
\end{equation}
we can further express $\ProbTwMB{1}{3}$ in terms of
$\ProbTwMB{1}{\oB}$ for $\oB\neq3$. In terms of the matrix of
coefficients, this amounts to subtracting every entry in
$\bfb''_{0,1}$ by $[\bfb''_{0,1}]_3$ and adding this specific
entry to $b''_{0,0}$. Writing this out explicitly, we get the
matrix of coefficients for a fourth equivalent inequality:
\begin{gather}\label{Eq:I3:Intermediate:MarginalSorted}
    b^{'''(I_3)}:\sim \left(
    \begin{array}{r||rrr|rrr}
            \cdot & -3 & -3 &  \cdot & \cdot & \cdot &  \cdot\\ \hline\hline
           -3 &  3 &  3 &  \cdot & 3 & \cdot &  \cdot\\
           -3 &  \cdot &  3 &  \cdot & 3 & 3 &  \cdot\\
            \cdot &  \cdot &  \cdot &  \cdot & \cdot & \cdot &  \cdot\\ \hline
            \cdot &  3 &  \cdot &  \cdot &-3 & \cdot &  \cdot\\
            \cdot &  3 &  3 &  \cdot &-3 &-3 &  \cdot\\
            \cdot &  \cdot &  \cdot &  \cdot & \cdot & \cdot &  \cdot\\
    \end{array}\right).
\end{gather}
What remains to be done now is to swap all of Bob's first and
second measurement outcomes, which gives a fifth equivalent
inequality with matrix of coefficients:
\begin{gather}\label{Eq:I3:Intermediate:Final}
    b^{''''(I_3)}:\sim 3\left(
    \begin{array}{r||rrr|rrr}
            \cdot & -1 & -1 &  \cdot & \cdot & \cdot &  \cdot\\ \hline\hline
           -1 &  1 &  1 &  \cdot & \cdot & 1 &  \cdot\\
           -1 &  1 &  \cdot &  \cdot & 1 & 1 &  \cdot\\
            \cdot &  \cdot &  \cdot &  \cdot & \cdot & \cdot &  \cdot\\ \hline
            \cdot &  \cdot &  1 &  \cdot & \cdot &-1 &  \cdot\\
            \cdot &  1 &  1 &  \cdot &-1 &-1 &  \cdot\\
            \cdot &  \cdot &  \cdot &  \cdot & \cdot & \cdot &  \cdot\\
    \end{array}   \right),
\end{gather}
\end{subequations}
which can be seen to be equivalent to the $I_{2233}$
inequality, Eq.~\eqref{Ineq:Prob:I2233}.

In exactly the same manner, we see that for $n=4$, we have
\begin{align*}
    b^{(I_4)}:&\sim\left(
    \begin{array}{r||rrrr|rrrr}
            -2    &  \cdot &  \cdot &  \cdot &  \cdot &  \cdot &  \cdot &  \cdot &  \cdot\\ \hline\hline
            \cdot &  1 & 1/3&-1/3& -1 &  1 & -1 &-1/3& 1/3\\
            \cdot & -1 &  1 & 1/3&-1/3& 1/3&  1 & -1 &-1/3\\
            \cdot &-1/3& -1 &  1 & 1/3&-1/3& 1/3&  1 & -1 \\
            \cdot & 1/3&-1/3& -1 &  1 & -1 &-1/3& 1/3&  1 \\\hline
            \cdot &  1 & -1 &-1/3& 1/3& -1 &  1 & 1/3&-1/3\\
            \cdot & 1/3&  1 & -1 &-1/3&-1/3& -1 &  1 & 1/3\\
            \cdot &-1/3& 1/3&  1 & -1 & 1/3&-1/3& -1 &  1 \\
            \cdot & -1 &-1/3& 1/3&  1 &  1 & 1/3&-1/3& -1
    \end{array}   \right),
\end{align*}
then, by zeroing the coefficients associated with
$\ProbTwJ{\sA}{4}{\sB}{\oB}$ and $\ProbTwJ{\sA}{\oA}{\sB}{4}$
via the no-signaling condition,
Eq.~\eqref{Eq:Dfn:No-signaling}, we get
\begin{align*}
    b^{'(I_4)}:&\sim\left(
    \begin{array}{r||rrrr|rrrr}
            -2    &-2/3&-2/3&-2/3&  2 &  \cdot &  \cdot &  \cdot &  \cdot \\ \hline\hline
            \cdot & 2/3& 2/3& 2/3& -2 &  2 &-2/3&-2/3&-2/3\\
            \cdot &-4/3& 4/3& 4/3&-4/3& 4/3& 4/3&-4/3&-4/3\\
            \cdot &-2/3&-2/3&  2 &-2/3& 2/3& 2/3& 2/3& -2 \\
            \cdot &  \cdot &  \cdot &  \cdot &  \cdot &  \cdot &  \cdot &  \cdot &  \cdot \\\hline
            \cdot &  2 &-2/3&-2/3&-2/3& -2 & 2/3& 2/3& 2/3\\
            \cdot & 4/3& 4/3&-4/3&-4/3&-4/3&-4/3& 4/3& 4/3\\
            \cdot & 2/3& 2/3& 2/3& -2 &-2/3&-2/3&-2/3&  2 \\
            \cdot &  \cdot &  \cdot &  \cdot &  \cdot &  \cdot &  \cdot &  \cdot &  \cdot
    \end{array}   \right)\\
    \Rightarrow b^{''(I_4)}:&\sim\left(
    \begin{array}{r||rrrr|rrrr}
            -2    &-2/3&-2/3&-2/3&  2 &  \cdot &  \cdot &  \cdot &  \cdot \\ \hline\hline
          -8/3& 8/3& 8/3& 8/3&  \cdot & 8/3&  \cdot &  \cdot &  \cdot \\
          -8/3&  \cdot & 8/3& 8/3&  \cdot & 8/3& 8/3&  \cdot &  \cdot \\
          -8/3&  \cdot &  \cdot & 8/3&  \cdot & 8/3& 8/3& 8/3&  \cdot \\
            \cdot &  \cdot &  \cdot &  \cdot &  \cdot &  \cdot &  \cdot &  \cdot &  \cdot \\\hline
            \cdot & 8/3&  \cdot &  \cdot &  \cdot &-8/3&  \cdot &  \cdot &  \cdot \\
            \cdot & 8/3& 8/3&  \cdot &  \cdot &-8/3&-8/3&  \cdot &  \cdot \\
            \cdot & 8/3& 8/3& 8/3&  \cdot &-8/3&-8/3&-8/3&  \cdot \\
            \cdot &  \cdot &  \cdot &  \cdot &  \cdot &  \cdot &  \cdot &  \cdot &  \cdot
    \end{array}   \right).
\end{align*}
Finally, by zeroing $\ProbTwMB{1}{4}$ using
Eq.~\eqref{Eq:MProbalities:Normalization} and swapping Bob's
measurement outcomes, $\oB\leftrightarrow n-\oB$, we end up
with an equivalent inequality with matrix of coefficients:
\begin{align*}
    b^{''''(I_4)}:&\sim\frac{8}{3}\left(
    \begin{array}{r||rrrr|rrrr}
            \cdot & -1 & -1 & -1 &  \cdot & \cdot & \cdot & \cdot &  \cdot\\ \hline\hline
           -1 &  1 &  1 &  1 &  \cdot & \cdot & \cdot & 1 &  \cdot\\
           -1 &  1 &  1 &  \cdot &  \cdot & \cdot & 1 & 1 &  \cdot\\
           -1 &  1 &  \cdot &  \cdot &  \cdot & 1 & 1 & 1 &  \cdot\\
            \cdot &  \cdot &  \cdot &  \cdot &  \cdot & \cdot & \cdot & \cdot &  \cdot\\ \hline
            \cdot &  \cdot &  \cdot &  1 &  \cdot & \cdot & \cdot &-1 &  \cdot\\
            \cdot &  \cdot &  1 &  1 &  \cdot & \cdot &-1 &-1 &  \cdot\\
            \cdot &  1 &  1 &  1 &  \cdot & -1&-1 &-1 &  \cdot\\
            \cdot &  \cdot &  \cdot &  \cdot &  \cdot & \cdot & \cdot & \cdot &  \cdot
    \end{array}   \right),
\end{align*}
$I_4$ is thus equivalent to the $I_{2244}$ inequality, c.f.
Eq.~\eqref{Ineq:Prob:I2244}.

More generally, we can prove that $I_n$ is equivalent to
$I_{22nn}$ by generalizing the above procedures. Firstly, we
note that the matrix of coefficients $b^{(I_n)}$ for the
$n$-outcome CGLMP inequality,
Eq.~\eqref{Ineq:Prob:CGLMP:Functional2}, is made up from blocks
of circulant matrices $\bfb_{\sA,\sB}$, with entries given by
\begin{subequations}\label{Eq:bIn:Original}
\begin{equation}
    [\bfb_{1,1}]_{\oA,\oB}=
    \left\{ \begin{array}{c@{\quad:\quad}l}
         1+\frac{2(\oA-\oB)}{n-1} & \oB\ge\oA\\
        -1+\frac{2(\oA-\oB-1)}{n-1} & \oB<\oA
        \end{array} \right.,
\end{equation}
\begin{equation}
    [\bfb_{1,2}]_{\oA,\oB}=[\bfb_{2,1}]_{\oA,\oB}=-[\bfb_{2,2}]_{\oA,\oB}
    =\left\{ \begin{array}{c@{\quad:\quad}l}
        -1+\frac{2(\oB-\oA-1)}{n-1} & \oB>\oA\\
         1+\frac{2(\oB-\oA)}{n-1} & \oB\le\oA
        \end{array} \right.,
\end{equation}
and marginal blocks $\bfb_{\sA,0}$, $\bfb_{0,\sB}$, $b_{0,0}$:
\begin{equation}\label{Eq:bIn:Marginal}
    \bfb_{1,0}=\bfb_{2,0}=\zero_n,\quad
    \bfb_{0,1}=\bfb_{0,2}=\zero_n\t,\quad b_{0,0}=-2.
\end{equation}
\end{subequations}
where $\zero_n$ is an $n\times 1$ null vector.

As in Eq.~\eqref{Eq:I3:Intermediate:NoA3}, we will now make use
of the no-signaling condition, Eq.~\eqref{Eq:Dfn:No-signaling},
to zero the coefficients associated with the joint
probabilities $\ProbTwJ{\sA}{n\,}{\sB}{\oB}$. This gives rise
to an equivalent inequality whose matrix of coefficients
$b^{'(I_n)}$ is related to the original one, $b^{(I_n)}$ by
\begin{gather*}
    \bfb'_{0,0}=\bfb_{0,0},\quad \bfb'_{\sA,0}=\bfb_{\sA,0},\quad \sA=1,2,\\
    [\bfb'_{0,\sB}]_{\oB}=[\bfb_{0,\sB}]_{\oB}+\sum_{\sA=1}^2
    [\bfb_{\sA,\sB}]_{n,\oB},\quad \sB=1,2,\quad\oB=1,2,\ldots,n, \\
    [\bfb'_{\sA,\sB}]_{\oA,\oB}=[\bfb_{\sA,\sB}]_{\oA,\oB}
    -[\bfb_{\sA,\sB}]_{n,\oB},\quad
    \sA,\sB=1,2,\quad\oA,\oB=1,2,\ldots,n.
\end{gather*}
Next, we will again make use of the no-signaling condition, but
to instead zero the coefficients associated with the joint
probabilities $\ProbTwJ{\sA}{\oA}{\sB}{n}$ [c.f.
Eq.~\eqref{Eq:I3:Intermediate:NoB3}]. This gives rise to
another equivalent inequality whose matrix of coefficients
$b^{''(I_n)}$ is related to the existing one, $b^{'(I_n)}$ by
\begin{gather*}
    \bfb''_{0,0}=\bfb'_{0,0},\quad \bfb''_{0,\sB}=\bfb'_{0,\sB},\quad \sB=1,2,\\
    [\bfb''_{\sA,0}]_{\oA}=[\bfb'_{\sA,0}]_{\oA}+\sum_{\sB=1}^2
    [\bfb'_{\sA,\sB}]_{\oA,n},\quad \sA=1,2,\quad\oA=1,2,\ldots,n, \\
    [\bfb''_{\sA,\sB}]_{\oA,\oB}=[\bfb'_{\sA,\sB}]_{\oA,\oB}
    -[\bfb'_{\sA,\sB}]_{\oA,n},\quad
    \sA,\sB=1,2,\quad\oA,\oB=1,2,\ldots,n.
\end{gather*}
Now, we will make use of the normalization of marginal
probabilities, Eq.~\eqref{Eq:MProbalities:Normalization}, to
zero the coefficients associated with the marginal
probabilities $\ProbTwMA{\sA}{n}$. This gives rise to another
equivalent inequality whose matrix of coefficients
$b^{'''(I_n)}$ is related to the existing one, $b^{''(I_n)}$ by
\begin{gather*}
    b'''_{0,0}=b''_{0,0}+\sum_{\sB=1}^2[\bfb''_{0,\sB}]_n,\\
    [\bfb'''_{0,\sB}]_{\oB}=[\bfb''_{0,\sB}]_{\oB}-[\bfb''_{0,\sB}]_n,
    \quad \sB=1,2,\quad \oB=1,2,\ldots,n,\\
    \bfb'''_{\sA,0}=\bfb''_{\sA,0},\quad\sA=1,2,\\
    \bfb'''_{\sA,\sB}=\bfb''_{\sA,\sB},\quad\sA,\sB=1,2.
\end{gather*}
More explicitly, it is easy to check that the matrix of
coefficients $b^{'''(I_n)}$ reads
\begin{subequations}
\begin{gather}
    [\bfb'''_{1,1}]_{\oA,\oB}=
    \left\{ \begin{array}{c@{\quad:\quad}l}
        \frac{2n}{n-1} & \oB\ge\oA, \oB\neq n\\
        0 & \oB\ge\oA, \oB=n\\
        0 & \oB<\oA,
        \end{array} \right.,\\
    [\bfb'''_{1,2}]_{\oA,\oB}=[\bfb'''_{2,1}]_{\oA,\oB}=-[\bfb'''_{2,2}]_{\oA,\oB}
    =\left\{ \begin{array}{c@{\quad:\quad}l}
        0 & \oB>\oA\\
        \frac{2n}{n-1} & \oB\le\oA, \oA\neq n,\\
        0 & \oB\le\oA, \oA= n,
        \end{array} \right.,\\
    [\bfb'''_{0,1}]_{\oB}=
    \left\{ \begin{array}{c@{\quad:\quad}l}
        -\frac{2n}{n-1} & \oB<n\\
         0 & \oB=n
        \end{array} \right.,\\
    \label{Eq:bIn:Marginal10d}
    [\bfb'''_{1,0}]_{\oA}=
    \left\{ \begin{array}{c@{\quad:\quad}l}
        -\frac{2n}{n-1} & \oA<n\\
         0 & \oA=n
        \end{array} \right.,\\
    \bfb'''_{2,0}=\zero_n,\quad
    \bfb'''_{0,2}=\zero_n\t,\quad b'''_{0,0}=0.
\end{gather}
\end{subequations}

Finally, by swapping Bob's measurement outcomes,
$\oB\leftrightarrow n-\oB$ for all $\oB$, $\sA$ and $\sB$, it
is readily seen that the matrix of coefficients for this fifth
equivalent inequality is related to that of $I_{22nn}$ by
\begin{equation}
    b^{''''(I_n)}=\frac{2n}{n-1}b^{(I_{22nn})},
\end{equation}
and hence the CGLMP inequality with $n$ outcomes,
Eq.~\eqref{Ineq:Prob:CGLMP:Functional2}, is equivalent to the
$I_{22nn}$ inequality. Clearly, since the two inequalities,
Eq.~\eqref{Ineq:Prob:CGLMP:Functional} and
Eq.~\eqref{Ineq:Prob:CGLMP:Functional2}, are identical up to
simple algebraic manipulations, the $I_{22nn}$ inequality must
also be equivalent to the CGLMP inequality written in the form
of Eq.~\eqref{Ineq:Prob:CGLMP:Functional}. In particular, if we
denote by $\Bell_{I_n}$ and $\Bell_{I_{22nn}}$, respectively,
the Bell operator derived from
Eq.~\eqref{Ineq:Prob:CGLMP:Functional} and
Eq.~\eqref{Ineq:Prob:I22nn:Functional}. Then, for any quantum
state $\rho$, the expectation values of these Bell operators
with respect to $\rho$ are related by
\begin{equation}
    \tr\left(\rho~\Bell_{I_n}\right)=
    \frac{2n}{n-1}\tr\left(\rho~\Bell_{I_{22nn}}\right)+2.
    \tag{\ref{Eq:In-I22nn}}
\end{equation}

\section{Quantum Correlations and Locally Causal Quantum States}
\label{App:Sec:Q.Cn:Classical}

\subsection{Convexity of Non-Bell-Inequality-Violating States}
\label{App:Sec:Convexity:LocalStates}

Here, we will prove that the set of quantum states not
violating a given Bell inequality in the sense of
Definition~\ref{Dfn:BIViolation} is convex. Let us denote by
$\Bell_k$ the Bell operator associated with a Bell inequality
$I_k:\SLHVBI{k}\le\bLHV{k}$ and $\NV{k}$ the set of quantum
states not violating $I_k$ via a standard Bell experiment. A
quantum state $\rho\in\NV{k}$ if
\begin{equation}
    \SqmBI{k}(\rho)\le \bLHV{k}.
\end{equation}
Clearly, this implies that for any local observables
constituting the Bell operator $\Bell_k$, we must have
\begin{equation}
    \tr\left(\rho~\Bell_k\right)\le \bLHV{k}.
\end{equation}

Now, let two quantum states $\rho_1,\rho_2\in\NV{k}$. Then, for
any convex combination of them, i.e.,
\begin{equation}
    \rho'\equiv p~\rho_1+(1-p)~\rho_2,\quad 0\le p\le1,
\end{equation}
we see that
\begin{align*}
    \tr\left(\rho'~\Bell_k\right)&=p~\tr\left(\rho_1~\Bell_k\right)
    +(1-p)~\tr\left(\rho_2~\Bell_k\right),\\
    &\le p\,\bLHV{k}+(1-p)\,\bLHV{k},\\
    &=\bLHV{k}.
\end{align*}
Since this is true for any local observables constituting
$\Bell_k$, we must have
\begin{equation}
    \SqmBI{k}(\rho')\le \bLHV{k}.
\end{equation}
That is, $\rho'$ is also a member of $\NV{k}$, and hence
$\NV{k}$ is a convex set. Moreover, since this is true for an
arbitrary Bell inequality, it follows that the set of NBIV
quantum states, $\NVg$ is also convex.

\section{Bounds on Quantum Correlations in Standard Bell
Experiments}\label{App:Sec:QuantumBounds}

\subsection{Bell-CH Inequality and Full Rank Projector}
\label{App:Sec:FullRank}

In this section, we will prove that the Bell-CH inequality with
only two possible outcomes cannot be violated if any of the
POVM elements involved is a full rank projector.\footnote{This
necessarily implies that the complementary POVM element is a
zero matrix.} For that matter, it suffices to show that in any
of these scenarios, the resulting Bell operator $\Bell_{\rm
CH}$ is strictly negative semidefinite (NSD), since the trace
of a positive semidefinite (PSD) matrix $\rho$ against an NSD
matrix cannot be positive.

Now, let us recall that for a two-outcome Bell experiment, the
Bell-CH operator can be written as
\begin{equation}
    \Bell_{\rm CH}
    =A_1^+\ten(B_2^+-B_1^-)-A_2^-\ten B_1^+-A_2^+\ten
B_2^+\tag{\ref{Eq:BellOperator:CH}}.
\end{equation}
Then, by utilizing the normalization of POVM elements,
Eq.~\eqref{Eq:POVM:Normalization}, we see that when
\begin{enumerate}
    \item  $B_1^+=\zero_{\dB\times\dB}$,
        $B_1^-=\unit_{\dB}$,
    \begin{align*}
        \Bell_{\rm CH}&= - A_1^+\ten B_2^- - A_2^+ \ten B_2^+ \le\zero_{\dA\dB\times\dA\dB};
    \end{align*}
    \item  $B_1^+=\unit_{\dB}$,
        $B_1^-=\zero_{\dB\times\dB}$,
    \begin{align*}
        \Bell_{\rm CH}&=A_1^+\ten B_2^+-A_2^-\ten\unit_{\dB}-A_2^+\ten B_2^+
        =-A_1^-\ten B_2^+-A_2^-\ten B_2^-\le\zero_{\dA\dB\times\dA\dB};
    \end{align*}
    \item  $B_2^+=\zero_{\dB\times\dB}$,
        $B_2^-=\unit_{\dB}$,
    \begin{align*}
        \Bell_{\rm CH}&= -A_1^+\ten B_1^--A_2^-\ten B_1^+\le\zero_{\dA\dB\times\dA\dB};
    \end{align*}
    \item  $B_2^+=\unit_{\dB}$,
        $B_2^-=\zero_{\dB\times\dB}$,
    \begin{align*}
        \Bell_{\rm CH}&= A_1^+\ten B_1^+-A_2^-\ten B_1^+-A_2^+\ten\unit_{\dB}
        = -A_1^-\ten B_1^+-A_2^+\ten B_1^- \le\zero_{\dA\dB\times\dA\dB}.
    \end{align*}
\end{enumerate}
Since the Bell-CH inequality is symmetrical with respect to
swapping the two parties, exactly the same argument can be
applied to show that the resulting Bell-CH operator is NSD if
any of Alice's POVM element is a full rank projector. Hence,
the Bell-CH inequality cannot be violated in a standard Bell
experiment involving at most two possible outcomes and where
one of the measurement devices always gives the same
measurement outcome.

\subsection{Derivation of Horodecki's Criterion using LB}
\label{App:Sec:HorodeckiCriterion}

In order to determine if a general two-qubit state $\rho$
violates the Bell-CH inequality,
Eq.~\eqref{Ineq:CH:Functional}, we will have to first obtain an
explicit expression for $\SqmCH(\rho)$. This can be done, for
example, by evaluating Eq.~\eqref{Eq:S:implicitA} which, in
turn, requires us to know the eigenvalues of
$\rho_{B^+_{\sB}}-\rho_{B^-_{\sB}}$, Eq.~\eqref{Eq:Dfn:rhoAB}
for all $\sB$.

In this regard, let us note that for the Bell-CH inequality, we always
have
\begin{subequations}\label{Eq:rhoB:Bell-CH}
\begin{gather}
    \rho_{B^+_1}-\rho_{B^-_1}=\trA\left\{\rho
    \left[\left(A^+_1-A^-_2\right)\ten\unit_{\dB}\right]\right\},
    \label{Eq:rhoAB1}\\
    \rho_{B^+_2}-\rho_{B^-_2}=\trA\left\{\rho\left[
    \left(A^+_1-A^+_2\right)\ten\unit_{\dB}\right]\right\},
    \label{Eq:rhoAB2}\\
    \sum_{\sB,\oB}\tr\left(\rho_{B^{\oB}_{\sB}}\right)=-1,
    \label{Eq:TrrhoAB}
\end{gather}
\end{subequations}
since $b^{++}_{12}=-b^{+-}_{11}=-b^{-+}_{21}=-b^{++}_{22}=1$
while all the other $\CoeffG=0$ [c.f.
Eq.~\eqref{Eq:BellOperator:Probabilities} and
Eq.~\eqref{Eq:BellOperator:CH}].

When Alice's choice of POVM is given by
Eq.~\eqref{Eq:TwoOutomceProjectors}, it follows from
Eq.~\eqref{Eq:TwoQubit:HS} that the above expressions can be
written more explicitly as
\begin{align*}
    \rho_{B_1^+}-\rho_{B_1^-}&=\quar\left[\bfr_\A\cdot(\hat{a}_1+\hat{a}_2)~\unit_2
    +\sum_{i,j=x,y,z}(\hat{a}_1+\hat{a}_2)_i[T]_{ij}\,\sigma_j\right],\\
    \rho_{B_2^+}-\rho_{B_2^-}&=\quar\left[\bfr_\A\cdot(\hat{a}_1-\hat{a}_2)~\unit_2
    +\sum_{i,j=x,y,z}(\hat{a}_1-\hat{a}_2)_i[T]_{ij}\,\sigma_j\right],
\end{align*}
which gives, respectively, eigenvalues
\begin{align}\label{Eq:Eig:rhoAB}
    \lambda^{\pm}_{1}=\half\left(\cos\theta~\hat{c}\cdot\bfr_\A\pm
    |\cos\theta| \,||T^{\mbox{\tiny T}}\hat{c}||\right),\quad
    \lambda^{\pm}_{2}=\half\left(\sin\theta~\hat{c}\,'\cdot\bfr_\A\pm
    |\sin\theta|\,||T^{\mbox{\tiny T}}\hat{c}\,'||\right)
\end{align}
where $\hat{c}\in\mathbb{R}^3$ and $\hat{c}\,'\in\mathbb{R}^3$
are orthogonal unit vectors defined via
\begin{equation}
    \hat{a}_1+\hat{a}_2\equiv 2\cos\theta\,\hat{c},\qquad
    \hat{a}_1-\hat{a}_2\equiv 2\sin\theta\,\hat{c}\,'.
\end{equation}
We can now write Eq.~\eqref{Eq:S:implicitA} as
\begin{equation}\label{Eq:Sqm:Two-qubit:Intermediate}
    \SqmCH(\rho, \hat{c}, \hat{c}\,',\theta)
    =\half\sum_{\sB=1}^2\sum_{\oB=\pm}
    \left|\lambda_{\sB}^{\oB}(\hat{c},\hat{c}\,',\theta)\right|-\half,
\end{equation}
which is to be maximized over all legitimate choices of
$\hat{c}$, $\hat{c}'$ and $\theta$ to give $\SqmCH(\rho)$.

Let us now consider the case in which $\SqmCH(\rho)$ is
obtained by choosing $\hat{c}$, $\hat{c}'$ and $\theta$ in
$\Sqm^{\mbox{\tiny (CH)}}(\rho, \hat{c}, \hat{c}\,',\theta)$
such that $\sgn(\lambda_{\sB}^+)\neq\sgn(\lambda_{\sB}^-)$ for
{\em all} $\sB$. In this case,
Eq.~\eqref{Eq:Sqm:Two-qubit:Intermediate} becomes
\begin{equation}
    \SqmCH(\rho, \hat{c}, \hat{c}\,',\theta )=
    \half\left(||T^{\mbox{\tiny T}}\hat{c}||\cos\theta
    + ||T^{\mbox{\tiny T}}\hat{c}\,'||\sin\theta\right)-\half,
    \quad \theta\in\left[0,\frac{\pi}{4}\right],
\end{equation}
where we have redefined $\theta$ such that it now falls within
0 and $\pi/4$. The maximization over $\theta$ can now be
carried out by choosing $\theta=\theta^*$ such that
$||T^{\mbox{\tiny T}}\hat{c}||\sin\theta^*=||T^{\mbox{\tiny T}}
\hat{c}\,'||\cos\theta^*$, i.e.,\footnote{That this choice is
of $\theta$ is always possible follows from the well-known fact
that all singular values of $T$ are less than or equal to one
(see, for example, pp. 1840 of
Ref.~\cite{R.Horodecki:PRA:1996}).}
\begin{equation}
    \SqmCH(\rho, \hat{c}, \hat{c}\,',\theta^* )=
    \half\sqrt{||T^{\mbox{\tiny T}}\hat{c}||^2+ ||T^{\mbox{\tiny
    T}}\hat{c}\,'||^2}-\half.
\end{equation}
From here, it suffices to choose $\hat{c}$ and $\hat{c}'$ as
the (orthonormal) eigenvectors of $T\,T\t$ corresponding to the
two largest eigenvalues. When arranged in descending order, the
$\idx{k}$ eigenvalue of $T\,T\t$, however, is just the square
of the $\idx{k}$ singular value of $T$, which we shall denote
by $\varsigma_k$. Hence, in this particular case, we have
\begin{equation}
    \SqmCH(\rho)=\half\sqrt{\varsigma_1^2+\varsigma_2^2}-\half.
\end{equation}

What about the other cases in which $\SqmCH(\rho)$ is obtained
by choosing $\hat{c}$, $\hat{c}'$ and $\theta$ in
$\Sqm^{\mbox{\tiny (CH)}}(\rho, \hat{c}, \hat{c}\,',\theta)$
such that $\sgn(\lambda_{\sB}^+)=\sgn(\lambda_{\sB}^-)$ for at
least one of the $\sB$'s? In these cases, it follows from our
discussion in Sec.~\ref{Sec:two-outcome} that for each of such
$\sB$'s, the corresponding pair of optimal $B^{\oB}_{\sB}$ is
given by $\{\zero,\unit_2\}$. However, as we have seen in
Appendix~\ref{App:Sec:FullRank}, if any of Alice's (or Bob's)
POVM element is a full rank projector, the corresponding Bell
operator is an NSD matrix, and hence cannot lead to a Bell-CH
inequality violation. Moreover, the best that one can do in
this case is to pick a classical strategy such that
\begin{equation}
    \SqmCH(\rho)=0,
\end{equation}
which is necessarily greater than
$\sqrt{\varsigma_1^2+\varsigma_2^2}/2-1/2$.

Therefore, for a general two qubit state $\rho$, we have
\begin{equation}
    \SqmCH(\rho)=\max\left\{0,\half\sqrt{\varsigma_1^2+\varsigma_2^2}-\half\right\}.
\end{equation}
Recall that a Bell-CH violation by $\rho$ is possible if and
only if $\SqmCH(\rho)>0$. Hence, a two-qubit state violates the
Bell-CH inequality if and only if
\begin{equation}
    \varsigma_1^2+\varsigma_2^2>1.\tag{\ref{Eq:HorodeckiCriterion}}
\end{equation}

\section{Bell-Inequality Violations by Quantum States}
\label{App:Sec:BellViolation}

\subsection{Bell-CH-violation for Pure Two-Qudit States}
\label{App:Sec:Bell-CH:Pure2Qudit:Measurements}

Here, we will provide more details about the intermediate
calculations leading to Eq.~\eqref{Eq:Sqm:CH:pure} and the
corresponding optimal measurements, i.e.,
$\left\{B_{\sB}^\pm\right\}_{\sB=1}^2$, that should be carried
out by Bob.

To begin with, we note from Eq.~\eqref{Eq:Alice-POVM} that
\begin{equation}\label{Eq:DiffAlicePOVM}
    A_1^+-A_2^\mp=\half\left(Z\pm X\right)=\half\left[
    \bigoplus_{i=1}^{\lfloor d/2\rfloor}X_\pm+(1\pm1)\,\Xi\right],\quad
\end{equation}
where
\begin{gather*}
    X_{\pm}\equiv\sigma_z\pm\sigma_x=\left(
    \begin{array}{rr}
        1 & \pm1 \\
        \pm1 & -1
    \end{array}\right),\\
    \left[\Xi\right]_{ij}=0\quad\forall\quad
    i,j\neq d,\qquad \left[\Xi\right]_{dd}\equiv\xi=d~{\rm mod}~2;
\end{gather*}
here and below, whenever $d$ is odd, we will assume that the
end product of the direct sum is appended with zero entries to
make the dimension of the resulting matrix $d\times d$.

From Eq.~\eqref{Eq:TwoQudit:SchmidtForm} and
Eq.~\eqref{Eq:rhoB:Bell-CH}, it then follows that
\begin{subequations}\label{Eq:rhoB:intermediate}
\begin{align}
    \rho_{B^+_1}-\rho_{B^-_1}&=\half\sum_{i,j=1}^{2\lfloor d/2\rfloor}
    c_ic_j {}_\A\bra{j}
    \left(\bigoplus_{n=1}^{\lfloor d/2\rfloor}X_+\right)
    \ket{i}_\A~\ket{i}_{\B}{}_{\B}\bra{j} +c_d^2\,\Xi,\nonumber\\
    &=\half\bigoplus_{n=1}^{\lfloor d/2\rfloor}  \left(
    \begin{array}{cc}
        c_{2n-1}^2 &  c_{2n-1}c_{2n} \\
        c_{2n-1}c_{2n} & -c_{2n}^2
    \end{array}\right) + c_d^2\,\Xi,
\end{align}
and
\begin{align}
    \rho_{B^+_2}-\rho_{B^-_2}=\half\bigoplus_{n=1}
    ^{\left\lfloor d/2\right\rfloor}
    \left(
    \begin{array}{cc}
        c_{2n-1}^2 & -c_{2n-1}c_{2n} \\
        -c_{2n-1}c_{2n} & -c_{2n}^2
    \end{array}\right),
\end{align}
\end{subequations}

Some further calculations show that both these matrices have
the following $2\lfloor d/2\rfloor$ eigenvalues
\begin{equation*}
    \lambda_{n,\pm}=\quar\left(c_{2n-1}^2-c_{2n}^2\pm\kappa_n
    \right),\quad n=1,2,\ldots,\left\lfloor \frac{d}{2}\right\rfloor,
\end{equation*}
where
\begin{equation*}
    \kappa_n\equiv\sqrt{(c_{2n-1}^2 +c_{2n}^2)^2+4c_{2n-1}^2c_{2n}^2}.
\end{equation*}
For each $n$, let us denote the eigenvectors of
$\rho_{B^+_{\sB}}-\rho_{B^-_{\sB}}$ corresponding to eigenvalue
$\lambda_{n,\pm}$ as $\ket{v^{\sB}_{n,\pm}}$, then, it can be
shown that these eigenvectors only have the following nonzero
entries
\begin{align*}
    [\ket{v^1_{n,\pm}}]_{2n-1}&=\eta_{n,\mp}\left(c_{2n-1}^2+c_{2n}^2\pm
    \kappa_n\right),\quad
    [\ket{v^1_{n,\pm}}]_{2n}=2\eta_{n,\mp}\,c_{2n-1}c_{2n},\\
    [\ket{v^2_{n,\pm}}]_{2n-1}&=\eta_{n,\mp}\left(c_{2n-1}^2+c_{2n}^2\pm
    \kappa_n\right),\quad
    [\ket{v^2_{n,\pm}}]_{2n}=-2\eta_{n,\mp}\,c_{2n-1}c_{2n}.
\end{align*}
where
\begin{equation*}
    \eta_{n,\pm}=\sqrt{\frac{\kappa_n\pm(c_{2n-1}^2+c_{2n}^2)}
    {8\,c_{2n-1}^2c_{2n}^2\kappa_n}}
\end{equation*}
is a normalization constant. When $d$ is odd,
$\rho_{B^+_1}-\rho_{B^-_1}$ and $\rho_{B^+_2}-\rho_{B^-_2}$,
respectively, also have the eigenvalue $c_d^2$ and 0. In this
case, the additional eigenvector of
$\rho_{B^+_{\sB}}-\rho_{B^-_{\sB}}$, denoted by
$\ket{v^{\sB}_d}$, where $\sB=1,2$, only has the following
nonzero entry $[\ket{v^{\sB}_d}]_d=1$.

Following the arguments presented in
Sec.~\ref{Sec:two-outcome}, we then know that  the
corresponding optimal measurements for Bob can be chosen to be
\begin{equation}\label{Eq:Bob-POVM}
    B_{\sB}^+=\sum_{n=1}^{\lfloor d/2\rfloor} \ketbra{v^{\sB}_{n,+}}
    +\xi\ketbra{v^{\sB}_d},\quad  B_{\sB}^-=\unit_{\dB}-B_{\sB}^+.
\end{equation}
Moreover, the corresponding expectation value of Bell operator
reads
\begin{align*}
    \eBellCH_{\ket{\Phi_d}}&=\half\sum_{\sB=1}^2\left|
    \left|\rho_{B^+_{\sB}}-\rho_{B^-_{\sB}}\right|\right|
    +\half\sum_{\sB}\sum_{\oB=\pm}\tr
    \left(\rho_{\POVMBg}\right),\\
    &=\quar\sum_{\sB=1}^2\sum_{n=1}^{\lfloor  d/2\rfloor}\kappa_n
    +\frac{\xi}{2}c_d^2-\half,\\
    &=\half\sum_{n=1}^{\lfloor  d/2\rfloor}
    \sqrt{(c_{2n-1}^2+c_{2n}^2)^2
    +4c_{2n}^2 c_{2n-1}^2}+\frac{\xi}{2}c_d^2-\half,
\end{align*}
where we have also made used of Eq.~\eqref{Eq:TrrhoAB} and the
fact that
\begin{equation*}
    c_{2n-1}^2-c_{2n}^2<\sqrt{(c_{2n-1}^2+c_{2n}^2)^2+4c_{2n-1}^2c_{2n}^2}.
\end{equation*}

\section{Nonstandard Bell Experiments and Hidden Nonlocality}

\subsection{Proof of Lemma~\ref{Lem:LMI->Separable}}
\label{App:Sec:ProofOfLemma}

\newcounter{tmp}
\setcounter{tmp}{\value{theorem}} \setcounter{theorem}{16}

For ease of reference, let us reproduce
Lemma~\ref{Lem:LMI->Separable} as follows:
\begin{lemma}
    Let $\Omega_x: [\Cd{2}] \otimes [\Cd{2}] \rightarrow
    [\H_{\A}\otimes \Cd{2}]\otimes[\H_{\B} \otimes \Cd{2}]$ be a family of
    maps, separable with respect to the partition denoted by the brackets.
    Let $\mu$ be a unit-trace, PSD matrix acting on
    $[\H_{\A}]\otimes[\H_{\B}]$ such that
    \begin{equation}\tag{\ref{principal}}
        \mu\t\! \otimes H_{\frac{\pi}{4}} - \int\dd{x}
        \ \Omega_x\!\left( H_{\theta_x} \right)
        \geq 0,
    \end{equation}
    where $H_\theta$ is defined in Eq.~\eqref{Eq:Dfn:Htheta}, then
    $\mu$ has to be separable.
\end{lemma}
\noindent In order to prove this Lemma, and therefore
Theorem~\ref{Thm:ObservableNonlocality}, it is necessary to use
the constraint that the maps $\Omega_x$ are separable. The
problem of characterizing the separable maps is hard in general
since it maps onto the separability problem for bipartite
states. However it turns out only to be necessary to determine
the set of separable maps that take Bell diagonal states to
Bell diagonal states and this can be done exactly
(Appendix~\ref{App:SLOCC}). In what follows, we will provide
the details for the proof of this Lemma.

\begin{proof}
The proof basically consists of three main steps. Firstly, we
will need to characterize the set of separable maps $\Omega_x$
that is relevant to Eq.~\eqref{principal}. Then, we will need
to determine the values of $\theta_x$ that are allowed by
the matrix inequality. Once we have characterized the set of
separable maps $\Omega_x$ and inputs $H_{\theta_x}$ that
satisfy the matrix inequality \eqref{principal}, it can further
be shown that $\mu$ is the result of a separable map acting
on a separable state, and hence separable.

Now, let us begin by characterizing the set of separable maps
$\Omega_x: [\Cd{2}] \ten [\Cd{2}] \rightarrow [\H_{\A}\ten
\Cd{2}]\ten[\H_{\B} \ten \Cd{2}]$ that satisfy the matrix
inequality \eqref{principal}. For future reference, we will
also refer to the first and second (output) qubit space
involved in $\Omega_x$ as $\H_{\A''}$ and $\H_{\B''}$
respectively. Now, recall that the Bell basis is defined as
\begin{subequations}
    \begin{align}
      \ket{\Phi_{^1_2}} &\equiv \frac{1}{\sqrt{2}}
      \left( \ket{0}\ket{0} \pm \ket{1}\ket{1} \right),
      \tag{\ref{Eq:BellBases12}}      \\
      \ket{\Phi_{^3_4}} &\equiv \frac{1}{\sqrt{2}}
      \left( \ket{0}\ket{1} \pm \ket{1}\ket{0} \right).
      \tag{\ref{Eq:BellBases34}}
    \end{align}
\end{subequations}
It is easy to show that the matrices $H_\theta$ defined in
Eq.~\eqref{Eq:Dfn:Htheta} are diagonal in this basis, i.e.,
\begin{subequations}
    \begin{equation}
        H_\theta=\sum_{i=1}^4 [N_\theta]_i~\Pi_i,
    \end{equation}
    where $\Pi_i\equiv\ketbra{\Phi_i}$ ($i=1,2,3,4$)
    are the Bell projectors and $[N_\theta]_i$ is the $\idx{i}$ components of the
    vector
    \begin{equation}\label{Eq:Dfn:Ntheta}
        N_\theta \equiv \left(
        \begin{array}{c}
          1-\cos\theta-\sin\theta \\
          1+\cos\theta-\sin\theta \\
          1-\cos\theta+\sin\theta \\
          1+\cos\theta+\sin\theta \\
        \end{array}\right).
    \end{equation}
\end{subequations}
For each value of $x$, let us now define the sixteen matrices
\begin{equation}\label{Eq:Dfn:omega}
    \omega_x^{ij} \equiv \tr_{\A'' \B''}\!\left[\left(\unit\ten\Pi_i\right)\,
    \Omega_x(\Pi_j)\right],\quad i,j=1,2,3,4,
\end{equation}
where the identity matrix $\unit$ acts on
$\H_{\A}\otimes\H_{\B}$ and $\Pi_i$ acts on
$\H_{\A''}\otimes\H_{\B''}$. Each $\omega_x^{ij}$ is the result
of a physical operation, and hence PSD. Projecting the {\em
lhs} of the matrix inequality \eqref{principal} using the four
Bell projectors $\Pi_i$, and taking the partial trace over
$\H_{\A''}\otimes\H_{\B''}$, we get
\begin{equation}\label{Eq:LMI:muTNi}
    \mu\t \left[N_{\frac{\pi}{4}}\right]_i - \int\dd{x}
    \sum_{j=1}^4\ \omega_x^{ij} [N_{\theta\!_x}]_j \geq 0,\quad i=1,2,3,4.
\end{equation}
We shall also define a $4\times4$ matrix $M_x$ whose $(i,j)$
component is given by the trace of the  corresponding
$\omega_x^{ij}$, i.e.,\footnote{Any of these $4\times4$
matrices is essentially the Jamio\l{k}owski state corresponding
to a separable, Bell-diagonal-preserving map written in the
tensored Bell basis (Appendix~\ref{App:SLOCC}).}
\begin{equation}\label{Eq:Dfn:Mx}
    [M_x]_{i,j}\equiv\tr~\omega_x^{ij}.
\end{equation}
Performing the trace on the {\em lhs} of the matrix inequality
\eqref{Eq:LMI:muTNi},  we obtain four inequalities which are
associated with each of the four components of
$N_{\frac{\pi}{4}}$,
\begin{equation}\label{Eq:4IneqN}
    N_{\frac{\pi}{4}} - \int\dd{x}
    \ M_x \cdot N_{\theta\!_x} \succeq \zero_4,
\end{equation}
where $\zero_4$ is the 4-dimensional null vector, and the
symbols $\cdot$ and $\succeq$ mean,  respectively, standard
matrix multiplication and component-wise inequality.

Consider the set of matrices $M$ that are generated by tracing
the {\em lhs} of Eq.~\eqref{Eq:Dfn:omega} when $\Omega_x:
[\Cd{2}] \ten [\Cd{2}] \rightarrow [\H_{\A}\ten
\Cd{2}]\ten[\H_{\B} \ten \Cd{2}]$ is any separable map. The
characterization of this set of matrices can be found in
Appendix~\ref{App:Sec:SLOCC:SeparableStates}. In particular,
let us denote by $\mathcal{D}$ and $\mathcal{G}$, respectively,
the convex hull of all matrices obtained by independently
permuting the rows and/or columns of $D_0$ and $G_0$, c.f.
Eq.~\eqref{Eq:D0&G0}. It then follows from
Definition~\ref{dfn:rhos}, Theorem~\ref{Thm:SeparableUUVV} and
Choi-Jamio{\l}kowski isomorphism
(Appendix~\ref{App:Sec:BellMaps}) that any matrix $M$ as
defined above can be written as
\begin{equation}\label{m}
    M=p\,D+q\,G,
\end{equation}
where $D\in \mathcal{D}$, $G\in \mathcal{G}$, and $p,q\geq 0$.
Then, any solution to the vector inequality~\eqref{Eq:4IneqN}
can be labeled by giving $(\theta_x,p_x,q_x,D_x,G_x)$.

Now, let us characterize the set of admissible solutions to the
vector inequality \eqref{Eq:4IneqN}. By using the fact that $G
\cdot N_\theta\succeq \zero_4$ for all $\theta$ and all $G\in
\mathcal{G}$, we can see that any solution of the vector
inequality \eqref{Eq:4IneqN} must satisfy
\begin{equation}\label{D}
    N_{\frac{\pi}{4}} \succeq \int\dd{x}
    \ p_x\, D_x \cdot N_{\theta\!_x}.
\end{equation}
Recall that this component-wise inequality entails four
inequalities. Adding them together  we obtain the condition
\begin{equation}\label{Eq:p:Normalization}
    \int\dd{x}~p_x \leq 4.
\end{equation}

Denote by $\N$ the set of all vectors obtained by permuting the
components of $N_\theta$, Eq.~\eqref{Eq:Dfn:Ntheta}, when
$\theta$ runs through $[0,\pi/4]$. With some thought, it is not
difficult to see that the convex hull of $\N$, denoted by
$\co{\N}$, is precisely the set of vectors that can be written
as the {\em rhs} of the vector inequality \eqref{D} under the
constraint given by Eq.~\eqref{Eq:p:Normalization}. We can then
write the first inequality of \eqref{D} as
\begin{equation} \label{N1}
    1-\sqrt{2} \geq [N]_1\ ,
\end{equation}
where $[N]_1$ is the first component of $N\in \co{\N}$. It is
easy to see that all  vectors $N\in\co{\N}$ satisfy the
converse inequality, namely, $1-\sqrt{2} \leq [N]_1$, and only
$N_{\frac{\pi}{4}}$ saturates it. Hence, the only admissible
solution for the {\em rhs} of  the vector inequality \eqref{D}
is $N_{\frac{\pi}{4}}$. Substituting this into the vector
inequality \eqref{Eq:4IneqN}, and again using Eq.~\eqref{m} and
Eq.~\eqref{Eq:p:Normalization}, we obtain $-\int\dd{x}\ q_x\,
G_x\cdot N_{\theta\!_x} \succeq \zero_4$. However, as mentioned
above, $G \cdot N_{\theta} \succeq \zero_4$ for all $\theta$
and all $G\in\mathcal{G}$, which implies that for any solution
to the vector inequality \eqref{Eq:4IneqN}, we must have $\int
\dd{x}\ q_x\, G_x \cdot N_{\theta\!_x}= \zero_4$. Therefore,
the vector inequality \eqref{Eq:4IneqN} may now be written as
\begin{equation}\label{Eq:LVI:Npi4M0}
    N_{\frac{\pi}{4}} - M_0\cdot N_{\frac{\pi}{4}} \succeq \zero_4,
\end{equation}
where $M_0$ is any doubly-stochastic matrix such that
\begin{equation} \label{Eq:DoublyStochasticM0}
    M_0\cdot N_{\frac{\pi}{4}}=N_{\frac{\pi}{4}}.
\end{equation}
With some thought, it can be shown that the form of
$N_{\frac{\pi}{4}}$ demands that  doubly-stochastic matrices
that satisfy Eq.~\eqref{Eq:DoublyStochasticM0} must have the
following form
\begin{equation}\label{M_0} M_0 = \left(
    \begin{array}{cccc}
        1 & \cdot & \cdot & \cdot \\
        \cdot & 1- \eta & \eta & \cdot \\
        \cdot & \eta & 1- \eta & \cdot \\
        \cdot & \cdot & \cdot & 1 \\
    \end{array}\right),
\end{equation}
where $\eta\in [0,1]$.

On the other hand, the vector inequality \eqref{Eq:LVI:Npi4M0}
and  Eq.~\eqref{Eq:DoublyStochasticM0} together imply that the
{\em lhs} of the former, and hence Eq.~\eqref{Eq:4IneqN} is
$\zero_4$. Since the four inequalities in Eq.~\eqref{Eq:4IneqN}
were obtained by taking the trace of the matrix inequality
\eqref{Eq:LMI:muTNi}, this further implies that the {\em lhs}
of the matrix inequality \eqref{Eq:LMI:muTNi} is traceless for
all $i$. The only positive matrix with zero trace is the null
matrix, therefore we must have
\begin{equation}\label{fin}
    \mu\t [N_{\frac{\pi}{4}}]_i =
    \sum_{j=1}^4\, \omega_0^{ij} [N_{\frac{\pi}{4}}]_j,\quad i=1,2,3,4,
\end{equation}
where $\omega_0$ is any $\omega_x$ that gives rise to $M_0$. By
the same  token, c.f. Eq.~\eqref{Eq:Dfn:Mx}, the pairs $(i,j)$
for which $[M_0]_{i,j}=0$ must have originated from
$\omega_0^{ij}$ which is a null matrix.

Finally, if we now add the equalities in Eq.~\eqref{fin}
corresponding to $i=2,3$, it follows from the definition of
$\omega_0^{ij}$ [Eq.~\eqref{Eq:Dfn:omega}] that
\begin{equation}\label{fin2}
    2~\mu\t = \tr_{\A'' \B''}~\left[\left(\unit\ten\Psi\right)
    \Omega_0(\Psi)\right],
\end{equation}
where $\Psi=\Pi_2+ \Pi_3$, and $\Omega_0$ is any $\Omega_x$
that gives  rise to $\omega_0$. From the PPT criterion of
separability~\cite{A.Peres:PRL:1996,MPR.Horodecki:PLA:1996},
one can easily check that the (unnormalized) two-qubit state
$\Psi$ is a separable state. Eq.~\eqref{fin2} implies that
$\mu\t$ is the output of a separable map applied to a separable
input state, and hence is a separable state as we have wanted
to prove.
\end{proof}

\setcounter{theorem}{\value{tmp}}

\chapter{Semidefinite Programming and Relaxations}

\section{Semidefinite Programs}\label{App:Sec:SDP}

A semidefinite program (SDP) is a convex optimization over
Hermitian
matrices~\cite{L.Vandenberghe:SR:1996,S.Boyd:Book:2004}. The
objective function depends linearly on the matrix variable (as
expectation values do in quantum mechanics for example) and the
optimization is carried out subjected to the constraint that
the matrix variable is positive semidefinite (PSD) and
satisfies various affine constraints. Any semidefinite program
may be written in the following {\it standard form}:
\begin{subequations}\label{Eq:SDP:Matrix}
\begin{align}
    &\text{maximize \ }-\text{tr}\left[ F_{0}Z\right],
    \label{Eq:SDP:Matrix:Obj}\\
    &\text{subject to\ }\quad\tr\left[ F_iZ\right] =c_i \quad \forall~i,
    \label{Eq:SDP:Matrix:Eq}\\
    &\text{ \ \ \ \ \ \ \ \ \ \ \ \ \ \ \ \ \ \ \ \ }Z\geq
    0,\label{Eq:SDP:Matrix:Ineq}
\end{align}\end{subequations}
where $F_0$ and all the $F_i$'s are Hermitian matrices and the
$c_i$ are real numbers that together specify the optimization;
$Z$ is the Hermitian matrix variable to be optimized.

An SDP also arises naturally in the {\em inequality form},
which seeks  to minimize a linear function of the optimization
variables $\bfx\in\mathbb{R}^n$, subjected to a linear matrix
inequality (LMI):
\begin{subequations}\label{Eq:SDP:Vector}
\begin{gather}
    \text{minimize \ \ \ \ }\quad \mathbf{x}\t\mathbf{c'}\qquad
    \label{Eq:SDP:Vector:Obj}\\
    \text{subject to \ \ }G_0+\sum_i[\bfx]_iG_i \geq0.
    \label{Eq:SDP:Vector:Ineq}
\end{gather}\end{subequations}
As in the standard form, $G_0$ and all the $G_i$'s are
Hermitian matrices, while $\mathbf{c'}$ is a real vector of
length $n$.

\section{Semidefinite Relaxation to Finding $\Sqm(\rho)$}
\label{Sec:semidefinite relaxation}

The global optimization problem of finding $\Sqm(\rho)$, either
in the form of Eq.~\eqref{Eq:GlobalOptimization:Correlation}
for a two-outcome Bell correlation inequality, or
Eq.~\eqref{Eq:GlobalOptimization:3322} for a two-outcome Bell
inequality for probabilities, is a QCQP. As was demonstrated in
Sec.~\ref{Sec:UB}, an upper bound on $\Sqm(\rho)$ can then be
obtained by considering the corresponding {\em
  Lagrange Dual}.

More generally, the global optimization problem of finding
$\Sqm(\rho)$ can be mapped to a real polynomial optimization
problem:\begin{subequations}\label{Eq:polynomial optimization}
\begin{align}
&\text{maximize \ } f_\text{obj}(\bfy),\label{Eq:Global.Obj}\\
&\text{subject to \ }
f_{{\rm eq},i}(\bfy)=0,\quad i=1,2,\ldots,N_{eq},\label{Eq:Global.Eq}\\
&\qquad\qquad\quad f_{{\rm ineq},j}(\bfy)\ge 0,\quad
j=1,2,\ldots,N_{ineq},\label{Eq:Global.Ineq}
\end{align}\end{subequations}
where $\bfy$ is a  vector of {\em real} variables formed by the
expansion coefficients of local observables $\{O_m\}$ in terms
of Hermitian basis operators.

By considering Positivstellensatz-based relaxations, a
hierarchy of upper bounds for $f_\text{obj}(\bfy)$ can be
obtained by solving appropriate SDPs (see, for example,
Ref.~\cite{P.A.Parrilo:MP:2003} and references therein). To see
this, let us first note that $\gamma$ will be an upper bound on
the constrained optimization problem \eqref{Eq:polynomial
optimization} if there exists a set of sum of squares (SOS)
$\mu_i(\bfy)$'s (i.e., nonnegative, real polynomials that can
be written as $\sum_j [h_j(\bfy)]^2$ with $h_j(\bfy)$ being
some real polynomials of $\bfy$), and  a set of real
polynomials $\nu_j(\bfy)$ such
that~\cite{P.A.Parrilo:MP:2003,S.Prajna:LNCIS:2005,K.Schmudgen:MA:1991}
\begin{align}
\gamma-f_\text{obj}(\bfy)=&\mu_0(\bfy)+\sum_j\nu_j(\bfy)f_{{\rm
eq},j}(\bfy)+\sum_i\mu_i(\bfy)f_{{\rm
    ineq},i}(\bfy)\nonumber\\
\label{Eq:G(x)}&+\sum_{i_1,i_2}\mu_{i_1,i_2}(\bfy)f_{{\rm
ineq},i_1}(\bfy)f_{{\rm
    ineq},i_2}(\bfy)+\ldots.
\end{align}
The relaxed optimization problem then consists of minimizing
$\gamma$ subjected to the above constraint. Clearly, at values
of $\bfy$ where the constraints are satisfied, $\gamma$ gives
an upper bound on $f_\text{obj}(\bfy)$. The auxiliary
polynomials $\nu_j(\bfy)$ and SOS $\mu_i(\bfy)$ are thus
analogous to the Lagrange multipliers in the relaxed
optimization problem.

For a fixed degree of the above expression, this relaxed
optimization problem can be cast as an SDP in the form of
Eq.~\eqref{Eq:SDP:Vector}~\cite{P.A.Parrilo:MP:2003}. For the
lowest order relaxation, the auxiliary polynomials
$\nu_j(\bfy)$ and SOS $\mu_i(\bfx)$ are chosen such that degree
of the expression in Eq.~\eqref{Eq:G(x)} is no larger than the
maximum degree of the set of polynomials
\begin{equation*}
f_\text{obj}(\bfy),f_{{\rm eq},1}(\bfy),\ldots,f_{{\rm eq},N_{\rm
    eq}}(\bfy),f_{{\rm ineq},1}(\bfy), \ldots,f_{{\rm ineq},N_{\rm ineq}}(\bfy);
\end{equation*}
for a QCQP with no inequality constraints, this amounts to
setting all the $\mu_i(\bfy)$ to zero and all the $\nu_j(\bfy)$
to numbers.

For higher order relaxation, we increase the degree of the
expression in Eq.~$\eqref{Eq:G(x)}$ by increasing the degree of
the auxiliary polynomials. At the expense of involving more
computational resources, a tighter upper bound on
$f_\text{obj}(\bfy)$ can then be obtained by solving the
corresponding SDP.

\subsection{Lowest Order Relaxation with Observables
of Fixed Trace}\label{App:Sec:Fixed-rank}

We have seen in Sec.~\ref{Sec:State-independent Bound} that a
direct implementation of the Lagrange dual to the optimization
problem given in Eq.~\eqref{Eq:GlobalOptimization:Correlation}
 --- disregarding the constraint given by \eqref{Eq:z} --- gives
rise to an upper bound on $\Sqm(\rho)$ that is apparently
state-independent. To obtain a tighter upper bound on
$\Sqm(\rho)$ using again the lowest order relaxation to
Eq.~\eqref{Eq:GlobalOptimization:Correlation}, we found it most
convenient to express the optimization problem in terms of the
{\em real} optimization variables,
\begin{equation}\label{Eq:y:dfn}
    y_{mn}\equiv \tr\left(O_m\sigma_n\right),\quad n=0,1,\ldots,d^2-1,
\end{equation}
which are just the expansion coefficients of each $O_m$ in
terms of a set of Hermitian basis operators
$\{\sigma_n\}_{n=0}^{d^2-1}$ satisfying
Eq.~\eqref{Eq:basis:Gell-Mann}. The constraint \eqref{Eq:z} can
then be taken care of by setting each $y_{m0}=z_m/\sqrt{d}$. It
is also expedient to express the density matrix $\rho$ in terms
of the same basis of Hermitian operators
\begin{subequations}\label{Eq:rho:expansion}
\begin{align}\label{Eq:rho:CoherenceRepn}
    \rho=\frac{\unit_{d}\ten\unit_d}{d^2}+\sum_{i=1}^{d^2-1}\left([\bfr_\A]_i\sigma_i\ten\sigma_0
    +[\bfr_\B]_i\sigma_0\ten\sigma_i\right)
    +\sum_{i,j=1}^{d^2-1}[R']_{ij}\sigma_i\ten\sigma_j
\end{align}
where
\begin{gather}
    [R']_{ij}=\tr\left(\rho~\sigma_i\ten\sigma_j\right),\label{Eq:R':Defn}\\
    [\bfr_\A]_i\equiv
    \tr(\rho~\sigma_i\ten\sigma_0),\quad[\bfr_\B]_j\equiv
    \tr(\rho~\sigma_0\ten\sigma_j);\label{Eq:r:Defn}
\end{gather}\end{subequations}
$\bfr_\A$, $\bfr_\B$ are simply the coherence vectors that have
been studied in the
literature~\cite{M.S.Byrd:PRA:2003,G.Kimura:PLA:2003}.

We will now incorporate the constraints \eqref{Eq:z} by
expressing the Lagrangian \eqref{Eq:Lagrangian} as a function
of the reduced set of variables
\begin{equation}
    \left(\bfy'\right)\t\equiv[y_{11}~y_{12}~\ldots~y_{1\,d^2-1}~y_{21}~\ldots
    y_{\mA+\mB\,d^2-1}],
\end{equation}
while all the $y_{m0}=z_m/\sqrt{d}$ are  treated as fixed
parameters of the problem. With this change in basis, and after
some patient algebra, the Lagrangian can be rewritten as
\begin{align}
    \mathcal{L}(\bfy',\lambda_{mn})&=\sum_{m=1}^{\mA+\mB}\lambda_{m0}
    \left(\sqrt{d}-\frac{z_m^2}{d\sqrt{d}}\right)+\sum_{\sA=1}^{\mA}
    \sum_{\sB=1}^{\mB}b_{\sA\sB}
    \frac{z_{\sA}z_{\sB+\mA}}{d^2}\nonumber\\
    &-\frac{1}{\sqrt{d}}({\bfl}-\bfr)\t\left(\bfy'\right)
    -\left(\bfy'\right)\t\Omega'\left(\bfy'\right),
    \label{Eq:Lagrangian:FixedRank}
\end{align}
where $\lambda_{mn}$ are defined in
Eq.~\eqref{Eq:Lambda:Expansion},
\begin{subequations}\label{Eq:l,r,t,etc.}
\begin{gather}
    \Omega'\equiv\half\left(
    \begin{array}{cc}
        \zero_{\mA(d^2-1)\times\mA(d^2-1)} & -b\ten R' \\
        -\left(b\ten R'\right)\t & \zero_{\mB(d^2-1)\times\mB(d^2-1)}\\
    \end{array}\right)
    +\bigoplus_{m=1}^{\mA+\mB} M_m,\nonumber\\
    {\bf l}\equiv\vvec{L},\qquad  \bfr\equiv\left(
    \begin{array}{c}
        {\bft}_\A\ten\bfr_\A\\
        {\bft}_\B\ten\bfr_\B\end{array}
    \right),
    \label{Eq:Omega,l,r:dfn}
\end{gather}
and for $i,j=1,2,\ldots,d^2-1$,
\begin{gather}
    [L]_{j,m}=2z_m\lambda_{mj},~[{\bf
    t}_\A]_{\sA}=\sum_{\sB=1}^{\mB}
    b_{\sA\sB}z_{\sB+\mA},~[{\bf t}_\B]_l=\sum_{\sA=1}^{\mA} b_{\sA\sB}z_{\sA},\nonumber\\
    M_m=\sum_{n=0}^{d^2-1}\lambda_{mn}P_n,\quad
    [P_n]_{i,j}=\half\tr\left(\sigma_n\left[\sigma_i,\sigma_j\right]_+\right);
\end{gather}\end{subequations}
$[\sigma_i,\sigma_j]_+\equiv \sigma_i\sigma_j
+\sigma_j\sigma_i$ is the anti-commutator of $\sigma_i$ and
$\sigma_j$.

As before, we now maximize the Lagrangian
\eqref{Eq:Lagrangian:FixedRank} over $\bfy'$ to obtain the
corresponding Lagrange dual function. The latter, however, is
unbounded above unless
\begin{equation}
    \left(
    \begin{array}{cc}
        -2t & \frac{1}{\sqrt{d}}\left({\bf l}\t-\bfr\t\right)\\
        \frac{1}{\sqrt{d}}\left({\bf l}-\bfr\right) & 2\Omega'
    \end{array}\right)\ge0, \label{Eq:Constraint:FixedRank}
\end{equation}
for some finite $t$. The convex optimization problem dual to
Eq.~\eqref{Eq:GlobalOptimization:Correlation} with fixed trace
for each observables is thus
\begin{align}
    &\text{minimize}\sum_{m=1}^{\mA+\mB}\lambda_{m0}
    \left(\sqrt{d}-\frac{z_m^2}{d\sqrt{d}}\right)
    +\sum_{\sA=1}^{\mA}\sum_{\sB=1}^{\mB}b_{\sA\sB}\frac{z_{\sA}z_{\sB+\mA}}{d^2}
    -t,\nonumber\\
    &\text{subject to \ \ }\left(
    \begin{array}{cc}
        -2t & \frac{1}{\sqrt{d}}\left({\bf l}\t-\bfr\t\right)\\
        \frac{1}{\sqrt{d}}\left({\bf l}-\bfr\right) & 2\Omega'
    \end{array}\right)\ge0.\label{Eq:SDP:FixedRank}
\end{align}

\subsection{Sufficient Condition for No-violation of the
Bell-CHSH Inequality}\label{App:Sec:UpperBound:Analytic}

To derive the semianalytic criterion
Eq.~\eqref{Eq:CHSHViolation:Criterion}, we now note that any
choice of $\{\lambda_{mn}\}_{n=0}^{d^2-1}$ that satisfy
constraint \eqref{Eq:Constraint:FixedRank} will provide an
upper bound on the corresponding $\SqmCHSH(\rho)$. In
particular, an upper bound can be obtained by setting
\begin{equation}
    \lambda_{mn}=\delta_{n0}\left[\lambda_\A\left(\delta_{m1}+
    \delta_{m2}\right)+\lambda_\B\left(\delta_{m3}+\delta_{m4}\right)\right],
\end{equation}
and solving for $\lambda_\A$, $\lambda_\B$ that satisfy the
constraint \eqref{Eq:Constraint:FixedRank}. With this choice of
the Lagrange multipliers, and for quantum states with vanishing
coherence vectors, the constraint
\eqref{Eq:Constraint:FixedRank} becomes
\begin{align}
    \left(
    \begin{array}{ccc}
    -2t & \zero_{2(d^2-1)}\t & \zero_{2(d^2-1)}\t \\
    \zero_{2(d^2-1)} & \frac{2\lambda_\A}{\sqrt{d}}\unit_2\ten\unit_{d^2-1} &
    -b\ten R' \\
    \zero_{2(d^2-1)} & -\left(b\ten R'\right)\t &
    \frac{2\lambda_\B}{\sqrt{d}}\unit_2\ten\unit_{d^2-1}
    \end{array}\right)\ge 0,
\end{align}
where $b$ and $R'$ are defined, respectively, in
Eq.~\eqref{Eq:b-CHSH:dfn} and Eq.~\eqref{Eq:R':Defn}. This, in
turn is equivalent to\begin{subequations}
\begin{gather}
    -t\ge0,\label{Eq:Constraint:FixedRank:t}\\
    \left(
    \begin{array}{cc}
    \frac{2\lambda_\A}{\sqrt{d}}\unit_2\ten\unit_{d^2-1} & -b\ten R'\\
    -\left(b\ten R'\right)\t & \frac{2\lambda_\B}{\sqrt{d}}\unit_2\ten
    \unit_{d^2-1}
    \end{array}\right)\ge 0.
    \label{Eq:Constraint:FixedRank:Reduced}
\end{gather}\end{subequations}
Using Schur's
complement~\cite{R.A.Horn:Book:1990,G.Strang:Book:1988} and
Eq.~\eqref{Eq:b-CHSH:dfn}, the
constraint~\eqref{Eq:Constraint:FixedRank:Reduced} can be
explicitly solved to give
\begin{equation*}
    \lambda_\A\lambda_\B\ge \half s_1^2 d,
\end{equation*}
where $s_1$ is the largest singular value of the matrix $R'$.
Substituting this and Eq.~\eqref{Eq:Constraint:FixedRank:t}
into Eq.~\eqref{Eq:SDP:FixedRank}, and after some algebra, we
see that $\SqmCHSH(\rho)$ for a quantum state $\rho$ with
vanishing coherence vectors cannot be greater  than
\begin{equation*}
    \max_{z_1,z_2,z_3,z_4}
    2\sqrt{2}s_1d\sqrt{\prod_{i=1}^2
    \frac{2d^2-z_{2i-1}^2-z_{2i}^2}{2d^2}}+\sum_{\sA,\sB=1}^2
    b_{\sA\sB}\frac{z_{\sA}z_{\sB+2}}{d^2}.
\end{equation*}
For $\rho$ to violate the Bell-CHSH inequality, we must have
this upper bound greater than the classical threshold value,
$\bLHV{\mbox{\tiny CHSH}}=2$, c.f. Eq.~\eqref{Ineq:Bell-CHSH}.
Hence a sufficient condition for $\rho$ to satisfy the
Bell-CHSH inequality is given by
Eq.~\eqref{Eq:CHSHViolation:Criterion}.

\section{Explicit Forms of Semidefinite Programs}

\subsection{SDP for the LB Algorithm}\label{App:Sec:SDP:LB}

Here, we provide an explicit form for the matrices $F_i$ and
constants $c_i$ that define the SDP used in the LB algorithm,
Eq.~\eqref{Eq:sdpiter}. By setting
\begin{gather*}
Z=\left(
\begin{array}{cccccc}
B^1_1 & \zero & \zero & \zero  & \zero & \zero\\
\zero & \ddots & \zero & \zero & \zero  & \zero \\
\zero & \zero & B_1^{\nB} & \zero  & \zero & \zero\\
\zero & \zero & \zero & B_2^1 &\zero & \zero\\
\zero & \zero & \zero & \zero &\ddots & \zero\\
\zero & \zero & \zero & \zero &\zero & B^{\nB}_{\mB}
\end{array}\right),
\end{gather*}
in Eq.~\eqref{Eq:SDP:Matrix}, we see that the inequality
constraint \eqref{Eq:SDP:Matrix:Ineq} of the SDP entails the
positive semidefiniteness of the POVM elements
$\left\{\left\{B^{\oB}_{\sB}\right\}_{\oB=1}^{\nB}\right\}_{\sB=1}^{\mB}$,
and hence Eq.~\eqref{Eq:SDPIter:ineq}. On the other hand, with
\begin{equation*}
F_0=-\left(
\begin{array}{cccccc}
\rho_{B^1_1} & \zero & \zero & \zero  & \zero & \zero\\
\zero & \ddots & \zero & \zero & \zero  & \zero \\
\zero & \zero & \rho_{B_1^{\nB}} & \zero  & \zero & \zero\\
\zero & \zero & \zero & \rho_{B_2^1} &\zero & \zero\\
\zero & \zero & \zero & \zero &\ddots & \zero\\
\zero & \zero & \zero & \zero &\zero & \rho_{B^{\nB}_{\mB}}
\end{array}\right),
\end{equation*}
where $\rho_{B^{\oB}_{\sB}}$ is defined in
Eq.~\eqref{Eq:Dfn:rhoAB}, the equality constraint
\eqref{Eq:SDP:Matrix:Eq}, together with appropriate choice of
$F_i$ and $c_i$, ensures that the normalization condition
\eqref{Eq:SDPIter:eq} is satisfied.

In particular, each $F_i$ is formed from a direct sum of
Hermitian basis operators. A convenient choice of such basis
operators is given by the traceless Gell-Mann matrices, denoted
by $\{\sigma_n\}_{n=1}^{d^2-1}$, supplemented by
\begin{subequations}\label{Eq:basis:Gell-Mann}
\begin{equation}
    \sigma_0=\frac{1}{\sqrt{d}}\unit_d,
\end{equation}
such that
\begin{equation}
    \tr\left(\sigma_n\sigma_{n'}\right)=\delta_{nn'}\quad\text{and}\quad
    \tr\left(\sigma_n\right)=\sqrt{d}\,\delta_{n0},
\end{equation}
\end{subequations}
where $d=\dB$ is the dimension of the state space that each
$B^{\oB}_{\sB}$ acts on. A typical $F_i$ then consists of $\nB$
diagonal blocks of $\sigma_n$ at positions corresponding to the
$\nB$ POVM elements $\{\POVMBg\}_{\oB=1}^{\nB}$ in $Z$ for a
fixed $\sB$. For instance, the set of $F_i$
\begin{equation*}
F_i=\left(
\begin{array}{cccccc}
\sigma_{i-1} & \zero & \zero & \zero  & \zero & \zero\\
\zero & \ddots & \zero & \zero & \zero  & \zero \\
\zero & \zero & \sigma_{i-1} & \mathbf{0}  & \zero & \zero\\
\zero & \zero & \mathbf{0} & \zero &\zero & \zero\\
\zero & \zero & \mathbf{0} & \zero &\ddots & \zero\\
\zero & \zero & \mathbf{0} & \zero &\zero & \zero
\end{array}\right),\quad 1\le i\le d^2,
\end{equation*}
together with $c_i=\sqrt{d}~\delta_{i1}$ entails the
normalization of $\{\POVMBg\}_{\oB=1}^{\nB}$, i.e.,
$\sum_{\oB=1}^{\nB} \POVMB{1}{\oB}=\unit_{\dB}$; the remaining
$(\mB-1)d^2$ $F_i$ are defined similarly and can be obtained by
shifting the nonzero diagonal blocks diagonally downward by
appropriate multiples of $\nB$ blocks. The SDP thus consists of
solving Eq.~\eqref{Eq:SDP:Matrix} for a $\mB\nB{d}\times
\mB\nB{d} $ Hermitian matrix $Z$ subjected to $d^2\mB$ affine
constraints.

\subsection{SDP for the UB Algorithm}

In analogy with the previous section, we will provide, in this
section, an explicit form for some of the SDPs used in the UB
algorithm. In particular, we find it expedient to express these
SDPs in the inequality form, Eq.~\eqref{Eq:SDP:Vector}, but for
convenience, we will use two indices $m$ and $n$
($m=1,2,\ldots,\mA+\mB$, $n=0,1,\ldots,d^2-1$), instead of the
single index $i$ [c.f. Eq.~\eqref{Eq:SDP:Vector:Ineq}] to label
the Hermitian matrices $G_{mn}$ and the components of the
vector $\bfc'$.  Throughout this section, $\sigma_n$ will refer
to a Hermitian basis operator satisfying
Eq.~\eqref{Eq:basis:Gell-Mann}.

\subsubsection{State-independent Bound}

Now, we will give the matrices $G_{mn}$ and constants
$[\bfc']_{mn}$ that define the SDP obtained from the lowest
order relaxation to
Eq.~\eqref{Eq:GlobalOptimization:Correlation} given by
Eq.~\eqref{Eq:SDP-relaxation:deg2} and
Eq.~\eqref{Eq:Lambda:Expansion}. To begin with, it is
straightforward to see that by setting
\begin{align*}
    [\bfx]_{mn}&=\lambda_{mn},\quad [{\bf c'}]_{mn}=\sqrt{d}\,\delta_{n0}
\end{align*}
in Eq.~\eqref{Eq:SDP:Vector}, we obtain the same objective
function as that in Eq.~\eqref{Eq:SDP-relaxation:deg2}, where
$\lambda_{mn}$ is defined in Eq.~\eqref{Eq:Lambda:Expansion}.
Next, if we further set [c.f. Eq.~\eqref{Eq:Dfn:Omega0} and
Eq.~\eqref{Eq:Dfn:Omega}]
\begin{align*}
    G_0&=\Omega_0=\half\left(
        \begin{array}{cc}
        \zero_{d^2\mA\times d^2\mA} & -b\ten R \\
        -b^T\ten R^\dag & \zero_{d^2\mB\times d^2\mB}\\
        \end{array}\right),
\end{align*}
where $b$ and $R$ are defined just after
Eq.~\eqref{Eq:Dfn:Omega0}, and
\begin{align*}
    G_{mn}&=\bigoplus_{k=1}^{m-1}\zero_{d^2\times d^2}
    \bigoplus\unit_d\ten\sigma_n
    \bigoplus_{k=m+1}^{\mA+\mB}\zero_{d^2\times d^2},
\end{align*}
then it can be seen that Eq.~\eqref{Eq:SDP:Vector:Ineq}
enforces the inequality constraint given in
Eq.~\eqref{Eq:SDP-relaxation:deg2}. The SDP corresponding to
Eq.~\eqref{Eq:SDP-relaxation:deg2}, which apparently gives rise
to a state-independent upper bound on $\Sqm(\rho)$, thus
consists of solving Eq.~\eqref{Eq:SDP:Vector} for
$d^2(\mA+\mB)$ real variables subjected to a matrix inequality
of dimension $d^2(\mA+\mB)\times d^2(\mA+\mB)$, and which is
linear in the $d^2(\mA+\mB)$ real variables.

\subsubsection{State-dependent Bound}

For the more refined SDP given by Eq.~\eqref{Eq:SDP:FixedRank},
which gives rise to a state-dependent upper bound on
$\Sqm(\rho)$, we will instead set
\begin{gather*}
    \bfx=\bfx_0\oplus t,\quad \bfc'=\bfc_0\oplus -1,\\
    [\bfx_0]_{mn}=\lambda_{mn},\quad
    [{\bf c_0}]_{mn}=\left(\sqrt{d}-\frac{z_m^2}{d\sqrt{d}}\right)\,\delta_{n0},
\end{gather*}
in Eq.~\eqref{Eq:SDP:Vector}, where $z_m$ is the trace of local
observables defined in Eq.~\eqref{Eq:z}. It is easy to see that
with the above choice of $\bfx$ and $\bfc'$,
Eq.~\eqref{Eq:SDP:Vector:Obj} gives, apart from a constant that
is immaterial to the optimization, the same objective function
as that in Eq.~\eqref{Eq:SDP:FixedRank}. Next, we will set
\begin{gather*}
    G_0=-\left(
        \begin{array}{ccc}
        0 & \frac{1}{\sqrt{d}}\left(\bft_\A\ten\bfr_\A\right)\t
        & \frac{1}{\sqrt{d}}\left(\bft_\B\ten\bfr_\B\right)\t \\
        \frac{1}{\sqrt{d}}\bft_\A\ten\bfr_\A  & \zero_{(d^2-1)\mA\times(d^2-1)\mA}
        & b\ten R' \\
        \frac{1}{\sqrt{d}}\bft_\B\ten\bfr_\B  & \left(b\ten R'\right)\t &
        \zero_{(d^2-1)\mB\times(d^2-1)\mB}\\
        \end{array}\right),  \\
    G_{mn}=\left[0\bigoplus_{k=1}^{m-1}\zero_{(d^2-1)\times(d^2-1)}\bigoplus P_n
    \bigoplus_{k=m+1}^{\mA+\mB}\zero_{(d^2-1)\times(d^2-1)}\right]
    + (1-\delta_{n0})\frac{2z_m}{\sqrt{d}}G'_{mn},
\end{gather*}
where $\bft_\A$, $\bft_\B$, $P_n$ are defined in
Eq.~\eqref{Eq:l,r,t,etc.}, $\bfr_\A$, $\bfr_\B$, $R'$ are
defined in Eq.~\eqref{Eq:rho:expansion} and $G'_{mn}$ is a
$[1+(d^2-1)(\mA+\mB)]\times[1+(d^2-1)(\mA+\mB)]$ matrix that is
zero everywhere except for the following entries:
\begin{equation*}
    [G'_{mn}]_{1,1+(m-1)(d^2-1)+n}=[G'_{mn}]_{1+(m-1)(d^2-1)+n,1}=1.
\end{equation*}
Finally, by setting
\begin{equation*}
    G_t=\left(
        \begin{array}{cc}
        -2 & \zero_{(d^2-1)(\mA+\mB)}\t \\
        \zero_{(d^2-1)(\mA+\mB)} & \zero_{(d^2-1)(\mA+\mB),(d^2-1)(\mA+\mB)}
        \end{array}\right),
\end{equation*}
in Eq.~\eqref{Eq:SDP:Vector}, which is the $G_i$ corresponding
to the variable $t$, it can be seen that
Eq.~\eqref{Eq:SDP:Vector:Ineq} enforces the matrix inequality
constraint given in Eq.~\eqref{Eq:SDP:FixedRank}. The SDP
corresponding to Eq.~\eqref{Eq:SDP:FixedRank} thus consists of
solving Eq.~\eqref{Eq:SDP:Vector} for $d^2(\mA+\mB)+1$ real
variables subjected to a matrix inequality of dimension
$[1+(d^2-1)(\mA+\mB)]\times [1+(d^2-1)(\mA+\mB)]$, and which is
linear in the $d^2(\mA+\mB)+1$ real variables.

\subsection{SDP for the Verification of Entanglement Witness}
\label{App:Sec:SDP:EW}

Here, we will show that, in the context of
Lemma~\ref{Lem:ProofOfWitnesses}, the search for a PSD $\Z$
satisfying Eq.~\eqref{Eq:Z:Dfn} is a semidefinite programming
feasibility
problem~\cite{L.Vandenberghe:SR:1996,S.Boyd:Book:2004}, i.e.,
an SDP whereby the objective function is some constant that is
independent of any optimization variables. In particular, we
will show that this SDP is readily written in the {\em standard
form}, Eq.~\eqref{Eq:SDP:Matrix}, but for convenience, we will
use two indices $i$ and $j$ instead of the single index $i$
[c.f. Eq.~\eqref{Eq:SDP:Matrix:Eq}] to label the Hermitian
matrices $F_{ij}$ and the constants $c_{ij}$.

Let us denote by $\{\sigma^{\A}_{i}\}_{i=0}^{\dA'^2-1}$ and
$\{\sigma^{\B}_{j}\}_{j=0}^{\dB'^2-1}$, respectively, a
complete set of Hermitian basis operators acting on
$\Pi_\A\,\HA^{\ten m}\,\Pi_\A$ and $\Pi_\B\,\HB^{\ten
n}\,\Pi_\B$ where $\Pi_\A$, $\Pi_\B$ are, respectively, the
projectors onto the symmetric subspace of $\HA^{\ten m}$ and
$\HB^{\ten n}$ and $\dA'$, $\dB'$ are the corresponding
dimensions of these symmetric subspaces. As before, a
convenient choice of such basis operators is given by the
orthonormal set which satisfies Eq.~\eqref{Eq:basis:Gell-Mann}.
Since both the {\em lhs} and {\em rhs} of Eq.~\eqref{Eq:Z:Dfn}
are Hermitian matrices, if the equation holds true, it follows
that for all $i$ and $j$ we must have
\begin{align*}
    \tr\left[\Pi_\A\ten\Pi_\B\,\left(\unit_{d_\A}^{\otimes  m-1}\ten
    Z_w\ten\unit_{d_\B}^{\ten n-1}\right)\,\Pi_\A\ten\Pi_\B
    \,\sigma^{\A}_{i}\ten\sigma^{\B}_{j}\right] &=
    \tr\left[\Pi_\A\ten\Pi_\B\,\Z^{\mbox{\tiny $T_s$}}\,\Pi_\A\ten\Pi_\B
    \,\sigma^{\A}_{i}\ten\sigma^{\B}_{j}\right],\nonumber\\
    \Rightarrow\tr\left[\left(\unit_{d_\A}^{\otimes  m-1}\ten
    Z_w\ten\unit_{d_\B}^{\ten n-1}\right)
    ~\sigma^{\A}_{i}\ten\sigma^{\B}_{j}\right] &=
    \tr\left[\Z^{\mbox{\tiny $T_s$}}~
    \sigma^{\A}_{i}\ten\sigma^{\B}_{j}\right],\nonumber\\
    \Rightarrow\tr\left[\left(\unit_{d_\A}^{\otimes  m-1}\ten
    Z_w\ten\unit_{d_\B}^{\ten n-1}\right)
    ~\sigma^{\A}_{i}\ten\sigma^{\B}_{j}\right] &=
    \tr\left[\Z~\left(\sigma^{\A}_{i}\ten\sigma^{\B}_{j}
    \right)^{\mbox{\tiny $T_s$}}\right].
\end{align*}
Moreover, it is easy to see that whenever this last expression
holds true for all $i$ and $j$, one can construct a PSD $\Z$
such that Eq.~\eqref{Eq:Z:Dfn} holds true. Hence, if we set
\begin{align*}
    F_0&=\zero_{\dA^m\dB^n\times\dA^m\dB^n},\quad
    F_{ij}=\left(\sigma^{\A}_{i}\ten\sigma^{\B}_{j}\right)^{\mbox{\tiny $T_s$}}, \\
    Z&=\Z,\quad c_{ij}=\tr\left[\left(\unit_{d_\A}^{\otimes  m-1}\ten
    Z_w\ten\unit_{d_\B}^{\ten n-1}\right)
    ~\sigma^{\A}_{i}\ten\sigma^{\B}_{j}\right],
\end{align*}
in Eq.~\eqref{Eq:SDP:Matrix}, we will have expressed the
problem of searching for a legitimate $\Z$ as a semidefinite
programming feasibility problem.

On the other hand, for numerical implementation of the above
SDP, it may be advantageous to formalize the above problem as
an ordinary SDP where $F_0$ is nonzero. For that purpose, one
sets, instead,
\begin{align*}
    F_0&=\left(\sigma^{\A}_{0}\ten\sigma^{\B}_{0}\right)^{\mbox{\tiny $T_s$}},\quad
    F_{ij}=\left(\sigma^{\A}_{i}\ten\sigma^{\B}_{j}\right)^{\mbox{\tiny $T_s$}}, \\
    c_{ij}&=\tr\left[\left(\unit_{d_\A}^{\otimes  m-1}\ten
    Z_w\ten\unit_{d_\B}^{\ten n-1}\right)
    ~\sigma^{\A}_{i}\ten\sigma^{\B}_{j}\right],
\end{align*}
in Eq.~\eqref{Eq:SDP:Matrix}, where now we have excluded
$F_{00}$ from the set of $F_{ij}$. With some thought, it is not
difficult to see that a legitimate $\Z$ that satisfies all the
constraints exists if and only if the optimum of the
optimization, $Z^*$ satisfies
\begin{equation}
    -\tr(F_0\,Z^*)\ge -\tr\left[\left(\unit_{d_\A}^{\otimes  m-1}\ten
    Z_w\ten\unit_{d_\B}^{\ten n-1}\right)
    ~\sigma^{\A}_{0}\ten\sigma^{\B}_{0}\right],
\end{equation}
in which case the desired $\Z$ can be constructed as
\begin{equation}
    \Z = Z^* +\left\{\tr\left[\left(\unit_{d_\A}^{\otimes  m-1}\ten
    Z_w\ten\unit_{d_\B}^{\ten n-1}\right)
    ~\sigma^{\A}_{0}\ten\sigma^{\B}_{0}\right]-\tr(F_0\,Z^*)\right\}\sigma^{\A}_{0}\ten\sigma^{\B}_{0}.
\end{equation}
Hence, the search for a legitimate $\Z$ can also be formalized
as an SDP which consists of solving Eq.~\eqref{Eq:SDP:Matrix}
for a $\dA^m\dB^n\times\dA^m\dB^n$ PSD matrix $Z$ subjected to
$(\dA'^2-1)(\dB'^2-1)-1$ affine constraints.

\backmatter

\chapter{List of Symbols}

The following list is neither exhaustive nor exclusive, but may
be helpful.
\begin{list}{}{%
\setlength{\labelwidth}{24mm}
\setlength{\leftmargin}{35mm}}

\item[$\ProbTwGJ$] Joint probability that the $\idx{\oA}$ and
    $\idx{\oB}$ experimental outcomes are observed,
    respectively, at Alice's and Bob's site given that she
    performs the $\idx{\sA}$ and he performs the $\idx{\sB}$
    measurement.

\item[$\ProbTwGMA$] The marginal probability that the
    $\idx{\oA}$ experimental outcome is observed at Alice's
    site given that she performs the $\idx{\sA}$ measurement.

\item[$\ProbTwGMB$] The marginal probability that the
    $\idx{\oB}$ experimental outcome is observed at Bob's
    site given that he performs the $\idx{\sB}$ measurement.

\item[$\Cp$] The set of probability vectors obeying the
    no-signaling conditions when Alice and Bob are allowed to
    perform, respectively, $\mA$ and $\mB$ alternative
    measurements and where each local measurement yields,
    correspondingly, one of $\nA$ and $\nB$ outcomes.

\item[$\polycl$] The set of classical probability vectors in
    $\Cp$; each member of $\polycl$ can be described with some
    LHVM.

\item[$\Qp$] The set of quantum probability vectors in $\Cp$;
    each member of $\Qp$ can be realized by some quantum
    strategy.

\item[$\POVMAg$] The POVM element associated with the
    $\idx{\oA}$ outcome of Alice's $\idx{\sA}$
    measurement

\item[$\POVMBg$] The POVM element associated with the
    $\idx{\oB}$ outcome of Bob's $\idx{\sB}$
    measurement

\item[$E(A_{\sA},B_{\sB})$] Correlation function associated
    with Alice measuring $A_{\sA}$ and Bob measuring $B_{\sB}$.

\item[$\B(\HA\ten\HB)$] Bounded operator acting on the Hilbert
    space $\HA\ten\HB$.

\item[$\H^{[k]}$] The Hilbert space associated with the $\idx{k}$ subsystem.

\item[$\MEd$] The $d$-dimensional maximally entangled state.

\item[$\rd{d\times d}$] The $d\times d$-dimensional maximally
    mixed state, i.e., $\rd{d\times
    d}=\frac{\unit_d\times\unit_d}{d^2}$.

\item[$\rw(p)$] The $d$-dimensional Werner state.

\item[$\rI(p)$] The $d$-dimensional isotropic state.

\item[$\SLHVBI{k}$] Functional form of the Bell inequality
    labeled by ``k".

\item[$\SqmBI{k}(\rho)$] Maximal expectation value of the Bell
    operator derived from the Bell inequality ``k" with
    respect to the quantum state $\rho$.

\item[$\Bell_k$] The Bell operator derived from the Bell
    inequality ``k".

\item[$\eBell{k}{\rho}$] Expectation value of the Bell operator
    $\Bell_k$ with respect to a quantum state $\rho$.

\item[$\lfloor a \rfloor$] The largest integer smaller than $a$

\item[{$[M]_{i,j}$}] The $(i,j)$ entry of a matrix $M$.

\item[$M\t$] The transpose of $M$.

\item[$M^{\mbox{\tiny $T_k$}}$] The partial transpose of $M$
    with respect to the $\idx{k}$ subsystem.

\item[$\tr\left(M\right)$] The trace of $M$.

\item[$\tr_\A\left(M\right)$] The partial trace of $M$ over subsystem $\A$.

\item[$\left|\left|M\right|\right|$] The trace norm of $M$,
    i.e., the sum of the absolute value of $M$'s eigenvalues.

\item[$\Pi$] Projector, i.e., $\Pi^2=\Pi$.

\item[$\unit_d$] The $d\times d$ identity matrix.

\item[$\zero$] The null operator/ zero matrix.

\item[$\zero_n$] The $n\times1$ null vector.

\item[$\zero_{d_\A\times d_\B}$] The $d_\A\times d_\B$ zero
    matrix.

\item[$\CSLOCC{CHSH}$] The set of quantum states not violating
    the Bell-CHSH inequality even after arbitrary local
    filtering operations.

\end{list}

\end{document}